%% file: paper.tex
\documentclass[12pt]{iopart}

\usepackage{graphicx}
\usepackage{booktabs}
\usepackage{amssymb}
\usepackage{bm}
\usepackage{multirow}
\usepackage{arydshln}
\usepackage[colorlinks,citecolor=blue,anchorcolor=red,menucolor=red,linkcolor=red,filecolor=red,runcolor=red,urlcolor=blue,frenchlinks=red]{hyperref}

\def\ov{\overline}
\newcommand{\dszero}{D_{s0}^*}
\newcommand{\dsone}{D_{s1}}

\graphicspath{{Figures/}} %

\begin{document}

\title{A review of the open charm and open bottom systems}

\author{Hua-Xing Chen$^1$, Wei Chen$^2$, Xiang Liu$^{3,4}$, Yan-Rui Liu$^5$ {\rm and} Shi-Lin Zhu$^{6,7,8}$}
\address{
$^1$School of Physics and Beijing Key Laboratory of Advanced Nuclear Materials and Physics, Beihang University, Beijing 100191, China \\
$^2$Department of Physics and Engineering Physics, University of Saskatchewan, Saskatoon, Saskatchewan, S7N 5E2, Canada \\
$^3$School of Physical Science and Technology, Lanzhou University, Lanzhou 730000, China \\
$^4$Research Center for Hadron and CSR Physics, Lanzhou University and Institute of Modern Physics of CAS, Lanzhou 730000, China \\
$^5$School of Physics and Key Laboratory of Particle Physics and Particle Irradiation (MOE), Shandong University, Jinan 250100, China \\
$^6$School of Physics and State Key Laboratory of Nuclear Physics and Technology, Peking University, Beijing 100871, China \\
$^7$Collaborative Innovation Center of Quantum Matter, Beijing 100871, China \\
$^8$Center of High Energy Physics, Peking University, Beijing 100871, China
}
\ead{hxchen@buaa.edu.cn, wec053@mail.usask.ca, xiangliu@lzu.edu.cn, yrliu@sdu.edu.cn {\rm and} zhusl@pku.edu.cn}
\vspace{10pt}
\begin{indented}
\item[]September 2016
\end{indented}

\begin{abstract}
Since the discovery of the first charmed meson in 1976,
many open-charm and open-bottom hadrons were observed.
In 2003 two narrow charm-strange states $D_{s0}^*(2317)$ and
$D_{s1}(2460)$ were discovered by the BaBar and CLEO Collaborations,
respectively. After that, more excited heavy hadrons were reported.
In this work, we review the experimental and theoretical progress in
this field.
\end{abstract}


\vspace{2pc}

\submitto{\RPP}
%
%
%
\tableofcontents

\input{section1.1.tex}

\input{section2.1.tex}
\input{section2.2.tex}
\input{section2.3.tex}

\input{section3.1.tex}
\input{section3.2.tex}

\input{section6.1.tex}
\input{section6.2.tex}

\input{section7.1.tex}

\input{section4.1.tex}
\input{section4.2.tex}

\input{section5.tex}

\vspace{0.2cm}

\section*{Acknowledgments}

We would like to express our gratitude to all the collaborators and
colleagues who contributed to the investigations presented here, in
particular to Dian-Yong Chen, Xiao-Lin Chen, Wei-Zhen Deng, Li-Sheng Geng,
Qi Huang, Atsushi Hosaka, Hong-Wei Ke, Takayuki Matsuki, Qiang Mao, Ning Li, Zhi-Gang Luo,
Cheng-Ping Shen, T. G. Steele, Qing-Tao Song, Yuan Sun, Zhi-Feng Sun, Jun-Zhang Wang,
Hao Xu, Jie-Sheng Yu, Bo Zhang, and Dan Zhou. We appreciate Bing Chen
from Anyang Normal University for the careful reading of the
manuscript and valuable suggestions. We thank Er-Liang Cui for
helping prepare some relevant documents. This project is supported
by the National Natural Science Foundation of China under Grants No.
11175073, No. 11222547, No. 11275115, No. 11475015, No. 11575008,
No. 11261130311, the 973 program, the Fundamental Research Funds for
the Central Universities, the National Program for Support of Top-notch Young Professionals, and the Natural
Sciences and Engineering Research Council of Canada (NSERC).

\vspace{0.2cm}

\section*{Author Contributions:}

S.L.Z. conceptualized this review and wrote the outline.
Y.R.L. wrote Sec.~1. H.X.C. wrote Sec.~2, 3 and 4. W.C. wrote Sec.~5 and 6.
X.L. and S.L.Z. wrote Sec.~7, systematically revised and finalized the manuscript.

\end{document}

%% file: section1.1.tex
\section{Introduction}
\label{sec1}

\subsection{From QED to QCD}
\label{sec1.0}

There are four fundamental interactions in nature: the
electromagnetic interaction, the weak interaction, the strong
interaction, and gravitation. The electromagnetic interaction occurs
between electrically charged particles, which is responsible for
electricity and magnetism as well as light. We encounter the
electromagnetic interaction and gravity every day.

Quantum Electrodynamics (QED) is the underlying theory of the
electromagnetic interactions. Its Lagrangian reads
\begin{eqnarray}\label{QED}
\mathfrak{L}^{QED} &=& \bar \psi \Big ( i \gamma^\mu D_\mu - m \Big ) \psi
- {1 \over 4} F_{\mu\nu} F^{\mu\nu} \, ,
\end{eqnarray}
where the gauge covariant derivative is defined as
\begin{eqnarray}
D_\mu &=& \partial_\mu + i e A_\mu \, .
\end{eqnarray}
Here, $\psi(x)$ denotes the electron/positron field, $A_\mu$ is the
electromagnetic field, $F_{\mu\nu}=\partial_\mu A_\nu-\partial_\nu A_\mu$, and $\gamma_\mu$ are Dirac matrices.

The electromagnetic interaction holds the electrons and protons
together inside a single atom and leads to the gross, fine, and
hyperfine structures of the line spectra. The electromagnetic
interactions between the atoms appear as various chemical bonds, which
bind the atoms to form the molecules and drive the chemical
reactions. The residual electromagnetic interaction (together with
some other sources) between the neutral molecules becomes the van
der Waals force, which plays a fundamental role in condensed matter
physics etc.

The strong interaction, another fundamental interaction that occurs
between quarks and gluons, is similar to the electromagnetic
interaction in some aspects. In particle physics, Quantum
Chromodynamics (QCD) is the underlying theory of strong interactions.
Its Lagrangian is similar to the QED one:
\begin{eqnarray}\label{QCD}
\mathfrak{L}^{QCD} &=& \bar \psi_i \Big ( i \gamma^\mu (D_\mu)_{ij} - m
\delta_{ij} \Big ) \psi_j - {1 \over 4} G_{\mu\nu}^a G^{\mu\nu}_a \,
\end{eqnarray}
with the covariant derivative, which has the definition
\begin{eqnarray}
(D_\mu)_{ij} &=& \partial_\mu \delta_{ij} - i g A_\mu^a T^a_{ij} \,
.
\end{eqnarray}
Here, $\psi_i(x)$ is the quark/antiquark field and $A_\mu^a$ is the gluon
field, both of which carry the color charge. $\gamma_\mu$ are Dirac
matrices and $T^a_{ij}={\lambda^a_{ij} / 2}$ are the generators of
the SU(3) gauge group. The tensor of the gluon field strength is defined
as $G^a_{\mu\nu}=\partial_\mu A^a_\nu-\partial_\nu A^a_\mu+gf^{abc}A^b_\mu A^c_\nu$,
where $f^{abc}$ denotes the antisymmetric structure constant of the group.

Similar to the electromagnetic interaction, the strong interaction
also has several ranges. On a larger scale about 1 to 3 fm, its
residual strong interaction between the nucleons becomes the nuclear
force, which binds the protons and neutrons into atomic nuclei. On
the smaller scale less than 1 fm, the strong QCD interaction
confines the quarks and gluons to form various color-singlet
hadrons. The gross, fine and hyperfine structures also exist in the
hadron spectra.

An ideal platform to study these structures is the heavy hadrons
containing one charm or bottom quark. In recent years there have
been significant experimental developments on these heavy
hadrons~\cite{Olive:2016xmw,Amhis:2016xyh}. We shall review all these
experimental progresses in Sec.~\ref{sec2}. We try to put them
together into an integrated whole to let the readers know the
current experimental status. We shall also review the experimental
information on the four-quark state candidate $X(5568)$, which was
recently reported by the D\O\, Collaboration~\cite{D0:2016mwd}, but not
confirmed by the LHCb and CMS
collaborations~\cite{Aaij:2016iev,CMS:2016X5568}.

However, different from QED, QCD is a non-Abelian quantum field
theory. Because of the difficulty in understanding the
nonperturbative nature of QCD at low energy, one has to rely on the
effective theoretical approaches to study hadron properties. Various
methods reflecting several aspects of QCD have been proposed, such
as the relativistic quark model, the constitute quark model, the
chiral quark model, the quark pair creation (QPC) model, the Regge
trajectory phenomenology, the chiral unitary model, the QCD sum
rule, and some effective Lagrangian theories/approaches, etc. Among
these models, the most famous one is the Godfrey-Isgur (GI)
relativized quark model~\cite{Godfrey:1985xj,Capstick:1986bm}, which
we shall pay particular attention to in the present review. Besides
these models, the coupled-channel effect and the screening effect
are sometimes important.

In this report we shall review the theoretical and experimental
progresses on open-flavored heavy hadrons containing the charm and
bottom quarks, and pay particular attention to their mass spectra
and theoretical interpretations. In this section we first give an
outline of the widely used quark level methods and hadron level
methods, and the detailed results will be presented separately in
the subsequent sections. In Sec.~\ref{sec3} we review the
conventional excited charmed and bottom mesons. In Sec.~\ref{sec6}
we review the conventional excited charmed and bottom baryons. In
Sec.~\ref{sec7} we review the doubly and triply heavy baryons. We
also refer to Ref.~\cite{Cheng:2015iom} for the recent theoretical
progresses about heavy baryons. In Sec.~\ref{sec4} we discuss
candidates of the exotic states, including the $D_{s0}^*(2317)$ and
$D_{s1}(2460)$ as well as the recently observed $X(5568)$. An
outlook and a brief summary will be given in Sec.~\ref{sec5}.

In this review we shall not discuss the top hadrons, because the top
quark decays weakly before it transforms into a
hadron~\cite{Kobayashi:1973fv,Abe:1995hr,Abachi:1994td}. For the
$B_c$ system, the readers may consult the
references~\cite{Chang:1991bp,Chang:1992bb,Chang:1992jb,Chang:1992pt,Chang:1994aw,Chang:2005bf,Chang:2005wd,Chang:2006xka,Chang:2007si,Ke:2010vx,Brambilla:2010cs,Wang:2011jt,Chang:2014jca}.

For the weak decays of the open-flavor mesons, there are
semileptonic decay process (e.g. $B^+\to D^*_2(2460)\ell^+\nu_\ell$)
and nonleptonic process (e.g. $B\to D^* D_{sJ}$). For the former
weak decays, the Isgur, Scora, Grinstein and Wise (ISGW) formalism is
applicable~\cite{Isgur:1988gb,Thomas:2005bu}. One factorizes the matrix element
into a leptonic and a hadronic part. The hadronic part can be
expanded with some form factors which contain the nonperturbative
strong interaction effects and can be evaluated with various methods
(see references in \cite{Hernandez:2006gt}). For the latter weak
process, one may use the factorization approximation
\cite{Ivanov:2006ni}. The leptonic decay is also possible. The study
procedure is very similar to that used for the charmonium decays
\cite{Novikov:1977dq}.

The heavy hadrons are also closely related to the studies of the
weak interaction and CP violation, which we omit. Interested readers
may consult the excellent reviews by the Heavy Flavor Averaging
Group (HFAG)~\cite{Amhis:2016xyh,Amhis:2014hma}, which report the
world averages of the measurements of their branching fractions,
lifetimes, neutral meson mixing parameters, semileptonic decay
parameters, CP violation parameters, and CKM matrix elements, etc.

We also note that there have accumulated huge experimental data in
hadron spectroscopy in the past decade, and the theoretical progress
is also significant. New phenomena on the higher hadrons provide us
a good opportunity to understand the strong interaction deeper.
There exist nice reviews for different types of hadrons in the
literature. For example, there are reviews on baryons
\cite{Cheng:2015iom,Crede:2013sze}, hybrid states
\cite{Meyer:2015eta}, heavy quark pentaquarks and tetraquarks
\cite{Chen:2016qju,Chen:2016heh}, exotic hadrons
\cite{Hosaka:2016pey,Richard:2016eis}, and heavy hadrons in nuclear
matter \cite{Hosaka:2016ypm}. See also reviews in
Refs.~\cite{Crede:2013sze,Ivanov:1998ms,Capstick:2000qj,Bali:2000gf,Aoki:2001ra,Colangelo:2004vu,Bugg:2004xu,Swanson:2006st,Rosner:2006jz,Zhu:2007wz,Klempt:2007cp,Li:2008ey,Klempt:2009pi,Drenska:2010kg,Oset:2016lyh,Oset:2016nvf}.

\subsection{Quark Model}
\label{sec1.0}

According to the conventional quark model (QM), the mesons are
composed of the quark-antiquark pair and baryons composed of
three quarks. Such a simple model has been very successful in
explaining hadron properties. However, recent progress on hadron
spectra is challenging the naive quark model
\cite{Olive:2016xmw,Chen:2016qju}. The challenges mainly come from
the hadrons containing heavy quarks, i.e. charmonium-like XYZ
mesons, $Q\bar{q}$-type mesons, and $Qqq$-type baryons ($Q$ denotes
the heavy charm/bottom quark, and $q$ denotes the light
up/down/strange quark). The presence of the heavy quark degrees of
freedom provides a useful handle to explore the candidates of the
exotic hadrons. For the excited states, more decay channels are
allowed and the coupled channel effects due to the nearby
hadron-hadron thresholds affect significantly the hadron properties.
For example, the low mass puzzle of the $D_{s0}^*(2317)$ is difficult to
understand if one does not consider the contributions from the $DK$
channel \cite{Lang:2014yfa}.

Up to now, all types of the QM mesons including $q\bar q$, $Q\bar q$
and $Q\bar Q$ have been found. But for the baryons, even the lowest
$QQq$ baryon ($\Xi_{cc}$) has not been confirmed, and no $QQQ$
baryon is observed at all. Although there are good candidates of
exotic hadrons beyond the QM assignment, e.g. the glueballs and
hybrid states, their confirmation is still on the way. The study of
hadron spectra helps us understand how the strong interaction binds
the quarks and gluons into matter fields and find out the relation
between QM and QCD.

In recent years, the development on experimental measurements makes
it possible to investigate excited hadrons. New open-flavored
hadrons ($H_Q$), especially the mesons, have been observed at $ee$,
$pp$, and $ep$ colliders. There, the produced hard heavy quark
becomes a softer heavy quark by emitting gluons or massive gauge
bosons and then fragments into $H_Q$ nonperturbatively. The hadrons
are usually detected in $B$ ($B_s$) decays or inclusive productions,
i.e. $e^+e^-\to Q\bar{Q}\to H_Q+X$, $pp\to H_Q+X$, $ep\to H_Q+X$.
Many interesting states were observed such as the charmed-strange
mesons $D_{s0}^*(2317)$ and $D_{s1}(2460)$, which are lower than the
QM prediction and were discussed widely in terms of the various
configurations like the molecule, tetraquark, and coupled channel
effect. Since the mass splittings between the higher states are
smaller than those of the lower states, different assignments
(orbital or radial excitation states) are possible and their nature
needs detailed investigations.

Before reviewing the widely used quark level methods and hadron
level methods, we would also like to note that the basic scales in
QCD are the $\Lambda_{QCD}$, the quark masses $m_q^\prime s$ and the
scale of chiral symmetry breaking $\Lambda_\chi$. Several symmetries
of QCD are hidden behind these scales. For example, in the limit
$m_{u,d,s}\to 0$, QCD has the chiral symmetry which is spontaneously
broken below the scale $\Lambda_{\chi}\sim 1$ GeV. The $c/b$ quark
is much heavier than the $u/d/s$ quark. Contrary to the chiral
symmetry, there is another symmetry in the heavy quark sector. In
the infinitely heavy limit of the heavy quark mass, the QCD
Lagrangian has a heavy quark symmetry which has two meanings: (1)
heavy quark flavor symmetry (HQFS) which is a symmetry for the
exchange of heavy quark flavors $b\leftrightarrow c$; and (2) heavy
quark spin symmetry (HQSS) which is a symmetry for the exchange of
heavy quark spins $\uparrow_Q\leftrightarrow\downarrow_Q$. This
spin-flavor symmetry plays a crucial role in understanding the
properties of hadrons containing heavy quark. Both the quark level and hadron
level investigations involve this important symmetry.

\subsubsection{Quark potential models}
\label{sec1.1.1} $\\$

The basic approach to study hadron spectra is the quark potential
model. Generally speaking, the potential includes the contributions
from the color Coulomb interaction, spin-orbit interaction,
spin-spin interaction, and quark confinement. The first three parts
result from the one-gluon-exchange force \cite{DeRujula:1975qlm}
between free quarks while the last part is added phenomenologically
to meet the fact that the quark interaction becomes stronger and
stronger with the increasing distance and thus no colored free quark
exists. Since the potential is not an experimental observable, any
versions of the potential model that can reproduce the hadron masses
are acceptable.

The confinement potential cannot be obtained analytically from QCD
now. There exist various types, e.g. the linear potential
\cite{Eichten:1978tg,Eichten:1979ms}, logarithmic
\cite{Quigg:1977dd}, power-law \cite{Martin:1980jx}, or
error-function \cite{Zhang:1992ag}. When considering the
electromagnetic properties, one needs the additional
one-photon-exchange interaction terms. Before the observation of the
exotic $D_{s0}^*(2317)$, $D_{s1}(2460)$, and $X(3872)$, the quark
model gives satisfactory descriptions for the hadron spectra except
a few exceptions, e.g. the Roper resonance and $\Lambda(1405)$. The
interpretation of these candidates of the exotic mesons requires the
important coupled channel effects. The quark potential model has to
be improved to account for the properties of the new hadrons.

The most famous potential is given in the Godfrey-Isgur (GI)
relativized quark model \cite{Godfrey:1985xj,Capstick:1986bm}. Its
Hamiltonian includes a relativistic kinetic term and a
momentum-dependent potential $V_{\rm eff}(\bm{p},r)$:
\begin{eqnarray}
\mathcal{H}\left(\textbf{p},\textbf{r}\right) =
\left(p^2+m_1^2\right)^{1/2} + \left(p^2+m_2^2\right)^{1/2} +
V_{\mathrm{eff}}\left(\textbf{p},\textbf{r}\right) \, .
\label{Sec11:EQGI}
\end{eqnarray}
The effective potential $V_{\mathrm{eff}}(\textbf{p},\textbf{r})$
contains two main ingredients: one is a short-distance interaction
of one-gluon-exchange, and the other is a long-distance interaction
of linear confining. The latter was firstly employed by the Cornell
group and later suggested by the lattice QCD. This potential
$V_{\mathrm{eff}}(\textbf{p},\textbf{r})$ can be obtained by the
on-shell $q\bar{q}$ scattering amplitudes in the center-of-mass
frame, and can be transformed to be the standard non-relativistic
potential $V_{\mathrm{eff}}(r)$:
\begin{eqnarray}
V_{\mathrm{eff}}(r)&\to&\sum_{i<j}\left(H^{\mathrm{conf}}_{ij}(r) +
H^{\mathrm{hyp}}_{ij}(r) + H^{\mathrm{SO(cm)}}_{ij}(r) +
H^{\mathrm{SO(tp)}}_{ij}(r)\right) \, .
\end{eqnarray}
The first term, $H^{\mathrm{conf}}_{ij}(r)$, is the spin-independent
potential, containing a constant term, a linear confining potential
and a one-gluon exchange potential:
\begin{eqnarray}
H^{\mathrm{conf}}_{ij}(r) &=&
-\left[\frac34c+\frac34br_{ij}-\frac{\alpha_s}{r_{ij}}\right]\bm{F}_i\cdot\bm{F}_j
\, .
\end{eqnarray}
The second term, $H^{\mathrm{hyp}}_{ij}(r)$, is the color-hyperfine
interaction:
\begin{eqnarray}
H^{\mathrm{hyp}}_{ij}(r)
&=&-\frac{\alpha_s}{m_im_j}\Bigg[\frac{8\pi}{3}\bm{S}_i\cdot\bm{S}_j\delta^3
(\bm r_{ij})
\\ \nonumber && ~~~~~~~~~~~~~~~ +\frac{1}{r_{ij}^3}\Big(\frac{3\bm{S}_i\cdot\bm r_{ij} \bm{S}_j\cdot\bm r_{ij}}{r_{ij}^2}-\bm{S}_i\cdot\bm{S}_j\Big)\Bigg] \bm{F}_i\cdot\bm{F}_j \, .
\end{eqnarray}
The third term $H^{\mathrm{SO(cm)}}_{ij}(r)$ is the color-magnetic
term, and the fourth term $H^{\mathrm{SO(tp)}}_{ij}(r)$ is the
Thomas-precession term, both of which are spin-orbit interactions:
\begin{eqnarray}
H^{\mathrm{SO(cm)}}_{ij}(r) &=&
-\frac{\alpha_s}{r_{ij}^3}\left(\frac{1}{m_i}+\frac{1}{m_j}\right)\left(\frac{\bm{S}_i}{m_i}+\frac{\bm{S}_j}{m_j}\right)\cdot
\bm{L}(\bm{F}_i\cdot\bm{F}_j) \, ,
\\
H^{\mathrm{SO(tp)}}_{ij}(r) &=& \frac{-1}{2r_{ij}}\frac{\partial
H^{\mathrm{conf}}}{\partial
r_{ij}}\Bigg(\frac{\bm{S}_i}{m^2_i}+\frac{\bm{S}_j}{m^2_j}\Bigg)\cdot
\bm{L} \, .
\end{eqnarray}
In these expressions, $b$ and $c$ are constants, $\bm{S}_i$ is the
spin operator for $i$-th quark, $\bm{L}=\bm{r}_{ij}\times
\bm{p}_i=-\bm{r}_{ij}\times\bm{p}_j$, and the relation between
$\bm{F}_i$ and the Gell-Mann matrix is
$\bm{F}_i=\bm{\lambda}_i/2\,(-\bm{\lambda}_i^*/2)$ for quarks
(antiquarks).

This model is embedded relativistic effects mainly in two ways. First,
a smearing function is introduced to incorporate the effects of an
internal motion inside a hadron and non-locality of interactions
between (anti)quarks. Secondly, a general formula of the potential
should depend on the center-of-mass momentum of the interacting
quarks, which effect can be taken into account by introducing the
momentum-dependent factors in the interactions. With the GI model,
one can get the hadron spectra as well as wave functions by
diagonalizing the Hamiltonian~(\ref{Sec11:EQGI}) with a simple
harmonic oscillator (SHO) basis through a variational method. More
details of the GI model can be found in
Refs.~\cite{Godfrey:1985xj,Capstick:1986bm}, and it was recently
updated in Refs.~\cite{Godfrey:2015dva,Godfrey:2016nwn}.

When one focuses mainly on the mass splittings of the hadrons with
the same quark content in the same multiplet, one may adopt the
chromomagnetic interaction and the above Hamiltonian can be reduced
to a simple form
\begin{eqnarray}
H&=&\sum_i m_i^{eff}+H_{CM}
\\ \nonumber &=& \sum_i m_i^{eff}-\sum_{i<j}\frac{C_{ij}}{m_im_j}\bm{F}_i\cdot\bm{F}_j\bm{S}_i\cdot\bm{S}_j,
\end{eqnarray}
where $m_i^{eff}$ is the effective quark mass incorporating the
effects from the kinetic energy, color-Coulomb, confinement, and so
on, the subscript $CM$ means the color magnetic, and the effective
coupling constant $C_{ij}$ is the average of the contact
interaction. With this simple chromomagnetic Hamiltonian, one may
discuss the possible partner states of a given hadron.

An alternate relativised quark model motivated by the potential
nonrelativistic QCD (pNRQCD) and lattice QCD can be found in
Refs.~\cite{Godfrey:2016nwn,Lakhina:2006vg,Lakhina:2006fy}. The form
for the spin-dependent interaction is consistent with that obtained
with the earlier approaches
\cite{Brown:1979ya,Eichten:1980mw,Gupta:1981pd,Gupta:1983we,Pantaleone:1985uf}.
One may organize the high-order corrections for the potential in a
systematic way. In a constituent quark model \cite{Segovia:2013wma},
the one loop correction for the quark potential is considered in
order to include different flavored quarks in a unified description.

In Refs. \cite{Galkin:1992ry,Ebert:1996ec}, a three-dimensional
relativistic QCD-motivated potential model was developed to discuss
the hadron properties. The wave function of the bound
quark-antiquark state satisfies the quasipotential equation
\begin{eqnarray}
\frac{1}{2\mu_R}(b^2(M)-\bm{p}^2)\psi_M(\bm{p})=\int\frac{d^3q}{(2\pi)^3}V(\bm{p},\bm{q};M)\psi_M(\bm{q}),
\end{eqnarray}
where $\mu_R=[M^4-(m_1^2-m_2^2)]/(4M^3)$ is the relativistic reduced
mass with $M$ ($m_{1,2}$) being the meson (quark) mass,
$b^2(M)=[M^2-(m_1+m_2)^2][M^2-(m_1-m_2)^2]/(4M^2)$, and the
quasipotential operator is defined by
\begin{eqnarray}
V(\bm{p},\bm{q};M)&=&\bar{u}_1(p)\bar{u}_2(-p)\Big[\frac43\alpha_s
D_{\mu\nu}(\bm{k})\gamma_1^\mu\gamma_2^\nu \label{sec11:EQRQM}
\\ \nonumber && +V_{conf}^V(\bm{k})\Gamma_1^\mu\Gamma_{2;\mu}+V_{conf}^S(\bm{k})\Big]u_1(q)u_2(-q) \, .
\end{eqnarray}
In the quasipotential, $\alpha_s$ is the QCD coupling constant,
$D_{\mu\nu}(\bm{k})$ is the gluon propagator in the Coulomb gauge
($\bm{k}=\bm{p}-\bm{q}$), $\gamma_\mu$ ($\Gamma_\mu(\bm{k})$) are the
Dirac matrices (effective long-range vertex), $u(p)$ denotes the quark
spinor, and the vector (scalar) confining potential in the
nonrelativistic limit reduces to the linear type $V_V(r)$
($V_S(r)$):
\begin{eqnarray}
V_{conf}^V(r) &=& (1-\epsilon) (Ar + B) \, ,
\\ \nonumber V_{conf}^S(r) &=& \epsilon (Ar + B) \, ,
\end{eqnarray}
where $\epsilon$ is the mixing coefficient, and these two equations
further produce
\begin{eqnarray}
V_{conf}(r) = V^V_{conf}(r) + V^S_{conf}(r) = Ar + B \, .
\end{eqnarray}
The structure of the spin-dependent interaction is also in agreement
with Ref. \cite{Eichten:1980mw}. One obtains the meson spectrum by
solving this quasipotential equation. This approach was applied to
the heavy-light mesons, doubly heavy baryons, heavy quarkonia, and
$B_c$ mesons in the following studies
\cite{Ebert:1997nk,Ebert:2002ig,Ebert:2002pp,Ebert:2009ua}. Another
quasipotential model was constructed with the spectator equation in
Ref. \cite{Zeng:1994vj}, where the spectator equation is one of a
class of three-dimensional reductions of the Bethe-Salpeter
equation. This relativistic quark model is a type of the covariant
extension of the GI model. One may consult
Refs.~\cite{Radford:2007vd,Radford:2009bs,Mukherjee:1993hb,Chang:1994un,Chang:2002vx,Chang:2003ua,Chang:2004im,Chang:2004br,Dai:1993kt,Dai:1993qr,Dai:1993np,Dai:1994hna,Chen:1992fq,Chen:1993us,Chao:1982yk,Qin:1987fc,Ding:1995he,Dai:1991xw,Dai:1992zg,Ding:1983en,Ding:1987iz,Wang:2005qx,Zhang:2010ur,Fu:2011zzo,Wang:2012pf,Wang:2012wk}
for the other potential approaches.

Because of the spontaneously broken chiral symmetry of QCD, the
massless Goldstone bosons appear below the scale $\Lambda_{\chi}$.
They correspond to the lowest pseudoscalar mesons. Between the
confinement scale $\Lambda_{QCD}$ and the chiral symmetry scale
$\Lambda_{\chi}$, an effective approach, the chiral quark model, was
proposed in Ref.~\cite{Manohar:1983md}. The fundamental fields are
quarks, gluons and pseudoscalar Goldstone bosons. When studying
hadron spectra with this model, one usually considers the
one-gluon-exchange interaction and one-pion-exchange interaction.
Similar to the linear sigma model, the exchange of the scalar mesons
can also be included. One may further include the vector meson
exchanges to compensate part of the contributions from the gluon
exchanges. From various physical considerations, different versions
of the chiral quark model have been
proposed~\cite{Zhang:1994pp,Zhang:1997ny,Dai:2003dz,Valcarce:1995dm,Vijande:2004he,Dziembowski:1996cv,Glantschnig:2004mu,Segovia:2011dg}.

\subsubsection{Coupled channel and screening effects}
\label{sec1.1.2} $\\$

The GI model has achieved a great success in describing the meson
spectrum~\cite{Godfrey:1985xj}. However, there still exist some
discrepancies between its predictions and recent experimental
observations. For example, the masses of the
$D_{s0}^*(2317)$~\cite{Aubert:2003fg,Besson:2003cp,Abe:2003jk,Aubert:2006bk},
$D_{s1}(2460)$~\cite{Besson:2003cp,Abe:2003jk,Aubert:2006bk,Aubert:2003pe},
and $X(3872)$~\cite{Choi:2003ue} deviate from those expected by
the GI model. These discrepancies are partly caused by coupled
channel
effects~\cite{Eichten:2004uh,vanBeveren:2003kd,Dai:2006uz,Liu:2009uz},
which appear to be the most important for the states lying the near
kinematic
thresholds~\cite{Godfrey:2015dva,Godfrey:2016nwn,Dai:2006uz}. For
example, the $D_{s0}^*(2317)$ is close to the $DK$ threshold and its
properties are affected significantly by the $DK$
channel~\cite{Lang:2014yfa,vanBeveren:2003kd,Dai:2006uz,Mohler:2013rwa}.
The coupled channel effects lower the bare mass of hadron in
QM.

There are actually two types of the coupled channel effects: (1)
without the quark pair creation or annihilation, and (2) with the
quark pair creation or annihilation. In the former case, e.g. the
mixing of the two $J^P=1^+$ $Q\bar{q}$ mesons or the mixing of the
$S$- and $D$-waves interactions for the deuteron, the formalism is
straightforward. In the latter case, one has to consider the
transitions between the 2-quark (3-quark) state and 4-quark
(5-quark) or more complicated Fock components by specifying the
quark pair creation mechanism, e.g. $^3P_0$ mechanism, and the
calculation is usually more complicated.

Besides the explicit inclusion of the created quark-antiquark pair
in the coupled channel calculation
\cite{Eichten:2004uh,vanBeveren:2003kd,Dai:2006uz,Liu:2009uz}, one
can also adjust the mass spectrum by screening the color charges at
the distance larger than about 1 fm~\cite{Born:1989iv}, where the
light quark-antiquark pairs are spontaneously created. This vacuum
polarization effect softens (or screens) the long distance linear
potential. For the low-lying hadrons, the creation of the
quark-antiquark pair may be neglected while its contribution affects
significantly the higher mass hadrons. This is the so-called
screening effect and has been confirmed by the unquenched Lattice
QCD and some holographic
models~\cite{Bali:2005fu,Armoni:2008jy,Bigazzi:2008gd,Namekawa:2011wt}.

Some literatures studied the meson mass spectrum by considering the
screening effect. Li and Chao adopted the screened potential to
compute the charmonium spectrum~\cite{Li:2009zu}. Li, Meng, and Chao
compared the charmonium spectra predicted by the coupled-channel
model and the screened potential model and found that the two models
have the similar global features in describing the charmonium
spectrum since they describe roughly the same
effect~\cite{Li:2009ad}. Mezzoir and Gonzalez investigated the
highly excited light mesons by flattening the linear potential $br$
above a certain saturation distance $r_s$~\cite{Mezoir:2008vx}. Song
{\it et al.} studied the charmed and charmed-strange meson systems
by considering the screening
effect~\cite{Song:2015nia,Song:2015fha}. Recently, Deng {\it et al.}
investigated systematically the mass spectrum of bottomonia and
charmonia with a nonrelativistic screened potential model in Refs.
\cite{Deng:2016ktl,Deng:2016stx}.

\begin{figure}[htb]
\centering
\includegraphics[width=0.6\textwidth]{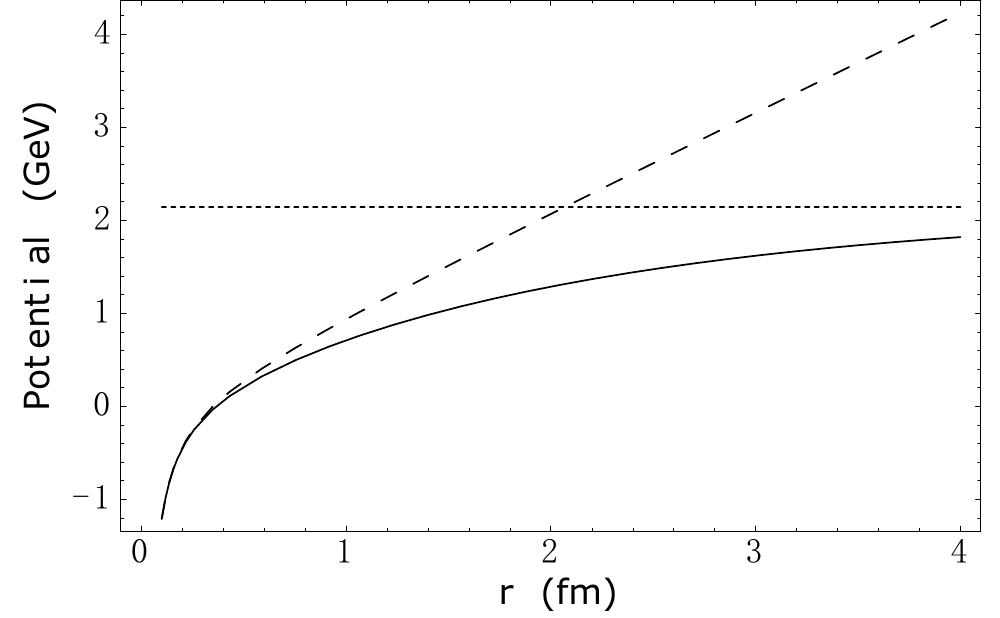}
\caption{The $r$ dependence of $V(r)=-\frac43\frac{\alpha_s}{r}+br$
(dashed line) and the screened potential
$V(r)=-\frac43\frac{\alpha_s}{r}+\frac{b(1-e^{-\mu r})}{\mu}$ (solid
line). The dotted line shows the asymptotic limit of the latter
potential. Taken from Ref.~\cite{Li:2009zu}.}
\label{fig:VvsVscr}
\end{figure}

To take into account the screening effects, one replaces the  linear
confining potential in Eq.~(\ref{Sec11:EQGI})
by~\cite{Laermann:1986pu,Chao:1992et,Ding:1993uy}
\begin{eqnarray}
br\to V^{\mathrm{scr}}(r)=\frac{b(1-e^{-\mu r})}{\mu} \, .
\label{sec11:EQscreen}
\end{eqnarray}
It is obvious that $V^{\mathrm{scr}}(r)$ behaves like a linear
potential $br$ at short distances ($r\ll \frac1\mu$) but approaches
to $\frac{b}{\mu}$ at long distances ($r\gg \frac1\mu$). A schematic
comparison for the potential $V(r)=-\frac43\frac{\alpha_s}{r}+br$
and $V(r)=-\frac43\frac{\alpha_s}{r}+\frac{b(1-e^{-\mu r})}{\mu}$ is
given in Fig. \ref{fig:VvsVscr}. The previous bare quark model is
sometimes called the quenched quark model, where the creation of a
quark-antiquark pair is not included.

To a large extent, the model with the screened potential is
equivalent to the unquenched quark model with the inclusion of the
quark-antiquark pair. Therefore, the screened model is sometimes
also denoted as the unquenched quark model. Because the potential is
revised at long distance, the mass spectra for the lower hadrons are
consistent with the quenched models. For the higher excitations of hadron, the
coupled channel effects due to the created quark pair shift the
hadron masses to a lower/higher position. It is not difficult to
understand, from Fig. \ref{fig:VvsVscr}, that the spectrum for
higher hadrons is shrunk with the screened potential, i.e.  the
flattened screened potential suppresses the masses of the higher
hadrons. Therefore, the screened potential effectively incorporates
the coupled channel effects. A detailed explanation of how to
introduce the screening effect into the GI model can be found in
Ref.~\cite{Song:2015nia}.

\subsubsection{Quark pair creation (QPC) models}
\label{sec1.1.3} $\\$

In studying the strong decays of the higher conventional hadrons or
studying their spectra with coupled channel effects, the creation of
a quark-antiquark pair is inevitable in quark models. The quantum
numbers of the quark pair depend on the QCD mechanism. Because of
the limitation of understanding the nonperturbative QCD, we have to
adopt some quark pair creation (QPC) models to study the decay
properties.

The widely used one was proposed by Micu in Ref.~\cite{Micu:1968mk}.
She assumed that the quark-antiquark pair was produced with the
vacuum quantum numbers ($J^{PC}=0^{++}$). Since the pair is in the
state $^{2S+1}L_J$ $=$ $^3P_0$, this model is called $^3P_0$ or TPZ
model. It was found that the observed partial widths were reasonably
explained although no explicit quark model wavefunctions were
assumed. In the following works by the Orsay group
\cite{LeYaouanc:1972vsx,LeYaouanc:1973ldf,LeYaouanc:1974cvx,LeYaouanc:1977fsz},
the model was developed by including explicit nonrelativistic wave
functions.

After the creation of the quark pair, the rearrangement with the
initial quarks gives an initial mock state. When calculating the
decay widths, one gets the decay amplitude through the combination
of the overlap integrals in spin, flavor, color, and orbital spaces
for the initial mock state and the final state. Up to now, this
formalism has been widely adopted in various processes. The relation
of this model to the QCD decay mechanisms was discussed for the open
flavor meson decays in Ref. \cite{Ackleh:1996yt}.

Another microscopic model for the strong decays was developed by the
Cornell group after the discovery of the
charmonium~\cite{Eichten:1978tg,Eichten:1979ms,Eichten:1975ag}. It
was assumed that the $q\bar{q}$ pair was produced from the linear
confining interaction. Since this model is mainly used in the heavy
quarkonium systems, we do not discuss it further.

\begin{figure*}[htb]
\centering
\includegraphics[width=0.6\textwidth]{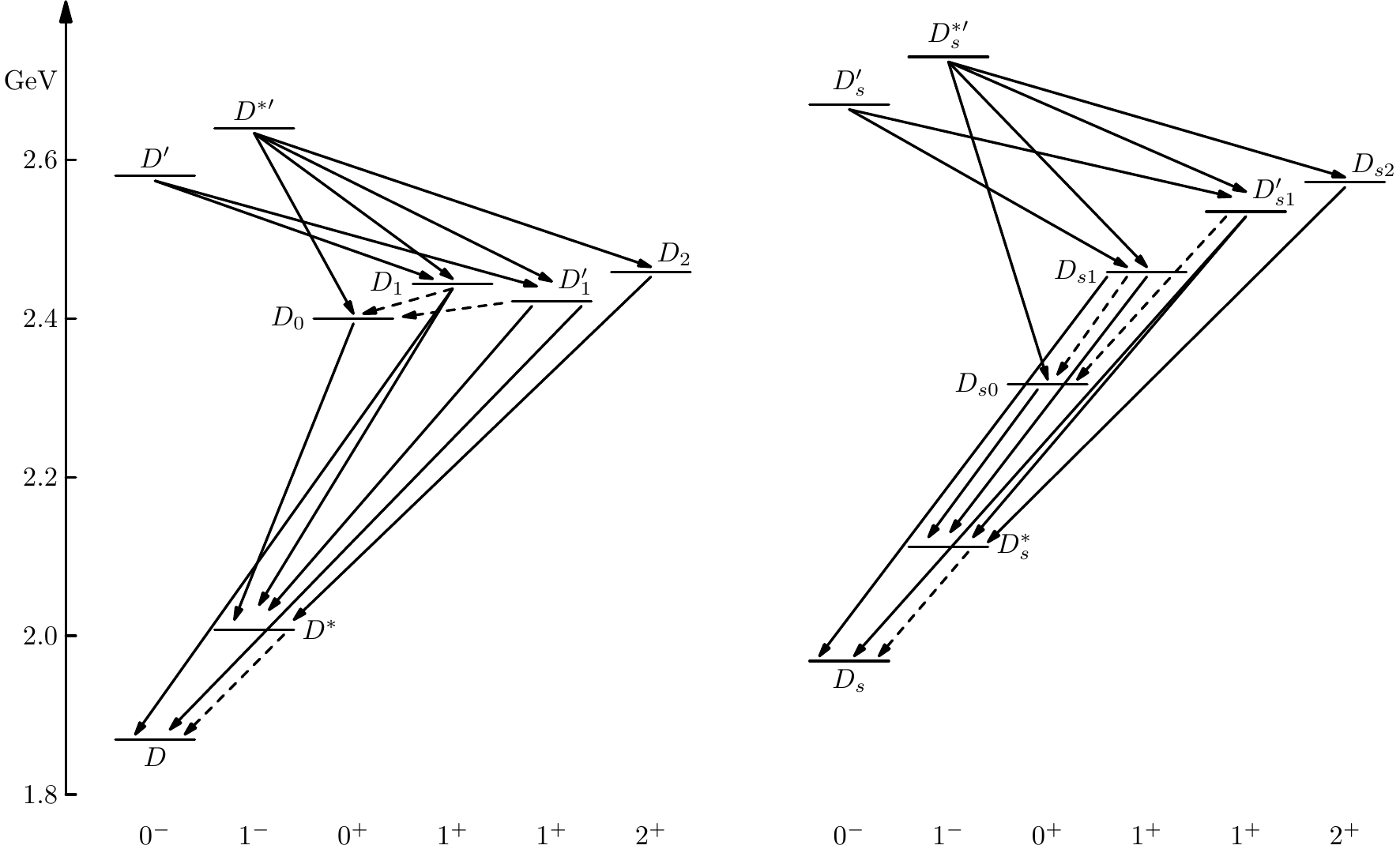}
\caption{Radiative decays of the charmed and charm-strange mesons.
The solid (dashed) line indicates the $E1$ ($M1$) transition.
}\label{DDs-rad-CS}
\end{figure*}

When the final states involve the soft pions, one may also use the
chiral quark model to study the strong decay properties. In this
model, the degrees of freedom are quarks, gluons and pseudoscalar
pions. One may also extend the model to incorporate the other scalar
and vector mesons. The coupling of the light mesons with the heavy
quarks is assumed to be weak. When discussing the heavy hadron
decays, one treats the heavy quark(s) as a spectator(s). The decay
amplitude relies on the coupling constants in the Lagrangian which
are easy to determine from the known data, e.g. $g_{\pi NN}$. Isgur
{\it et al.} used this formalism to discuss the hadron decays
\cite{Godfrey:1985xj,Copley:1979wj}. The electromagnetic decays can
also be discussed with the quark-photon interaction in the spectator
method. When discussing $E1$ and $M1$ decays, one may consult
Refs.~\cite{Novikov:1977dq,Eichten:1978tg,Godfrey:2016nwn} for
details. In Fig. \ref{DDs-rad-CS}, we show the radiative decays of the charmed and
charm-strange mesons, where the results are taken from Ref.~\cite{Close:2005se}.

\subsubsection{QCD potential}
\label{sec1.1.4} $\\$

In the above discussions we have introduced several quark potential
models, containing various potentials among quarks and gluons.
However, all these potentials are ``phenomenological'', and a complete derivation of the potential
from QCD still seems to be difficult. The QCD potential has been
studied using various methods equivalent to QCD, such as the lattice QCD
and nonrelativistic QCD (NRQCD), and we refer interested readers to
Refs.~\cite{Brambilla:2010cs,Bali:2000gf,Aoki:2001ra,Schroder:1998vy,Peter:1997me,Brambilla:2000gk,Pineda:2000sz,Brambilla:2009cd,Brambilla:1999xf,Brambilla:2004jw,Voloshin:2007dx,Lucker:2011dq,Murano:2011nz,Pineda:2011dg,Brambilla:2014jmp}
for more information.

These two types of potentials (fox example, the potentials used in
quark potential models and the static potential observed by the
Wilson loops in lattice QCD) are not exactly the same. In the
present view we shall pay more attention to the quark potential
models, which have been successfully applied to study heavy hadron
spectra, but note that the QCD potential is also important to
understand properties of hadrons such as the heavy quarkonium
states. Particularly, a method to calculate the quark and anti-quark
potential at finite quark masses in lattice QCD was proposed in
Ref.~\cite{Kawanai:2011xb}, which was latter applied in
Refs.~\cite{Kawanai:2011jt,Kawanai:2015tga,Nochi:2016wqg} to discuss
the potential description of the charmonia and charmed-strange
mesons. See the Tables~VI and IX of Ref.~\cite{Kawanai:2015tga},
where the authors compared the mass spectra of the charmonia and
charmed-strange mesons derived from lattice QCD and quark potential
models~\cite{Barnes:2005pb,Lucha:2003gs}.

\subsection{Regge trajectories}
\label{sec1.5}

Another widely used approach in studying the hadron spectra is the
well-known Regge
theory~\cite{Regge:1959mz,Regge:1960zc,Collins:1971ff}, which
preceded the QCD. The Regge theory was based on Lorentz invariance,
unitary and analyticity of the scattering matrix and has nothing
with quarks and gluons. It is a successful fundamental theory of
strong interactions at very high energies and still an indispensable
tool in phenomenological studies. The widely used notation in hadron
physics is the Regge trajectory.

There are two types of solutions of the Schr\"odinger equation for a
potential: bound state problem and scattering problem. It is well
known, e.g. with the Coulomb potential, that one may derive the
solution of the scattering problem ($E>0$) from that of the bound
state problem ($E<0$) by treating the principal quantum number $n$
as an imaginary number. In the scattering problem, the angular
momentum $\ell$ is usually taken as a physically meaningful integer
value. In fact, this quantum number from the quantization has little
to do with the interaction forces and is simply a parameter of the
Schr\"odinger equation. By considering the unphysical complex angular
momentum $\ell$, Regge studied the analytical properties of the
scattering amplitude in Refs. \cite{Regge:1959mz,Regge:1960zc}. He
found that the singularities of the amplitude in the complex $\ell$
plane are (Regge) poles which correspond to bound states or
resonances for physical angular momenta.

In a $2\to 2$ scattering process, the poles are determined by an
equation like $\alpha(t)=\ell$ with $\alpha(t)$ being a Regge
trajectory. The $t$ channel Mandelstam variable is used because
Regge poles generally arise in this channel. An interesting
observation in Regge theory is that $\alpha(t)$ is approximately a
linear function of $t$: $\alpha(t)\approx\alpha(0)+t\alpha'$. With
the crossing symmetry, one may relate the Regge poles to the
existing $s$ channel hadrons. Therefore, each hadron can be viewed
as a Regge particle or a Reggeon. Then the mass square of a hadron
is linearly related to the angular momentum ($\ell\to J$ once spin is
considered). The Regge's original work does not involve a confining
potential. Chew and Frautschi applied the theory to the case of
strong interactions and found mesons and baryons lie on linear
trajectories of the $(J, M^2)$ plane \cite{Chew:1961ev,Chew:1962eu}.
Hadrons having the same internal quantum numbers are on the same
trajectory. Thus, the Chew-Frautschi plot of Regge trajectories
provides a useful way of hadron classification.

After QCD as the fundamental theory of strong interactions was
established, there were lots of studies to understand the Regge
trajectory and the Regge particles with the quark-gluon
interactions~\cite{Bali:2000gf,Bugg:2004xu,Klempt:2007cp,Lucha:1991vn}.
Among them, the most simple and straightforward explanation towards
the linear Regge trajectories was probably the one proposed by Nambu
in 1978~\cite{Nambu:1974zg,Nambu:1978bd}, where the quark and
antiquark are assumed to be tied by the gluon flux tube. He further
assumed it to be a uniform flux tube, and the light quarks rotating
at its ends move at the speed of light at radius $R$. The mass
originating from this flux tube can be evaluated to be
\begin{eqnarray}
M = 2 \int_0^R {\sigma dr \over \sqrt{1 - v^2(r)}} = \pi \sigma R \,
,
\end{eqnarray}
where $\sigma$ is the string tension, i.e., the mass density per
unit length. The angular momentum of this flux tube can also be
evaluated to be
\begin{eqnarray}
J = 2 \int^R_0 {\sigma r v(r) dr \over \sqrt{1 - v^2(r)}} = {\pi
\sigma R^2 \over 2} + c^\prime \, .
\end{eqnarray}
Hence, $J$ and $M^2$ can be linearly related
\begin{eqnarray}
J = {M^2 \over 2 \pi \sigma} + c^{\prime\prime} \, ,
\end{eqnarray}
where $c^\prime$ and $c^{\prime\prime}$ are both constants. See also
explanations in Refs.
\cite{Bali:2000gf,Bugg:2004xu,Klempt:2007cp,Afonin:2007jd}.

If the separation between the quark and antiquark is larger than the
flux tube size, an updated picture was developed in
Ref.~\cite{Baker:2002km}, assuming that a linear potential acts to
confine the quarks in hadrons. Then one can arrive at more general
linear Regge trajectories
\begin{eqnarray}
M^2 = \alpha J + \beta n + c \, ~~~{\rm with}~~~ \alpha = \beta \,
, \label{eq:regge}
\end{eqnarray}
where $M$ is the mass, $J$ is the spin, $n$ is the principle quantum
number, $\alpha$ and $\beta$ are slopes, and $c$ is a constant. See
also discussions in Ref.~\cite{Zwiebach:2004tj}. This linear
behaviour is also expected by the dual
amplitudes~\cite{Ademollo:1969nx} and the AdS/QCD
model~\cite{Polchinski:2001tt,Brodsky:2006uq,Karch:2006pv,Forkel:2007cm}.
It is supported and can be applied to study various hadron spectra,
such as the light non-strange
mesons~\cite{Anisovich:2000kxa,Afonin:2006vi,Afonin:2007aa,Shifman:2007xn}.
As an example, the light vector mesons $\rho$, $\omega$,
$f_2(1270)$, $a_2(1320)$, $\omega_3(1670)$, $\rho_3(1690)$,
$a_4(2040)$, $f_4(2050)$, $\rho_5(2350)$, $a_6(2450)$, and
$f_6(2510)$ altogether compose one Regge trajectory, as shown in
Fig.~5 of Ref.~\cite{Klempt:2007cp}. We refer interested readers to
read
Refs.~\cite{Bugg:2004xu,Brisudova:2003dj,deTeramond:2005su,Bigazzi:2004ze,Glozman:2007ek,Guo:2008he}
for more information.

An important feature of Eq.~(\ref{eq:regge}) is that the slopes of
orbital and radial trajectories are almost equal ($\alpha = \beta$).
However, for heavy mesons it was found in Ref.~\cite{Ebert:2009ua}
that the $\alpha$ values are systematically smaller than the $\beta$
ones. See also
Refs.~\cite{Afonin:2007jd,Chan:1971gp,Kawarabayashi:1969yd,Kaidalov:1980bq,Burakovsky:1997vm,Brisudova:1999ut,Li:2004gu,Zhang:2004cd,Zhen:2013dca}
for more discussions.

In reality, the linear property is just the leading order phenomenon
in the Regge theory. Various studies have shown that the Regge
trajectories can be
nonlinear~\cite{Brisudova:1999ut,Tang:1992rp,Brisudova:1998wq,Tang:2000tb,PandoZayas:2003yb,Afonin:2004yb,Li:2016xbl}.
Although the string-like models or semirelativistic potential models
may give a linear relation between the hadron mass squared and its
quantum numbers, the universal behavior for both light and heavy
quark systems is difficult to obtain in a natural way
\cite{Afonin:2013hla,Afonin:2014nya}. The study of Regge
trajectories of hadron spectra is helpful to understand the strong
interactions from a different viewpoint and the open flavor hadrons
are good objects for such studies. We shall briefly discuss Regge
trajectories for both the heavy mesons and heavy baryons in
Sec.~\ref{sec3} and Sec.~\ref{sec6}, respectively.

\subsection{Heavy quark symmetry and effective Lagrangians}
\label{sec1.2}

One may use the heavy quark symmetry to study hadron properties by
constructing effective field theories (EFTs) at both the quark level
and hadron level. The well known EFTs for the heavy hadrons are
heavy quark effective theory (HQET), NRQCD,
and heavy hadron effective theories. These effective theories are
always related with some scales which constrain their application
region. Contrary to the phenomenological models, the Lagrangians in
effective theories are constructed according to QCD symmetries and
the interaction terms are organized systematically by some expansion
parameter(s). The effective theories are model-independent.

In the limit $m_Q\to\infty$, the one-gluon exchange (OGE)
interaction between a heavy quark and another light quark is
independent of the heavy quark mass, which results in the heavy
quark flavor symmetry and heavy quark spin symmetry. This can be
easily seen from the quark model potential in Eq.
(\ref{Sec11:EQGI}). Now the OGE potential contains only the Coulomb
part and the spin-orbit part of the light quarks. In other words,
the heavy quark looks like a static color source for the light
quarks. This picture is similar to the atom system and therefore the
hadron properties are mainly determined by the light degree of
freedom (d.o.f.). In this limit, the momentum of a heavy quark may
be written as $p_{Q\mu}=m_Qv_\mu+k_\mu$, where the four-velocity
$v_\mu$ is fixed and the residue momentum $k_\mu\sim\Lambda_{QCD}$.
The typical momentum of the light d.o.f. is also the scale of
$\Lambda_{QCD}$. For convenience, one may revise the description for
the heavy quark interaction by eliminating the large parameter $m_Q$
in order that the obtained effective field theory (HQET
\cite{Georgi:1990um}) describes physics below the scale $m_Q$. Its
Lagrangian can be expanded in $1/m_Q$.

In the QCD Lagrangian, one may replace the original quark field
$Q(x)$ by the velocity-dependent fields $H_v(x)$ and $h_v(x)$:
$Q(x)=e^{-im_Qv\cdot x}[H_v(x)+h_v(x)]$, where the rescaled fields
$H_v(x)$ and $h_v(x)$ satisfy
\begin{eqnarray}
\frac{1+v\!\!\!/}{2}H_v=H_v,\quad \frac{1-v\!\!\!/}{2}h_v=h_v.
\end{eqnarray}
Substituting these two fields into the QCD Lagrangian for the heavy
quark sector ${\cal L}^{(Q)}_{QCD}=\bar{Q}(iD\!\!\!\!/-m_Q)Q$, one
gets
\begin{eqnarray}
{\cal L}^{(Q)}_{QCD}&=&\bar{H}_v(iv\cdot D)H_v-\bar{h}_v(iv\cdot
D+2m_Q)h_v +\bar{H}_viD\!\!\!\!/ \,h_v+\bar{h}_viD\!\!\!\!/\, H_v,
\end{eqnarray}
where $h_v$ corresponds obviously to an excitation with the mass
$2m_Q$ (partly related to the antiquark d.o.f.). Since $2m_Q$ is the
energy for the creation of a heavy quark-antiquark pair and is above
the scale that HQET is applicable, this field can be ``integrated
out''. Therefore, the field $H_v$ annihilates a heavy quark but does
not create the corresponding antiquark. Then the resulting HQET
Lagrangian containing only $H_v$ is organized like
\begin{eqnarray}
{\cal L}_{HQET}={\cal L}_0+\frac{1}{m_Q}{\cal
L}_1+\frac{1}{m_Q^2}{\cal L}_2+\cdots,
\end{eqnarray}
where the leading Lagrangian ${\cal L}_0=\bar{H}_v(iv\cdot D)H_v$
has the exact heavy quark spin-flavor symmetry. The effects of $h_v$
are encoded in the coefficients of this EFT. If one studies the
system containing heavy antiquarks, the antiquark can be treated in
a similar way. Now, one projects away quarks. A more detailed
discussion is given in Ref. \cite{Georgi:1991mr}.

HQET is applicable to hadrons containing only one heavy quark. For
hadrons containing two or more heavy quarks, the correct EFT is
NRQCD \cite{Caswell:1985ui,Bodwin:1994jh}. In HQET, the leading
order Lagrangian does not depend on $m_Q$ while the kinetic energy
of the heavy quark $k^2/2m_Q$ is treated as a $1/m_Q$ correction.
For a heavy quarkonium system, the kinetic energy is needed to
stabilize a $Q\bar{Q}$ meson and cannot be treated as a
perturbation. Then the leading term in NRQCD expansion does not
conserve both HQFS and HQSS. In this frame, the energy and
three-momentum of the heavy quarks scale in a different way but
their ultraviolet (UV) cutoffs are considered to be the same
$\nu_{NR}$. The order relation of the scales is $\Lambda_{QCD}\sim
m_Qv^2\ll m_Qv\ll \nu_{NR}\ll m_Q$, where $v\ll c$ is the velocity
of the heavy quark in the meson rest frame. The
$v\sim\alpha_s(m_Qv)$ and the relativistic corrections of order
$(v^2)^n$ is more important than the perturbation corrections of
order $\alpha_s^{2n}(m_Q)$ \cite{Bodwin:1994jh}. In NRQCD, the
Lagrangian or effective operators are formulated with the expansion
parameter $v/c$ and the contributions from the hard scale ($\sim
m_Q$) are integrated out. Therefore, the Lagrangian in NRQCD is a
reorganized HQET Lagrangian. Now the heavy quark kinetic energy is
of leading order.

When applying this frame to the decay or production processes, the
decay width or production cross section can be factorized into the
short-range coefficients and long-range matrix elements. The former
part (the coefficients of the NRQCD effective operators) can be
computed perturbatively in the expansion of $\alpha_s(m_Q)$ while
the latter part needs to be evaluated nonperturbatively. If one
studies physics below the scale $m_Qv$, one obtains the potential
NRQCD (pNRQCD) by integrating out the relevant d.o.f. further
\cite{Pineda:1997bj}. More detailed discussions about EFTs for the
$Q\bar{Q}$ system can be found in Ref. \cite{Brambilla:2004jw}.
Analogous to the NRQCD for the heavy quarkonium, one may construct
effective theories for $QQq$ and $QQQ$ baryons
\cite{Brambilla:2005yk,Fleming:2005pd}. The NRQCD framework is
applicable to the production of such
baryons~\cite{Ma:2003zk,Li:2007vy,Yang:2007ep,Chen:2011mb,Jiang:2013ej,Yang:2014ita}.

The idea of hadron EFT is based on hadron classifications according
to HQSS. To be specific, we focus on the $Q\bar{q}$ mesons first.
Let $L$, $s_Q$, $s_q$, $S$, $j$, and $J$ denote the orbital angular
momentum, heavy quark spin, light quark spin, total spin, total
angular momentum of light d.o.f., and total angular momentum,
respectively. Now, the usual coupling type
$\vec{J}=\vec{L}+(\vec{s}_Q+\vec{s}_q)_S$ is reduced to the type
$\vec{J}=\vec{s}_Q+(\vec{L}+\vec{s}_q)_j$. In the latter coupling
type, the interaction mediator is the chromo-electric gluon, similar
to a photon. As a result of the heavy-quark-independent interaction,
the two mesons with the same $(L,j)$ form a degenerate doublet. The
mostly mentioned doublets are the low-lying $(0^-,1^-)=(D,D^*)$,
$(0^+,1^+)=(D_0^*,D_1^\prime)$, and $(1^+,2^+)=(D_1,D_2^*)$. With
the observation of more and more $Q\bar{q}$ mesons, one may also
find possible doublets with radial excitations
\cite{Colangelo:2010te}.

Similar to the $Q\bar{q}$ mesons, the $Qqq$ baryons also form spin
doublets (the lowest state $\Lambda_Q$ forms a spin singlet since
$s_{qq}=0$). Now, the Pauli principle works for light quarks and the
relation between flavor, spin, and orbital angular momentum must be
considered. Therefore the ground state doublet ($s_{qq}=1$) is
$(\frac12^+,\frac32^+)$ in the sextet representation of flavor
$SU(3)$. Up to now, quite a few excited charmed baryons have been
observed experimentally. Cheng and Chua discussed their strong
decays in Refs. \cite{Cheng:2006dk,Cheng:2015naa}.

Contrary to the $Qqq$ baryons, the Pauli principle works for the
heavy quarks in the $QQq$ baryons. For the $QQq$ ground state, the
spin of the heavy diquark is 1 since their color representation is
the antisymmetric $\bar{3}_c$. The resulting doubly heavy baryons
also form a degenerate doublet $(\frac12^+,\frac32^+)$. The $QQq$
baryons and $\bar{Q}q$ mesons have identical configurations of the
light d.o.f. Their properties are related by the doubly heavy
diquark-antiquark symmetry in the heavy quark limit
\cite{Fleming:2005pd,Savage:1990di,Cohen:2006jg}. For example, one
can predict the relation for the hyperfine splittings between the
ground baryons and ground mesons
$m_{J=\frac32}-m_{J=\frac12}=\frac34(m_{P^*}-m_P)$ once symmetry
breaking effects are considered. Similar relations for the $Qqq$
baryons and the proposed $T_{QQ}$ ($\bar{Q}\bar{Q}qq$) mesons can
also be predicted.

The heavy quark symmetry controls the transformation of the heavy
quark sector of the open-flavor hadrons. The transformation of the
light quark sector is determined by the chiral symmetry. Both of
them are very useful in the study of the properties of heavy
hadrons. In studying $D_{s0}^*(2317)$ and $D_{s1}(2460)$, Bardeen,
Eichten, and Hill proposed to assign them into the $(0^+,1^+)$
doublet which belongs to a chiral multiplet together with the lowest
$(0^-,1^-)$ doublet \cite{Bardeen:2003kt}. In their analysis, the
$Q\bar{q}$ system is viewed as a constituent quark which is tethered
by the heavy quark. In the case without the heavy quark, the
chirally symmetric phase of QCD needs the massless quark and the
symmetry spontaneously breaking results in the massless pions. But
now, the confinement forces the ``tethered'' constituent quark
states in the chirally symmetric phase to become the parity-doubled
bound states. They transform as the linear representations of the
light quark chiral symmetry. Once the chiral phase is spontaneously
broken, a mass gap between the degenerate parity partners appears.
Its value is determined by the couplings with the soft pions. The
situation in the $QQq$ baryons is similar. In fact, the chiral
partner structures also exist in the other systems. In Ref.
\cite{Jido:2001nt}, Jido, Oka and Hosaka explored chiral assignments
for the $qqq$ baryons, and further studies can be found in
Refs.~\cite{Chen:2008qv,Chen:2009sf,Chen:2010ba,Chen:2011rh,Dmitrasinovic:2016hup,Chen:2012vs,Chen:2012ex,Chen:2012ut,Chen:2013jra,Chen:2013gnu}
for baryon and tetraquark fields with $SU_L(3) \times SU_R (3)$
chiral symmetry. The chiral structure of the $Qqq$ baryons was
discussed in Ref. \cite{Liu:2011xc}.

When constructing effective theories at hadron level, the existence
of the multiplets is helpful to reduce the number of independent
interaction terms. The Lagrangians are required to be invariant with
respect to the transformations of the QCD symmetries such as heavy
quark symmetry and chiral symmetry. For the interactions of the
pions with the ground hadrons with one heavy quark, Yan et al. have
obtained the leading chiral order Lagrangians in Ref.
\cite{Yan:1992gz}. By using the HQSS, one reduces the number of
independent coupling constants, two for the meson case and six for
the baryon case, to one and two, respectively. The basic procedure
for the reduction to heavy hadron effective theories is similar to
that to HQET (heavy baryon chiral perturbation theory has the same
spirit) except for some technical details. The resulting Lagrangians
may be expressed in a compact form:
\begin{eqnarray}
{\cal L}_{meson}&=&tr[H(iv\cdot
D)\bar{H}]+g\,tr(H\gamma_5A\!\!\!/\bar{H})\, ,
\\ \nonumber
{\cal L}_{baryon}&=&\frac12tr[\bar{B}_{\bar3}(iv\cdot
D)B_{\bar3}]-tr[\bar{S}^\alpha(iv\cdot D-\Delta_B)S_\alpha]
\\ \nonumber
&&+[g_4\,gtr(\bar{S}^\mu A_\mu B_{\bar
3})+h.c.]+\frac32g_1(iv_\kappa)\epsilon^{\mu\nu\lambda\kappa}tr[\bar{S}_\mu
A_\nu S_\lambda]\, ,
\end{eqnarray}
where $D_\mu$ is the covariant derivative containing the chiral
connection, $A_\mu$ is the axial field in chiral perturbation
theory, $H$ denotes the rescaled $Q\bar{q}$ doublet mesons,
$\bar{H}=\gamma^0H^\dag \gamma^0$, $B_{\bar 3}$ ($S_\mu$) denotes
the rescaled flavor-antitriplet (sextet) baryons, $\bar{B}=B^\dag
\gamma^0$, and $\Delta_B=M_6-M_{\bar 3}$ is the baryon mass
difference between the two flavor multiplets in both heavy quark and
chiral limit. By extending the ground state heavy mesons to the
excited mesons, pions to the other light mesons, and by including an
additional hidden local
symmetry~\cite{Bando:1984ej,Bando:1985rf,Casalbuoni:1992gi,Meissner:1986ka,Bando:1987br},
one gets more effective Lagrangians in
Refs.~\cite{Liu:2011xc,Casalbuoni:1996pg}. The effective Lagrangians
for the $QQq$ baryons with pions can be found in
Refs.~\cite{Hu:2005gf,Ma:2015lba,Ma:2015cfa}.

The constructed effective Lagrangians may be used to study strong
decays, hadron productions, and hadron-hadron interactions. One may
also study radiative and (special) weak decays once transformations
for relevant external sources are included appropriately
\cite{Cheng:1992xi,Cho:1994vg,Cheng:1992ff}. For the semileptonic
$B$ decays, a convenient approach is to parameterize the matrix
elements with Isgur-Wise functions \cite{Isgur:1989vq,Isgur:1989ed}
by using the trace
formalism~\cite{Falk:1990yz,Falk:1991nq,Leibovich:1997tu,Leibovich:1997em,Cheng:1993ah}.

When studying coupled channel effects at hadron level, the quark
number conserving case, e.g. the hadron-hadron bound state problem
or scattering problem, is not difficult to deal with. For those
quark number changing cases, the hadron level calculation is much
easier than the quark level calculation because the quark pair
creation mechanism is hidden in the effective coupling terms and one
does not need to consider the microscopic details. There are
investigations on the hadron masses which are affected by the
virtual hadron loops. The quark fluctuation effects may be partly
considered in this way \cite{Barnes:2007xu}. After the creation of a
quark-antiquark pair and the formation of the new quark bound states
in the decay process, the residual strong interaction between new
hadrons probably plays an important role in understanding the
properties of the initial hadron. The final state interaction (FSI)
through the rescattering mechanism in a hadron loop plays a similar
role.

The heavy quark masses $m_c\approx 1270$ MeV, $m_b\approx 4660$ MeV
\cite{Olive:2016xmw}. Compared with the strange quark mass
$m_s\approx 95$ MeV, the charm quark is not so heavy. The recoil
corrections for the charmed hadrons are sometimes important, which
can be systematically included with high order
Lagrangians~\cite{Cheng:1993gc,Kitazawa:1993bk,Boyd:1994pa,Manohar:1997qy}.

\subsection{QCD sum rules}
\label{sec1.4}

Based on the heavy quark effective theory
(HQET)~\cite{Falk:1990yz,Grinstein:1990mj,Eichten:1989zv}, one can
use the method of QCD sum
rules~\cite{Shifman:1978bx,Reinders:1984sr,Nielsen:2009uh,Colangelo:2000dp} to study heavy hadrons.
This method has been successfully applied to study the
$S/P/D/F$-wave heavy
mesons~\cite{Dai:1993kt,Bagan:1991sg,Neubert:1991sp,Neubert:1993mb,Broadhurst:1991fc,Ball:1993xv,Huang:1994zj,Colangelo:1991ug,Colangelo:1992kc,Dai:1996yw,Dai:1996qx,Dai:2003yg,Colangelo:1998ga,Zhou:2014ytp,Zhou:2015ywa}
and the $S/P/D$-wave heavy
baryons~\cite{Shuryak:1981fza,Grozin:1992td,Bagan:1993ii,Dai:1995bc,Dai:1996xv,Groote:1996em,Zhu:2000py,Lee:2000tb,Huang:2000tn,Wang:2003zp,Liu:2007fg,Chen:2015kpa,Mao:2015gya,Chen:2016phw}.
We note that there are also some investigations using the method of
QCD sum rules in full
QCD~\cite{Bagan:1992tp,Bagan:1991sc,Duraes:2007te,Wang:2007sqa}.
Besides the heavy hadrons, this method has also been successfully
applied to study the exotic
hadrons~\cite{Chen:2006hy,Chen:2006zh,Chen:2007xr,Chen:2008ej,Chen:2008qw,Chen:2009gs,Chen:2010jd,Chen:2010ze,Chen:2013zia,Chen:2014vha,Chen:2015fwa,Chen:2015fsa,Chen:2015ata,Chen:2015moa,Chen:2016otp,Chen:2016ymy}.

In this section we briefly introduce the method of QCD sum rules
within HQET. Readers may consult
Refs.~\cite{Zhou:2014ytp,Zhou:2015ywa,Liu:2007fg,Chen:2015kpa,Mao:2015gya,Chen:2016phw}
for detailed discussions. In the QCD sum rule studies, one first
constructs the interpolating fields which couple to the physical
states. For example, the current
\begin{eqnarray}
J_{\Xi_c} &=& i \epsilon_{abc} \Big ( [\mathcal{D}_t^{\mu} u^{aT}] C
\gamma_t^\nu s^b -  u^{aT} C \gamma_t^\nu [\mathcal{D}_t^{\mu} s^b]
\Big ) \sigma_t^{\mu\nu} h_v^c \, , \label{eq:current15}
\end{eqnarray}
has the quark contents $usc$, the spin-parity quantum number $J^P =
1/2^-$, and one explicit orbital excitation, so it can couple to the
$P$-wave $\Xi_c(1/2^-)$ state through
\begin{eqnarray}
\langle 0| J_{\Xi_c}(x) | \Xi_c(1/2^-) \rangle = f_{\Xi_c}
u_{\Xi_c}(x) \, .
\end{eqnarray}
In Eq.~(\ref{eq:current15}): $a$, $b$ and $c$ are color indices, and
$\epsilon_{abc}$ is the totally antisymmetric tensor; $D^\mu =
\partial^\mu - i g A^\mu$ is the gauge-covariant derivative;
$\gamma_t^\mu = \gamma^\mu - v\!\!\!\slash v^\mu$,
$g_t^{\alpha_1\alpha_2}=g^{\alpha_1\alpha_2} - v^{\alpha_1}
v^{\alpha_2}$, and $D^\mu_t = D^\mu - (D \cdot v) v^\mu$ are the
transverse matrices and derivative; $C$ is the charge-conjugation
operator; the superscript $T$ represents the transpose of the Dirac
indices only; $q(x)$ is the light quark field at location $x$, and
it can be either $u(x)$ or $d(x)$ or $s(x)$; $h_v(x)$ is the heavy
quark field, and $v$ is its velocity.

One can use $J_{\Xi_c}$ to construct the two-point correlation
function
\begin{eqnarray}
{1 + v\!\!\!\slash \over 2} \Pi_{\Xi_c}(\omega) &=& i \int d^4 x
e^{i k x} \langle 0 | T[J_{\Xi_c}(x) \bar J_{\Xi_c}(0)] | 0 \rangle
\, , \label{sec1:pi}
\end{eqnarray}
where $\omega$ is the external off-shell energy, $\omega = v \cdot
k$. This two-point correlation function can be written at the hadron
level as
\begin{eqnarray}
\Pi_{\Xi_c}(\omega) = {f_{\Xi_c}^{2} \over
\overline{\Lambda}_{\Xi_c} - \omega} + \mbox{higher states} \, ,
\label{sec1:pole}
\end{eqnarray}
where
\begin{eqnarray}
\overline{\Lambda}_{\Xi_c} \equiv \lim_{m_Q \rightarrow \infty}
(m_{\Xi_c} - m_Q) \, ,
\end{eqnarray}
with $m_{\Xi_c}$ the mass of the lowest-lying heavy baryon state to
which $J_{\Xi_c}(x)$ couples.

The two-point correlation function (\ref{sec1:pi}) can also be
calculated at the quark-gluon level using the method of operator
product expansion (OPE). After performing the Borel transformation
at both the hadron and quark-gluon levels, one can obtain the mass
of the heavy baryons at the leading order in the $1/m_Q$ expansion
\begin{equation}
\overline{\Lambda}_{\Xi_c}(\omega_c, T) =
\frac{\frac{\partial}{\partial(-1/T)}\Pi_{\Xi_c}(\omega_c,
T)}{\Pi_{\Xi_c}(\omega_c, T)} \, .
\end{equation}

The mass at the order ${\mathcal O}(1/m_Q)$ can be evaluated based
on the HQET Lagrangian:
\begin{eqnarray}
\mathcal{L}_{\rm eff} = \overline{h}_{v}iv\cdot Dh_{v} +
\frac{1}{2m_{Q}}\mathcal{K} + \frac{1}{2m_{Q}}\mathcal{S} \, ,
\end{eqnarray}
where $\mathcal{K}$ is the operator of the nonrelativistic kinetic
energy
\begin{eqnarray}
\mathcal{K} &=& \overline{h}_{v}(iD_{t})^{2}h_{v} \, ,
\end{eqnarray}
and $\mathcal S$ is the Pauli term describing the chromomagnetic
interaction
\begin{eqnarray}
\mathcal{S} &=& \frac{g}{2} C_{mag} (m_{Q}/\mu) \overline{h}_{v}
\sigma_{\mu\nu} G^{\mu\nu} h_{v} \, .
\end{eqnarray}
Here $C_{mag} (m_{Q}/\mu) = [ \alpha_s(m_Q) / \alpha_s(\mu)
]^{3/\beta_0}$ with $\beta_0 = 11 - 2 n_f /3$.

One can rewrite Eq.~(\ref{sec1:pole}) up to the order ${\mathcal
O}(1/m_Q)$ as
\begin{eqnarray}
\Pi(\omega)_{pole} &=& \frac{(f+\delta
f)^{2}}{(\overline{\Lambda}+\delta m)-\omega}
\\ \nonumber &=& \frac{f^{2}}{\overline{\Lambda}-\omega}-\frac{\delta mf^{2}}{(\overline{\Lambda}-\omega)^{2}}+\frac{2f\delta f}{\overline{\Lambda}-\omega} \, ,
\end{eqnarray}
where $\delta m$ is the correction at the order ${\mathcal
O}(1/m_Q)$. It can be evaluated using the three-point correlation
functions:
\begin{eqnarray}
\delta_{O}\Pi_{\Xi_c}(\omega , \omega^\prime) &=& i^{2} \int
d^{4}xd^{4}ye^{ik\cdot x-ik^\prime\cdot
y}\times\langle0|T[J_{\Xi_c}(x) O(0) \bar J_{\Xi_c}(y)]|0\rangle \,
, \label{sec1:nextpi}
\end{eqnarray}
where $\omega = v \cdot k$, $\omega^\prime = v \cdot k^\prime$, and
$O = \mathcal{K}$ or $\mathcal{S}$. These three-point correlation
functions can be written at the hadron level as
\begin{eqnarray}
\delta_{\mathcal{K}}\Pi_{\Xi_c}(\omega,\omega^\prime) &=&
\frac{f_{\Xi_c}^{2}K_{\Xi_c}}{(\overline{\Lambda}_{\Xi_c}-\omega)(\overline{\Lambda}_{\Xi_c}-\omega^\prime)}
+ \cdots  \, , \label{sec1:defK}
\\ \delta_{\mathcal{S}}\Pi_{\Xi_c}(\omega,\omega^\prime) &=& \frac{d_{M}f_{\Xi_c}^{2}\Sigma_{\Xi_c}}{(\overline{\Lambda}_{\Xi_c}-\omega)(\overline{\Lambda}_{\Xi_c}-\omega^\prime)} + \cdots  \, ,
\label{sec1:defS}
\end{eqnarray}
where
\begin{eqnarray}
\nonumber K_{\Xi_c} &\equiv& \langle \Xi_c(1/2^-)
|\overline{h}_{v}(iD_{\bot})^{2}h_{v}| \Xi_c(1/2^-) \rangle \, ,
\\ \nonumber d_{M}\Sigma_{\Xi_c} &\equiv& \langle \Xi_c(1/2^-) | {g\over2} \overline{h}_{v}\sigma_{\mu\nu}G^{\mu\nu}h_{v}| \Xi_c(1/2^-) \rangle \, ,
\\ d_{M} &\equiv& d_{j,j_{l}} \, ,
\\ \nonumber d_{j_{l}-1/2,j_{l}} &=& 2j_{l}+2\, ,
\\ \nonumber d_{j_{l}+1/2,j_{l}} &=& -2j_{l} \, .
\end{eqnarray}

The three-point correlation functions (\ref{sec1:nextpi}) can also
be calculated at the quark-gluon level using the method of operator
product expansion. After performing a double Borel transformation
for $\omega$ and $\omega^\prime$ at both the hadron and quark-gluon
levels, one can evaluate $K_{\Xi_c}$ and $\Sigma_{\Xi_c}$ and
further obtain the mass of the heavy baryons at the order ${\mathcal
O}(1/m_Q)$:
\begin{eqnarray}
\delta m_{\Xi_c} = -\frac{1}{2m_{Q}}(K_{\Xi_c} +
d_{M}C_{mag}\Sigma_{\Xi_c} ) \, .
\end{eqnarray}

\subsection{Unsettled issues}
\label{sec1.3}

\begin{table*}
\centering \caption{The mass differences between the charmed-strange
and charmed mesons for the first three doublets. The masses are in
units of MeV \cite{Olive:2016xmw}.}\label{sec1:issues}
\begin{tabular}{cccccc}\hline \hline
$J^P$&Meson&Mass&Meson&Mass&Difference\\ \hline
$0^-$&$D^{0(\pm)}$& 1864.83 (1869.58)&$D_s^{\pm}$&1968.27&103.44 (98.69)\\
$1^-$&$D^{*0(\pm)}$&2006.85 (2010.26)&$D_s^{*\pm}$&2112.1&105.25
(101.84)\\\hline
$0^+$&$D_0^*(2400)^{0(\pm)}$&2318 (2351)&$D_{s0}^*(2317)^{\pm}$&2317.7&$-0.3$ ($-33.3$)\\
$1^+$&$D_1(2430)^{0}$ & 2427
&$D_{s1}(2460)^{\pm}$&2459.5&32.5\\\hline
$1^+$&$D_1(2420)^{0(\pm)}$&2420.8 (2423.2)&$D_{s1}(2536)^{\pm}$&2535.10&114.3 (111.9)\\
$2^+$&$D_2^*(2460)^{0(\pm)}$&2460.57
(2465.4)&$D_{s2}^*(2573)$&2569.1&108.53 (103.7)\\ \hline \hline
\end{tabular}
\end{table*}

There are two puzzles in the charm meson sector. The first one is
the low mass puzzle of the $D_{s0}^*(2317)$ and the $D_{s1}(2460)$
states, both of which were observed in 2003
\cite{Aubert:2003fg,Besson:2003cp}. Their masses (widths) are much
lower (narrower) than the QM predictions
\cite{Godfrey:1985xj,Godfrey:1986wj,DiPierro:2001dwf}. Although
there exist lots of discussions of these two states in the
literature, there are still open questions in understanding their
underlying structure.

The second puzzle is related to the degeneracy of the charmed mesons
and the charmed-strange mesons in the $(0^+,1^+)$ doublet. In Table
\ref{sec1:issues}, we have collected the masses of the S-wave and
P-wave charmed mesons and charmed-strange mesons in the $(0^-,1^-)$,
$(0^+,1^+)$ and $(1^+,2^+)$ doublets. The mass difference between
the charmed-strange and charmed mesons with the same quantum number
in the $(0^-,1^-)$ and $(1^+,2^+)$ doublets is around $105$ MeV,
which is the mass of the strange quark. However, the charmed mesons
and charmed-strange mesons in the $(0^+,1^+)$ doublet are almost
degenerate. Such a feature is very puzzling.

The surprisingly low masses of the $D_{s0}^*(2317)$ and
$D_{s1}(2460)$ states may arise from the coupled channel effects.
The S-wave $DK$ and $D^*K$ scattering states couple to the bare
quark model $c\bar s$ state strongly and lower the quark model
spectrum. Such a picture was supported by a recent lattice
simulation \cite{Lang:2014yfa}. However, one has to answer why the
S-wave $D\pi$ and $D^*\pi$ scattering states do NOT affect the
charmed mesons in the $(0^+,1^+)$ doublet. Is such an effect
amplified by the small mass difference between the $D_{s0}^*(2317)$
and $D_{s1}(2460)$ mesons and the $DK$ and $D^*K$ thresholds?

For comparison, the $\Lambda(1405)$ may be a state with similar
dynamics. Recall that the $\Lambda(1405)$ is a $\pi\Sigma$ resonance
and lies slightly below the $\bar{K}N$ threshold. The S-wave
$\bar{K}N$ scattering states couple to the bare P-wave $uds$ baryon
state in the quark model. Fifty years have passed since its
observation. People are still wondering whether the $\Lambda(1405)$
is a $\bar{K}N$ molecule or a pentaquark state or a three-quark
baryon affected by the coupled channel effects or there exist two
poles around 1405 MeV \cite{Hyodo:2015rnm}.

%% file: section2.1.tex
\section{Experimental progress on heavy hadrons}
\label{sec2}

The charm quark (or shortly denoted as the $c$ quark) is an
elementary fermion with spin 1/2. It has an electric charge of
$+2/3$ $e$ and a pole mass of $1.29^{+0.05}_{-0.11}$ GeV. The charm
quark was predicted in the GIM mechanism by Glashow, Iliopoulos, and
Maiani in 1970~\cite{Glashow:1970gm}. Four years later on 11
November 1974, the $J/\psi$ meson was discovered independently by
two research groups, one at the Brooklyn National Laboratory headed
by Ting~\cite{Aubert:1974js} and the other at the Stanford Linear
Accelerator Center headed by Richter~\cite{Augustin:1974xw}. After
that, lots of charmed particles were discovered/observed, including:
\begin{enumerate}

\item the charmed mesons, containing a charm quark and an up or
down antiquark, which will be reviewed in Sec.~\ref{sec2.1};

\item the charmed-strange mesons, containing a charm quark and
a strange antiquark, which will be reviewed in Sec.~\ref{sec2.2};

\item various singly charmed baryons, composed of a charm quark
and two light (up, down, or strange) quarks, which will be reviewed in Sec.~\ref{sec2.5};

\item the doubly charmed baryons, composed of two charm quarks
and one light quark, which will be reviewed in Sec.~\ref{sec2.7}.

\end{enumerate}
There were also many charmonium states observed in various
experiments, which we shall not discuss in this review.

Three years later in 1977, the bottom quark was discovered. The
bottom quark (or shortly denoted as the $b$ quark, also known as the
beauty quark) is an elementary fermion with spin 1/2. It has an
electric charge of $-1/3$ $e$ and a pole mass of
$4.18^{+0.03}_{-0.03}$ GeV. The bottom quark was proposed by
Kobayashi and Maskawa to explain CP violation in
1973~\cite{Kobayashi:1973fv}. In 1977, the $\Upsilon(1S)$, was
discovered at the Fermilab~\cite{Herb:1977ek}. After that, lots of
bottom particles were discovered/observed, including:
\begin{enumerate}

\item the bottom mesons, containing a bottom quark and an up
or down antiquark, which will be reviewed in Sec.~\ref{sec2.3};

\item the bottom-strange mesons, containing a bottom quark and
a strange antiquark, which will be reviewed in Sec.~\ref{sec2.4};

\item various singly bottom baryons, composed of a bottom quark
and two light quarks, which will be reviewed in Sec.~\ref{sec2.6}.

\end{enumerate}
Again, we shall not discuss the bottomonium states in this review.
We shall neither discuss the $B_c$ mesons, containing a bottom quark
and a charm antiquark or a bottom antiquark and a charm quark (see
Refs.~\cite{Chang:1992bb,Chang:1992jb,Chang:1992pt,Brambilla:2010cs,Ackerstaff:1998zf,Abe:1998wi,Aad:2014laa,Godfrey:2004ya,Brambilla:2004wf}
and related references).

We shall separately review experimental progresses on the heavy
meson/baryon states in the following, and a short summation is
given here:
\begin{enumerate}

\item The $1S$ and $1P$ charmed and charm-strange mesons have been
well established, while the higher states starting from the
$D(2550)$ and $D_{sJ}^\ast(2632)$ require more efforts.

\item All the $1S$ charmed baryons have been well established,
and the lowest-lying orbitally excited charmed states
$\Lambda_c(2595)$ of $J^P=1/2^-$, $\Lambda_c(2625)$ of $J^P=3/2^-$,
$\Xi_c(2790)$ of $J^P=1/2^-$, and $\Xi_c(2815)$ of $J^P=3/2^-$ have
also been well observed and complete the two $SU(3)$
$\mathbf{\bar3}$ multiplets, while all the higher states are not
well established.

\item The $1S$ bottom and bottom-strange mesons have been well
established, while more studies on the higher states starting from
the $B^*_J(5732)$ and $B^*_{sJ}(5850)$ are necessary.

\item All the $1S$ bottom baryons have been observed, except
the $\Omega_b^*$ of $J^P = 3/2^+$, but not all of them are well understood/established.
Besides them, there are only two excited bottom baryons observed in
experiments, the $\Lambda_b(5912)^0$ of $1/2^-$ and the
$\Lambda_b(5920)^0$ of $3/2^-$, which probably belong to the $SU(3)$
$\mathbf{\bar 3}_F$ multiplet of $J^P=1/2^-$ and $3/2^-$.

\end{enumerate}

\subsection{The charmed mesons}
\label{sec2.1}

\renewcommand{\arraystretch}{1.4}
\begin{table*}[htb]
\scriptsize
\caption{Experimental information of the observed charmed mesons.
The $1S$ charmed states ($D$ and $D^*$) and the $1P$ charmed states
($D_0^*(2400)$, $D_1(2420)$, $D_1(2430)$, and $D_2^*(2460)$) have
been well established, so we only list their averaged masses and
widths from PDG~\cite{Olive:2016xmw} together with the experiments
which first observed these states. However, the higher states
starting from the $D(2550)$ are not well established, so we list all
the relevant experiments together with their observed masses, widths,
and decay modes therein. \label{sec21:Dmeson} } \centering
\begin{tabular}{cccccc}
\toprule[1pt] State & $J^P$ & Mass (MeV) & Width (MeV) & Experiments
& Observed Modes
\\ \midrule[1pt]
$D^0$ & $0^-$ & $1864.83 \pm 0.05$ & $(410.1 \pm 1.5) \times
10^{-15}$ s & Mark I~\cite{Goldhaber:1976xn} & $K\pi$ and $K 3\pi$
\\
$D^\pm$ & $0^-$ & $1869.58 \pm 0.09$ & $(1040 \pm 7) \times
10^{-15}$ s & Mark I~\cite{Peruzzi:1976sv} & $K 2\pi$
\\ \hline
$D^{*0}$ & $1^-$ & $2006.85 \pm 0.05$ & $<2.1$ & Mark
I~\cite{Goldhaber:1977qn} & $e^+e^- \rightarrow D D^*$
\\
$D^{*\pm}$ & $1^-$ & $2010.26 \pm 0.05$ & $(83.4 \pm 1.8) \times
10^{-3}$ & Mark I~\cite{Goldhaber:1977qn} & $e^+e^- \rightarrow D
D^*$
\\ \hline
$D_{0}^*(2400)^0$ & $0^+$ & $2318 \pm 29$ & $267\pm40$ &
Belle~\cite{Abe:2003zm} & $D^+ \pi^-$
\\
$D_{0}^*(2400)^\pm$ & $0^+$ & $2351 \pm 7$ & $230 \pm 17$ & FOCUS~\cite{Link:2003bd} & $D^0 \pi^-$
\\ \hline
$D_{1}(2420)^0$ & $1^+$ & $2420.8 \pm 0.5$ & $31.7 \pm 2.5$ &
ARGUS~\cite{Albrecht:1985as} & $D^{*\pm}\pi^{\mp}$
\\
$D_{1}(2420)^\pm$ & $1^+$ & $2423.2 \pm 2.4$ & $25 \pm 6$ &
TPS~\cite{Anjos:1988uf} &$D^{*0} \pi^+$
\\ \hline
$D_{1}(2430)^0$ & $1^+$ & $2427\pm26\pm25$ & $384^{+107}_{-75}\pm75$
& Belle~\cite{Abe:2003zm} &$D^{*+} \pi^-$
\\ \hline
$D_{2}^*(2460)^0$ & $2^+$ & $2460.57 \pm 0.15$ & $47.7 \pm 1.3$ &
TPS~\cite{Anjos:1988uf} & $D^+\pi^-$
\\
$D_{2}^*(2460)^\pm$ & $2^+$ & $2465.4 \pm 1.3$ & $46.7 \pm 1.2$ &
ARGUS~\cite{Albrecht:1989gb} & $D^0\pi^+$
\\ \midrule[1pt]
$D(2550)^0$ & \multirow{2}{*}{$0^-$} & $2539.4\pm4.5\pm6.8$  &
$130\pm12\pm13$ & BaBar \cite{delAmoSanchez:2010vq}&  $D^* \pi$
\\
$D_J(2580)^0$                       && $2579.5\pm3.4\pm5.5$ &
$177.5\pm17.8\pm46.0$ & LHCb \cite{Aaij:2013sza}&  $D^* \pi$
\\ \hline
\multirow{3}{*}{$D_1^*(2600)^0$} & \multirow{3}{*}{$1^-$} &
$2608.7\pm2.4\pm2.5$ & $93\pm6\pm13$ & BaBar
\cite{delAmoSanchez:2010vq}&  $D^{(*)} \pi$
\\
                                                       && $2649.2\pm3.5\pm3.5$  & $140.2\pm17.1\pm18.6$ & LHCb \cite{Aaij:2013sza}&  $D^{*} \pi$
\\
                                                       && $ 2681.1 \pm 5.6 \pm 4.9 \pm 13.1$  & $186.7 \pm 8.5 \pm 8.6 \pm 8.2$ & LHCb \cite{Aaij:2016fma}&  $D \pi$
\\ \hline
$D(2750)^0$ & $?^?$ & $2752.4\pm1.7\pm2.7$ & $71\pm6\pm11$ & BaBar
\cite{delAmoSanchez:2010vq} &  $D^{*} \pi$
\\
$D_J(2740)^0$ & $2^-$ & $2737.0\pm3.5\pm11.2$ & $73.2\pm13.4\pm25.0$ &
LHCb \cite{Aaij:2013sza} &  $D^{*} \pi$
\\ \hdashline[2pt/2pt]
$D^*(2760)^0$ & $?^?$ & $2763.3\pm2.3\pm2.3$ & $60.9\pm5.1\pm3.6$ &
BaBar \cite{delAmoSanchez:2010vq} &  $D \pi$
\\
$D_J^*(2760)^0$ & $?^?$ & $2760.1 \pm 1.1 \pm 3.7$ & $74.4 \pm 3.4 \pm
19.1$ & LHCb \cite{Aaij:2013sza} &  $D \pi$
\\ \hdashline[2pt/2pt]
$D^*_1(2760)^0$ & $1^-$ & $2781 \pm 18 \pm 11 \pm 6$ & $177 \pm 32
\pm 20 \pm 7$ & LHCb \cite{Aaij:2015vea} & $D^+ \pi^-$
\\
$D^*_3(2760)^-$ & $3^-$ & $2798 \pm 7 \pm 1 \pm 7$ & $105 \pm 18 \pm
6 \pm 23$ & LHCb \cite{Aaij:2015sqa} & $\bar D^0 \pi^-$
\\
$D^*_3(2760)^0$ & $3^-$ & $ 2775.5 \pm 4.5 \pm 4.5 \pm 4.7$ & $95.3
\pm 9.6 \pm 7.9 \pm 33.1$ & LHCb \cite{Aaij:2016fma} & $D^+ \pi^-$
\\ \hline
$D_J(3000)^0$ & $?^?$ & $2971.8\pm8.7$ & $188.1\pm44.8$ &LHCb
\cite{Aaij:2013sza}&  $D^{*} \pi$
\\
$D_J^*(3000)^0$ & $?^?$ & $3008.1\pm4.0$  & $110.5\pm11.5$ &LHCb
\cite{Aaij:2013sza}&  $D \pi$
\\
$D_2^*(3000)^0$ & $2^+$ & $ 3214 \pm 29 \pm 33 \pm 36$  & $ 186 \pm
38 \pm 34 \pm 63$ &LHCb \cite{Aaij:2016fma}&  $D^+ \pi^-$
\\ \bottomrule[1pt]
\end{tabular}
\end{table*}

In this subsection we review the charmed mesons. Their experimental
information is listed in Table~\ref{sec21:Dmeson}. The $1S$ charmed
states ($D$ and $D^*$) and the $1P$ charmed states ($D_0^*(2400)$,
$D_1(2420)$, $D_1(2430)$ and $D_2^*(2460)$) have been well
established, completing one $S$-wave doublet $(0^-, 1^-)$ and two
$P$-wave doublets $(0^+, 1^+)$ and $(1^+, 2^+)$. Hence, we only list
in Table~\ref{sec21:Dmeson} their averaged masses and widths from
PDG~\cite{Olive:2016xmw} together with the experiments which first
observed them. However, the higher states starting from the
$D(2550)$ are not well established, so we list all the relevant
experiments. We note that the charmed state $D^\ast(2640)$ was
reported by the DELPHI Collaboration~\cite{Abreu:1998vk} but not
confirmed in any other experiments, so we do not include this state
in our review of this work.

Sometimes we use the words ``natural parity states'', labelled as
$D_J^*({\rm Mass})$, to denote the states having $P=(-1)^J$, such as
$J^P = 0^+$, $1^-$, $2^+$, etc.; we also use ``unnatural parity
states'', labelled as $D_J({\rm Mass})$, to denote the states having
$P=(-1)^{J+1}$, such as $J^P = 0^-$, $1^+$, $2^-$, etc.

\subsubsection{$D$ and $D^*$.}

\begin{figure}[htb]
\begin{center}
\includegraphics*[width=0.48\textwidth]{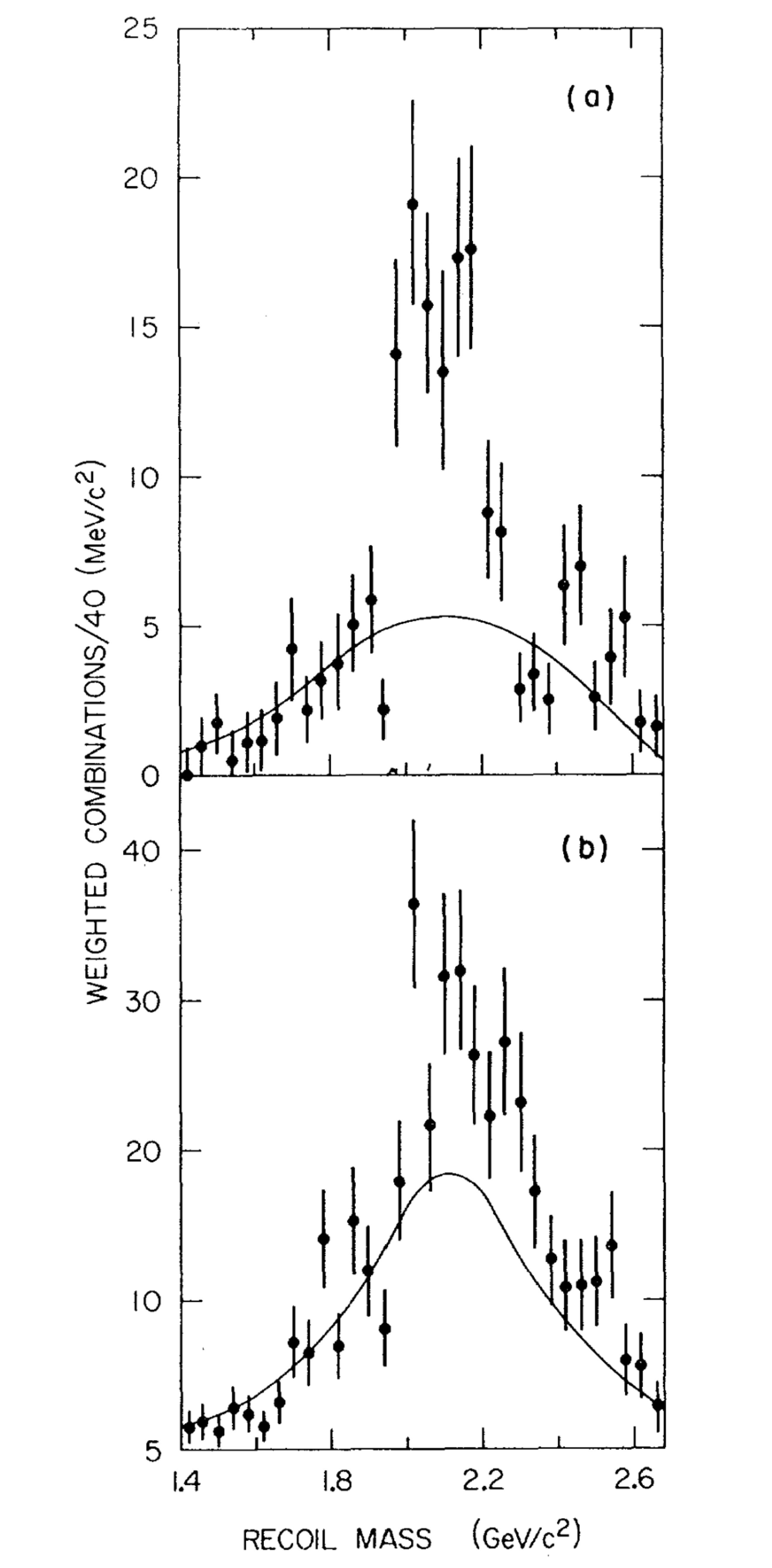}
\end{center}
\caption{
Mass distributions for the (a) $K \pi$ and (b) $K 3\pi$, where the peaks correspond to the $D$ meson.
Taken from Mark I~\cite{Goldhaber:1976xn}. }
\label{sec211:MARK}
\end{figure}

The lowest-lying charmed mesons, $D^0$ and $D^\pm$ of $J^P = 0^-$,
were observed in
1976~\cite{Goldhaber:1976xn,Peruzzi:1976sv,Wiss:1976gd}, as shown in
Fig.~\ref{sec211:MARK}. Their properties are known very
well~\cite{Olive:2016xmw}: the $D^0$ meson has a mass $1864.83 \pm 0.05$ MeV and a mean life $(410.1 \pm 1.5) \times 10^{-15}$ s; the
$D^\pm$ meson has a mass $1869.58 \pm 0.09$ MeV and a mean life
$(1040 \pm 7) \times 10^{-15}$ s; hundreds of their decay modes have
been observed in experiments, where the Cabibbo-allowed process $c
\rightarrow s W^+$ is preferred.

One year later in 1977, the lowest-lying vector charmed mesons,
$D^{*0}$ and $D^{*\pm}$ of $J^P = 1^-$, were
observed~\cite{Goldhaber:1977qn,Nguyen:1977kk,Peruzzi:1977ms}. Their
properties are also known very well~\cite{Olive:2016xmw}: the
$D^{*0}$ meson has a mass $2006.85 \pm 0.05$ MeV and the upper limit
of its decay width is $2.1$ MeV; the $D^{*\pm}$ meson has a mass
$2010.26 \pm 0.05$ MeV and a width $83.4 \pm 1.8$ keV; the $D^{*0}$
meson mainly decays into $D^0 \pi^0$ and $D^0 \gamma$, with
fractions $(64.7 \pm 0.9)\%$ and $(35.3 \pm 0.9)\%$, respectively;
the $D^{*\pm}$ meson decays into $D^0 \pi^\pm$, $D^\pm \pi^0$, and
$D^\pm \gamma$, with fractions $(67.7 \pm 0.5)\%$, $(30.7 \pm
0.5)\%$, and $(1.6 \pm 0.4)\%$, respectively.

\subsubsection{$D_0^*(2400)$, $D_1(2420)$, $D_1(2430)$, and $D_2^*(2460)$.}

\begin{figure}[htb]
\begin{center}
\includegraphics*[width=0.6\textwidth]{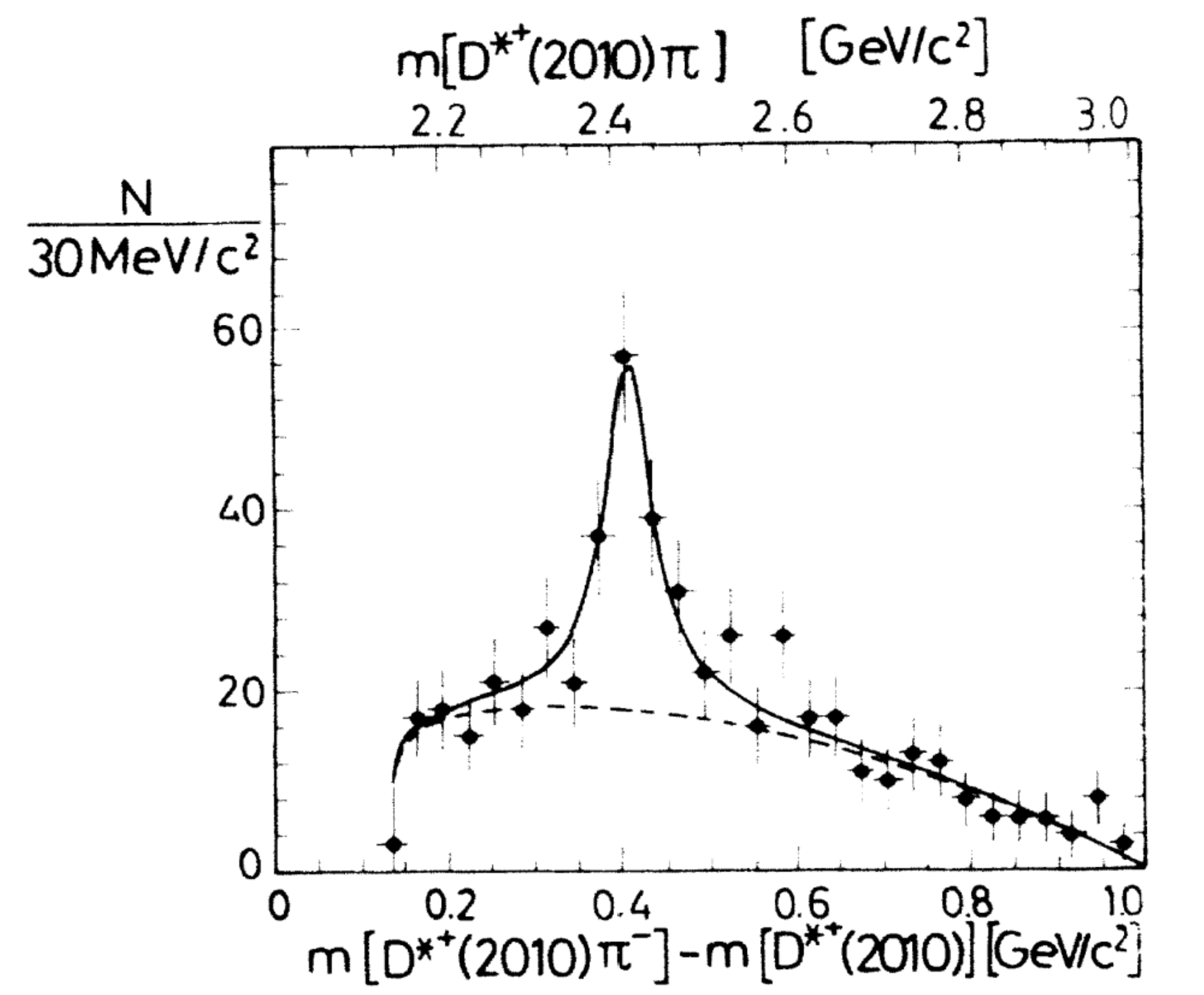}
\end{center}
\caption{The mass difference distribution $m(D^{*+}(2010) \pi^-)
- m(D^{*+}(2010))$, where the signal corresponds to the $D_1(2420)^0$.
Taken from ARGUS~\cite{Albrecht:1985as}. }
\label{sec211:ARGUS}
\end{figure}

Eight years later after the discovery of the $D$ and $D^*$ mesons,
the first orbitally excited charmed meson was observed. In
1985 the $P$-wave charmed meson $D_1(2420)^0$ of $J^P = 1^+$ was
reported in the $D^{*\pm} \pi^{\mp}$ invariant mass distribution by
the ARGUS Collaboration~\cite{Albrecht:1985as}, as shown in
Fig.~\ref{sec211:ARGUS}. Its charged partner, the $D_1(2420)^\pm$ of
$J^P = 1^+$, was observed in the $D^0 \pi^+$ invariant mass
distribution by the TPS Collaboration~\cite{Anjos:1988uf}. The
$D_1(2420)$ has been confirmed by many
experiments~\cite{Abe:2003zm,delAmoSanchez:2010vq,Abreu:1998vk,Albrecht:1988dj,Avery:1989ui,Frabetti:1993vv,Avery:1994yc,Ackerstaff:1997vc,Abe:2004sm,Abulencia:2005ry,Chekanov:2008ac,Aubert:2008zc,Abramowicz:2012ys}.
Its properties are known very well, as listed in
Table~\ref{sec21:Dmeson}. The widths of the $D_1(2420)^0$ and
$D_1(2420)^\pm$ are around 25 MeV. They have a partner $D_1(2430)$
with the similar mass but much larger width. The $D_1(2430)^0$ of
$J^P=1^+$ was observed in 2003 by the Belle Collaboration in the
$B^- \rightarrow \pi^- D_1(2430)^0 (\rightarrow D^{*+} \pi^-)$
process~\cite{Abe:2003zm}. In 2006, the BaBar Collaboration studied
the $D^*\pi$ invariant mass spectrum and confirmed the
$D_1(2430)^0$~\cite{Aubert:2006zb}. The charged partner of the $D_1(2430)^0$ has
not been observed yet.

Besides the $D_1(2430)^0$, Belle announced the observation of
another broad state, the $P$-wave charmed meson $D^*_0(2400)^0$ of
$J^P=0^+$~\cite{Abe:2003zm}, which was confirmed by the FOCUS, BaBar,
and Belle
experiments~\cite{Link:2003bd,Aubert:2009wg,Matvienko:2015gqa}. Its
charged partner $D^*_0(2400)^\pm$ was also observed by the FOCUS
Collaboration~\cite{Link:2003bd} and confirmed by the LHCb
Collaboration~\cite{Aaij:2015sqa}.

The TPS Collaboration observed another $P$-wave charmed meson with
either $J^P=0^+$ or $2^+$ in the $D^+\pi^-$ invariant mass
distribution~\cite{Anjos:1988uf}. This observation was confirmed by
many other experiments~\cite{Olive:2016xmw} and their angular
momentum analysis suggests the $J^P=2^+$ assignment to this state
\cite{Albrecht:1988dj}. Now this resonance is denoted as
$D_2^*(2460)^0$ of $J^P=2^+$. Its charged partner $D_2^*(2460)^\pm$
was observed by the ARGUS Collaboration in the $D^0\pi^+$
channel~\cite{Albrecht:1989gb} and confirmed in many other
experiments~\cite{Link:2003bd,delAmoSanchez:2010vq,Abreu:1998vk,Frabetti:1993vv,Avery:1994yc,Ackerstaff:1997vc,Abe:2004sm,Abulencia:2005ry,Chekanov:2008ac,Aubert:2008zc,Abramowicz:2012ys,Aubert:2009wg,Asratian:1995wx,Albrecht:1989pa}.

\subsubsection{$D(2550)$ and $D_J(2580)$.}

\begin{figure}[htb]
\begin{center}
\includegraphics*[width=0.48\textwidth]{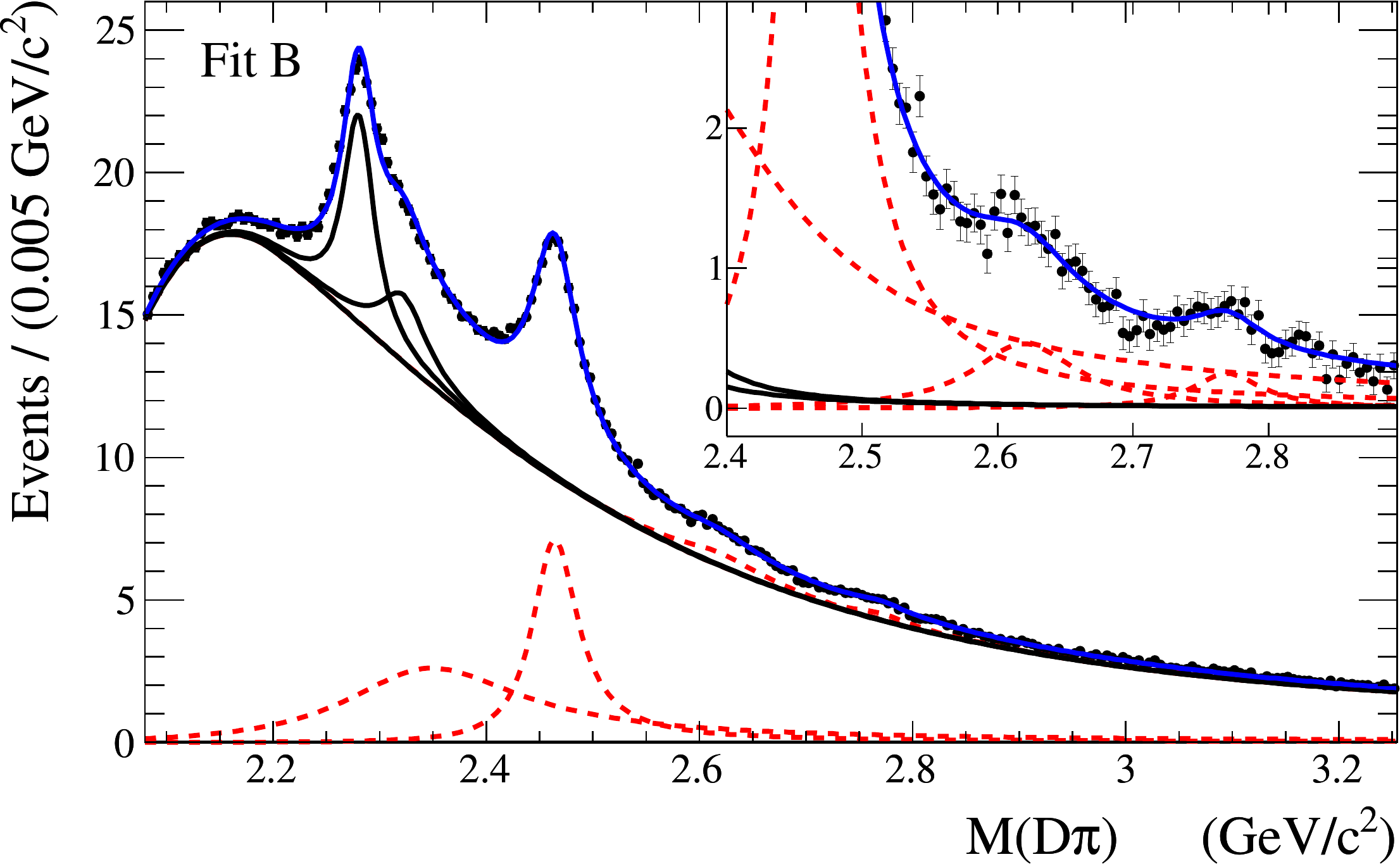}
\includegraphics*[width=0.48\textwidth]{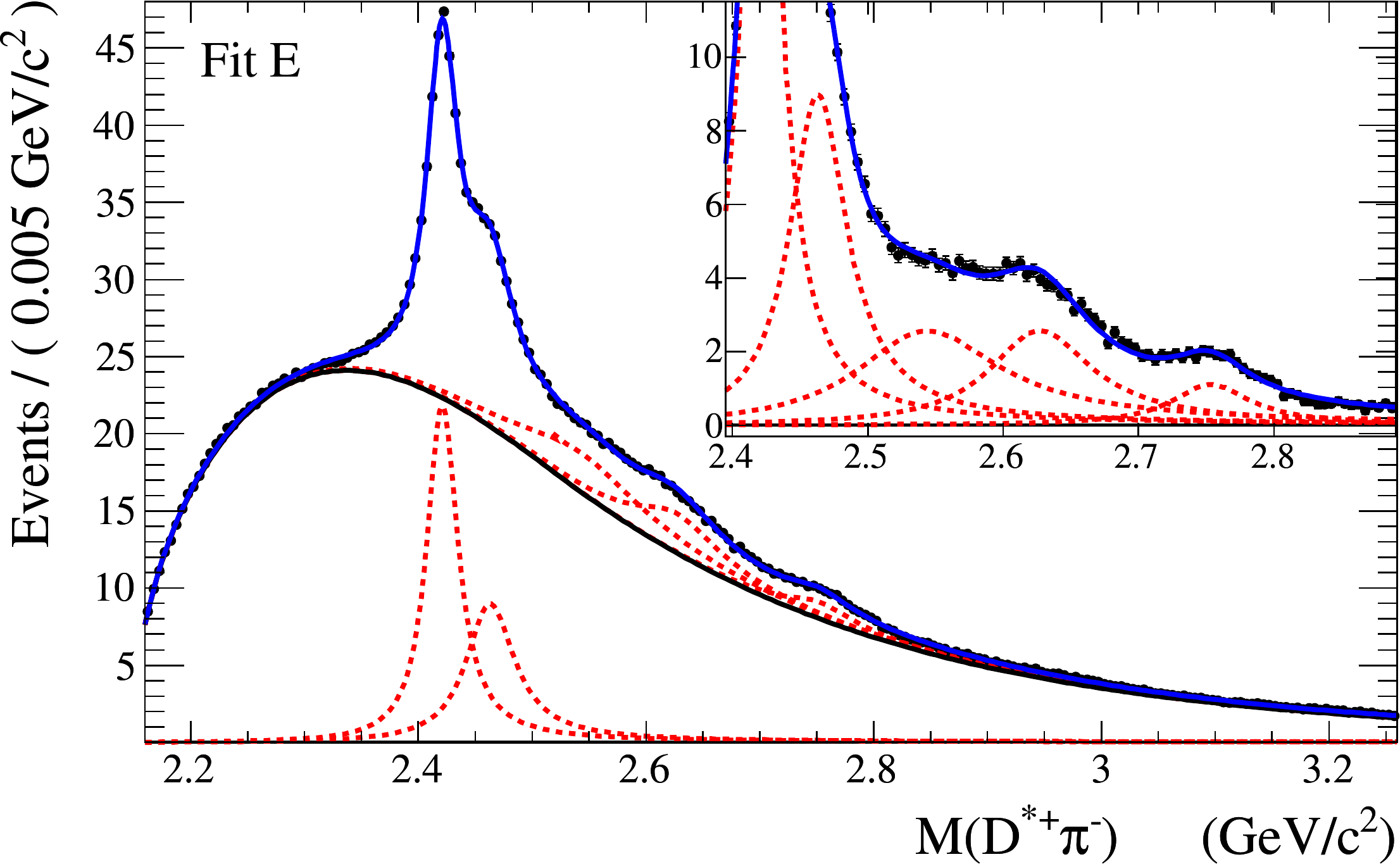}
\end{center}
\caption{(Color online) Mass distributions for the $D^0 \pi^+$ (left)
and $D^{*+} \pi^-$ (right).
Black points are experimental data, and the red dotted curves are the
signal components: in the left panel the signals above 2.4 GeV are
due to the $D_2^*(2460)^+$, $D^*(2600)^+$, and $D^*(2760)^+$; in the
right panel the signals above 2.4 GeV are due to the $D_1(2420)^0$,
$D_2^*(2460)^0$, $D(2550)^0$, $D^*(2600)^0$, and $D(2750)^0$.
Taken from BaBar~\cite{delAmoSanchez:2010vq}. }
\label{sec211:Babar}
\end{figure}

\begin{figure}[htb]
\begin{center}
\includegraphics*[width=0.48\textwidth]{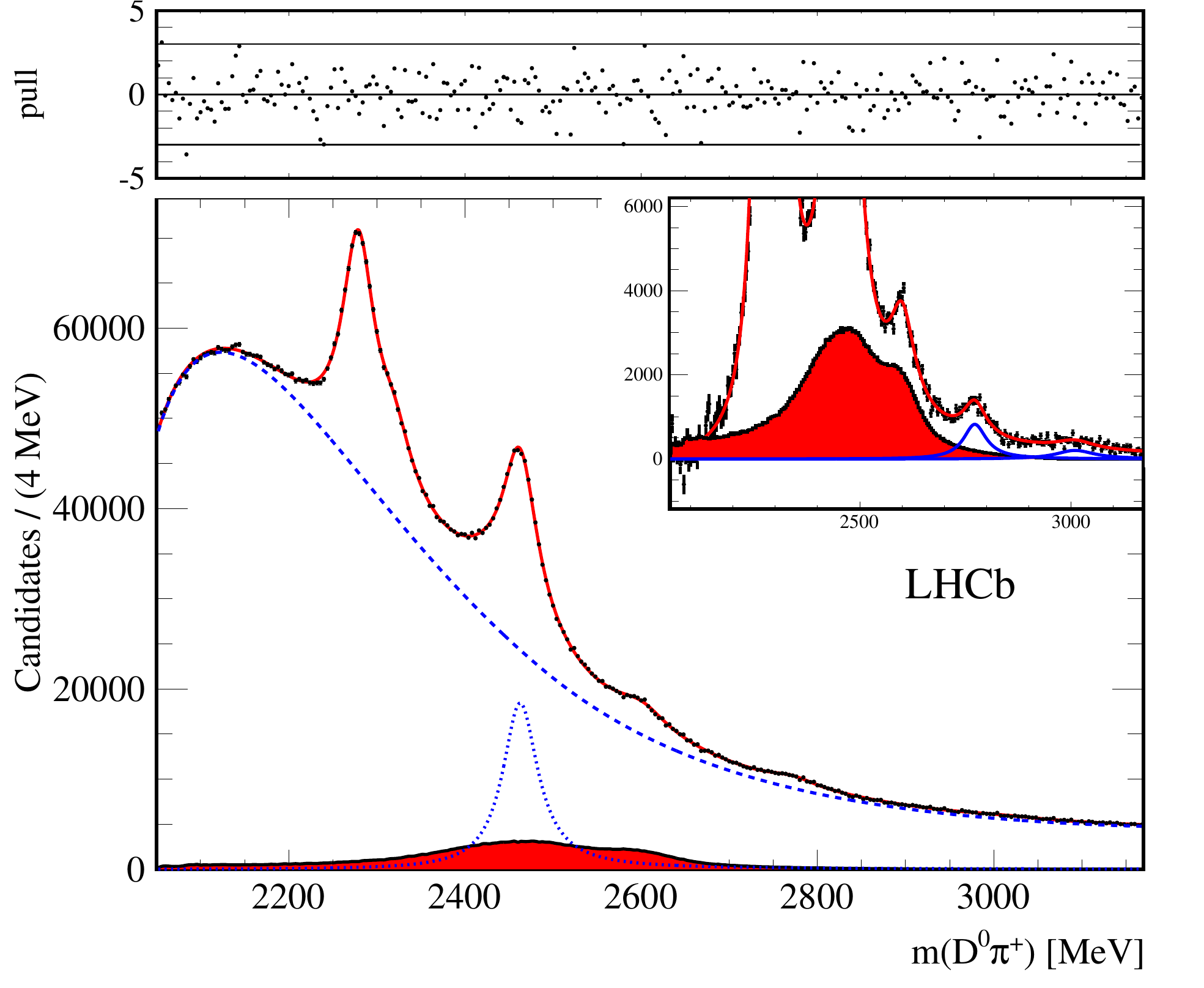}
\includegraphics*[width=0.48\textwidth]{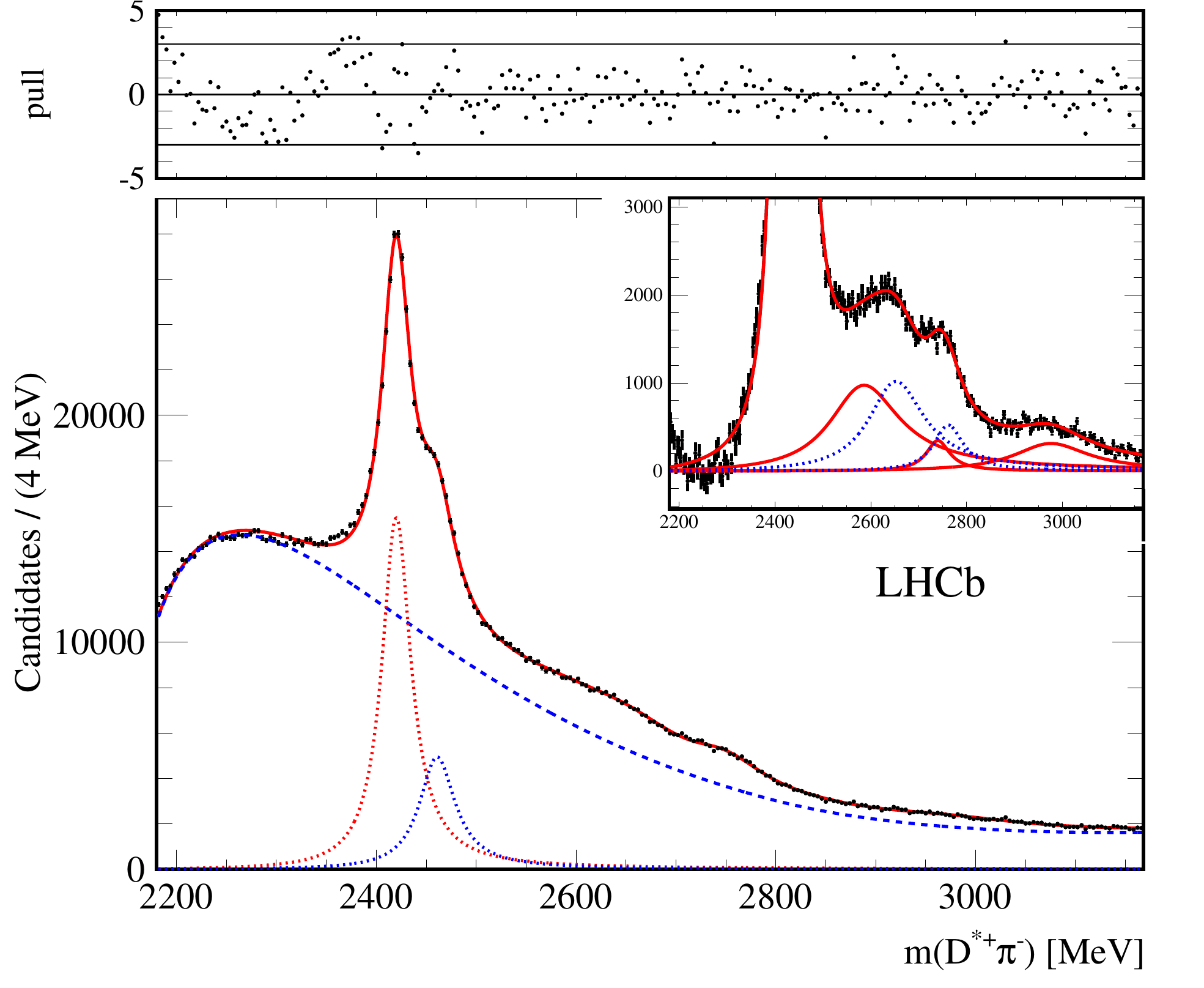}
\end{center}
\caption{(Color online) Mass distributions for the $D^0 \pi^+$ (left) and $D^{*+} \pi^-$ (right).
Black points are experimental data, and the red/blue dotted/full curves are the
signal components.
In the left panel: the dotted blue line is due to the $D_2^*(2460)^+$;
in the inset the full blue curves are due to the $D^*_J(2760)^+$ and $D^*_J(3000)^+$.
In the right panel: the dotted lines are due to the $D_1(2420)^0$ (red) and $D^*_2(2460)^0$ (blue);
in the inset the full red lines are due to the $D_J(2580)^0$, $D_J(2740)^0$, and $D_J(3000)^0$,
and the dotted blue lines are due to the $D^*_J(2650)^0$ and $D^*_J(2760)^0$.
Taken from LHCb~\cite{Aaij:2013sza}.
}
\label{sec211:LHCb}
\end{figure}

In 2010, the $D(2550)$ was observed by the BaBar Collaboration in the
$D^{*} \pi$ mass distribution in the inclusive $e^+ e^- \rightarrow c
\bar c$ interactions, as shown in the right panel of
Fig.~\ref{sec211:Babar}. Its mass and width were measured to be
$M=2539.4\pm4.5\pm6.8$ MeV and $\Gamma=130\pm12\pm13$ MeV,
respectively~\cite{delAmoSanchez:2010vq}. The $D(2550)$ is suggested
to be a candidate for $D(2^1S_0)$ by the helicity distribution
analysis~\cite{delAmoSanchez:2010vq}.

In 2013, an unnatural parity state $D_J(2580)$ was found in the $D^*
\pi$ invariant mass spectrum by the LHCb Collaboration through the
process $pp\rightarrow D \pi X$, as shown in the right panel of
Fig.~\ref{sec211:LHCb}. Its mass and width were measured to be
$M=2579.5\pm3.4\pm5.5$ MeV and $\Gamma=177.5\pm17.8\pm46.0$ MeV,
respectively~\cite{Aaij:2013sza}.

Since the resonance parameters of the $D_J(2580)$ are similar to
those of the $D(2550)$, they can be regarded as the same state.
Moreover, the LHCb results~\cite{Aaij:2013sza} are consistent with
the BaBar assignment~\cite{delAmoSanchez:2010vq} that it is a
$D(2^1S_0)$ state.

\subsubsection{$D^*(2600)$, $D^*_J(2650)$, and $D_1^*(2680)^0$.}

The BaBar Collaboration reported another resonance $D^*(2600)$ in
the $D^{(*)} \pi$ mass distribution~\cite{delAmoSanchez:2010vq}, as
shown in both the left and right panels of Fig.~\ref{sec211:Babar}.
This state has the mass $M=2608.7\pm2.4\pm2.5$ MeV and width
$\Gamma=93\pm6\pm13$ MeV and was regarded as a radial excitation of
the $D^*$ by the helicity distribution analysis. They also measured
the ratio \cite{delAmoSanchez:2010vq}
\begin{eqnarray}
\frac{\mathcal{B}(D^{*0}(2600)\rightarrow
D^+\pi^-)}{\mathcal{B}(D^{*0}(2600)\rightarrow
D^{*+}\pi^-)}=0.32\pm0.02\pm0.09 \, . \label{sec2:26000}
\end{eqnarray}
Later in the LHCb experiment~\cite{Aaij:2013sza}, a natural parity
state $D^*_J(2650)$ was found in the $D^* \pi$ invariant mass
spectrum, as shown in the right panel of Fig.~\ref{sec211:LHCb}.
Its mass and width were measured to be $2649.2\pm3.5\pm3.5$ MeV and
$140.2\pm17.1\pm18.6$ MeV, respectively. It is also tentatively
identified as a $J^P=1^-$ state, the radial excitation of the $D^*$.

Since the resonance parameters of the $D^*(2600)$ and $D^*_J(2650)$
are similar to each other, they are probably the same state. We use
$D_1^*(2600)$ to denote them together, as listed in
Table~\ref{sec21:Dmeson}.

Note: in a recent experiment reported by the LHCb
Collaboration~\cite{Aaij:2016fma}, a similar structure
$D_1^*(2680)^0$ was observed, which has parameters close to those
measured for the $D^*_J(2650)$:
\begin{eqnarray}
M_{D_1^*(2680)^0}&=& 2681.1 \pm 5.6 \pm 4.9 \pm 13.1 \,\mathrm{MeV}
\, ,
\\ \nonumber
\Gamma_{D_1^*(2680)^0}&=& 186.7 \pm 8.5 \pm 8.6 \pm 8.2
\,\mathrm{MeV} \, .
\end{eqnarray}

\subsubsection{$D(2750)$, $D_J(2740)$, $D^*(2760)$, $D^*_J(2760)$, $D^*_1(2760)^0$, and $D_3^*(2760)^-$.}

\begin{figure}[htb]
\begin{center}
\includegraphics*[width=0.48\textwidth]{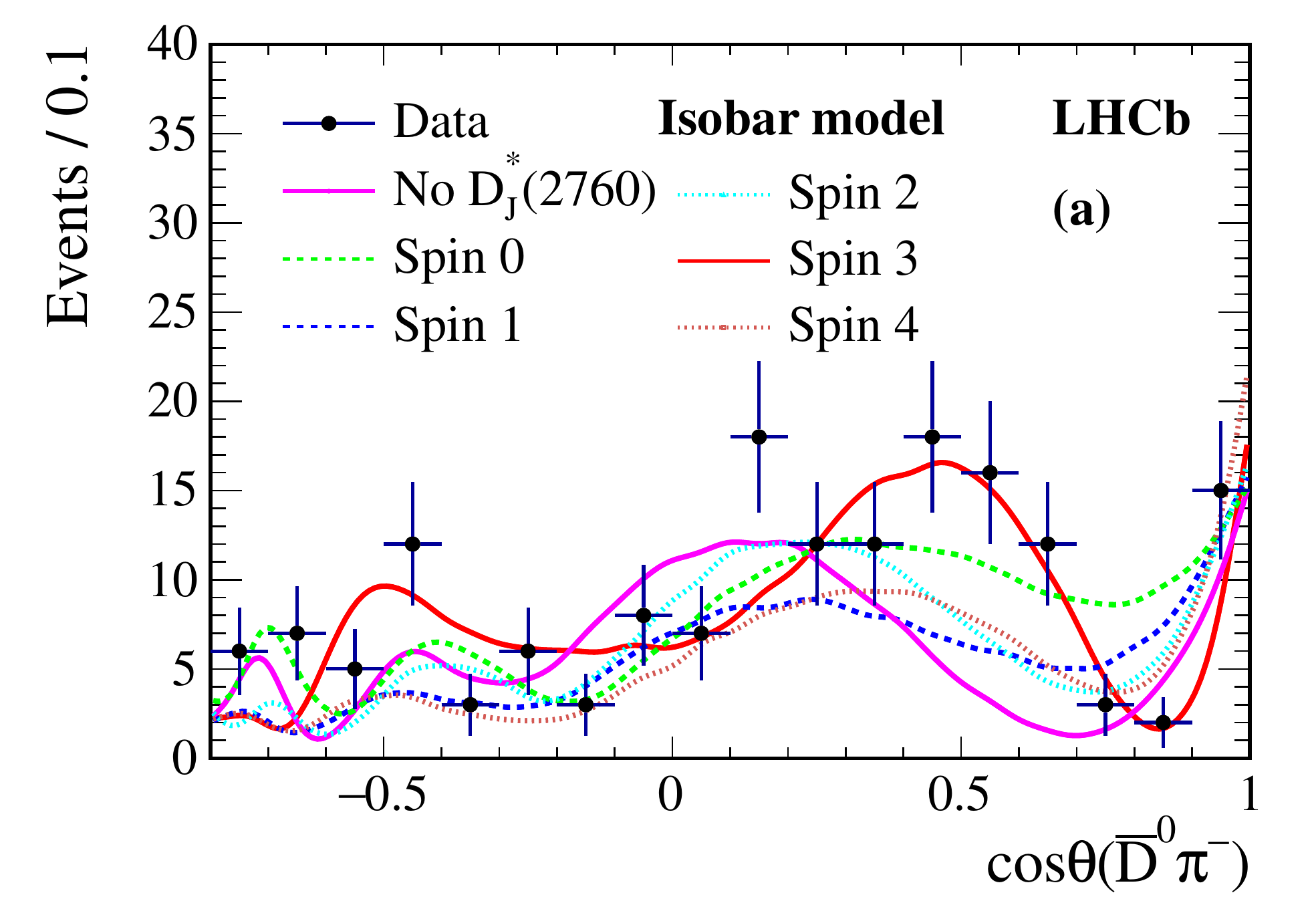}
\includegraphics*[width=0.48\textwidth]{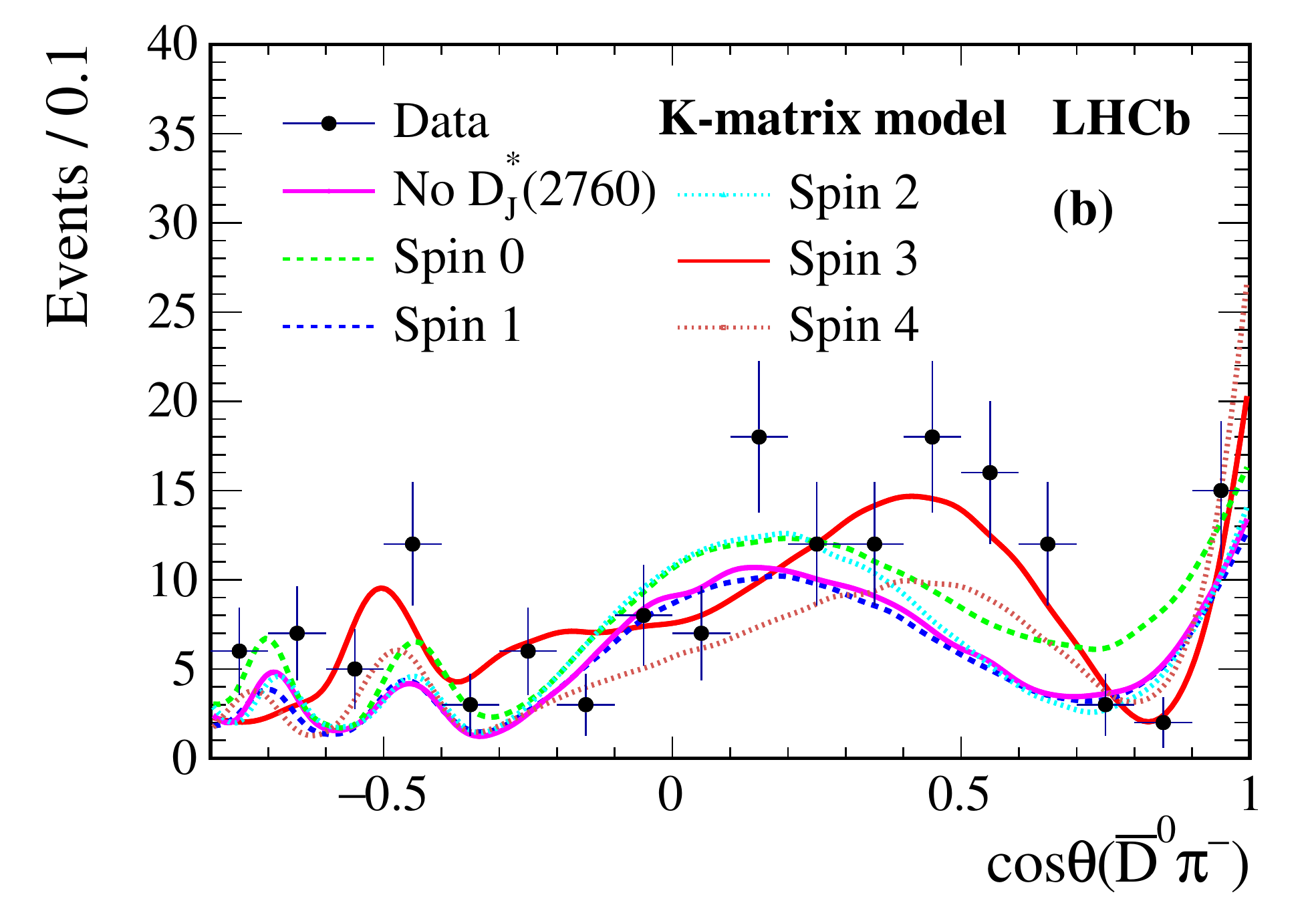}
\end{center}
\caption{(Color online) Cosine of the helicity angle distributions
in the range $7.4$~GeV$^2<m^2( \bar D^0 \pi^-) < 8.2$~GeV$^2$ for the
Isobar model (left) and the $K$-matrix model (right).
Black points are experimental data, which are fitted with different spin hypotheses of the $D^*_J(2760)^-$ as detailed in the legend.
Taken from LHCb~\cite{Aaij:2015sqa}. }
\label{sec211:LHCb2}
\end{figure}

The charmed mesons in the energy region around 2.75 GeV are slightly
confusing. There have been five measurements performed by the BaBar
and LHCb
Collaborations~\cite{delAmoSanchez:2010vq,Aaij:2013sza,Aaij:2015vea,Aaij:2015sqa}
and we review them in the following.

The $D^*(2760)$ was first observed by the BaBar Collaboration in the
$D \pi$ invariant mass spectrum~\cite{delAmoSanchez:2010vq}, as
shown in the left panel of Fig.~\ref{sec211:Babar}. It has the mass
and width $2763.3\pm2.3\pm2.3$ MeV and $60.9\pm5.1\pm3.6$ MeV,
respectively. It can be assigned as a $D$-wave charmed meson since
its mass is consistent with the theoretical prediction of the GI
model \cite{Godfrey:1985xj}. Later, LHCb announced the observation
of a natural parity state $D^*_J(2760)$ also in the $D \pi$
invariant mass spectrum~\cite{Aaij:2013sza}, as shown in the left
panel of Fig.~\ref{sec211:LHCb}. This state has the mass $2760.1 \pm
1.1 \pm 3.7$ MeV and width $74.4 \pm 3.4 \pm 19.1$ MeV. The
$D^*(2760)$ and $D^*_J(2760)$ can be regarded as the same state
since they have similar masses and widths and were observed in the
same decay modes.

Besides the $D^*(2760)$, another state $D(2750)$ was also observed
by the BaBar Collaboration but in the $D^* \pi$ mass spectrum, as
shown in the right panel of Fig.~\ref{sec211:Babar}, where its mass
and width were measured to be $M=2752.4\pm1.7\pm2.7$ MeV and
$\Gamma=71\pm6\pm11$ MeV, respectively \cite{delAmoSanchez:2010vq}.
Although the $D(2750)$ can be a good candidate of a $D$-wave charmed
meson according to the mass spectrum analysis of the GI
model~\cite{Godfrey:1985xj}, the helicity distribution analysis of
the $D(2750)$ didn't support the $D(1^3D_1)$ and $D(1^3D_3)$
assignments \cite{delAmoSanchez:2010vq}. BaBar also gave the ratio
\cite{delAmoSanchez:2010vq}
\begin{eqnarray}
\frac{\mathcal{B}(D^*(2760)^0\rightarrow
D^+\pi^-)}{\mathcal{B}(D(2750)^0\rightarrow
D^{*+}\pi^-)}=0.42\pm0.05\pm0.11.
\end{eqnarray}
Another similar unnatural parity state, the $D_J(2740)$, was found
by the LHCb Collaboration in the $D^* \pi$ mass spectrum, as shown
in the right panel of Fig.~\ref{sec211:LHCb}. It has the mass
$M=2737.0\pm3.5\pm11.2$ MeV, width $73.2\pm13.4\pm25.0$ MeV, and
spin-parity quantum numbers $J^P=2^-$~\cite{Aaij:2013sza}. Due to the
similarity between the $D(2750)$ and the $D_J(2740)$, they are
possibly the same state.

Recently in 2015, a spin-1 state, $D^*_1(2760)^0$, was observed in
the channel of $B^- \rightarrow D^{*0}_1 \pi^+ \rightarrow D^+ \pi^-
\pi^+$ by the LHCb Collaboration~\cite{Aaij:2015vea}. It is a
possible charmed meson with $J^P = 1^-$. Then the $D_3^*(2760)^-$
was reported in the similar process of $B^0 \rightarrow D^{*-}_3
\pi^+ \rightarrow \bar D^0 \pi^- \pi^+$ also by the LHCb
Collaboration~\cite{Aaij:2015sqa}. Its spin-parity were determined to
be $J^P = 3^-$. We show the LHCb results related to the
$D_3^*(2760)^-$ in Fig.~\ref{sec211:LHCb2}. Its neutral partner
$D_3^*(2760)^0$ was confirmed in a recent LHCb
experiment~\cite{Aaij:2016fma}. These new experimental results
provide interesting information for the D-wave charmed mesons in the
energy region around 2.75 GeV.

\subsubsection{$D_J(3000)$, $D^*_J(3000)$, and $D^*_2(3000)^0$.}

The LHCb Collaboration observed the unnatural parity state
$D_J(3000)$ in the $D^* \pi$ invariant mass
spectrum~\cite{Aaij:2013sza}, as shown in the right panel of
Fig.~\ref{sec211:LHCb}. Its resonance parameters are
\begin{eqnarray}
M=2971.8\pm8.7 \,\mathrm{MeV},\quad \Gamma=188.1\pm44.8
\,\mathrm{MeV}.
\end{eqnarray}
Another natural parity state $D^*_J(3000)$ was also reported by LHCb
but in the $D \pi$ invariant mass spectrum~\cite{Aaij:2013sza}, as
shown in the left panel of Fig.~\ref{sec211:LHCb}, which has
\begin{eqnarray}
M=3008.1\pm4.0 \,\mathrm{MeV},\quad \Gamma=110.5\pm11.5
\,\mathrm{MeV} .
\end{eqnarray}
Recently, the LHCb experiment observed another structure
$D^*_2(3000)^0$ in this energy region~\cite{Aaij:2016fma}. Its mass
and decay width were measured to be:
\begin{eqnarray}
M_{D^*_2(3000)^0}&=& 3214 \pm 29 \pm 33 \pm 36 \,\mathrm{MeV} \, ,
\\ \nonumber
\Gamma_{D^*_2(3000)^0}&=&  186 \pm 38 \pm 34 \pm 63 \,\mathrm{MeV}
\, .
\end{eqnarray}

There are many possible interpretations for the $D_J(3000)$ and
$D^*_J(3000)$. See Sec.~\ref{sec3.1} for more theoretical
discussions.

\subsection{The charmed-strange mesons}
\label{sec2.2}

\renewcommand{\arraystretch}{1.4}
\begin{table*}[htb]
\scriptsize
\caption{Experimental information of the observed charmed-strange
mesons. the $1S$ charmed-strange states ($D_s$, $D_s^*$) and the
$1P$ ones ($D_{s0}^*(2317)$, $D_{s1}(2460)$, $D_{s1}(2536)$, and
$D_{s2}^*(2573)$) are well established, so we only list their
averaged masses and widths from PDG~\cite{Olive:2016xmw} together
with the experiments which first observed them. However, the higher
states starting from the $D_{sJ}^\ast(2632)$ are not well
established, so we list all the relevant experiments together with
their observed masses, widths, and decay modes therein.
\label{sec22:Dsmeson} } \centering
\begin{tabular}{cccccc}
\toprule[1pt] State & $J^P$ & Mass (MeV) & Width (MeV) & Experiments
& Observed Modes
\\ \midrule[1pt]
$D_s$ & $0^-$ & $1968.27 \pm 0.10$ & $(500 \pm 7) \times 10^{-15}$ s
& DASP~\cite{Brandelik:1977fg} & $\eta \pi^\pm$
\\
$D_s^{\ast}$ & $1^-$ & $2112.1 \pm 0.4$& $<1.9$ &
DASP~\cite{Brandelik:1977fg} & $D_s \gamma$
\\ \hline
$D_{s0}^\ast(2317)$ & $0^+$ & $2317.7\pm0.6$ & $<3.8$  &
BaBar~\cite{Aubert:2003fg} & $D_s^+ \pi^0$
\\
$D_{s1}(2460)$ & $1^+$ & $2459.5\pm0.6$ & $<3.5$  &
CLEO~\cite{Besson:2003cp} & $D_s^{*+} \pi^0$
\\
$D_{s1}(2536)$ & $1^+$ & $2535.10 \pm 0.06$  & $0.92\pm0.03\pm0.04$  &
ITEP\&SERP~\cite{Asratian:1987rb} & $D_s^{*+} \gamma$
\\
$D_{s2}^\ast(2573)$ & $2^+$ & $2569.1 \pm 0.8$  & $16.9 \pm 0.8$  &
CLEO~\cite{Kubota:1994gn}&$D^0 K^+$
\\ \midrule[1pt]
$D_{sJ}^\ast(2632)$ & $?^?$ &$2632.5\pm1.7\pm5.0 $ & $<17$ &
SELEX~\cite{Evdokimov:2004iy} & $D K$
\\ \hline
\multirow{5}{*}{$D_{s1}^\ast(2700)$} & \multirow{5}{*}{$1^-$} &
$2688\pm4\pm3$  & $112\pm7\pm36$ & BaBar~\cite{Aubert:2006mh} &$D K$
\\
                                                             && $ 2708\pm9^{+11}_{-10}$ & $ 108\pm23^{+36}_{-31}$ & Belle~\cite{Brodzicka:2007aa} & $D K$
\\
                                                             && $ 2710\pm2^{+12}_{-7}$ & $ 149\pm7^{+39}_{-52}$ & BaBar~\cite{Aubert:2009ah} & $D^{(*)} K$
\\
                                                             && $ 2709.2\pm1.9\pm4.5$ & $ 115.8\pm7.3\pm12.1$ & LHCb~\cite{Aaij:2012pc} & $D K$
\\
                                                             && $2699^{+14}_{-7}$     & $127^{+24}_{-19}$ & BaBar~\cite{Lees:2014abp} & $D K$
\\ \hline
\multirow{3}{*}{$D_{sJ}^*(2860)$} & \multirow{3}{*}{$?^?$} &
$2856.6\pm1.5\pm5.0$ & $47\pm7\pm10$ & BaBar~\cite{Aubert:2006mh} &
$D K$
\\
                                                           && $2862\pm2^{+5}_{-2}$ &$48\pm3\pm6$ & BaBar~\cite{Aubert:2009ah} & $D^{(*)} K$
\\
                                                           && $2866.1\pm1.0\pm6.3$ & $69.9\pm3.2\pm6.6$ & LHCb~\cite{Aaij:2012pc} & $D K$
\\ \hdashline[2pt/2pt]
$D_{s3}^\ast(2860)$ & $3^-$ & $2860.5\pm2.6\pm2.5\pm6.0$ &
$53\pm7\pm4\pm6$ & LHCb~\cite{Aaij:2014xza,Aaij:2014baa} &
$\bar{D}^0K^-$
\\
$D_{s1}^\ast(2860)$ & $1^-$ & $2859\pm12\pm6\pm23$ &
$159\pm23\pm27\pm72$ & LHCb~\cite{Aaij:2014xza,Aaij:2014baa} &
$\bar{D}^0K^-$
\\ \hline
$D_{sJ}(3040)$ & $?^?$ & $3044\pm8^{+30}_{-5}$ &
$239\pm35^{+46}_{-42}$ & BaBar~\cite{Aubert:2009ah} & $D^*K$
\\ \bottomrule[1pt]
\end{tabular}
\end{table*}

In this subsection we review the charmed-strange mesons. Their
experimental information is listed in Table~\ref{sec22:Dsmeson}.
Similar to the charmed mesons, the $1S$ charmed-strange states
($D_s$, $D_s^*$) and the $1P$ ones ($D_{s0}^*(2317)$,
$D_{s1}(2460)$, $D_{s1}(2536)$, and $D_{s2}^*(2573)$) are well
established, completing one $S$-wave doublet $(0^-, 1^-)$ and two
$P$-wave doublets $(0^+, 1^+)$ and $(1^+, 2^+)$. Hence, we only list
in Table~\ref{sec22:Dsmeson} their averaged masses and widths from
PDG~\cite{Olive:2016xmw} together with the experiments which first
observed them. The higher states starting from the $D^*_{sJ}(2632)$
are not well established, and we list all the relevant experiments.

We note that the observed masses of the $D^*_{s0}(2317)$ and
$D_{s1}(2460)$ are far lower than the corresponding results
calculated using the GI model~\cite{Godfrey:1986wj}. There are many
different perspectives on their nature, which we shall review in
Sec.~\ref{sec4}.

\subsubsection{$D_s$ and $D_s^*$.}

The lowest-lying charmed-strange mesons, $D_s^\pm$ of $J^P = 0^-$,
was observed in 1977 by the DASP
Collaboration~\cite{Brandelik:1977fg}. The lowest-lying vector
charmed-strange mesons, $D_s^{*\pm}$ of $J^P = 1^-$, was observed in
the same experiment~\cite{Brandelik:1977fg}. Their properties are
known very well~\cite{Olive:2016xmw}: the $D_s^\pm$ has a mass
$1968.27 \pm 0.10$ MeV and a mean life $(500 \pm 7) \times 10^{-15}$
s; the $D_s^{*\pm}$ has a mass $2112.1 \pm 0.4$ MeV and the upper
limit of its width is 1.9 MeV; hundreds of decay modes of the
$D_s^\pm$ have been observed in experiments, while the $D_s^{*\pm}$
mainly decays into $D_s^\pm \gamma$ and $D_s^\pm \pi^0$, with
fractions $(93.5 \pm 0.7) \%$ and $(5.8\pm0.7) \%$, respectively.

\subsubsection{$D_{s1}(2536)$ and $D_{s2}^\ast(2573)$.}

Before 2003, there are only two good candidates for the $1P$
charmed-strange mesons, the $D_{s1}(2536)$ of $J^P=1^+$ and the
$D_{s2}^\ast(2573)$ of $J^P=2^+$.

In 1987, the $D_{s1}(2536)$ was first observed by analyzing the
$D_s^\ast \gamma$ invariant mass spectrum in the $\bar{\nu} N$
scattering process~\cite{Asratian:1987rb}. There, its measured mass
is $2535 \pm 28$ MeV. Later in 1989, it was observed in the $D^{\ast
+} K^0$ mass spectrum by the ARGUS
Collaboration~\cite{Albrecht:1989yi}, where its mass and width were
measured to be $M=2536 \pm 0.6 \pm 2.0$ MeV and $\Gamma<4.6$ MeV,
respectively. In 1993, the CLEO Collaboration measured the following
ratio~\cite{Alexander:1993nq}
\begin{eqnarray}
\frac{\Gamma(D_{s1}(2536) \to D_s^\ast \gamma)}{ \Gamma(D_{s1}(2536)
\to D^\ast K)} < 0.42.
\end{eqnarray}
The $D_{s1}(2536)$ has been confirmed by many other groups in
several different
channels~\cite{Avery:1989ui,Frabetti:1993vv,Chekanov:2008ac,Asratian:1993zx,Heister:2001nj,Aubert:2007rva,Abazov:2007wg,Lees:2011um,Belle:2011ad,Aaij:2012mra}.
Its mass value and narrow width are consistent with the theoretical
expectation that it is a charmed-strange meson of $J^P=1^+$ in the
$(1^+, 2^+)$ doublet \cite{Godfrey:1986wj}.

In 1994, the $D_{s2}^\ast(2573)$ was first observed by the CLEO
Collaboration in the $D^0K^+$ invariant mass spectrum
\cite{Kubota:1994gn}, where its mass and width were measured to be
$M=2573^{+1.7}_{-1.6} \pm 0.8 \pm 0.5$ MeV and $\Gamma =16^{+5}_{-4}
\pm3$ MeV, respectively. In addition, the following upper limit was
given
\begin{eqnarray}
\frac{\mathcal{B}(D_{s2}^\ast(2573)^{+} \to D^{\ast 0}
K^+)}{\mathcal{B}(D_{s2}^\ast(2573)^{+} \to D^{ 0} K^+) } <0.33,
\end{eqnarray}
A similar branching ratio was recently measured by the LHCb
Collaboration to be~\cite{Aaij:2016utb}
\begin{eqnarray}
\frac{\mathcal{B}(D_{s2}^\ast(2573)^{+} \to D^{\ast +}
K_S^0)}{\mathcal{B}(D_{s2}^\ast(2573)^{+} \to D^{+} K_S^0) } = 0.044
\pm 0.005 \pm 0.011 \, .
\end{eqnarray}
The $D_{s2}^\ast(2573)$ has also been confirmed by many other
experiments~\cite{Evdokimov:2004iy,Aubert:2006mh,Heister:2001nj,Aaij:2011ju,Albrecht:1995qx}.

\subsubsection{$D_{s0}^\ast(2317)$ and $D_{s1}(2460)$.}

\begin{figure}[htb]
\begin{center}
\includegraphics*[width=0.48\textwidth]{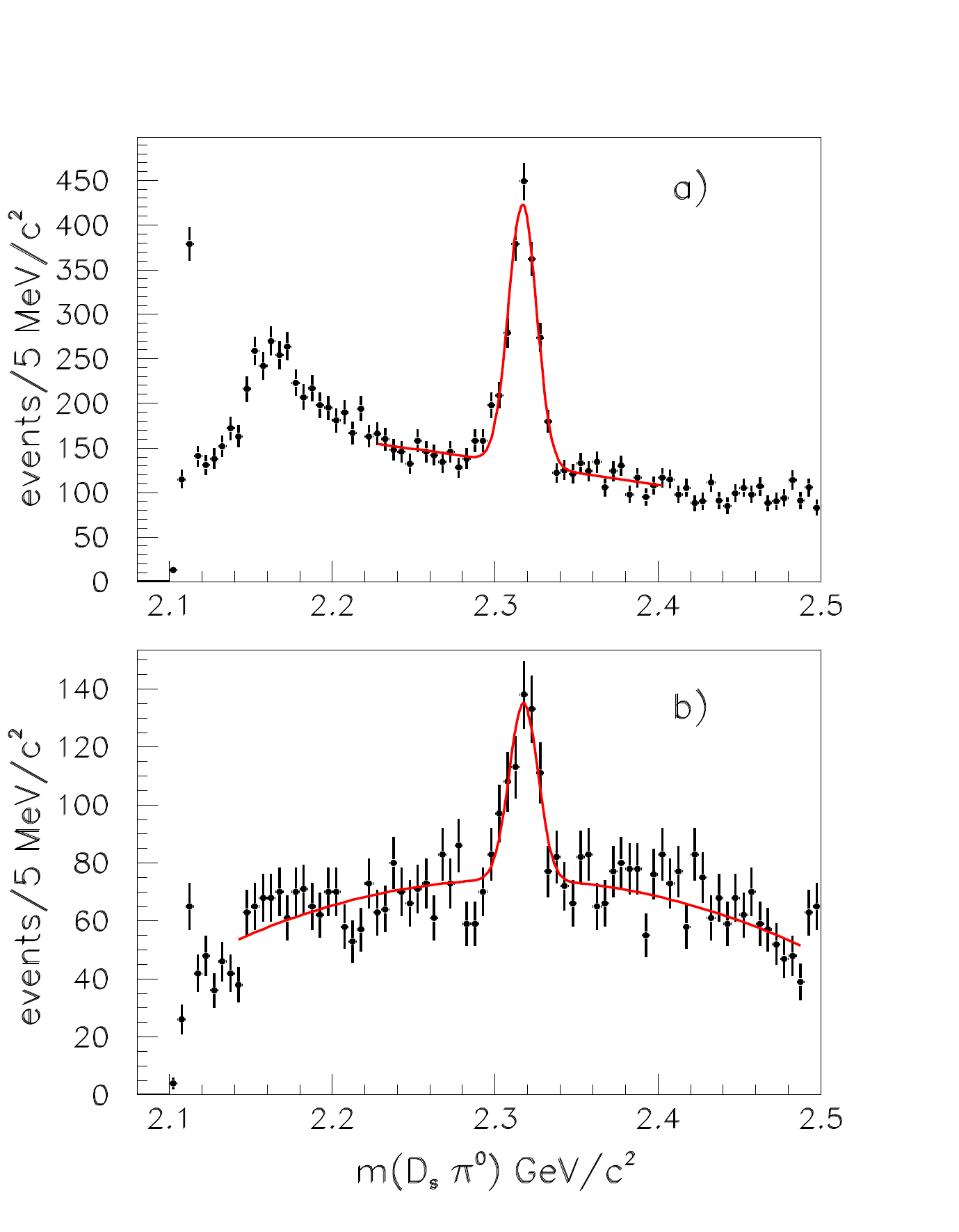}
\end{center}
\caption{(Color online) The $D^+_s \pi^0$ mass distribution for (a)
the $D^+_s$ decay into $K^+ K^- \pi^+$, and (b) the $D^+_s$ decay into $K^+ K^- \pi^+ \pi^0$.
The signals correspond to the $D_{s0}^*(2317)$.
Taken from BaBar~\cite{Aubert:2003fg}. } \label{sec212:Babar1}
\end{figure}

In 2003, a new charmed-strange meson $D_{s0}^*(2317)$ was observed by
the BaBar Collaboration in the $D_s^+ \pi^0$ invariant mass
distribution in the $B$ decay process~\cite{Aubert:2003fg}, as shown
in Fig.~\ref{sec212:Babar1}. Its mass is about 2.32 GeV.
Later in confirming this state, the CLEO Collaboration observed another narrow charmed-strange state $D_{s1}(2460)$~\cite{Besson:2003cp}.
The observed
masses of these two states are far lower than the corresponding
predictions from the GI model~\cite{Godfrey:1986wj}. These two
puzzling states quickly become the two superstars in the heavy meson
family. We shall review the relevant theoretical studies in
Sec.~\ref{sec4.1} carefully.

The $D_{s0}^*(2317)$ was also confirmed in the Belle and BaBar
experiments~\cite{Abe:2003jk,Aubert:2006bk,Aubert:2003pe}. In
addition, the following ratio was measured in the CLEO
experiment~\cite{Besson:2003cp}
\begin{eqnarray}
\frac{\mathcal{B}(D_{s0}^\ast(2317)^+ \to D_s^{\ast +}
\gamma)}{\mathcal{B}(D_{s0}^\ast(2317)^+ \to D_s^{ +} \pi^0)} <0.059
\, .
\end{eqnarray}
This upper bound was later measured by the Belle and BaBar
Collaborations to be 0.18~\cite{Abe:2003jk} and
0.16~\cite{Aubert:2006bk}, respectively.

\begin{figure}[htb]
\begin{center}
\includegraphics*[width=0.48\textwidth]{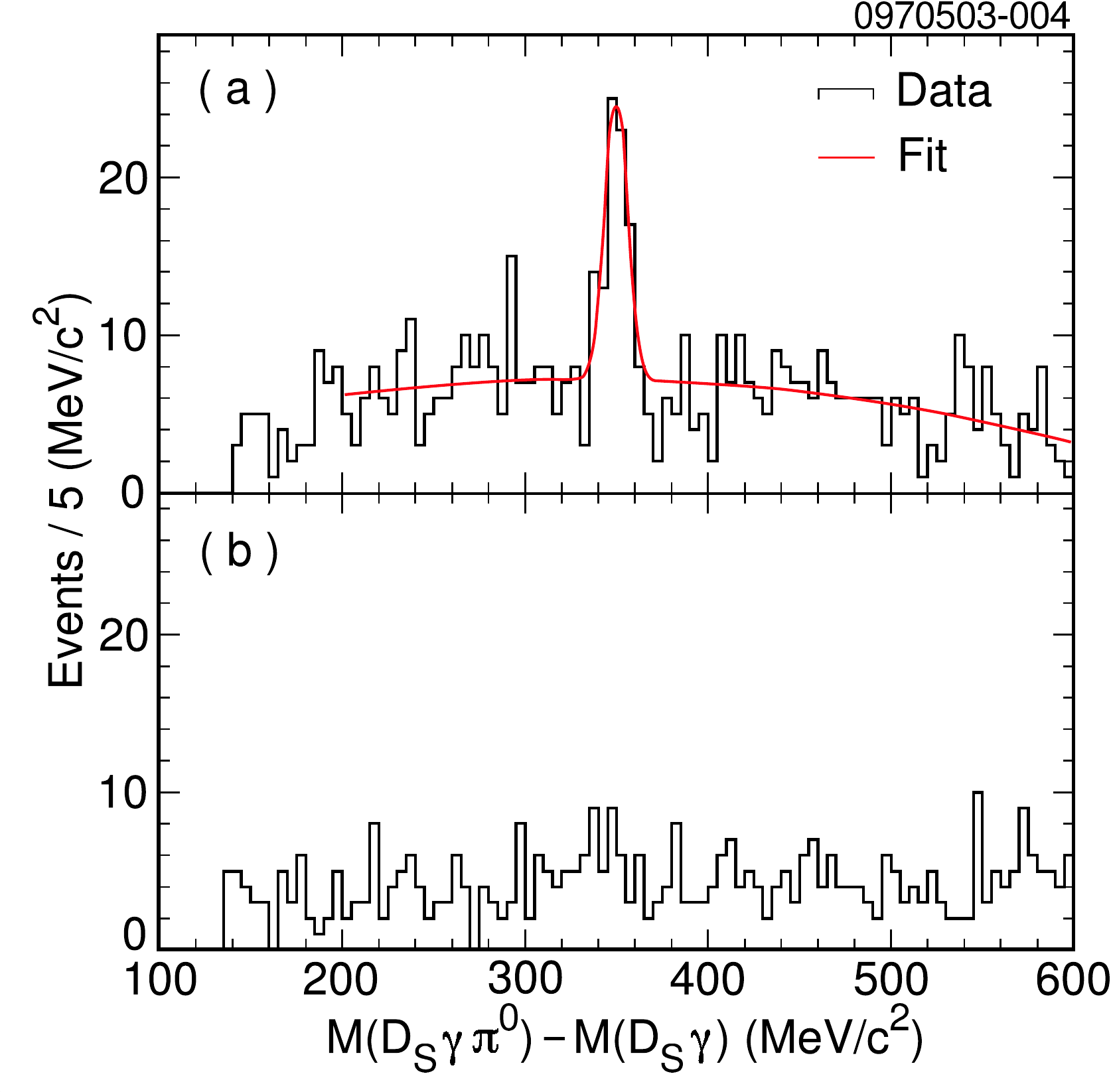}
\end{center}
\caption{(Color online) The mass difference distribution $\Delta
M(D^*_s \pi^0) = M(D_s \gamma \pi^0) - M(D_s \gamma)$, where (a) the
$D_s \gamma$ system is consistent with $D^*_s$ decay, and (b) the $D_s \gamma$ system further satisfies
$20.8~{\rm MeV} < |\Delta M(D_s \gamma) - 143.9~{\rm MeV} | <
33.8~{\rm MeV}$ with $\Delta M(D_s \gamma) = M(D_s \gamma) -
M(D_s)$.
The signal corresponds to the $D_{s1}(2460)$.
Taken from CLEO~\cite{Besson:2003cp}. } \label{sec212:CLEO}
\end{figure}

The $D_{s1}(2460)$ of $J^P = 1^+$ was first observed by the CLEO
Collaboration~\cite{Besson:2003cp} in the $D_s^{\ast +} \pi^0$
invariant mass spectrum, as shown in Fig.~\ref{sec212:CLEO}. Its
mass splitting with respect to $D_s$ was measured to be $350 \pm 1.2
\pm 1.0$ MeV and its width was given to be $\Gamma<7$ MeV at $90\%$
C.L. Later, the $D_{s1}(2460)$ was confirmed by the Belle and BaBar
experiments~\cite{Abe:2003jk,Aubert:2006bk,Aubert:2003pe}. Its mass,
narrow width, and decay properties are all consistent with those of
the $1^+$ charmed-strange meson. Particularly, the $D_{s1}(2460)$ is
above the $DK$ threshold, but it has a narrow width and does not
decay into $DK$, providing additional evidence that it has the
spin-parity quantum numbers $J^P=1^+$.

\subsubsection{$D_{sJ}^\ast(2632)$.}

In 2004, the SELEX Collaboration reported the observation of a
narrow charmed-strange meson $D_{sJ}^\ast(2632)$ in two decay modes,
$D^+_s \eta$ and $D^0 K^+$~\cite{Evdokimov:2004iy}. Its mass and
width were measured to be $2632.5 \pm 1.7$ MeV and $< 17$ MeV,
respectively. They also measured the relative branching ratio
$\Gamma(D^0 K^+)/\Gamma(D^+_s \eta)= 0.14 \pm 0.06$.

However, the following CLEO, BaBar, and FOCUS experiments (see
Ref.~\cite{Aubert:2004ku} and Refs.~[19,21] of
Ref.~\cite{Evdokimov:2004iy}) all reported negative results in their
search of $D_{sJ} (2632)$. The BaBar experiment searched for the
$D_{sJ}^\ast(2632)$ in the $e^+e^- \to c\bar c$ collisions and they
found no evidence of this state in the inclusive production of
$D^+_s \eta$, $D^0 K^+$, or $D^{+*}K_S$.

\subsubsection{$D_{s1}^\ast(2700)$.}

\begin{figure}[htb]
\begin{center}
\includegraphics*[width=1.0\textwidth]{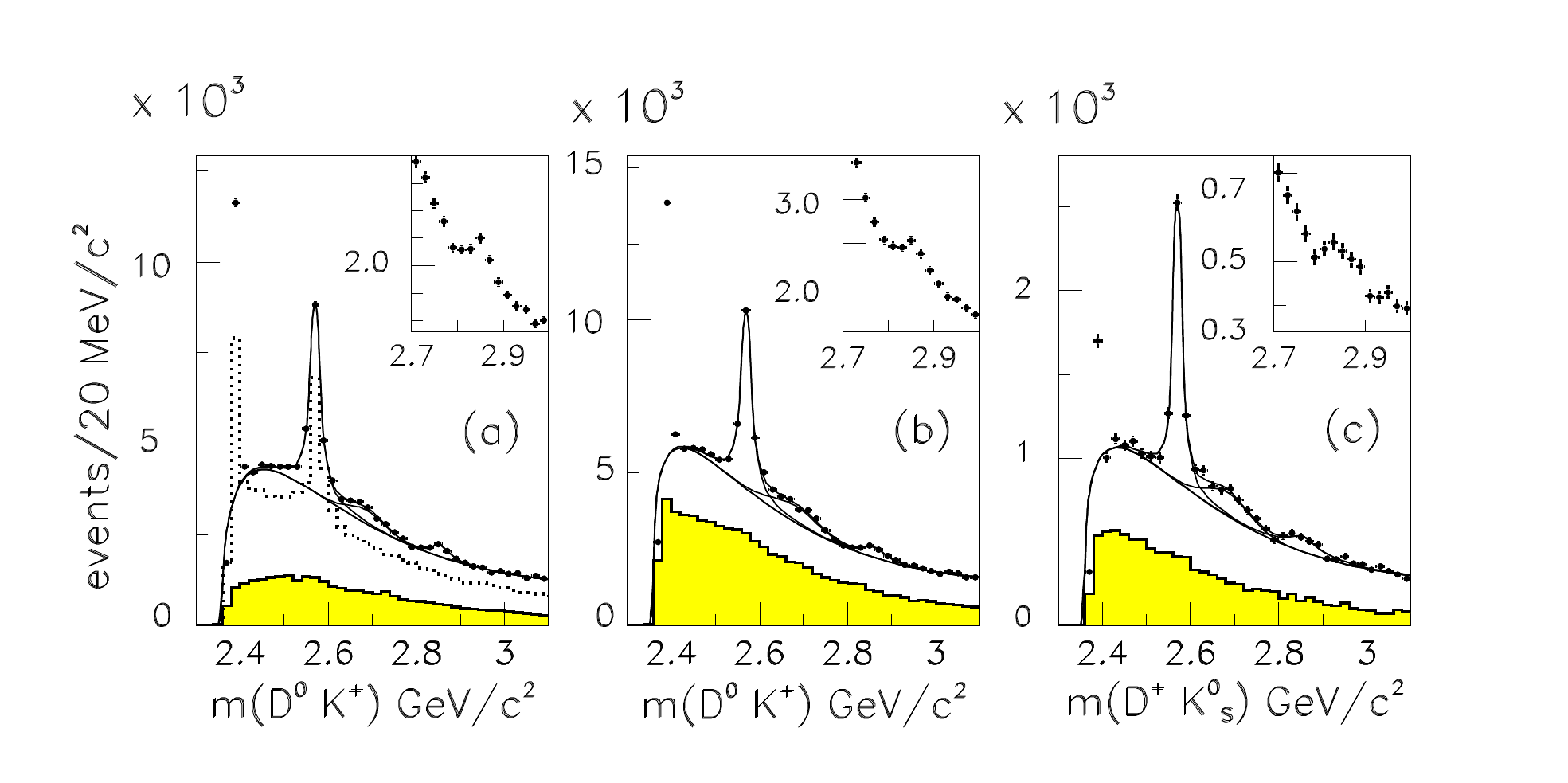}
\end{center}
\caption{(Color online) The $DK$ invariant mass distributions for
the (a) $D^0_{K^- \pi^+} K^+$, (b) $D^0_{K^- \pi^+ \pi^0} K^+$, and (c)
$D^+_{K^- \pi^+ \pi^+} K^0_s$. There are three structures: a
prominent narrow signal is due to the $D_{s2}(2573)^+$, a broad
structure peaking at a mass of approximately 2.7 GeV is identified
as the $D_{s1}^\ast(2700)$, and an enhancement around 2.86 GeV is
identified as the $D_{sJ}(2860)^+$. Taken from
BaBar~\cite{Aubert:2006mh}. } \label{sec212:Babar2}
\end{figure}

In 2006, a broad structure was observed by the BaBar Collaboration
in the $DK$ invariant mass spectrum, which was later named as
$D_{s1}^\ast(2700)$~\cite{Aubert:2006mh}, as shown in
Fig.~\ref{sec212:Babar2}. Its mass and width were measured to be
$M=2688 \pm 4 \pm3$ MeV and $\Gamma=112\pm 7 \pm 36$ MeV,
respectively. This state was confirmed in the following Belle and
LHCb experiments in the same
channel~\cite{Brodzicka:2007aa,Aaij:2012pc}, where its spin and
parity were determined to be $J=1$ and $P=-$ by the helicity angle
distribution and its decay to two pseudoscalar mesons, respectively.

The BaBar Collaboration also reported the $D^\ast K$ decay mode of
the $D_{s1}^*(2700)$ and measured the following
ratio~\cite{Aubert:2009ah}
\begin{eqnarray}
\frac{\mathcal{B}(D_{s1}^\ast(2700) \to D^\ast
K)}{\mathcal{B}(D_{s1}^\ast(2700) \to DK)} = 0.91 \pm 0.13 \pm 0.12
\, .
\label{sec21:ds12700ratio}
\end{eqnarray}
Later in 2014, the BaBar Collaboration further studied the
$D_{s1}^*(2700)$ in the $B^0 \to D^-D^0K^+$ and $B^+ \to \bar
D^0D^0K^+$ decays~\cite{Lees:2014abp} and measured its mass and
width to be $2699^{+14}_{-7}$ MeV and $127^{+24}_{-19}$ MeV,
respectively.

\subsubsection{$D_{sJ}^\ast(2860)$, $D_{s1}^\ast(2860)$, and $D_{s3}^\ast(2860)$.}

In 2006, an enhancement around 2.86 GeV was observed by the BaBar
Collaboration in the $DK$ invariant mass
spectrum~\cite{Aubert:2006mh}, as shown in Fig.~\ref{sec212:Babar2}.
This is the $D_{sJ}^\ast(2860)$, whose mass and width were measured
to be $M=2856.6 \pm 1.5\pm 5.0$ MeV and $\Gamma=48 \pm 7\pm10$ MeV,
respectively. It was confirmed in the $D^\ast K $ mode by the BaBar
Collaboration~\cite{Aubert:2009ah} as well as in the same $DK$ mode
by the LHCb Collaboration~\cite{Aaij:2012pc}. The following ratio
was measured by BaBar at the same time~\cite{Aubert:2009ah}
\begin{eqnarray}
\frac{\mathcal{B}(D_{sJ}^\ast(2860)^+ \to D^\ast K
)}{\mathcal{B}(D_{sJ}^\ast(2860)^+ \to D K )} = 1.10 \pm 0.15 \pm
0.19 \, .
\end{eqnarray}

\begin{figure}[htb]
\begin{center}
\includegraphics*[width=0.6\textwidth]{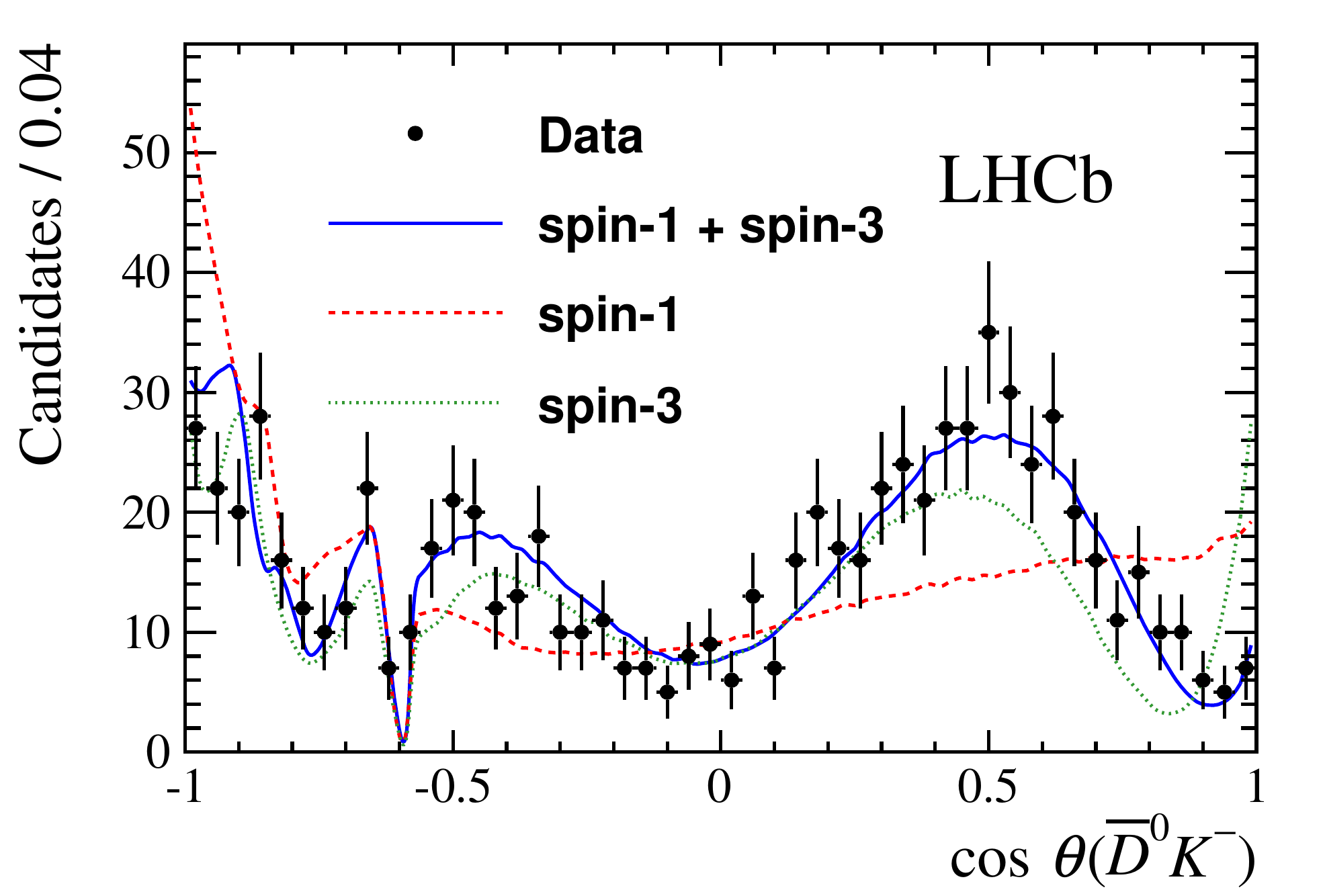}
\end{center}
\caption{(Color online) Cosine of the helicity angle of the $\bar
D^0K^-$ system, for 2.77~GeV~$< m(\bar D^0K^-) <$~2.91~GeV.
Black points are experimental data, which are fitted with different spin hypotheses of the $D_{sJ}^\ast(2860)$ as detailed in the legend.
Taken from LHCb~\cite{Aaij:2014xza}. } \label{sec212:LHCb}
\end{figure}

In 2014, the LHCb Collaboration further studied the structure around
2.86 GeV in the $B^0_s \to \bar{D}^0K^- \pi^+$
decay~\cite{Aaij:2014xza,Aaij:2014baa}. The amplitude analysis of
this decay indicates that this structure actually contains two
components, the $D_{s1}^\ast(2860)$ of $J^P = 1^-$ and the
$D_{s3}^\ast(2860)$ of $J^P = 3^-$, as shown in
Fig.~\ref{sec212:LHCb}. Their resonance parameters were measured to
be
\begin{eqnarray}
 \nonumber M_{D_{s1}^\ast(2860)}&=&2859 \pm 12\pm 6 \pm 23\,\mathrm{MeV} \, ,
\\
\Gamma_{D_{s1}^\ast(2860)}&=&159 \pm 23\pm 27 \pm 72\,\mathrm{MeV}
\, ,
\\ \nonumber
M_{D_{s3}^\ast(2860)}&=&2860.5 \pm 2.6\pm 2.5 \pm 6.0\,\mathrm{MeV}
\, ,
\\ \nonumber
\Gamma_{D_{s3}^\ast(2860)}&=&53 \pm 7\pm 4 \pm 6\,\mathrm{MeV} \, .
\end{eqnarray}
Comparing their widths, one finds that the $D_{s3}^\ast(2860)$ and the previously
observed
$D_{sJ}^\ast(2860)$~\cite{Aubert:2006mh,Aubert:2009ah,Aaij:2012pc}
may be the same state and the $D_{s1}^\ast(2860)$ may be a
different state.

\subsubsection{$D_{sJ}(3040)$.}

\begin{figure}[htb]
\begin{center}
\includegraphics*[width=1.0\textwidth]{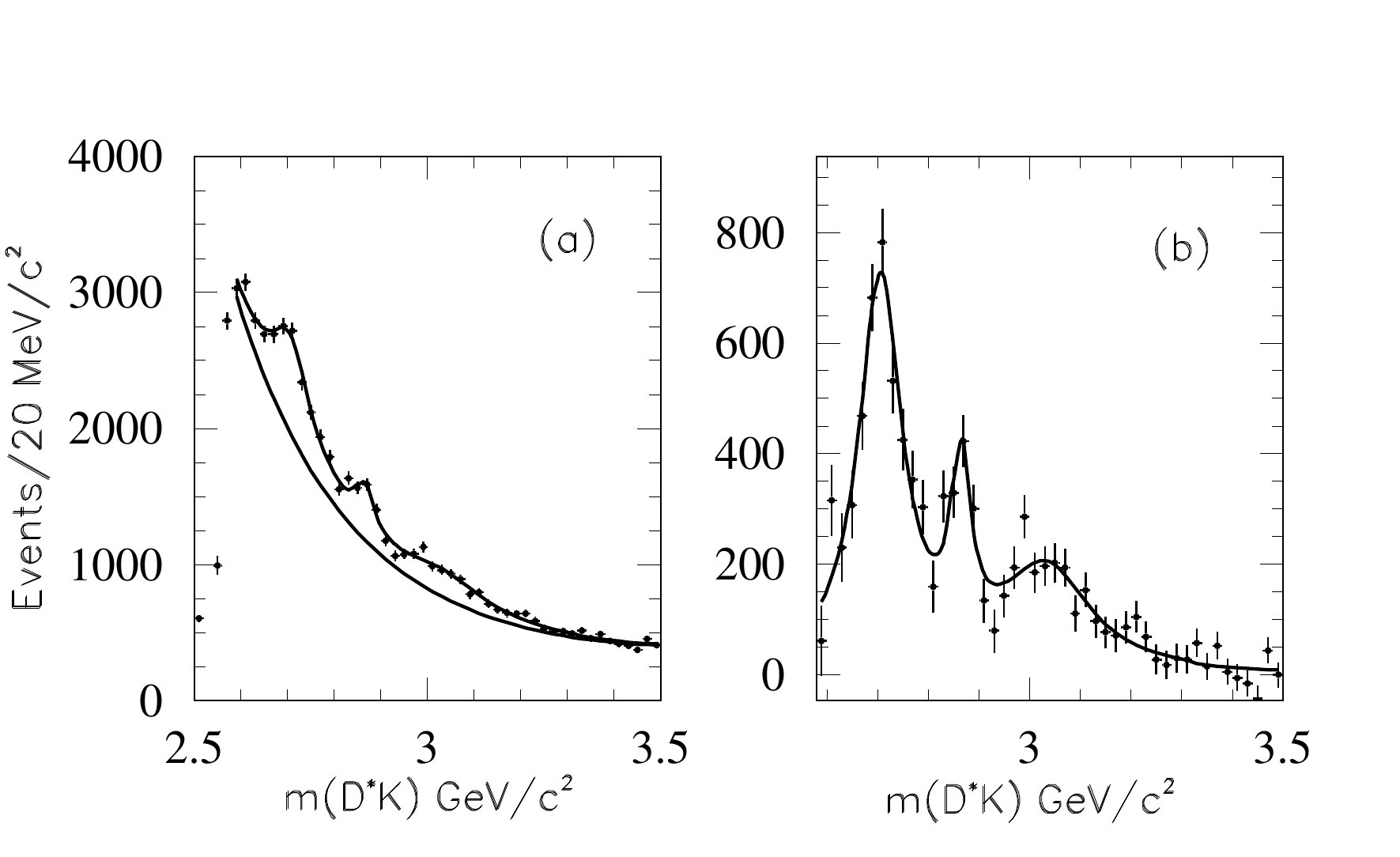}
\end{center}
\caption{The $D^*K$ invariant mass spectrum (a) with the background, and (b) after subtraction of the fitted
background. The signals correspond to the $D^*_{s1}(2710)^+$, $D^*_{sJ}(2860)^+$, and $D_{sJ}(3040)^+$. Taken from BaBar~\cite{Aubert:2009ah}. }
\label{sec212:Babar3}
\end{figure}

In 2009, the BaBar experiment observed a new broad structure in the
$D^\ast K$ invariant mass distribution~\cite{Aubert:2009ah}, as
shown in Fig.~\ref{sec212:Babar3}. They also confirmed the
$D_{s1}^\ast(2700)$ and $D_{sJ}^\ast(2860)$. The mass and width of
this new state $D_{sJ}(3040)$ were measured to be $M=3044 \pm
8^{+30}_{-5}$ MeV and $\Gamma=239 \pm 35^{+46}_{-42}$ MeV,
respectively. The negative result of its decay into $DK$ suggests
its unnatural parity quantum numbers.

\subsection{The bottom mesons}
\label{sec2.3}

\renewcommand{\arraystretch}{1.4}
\begin{table*}[htb]
\scriptsize
\caption{Experimental information of the observed bottom mesons.
Only the $1S$ bottom states ($B$ and $B^*$) are well established,
completing one $S$-wave doublet $(0^-, 1^-)$. Accordingly, we only
list their averaged masses and widths from PDG~\cite{Olive:2016xmw}
together with the experiments which first observed them. However,
the higher states starting from the $B^*_J(5732)$ are not well
established, so we list all the relevant experiments together with
their observed masses, widths, and decay modes therein.
\label{sec23:Bmeson} } \centering
\begin{tabular}{cccccc}
\toprule[1pt] State & $J^P$ & Mass (MeV) & Width (MeV) & Experiments
& Observed Modes
\\ \midrule[1pt]
$B^0$ & $0^-$ & $5279.62 \pm 0.15$ & $(1520 \pm 4) \times 10^{-15}$
s & CLEO~\cite{Behrends:1983er} & $D \pi \pi$\&$D^{*} \pi$
\\
$B^\pm$ & $0^-$ & $5279.31 \pm 0.15$ & $(1638 \pm 4) \times
10^{-15}$ s & CLEO~\cite{Behrends:1983er} & $D \pi$\&$D^{*} \pi \pi$
\\ \hline
$B^{*}$ & $1^-$ &  $5324.65 \pm 0.25$ & -- & CUSB~\cite{Han:1985uv}
& $B \gamma$
\\ \midrule[1pt]
\multirow{6}{*}{$B^*_J(5732)$} & \multirow{6}{*}{$?^?$} & $5681 \pm
11$  & $116 \pm 24$ & OPAL~\cite{Akers:1994fz} & $B^{(*)} \pi$
\\
                                                             && $5732 \pm 5 \pm 20$ & $145 \pm 28$ & DELPHI~\cite{Abreu:1994hj} & $B^{(*)} \pi$
\\
                                                             && $5695^{+17}_{-19}$ & -- & ALEPH~\cite{Barate:1998cq} & $B^{(*)} \pi$
\\
                                                             && $5713 \pm 2$ & $31 \pm 7$ & L3~\cite{Acciarri:1999jx} & $B \pi$
\\
                                                             && $5710 \pm 20$ & -- & CDF~\cite{Affolder:1999cx} & $B \pi$
\\
                                                             && $5738{^{+5}_{-6}\pm7}$ & $18{^{+15}_{-13}}{^{+29}_{-23}}$ & OPAL~\cite{Abbiendi:2000zv} & $B^{*} \pi$
\\ \hline
\multirow{4}{*}{$B_1(5721)^0$} & \multirow{4}{*}{$1^+$} & $5720.6
\pm 2.4 \pm 1.4$ & -- & D\O\,~\cite{Abazov:2007vq} & $B^{(*)+} \pi^-$
\\
&& $5725.3{^{+1.6}_{-2.2}}{^{+1.4}_{-1.5}}$ & -- &
CDF~\cite{Aaltonen:2008aa} & $B^{(*)+} \pi^-$
\\
&& $5726.6 \pm 0.9 {^{+1.1}_{-1.2}} \pm 0.4$ & -- &
CDF~\cite{Aaltonen:2013atp} & $B^{(*)+} \pi^-$
\\
&& $5727.7 \pm 0.7 \pm 1.4 \pm 0.17 \pm 0.4$ & $30.1 \pm 1.5 \pm
3.5$ & LHCb~\cite{Aaij:2015qla} & $B^{(*)+} \pi^-$
\\ \hdashline[2pt/2pt]
\multirow{2}{*}{$B_1(5721)^+$} & \multirow{2}{*}{$1^+$} & $5727 \pm
3 {^{+1}_{-3}} \pm 2$ & -- & CDF~\cite{Aaltonen:2013atp} & $B^{(*)0}
\pi^+$
\\
                                                           && $5725.1 \pm 1.8 \pm 3.1 \pm 0.17 \pm 0.4$ & $29.1 \pm 3.6 \pm 4.3$ & LHCb~\cite{Aaij:2015qla} & $B^{(*)0} \pi^+$
\\ \hline
\multirow{4}{*}{$B^*_2(5747)^0$} & \multirow{4}{*}{$2^+$} & $5746.8
\pm 2.4 \pm 1.7$ & -- & D\O\,~\cite{Abazov:2007vq} & $B^{(*)+} \pi^-$
\\
                                                           && $5740.2{^{+1.7}_{-1.8}}{^{+0.9}_{-0.8}}$ & $22.7{^{+3.8}_{-3.2}}{^{+3.2}_{-10.2}}$ & CDF~\cite{Aaltonen:2008aa} & $B^{(*)+} \pi^-$
\\
                                                           && $5736.7 \pm 1.2 {^{+0.8}_{-0.9}} \pm 0.2$ & -- & CDF~\cite{Aaltonen:2013atp} & $B^{(*)+} \pi^-$
\\
                                                           && $5739.44 \pm 0.37 \pm 0.33 \pm 0.17$ & $24.5 \pm 1.0 \pm 1.5$ & LHCb~\cite{Aaij:2015qla} & $B^{(*)+} \pi^-$
\\ \hdashline[2pt/2pt]
\multirow{2}{*}{$B^*_2(5747)^+$} & \multirow{2}{*}{$2^+$} & $5736.9
\pm 1.2 {^{+0.3}_{-0.9}} \pm 0.2$ & -- & CDF~\cite{Aaltonen:2013atp}
& $B^{(*)0} \pi^+$
\\
                                                           && $5737.20 \pm 0.72 \pm 0.40 \pm 0.17$ & $ 23.6 \pm 2.0 \pm 2.1$ & LHCb~\cite{Aaij:2015qla} & $B^{(*)0} \pi^+$
\\ \hline
$B_J(5840)^0$ & $?^?$ & $5862.9 \pm 5.0 \pm 6.7 \pm 0.2$ & $127.4
\pm 16.7 \pm 34.2$ & LHCb~\cite{Aaij:2015qla} & $B \pi$
\\ \hdashline[2pt/2pt]
$B_J(5840)^+$ & $?^?$  & $5850.3 \pm 12.7 \pm 13.7 \pm 0.2$ & $224.4
\pm 23.9 \pm 79.8$ & LHCb~\cite{Aaij:2015qla} & $B \pi$
\\ \hline
$B(5970)^0$ & $?^?$ & $5978 \pm 5 \pm 12$ & -- &
CDF~\cite{Aaltonen:2013atp} & $B^+ \pi^-$
\\
$B_J(5960)^0$ & $?^?$  & $5969.2 \pm 2.9 \pm 5.1 \pm 0.2$ & $82.3
\pm 7.7 \pm 9.4$ & LHCb~\cite{Aaij:2015qla} & $B^+ \pi^-$
\\ \hdashline[2pt/2pt]
$B(5970)^+$ & $?^?$ & $5961 \pm 5 \pm 12$ & -- &
CDF~\cite{Aaltonen:2013atp} & $B^0 \pi^+$
\\
$B_J(5960)^+$ & $?^?$  & $5964.9 \pm 4.1 \pm 2.5 \pm 0.2$ & $63.0
\pm 14.5 \pm 17.2$ & LHCb~\cite{Aaij:2015qla} & $B^0 \pi^+$
\\ \bottomrule[1pt]
\end{tabular}
\end{table*}

In this subsection we review the bottom mesons. Their experimental
information is listed in Table~\ref{sec23:Bmeson}. Different from
the charmed and charmed-strange mesons, only the $1S$ bottom states
($B$ and $B^*$) are well established, completing one $S$-wave
doublet $(0^-, 1^-)$. Accordingly, we only list their averaged
masses and widths from PDG~\cite{Olive:2016xmw} together with the
experiments which first observed them. However, the higher states
starting from the $B^*_J(5732)$ are not well established, so we list
all the relevant experiments together with their observed masses,
widths, and decay modes therein.

\subsubsection{$B$ and $B^*$.}

\begin{figure}[htb]
\begin{center}
\includegraphics*[width=0.48\textwidth]{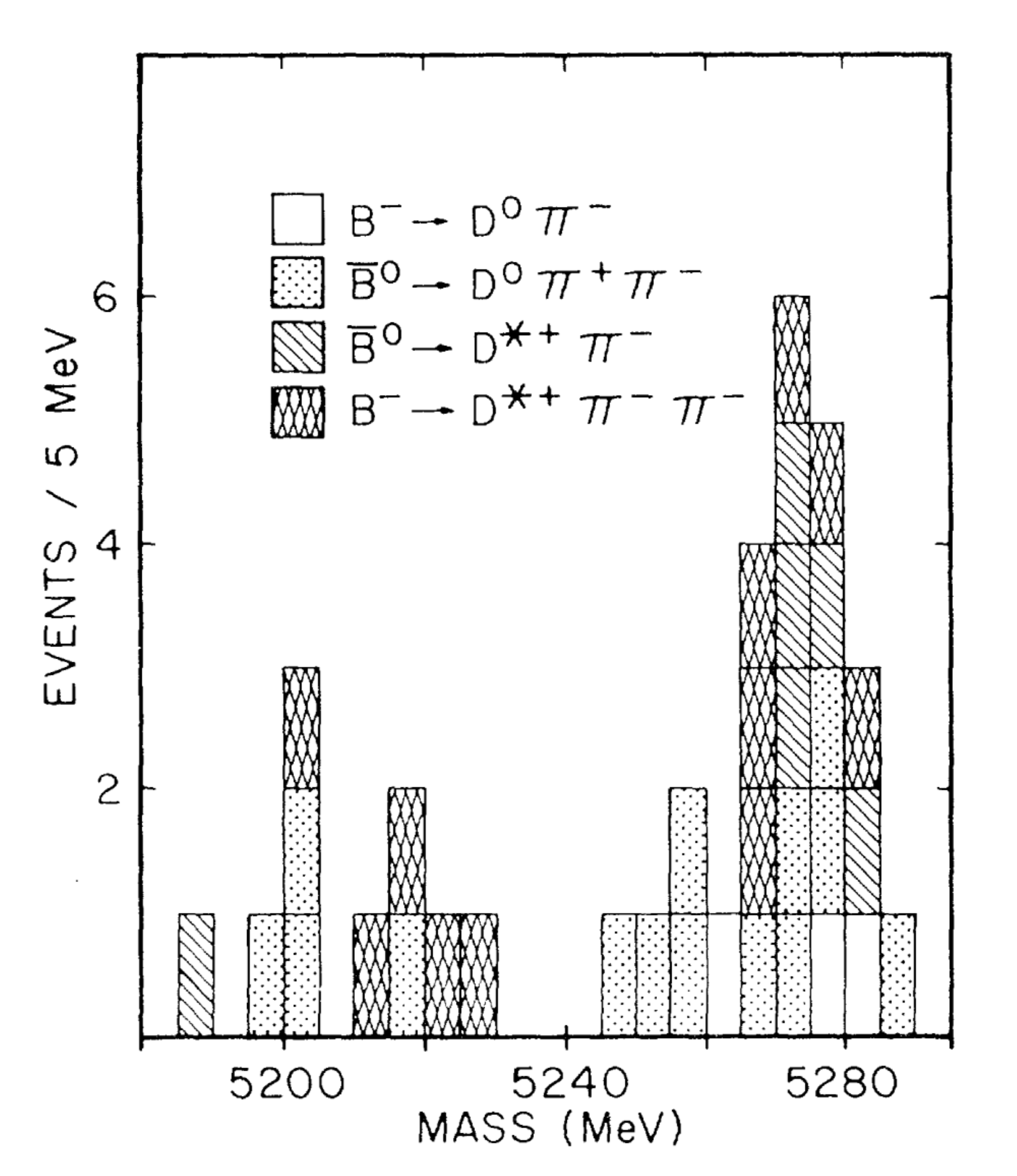}
\end{center}
\caption{Mass distribution of the $B$ meson candidates. Taken from
CLEO~\cite{Behrends:1983er}. } \label{sec213:CLEO}
\end{figure}

The lowest-lying bottom mesons, $B^0$ and $B^\pm$ of $J^P = 0^-$,
were observed in 1983 by the CLEO
Collaboration~\cite{Behrends:1983er}, as shown in
Fig.~\ref{sec213:CLEO}. Their properties are known very
well~\cite{Olive:2016xmw}: the $B^0$ meson has a mass $5279.62 \pm 0.15$ MeV and a mean life $(1520 \pm 4) \times 10^{-15}$ s; the
$B^\pm$ meson has a mass $5279.31 \pm 0.15$ MeV and a mean life
$(1638 \pm 4) \times 10^{-15}$ s; hundreds of their decay modes have
been observed in experiments, where the Cabibbo-allowed process $b
\rightarrow c W^-$ is preferred.

Two years later in 1985, the lowest-lying vector bottom meson,
$B^{*}$ of $J^P = 1^-$, was observed~\cite{Han:1985uv}. Its
existence has been confirmed by many following experiments, but its
properties were not well measured. The best measured quantity is the
mass difference between $B$ and $B^*$, that is $m_{B^*} - m_B =
45.18 \pm 0.23$ MeV~\cite{Olive:2016xmw}. In 2012, the LHCb
experiment measured the $B^{*+}$ mass to be $m_{B^{*+}} = 5324.26
\pm 0.30 \pm 0.23 \pm 0.17$ MeV~\cite{Aaij:2012uva}. The $B^{*}$
meson mainly decays into $B \gamma$.

\subsubsection{$B^*_J(5732)$, $B_1(5721)^0$, and $B^*_2(5747)^0$.}

\begin{figure}[htb]
\begin{center}
\includegraphics*[width=0.48\textwidth]{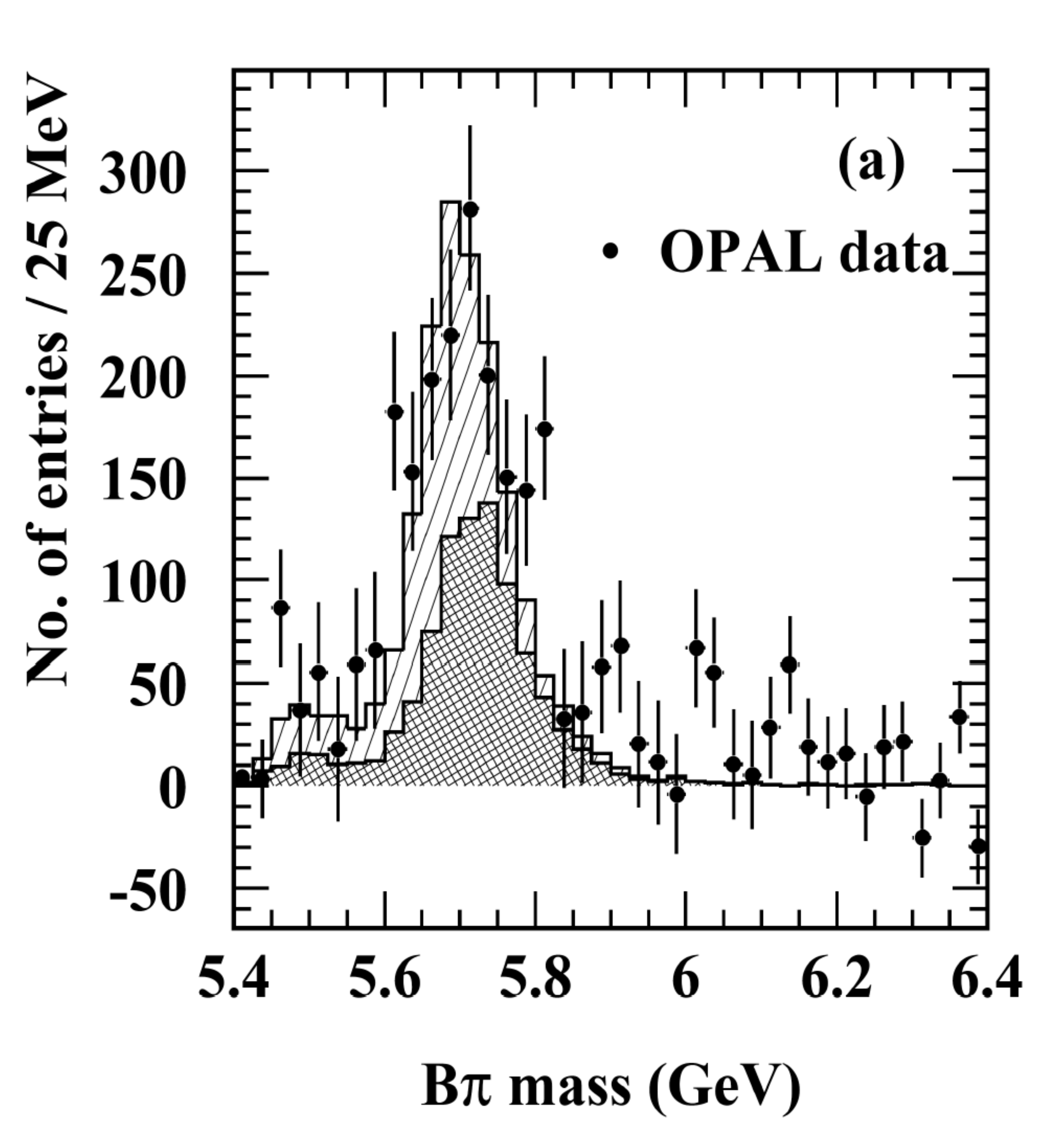}
\end{center}
\caption{The $B\pi$ invariant mass distribution,
where the signal corresponds to the $B^*_J(5732)$.
Taken from OPAL~\cite{Akers:1994fz}. }
\label{sec213:OPAL}
\end{figure}

In 1994, the first orbitally excited bottom meson, the $B^*_J(5732)$, was
observed by the OPAL detector at LEP in the $B^{(*)+} \pi^-$
invariant mass distribution~\cite{Akers:1994fz}, as shown in
Fig.~\ref{sec213:OPAL}. This observation was confirmed by the
following DELPHI, ALEPH, L3, and CDF
experiments~\cite{Abreu:1994hj,Barate:1998cq,Acciarri:1999jx,Affolder:1999cx,Buskulic:1995mt}.
All the experimental information is listed in
Table~\ref{sec23:Bmeson}. In 2000, the OPAL Collaboration further
studied this state in the $B^{*} \pi$ invariant mass
distribution~\cite{Abbiendi:2000zv}. Its mass and width were
measured to be $5738{^{+5}_{-6}\pm7}$ MeV and
$18{^{+15}_{-13}}{^{+29}_{-23}}$ MeV, respectively. They also
measured its branching ratio decaying into $B^* \pi (X)$ to be
\begin{eqnarray}
\mathcal{B}(B^*_J \to B^* \pi (X)) = 0.85 {^{+0.26}_{-0.27} \pm
0.12} \, ,
\end{eqnarray}
where $(X)$ refers to decay modes with or without additional
accompanying decay particles. Based on the heavy quark symmetry,
they determined its branching ratio decaying into $B^* \pi$ to be
\begin{eqnarray}
\mathcal{B}(B^*_J \to B^* \pi) = 0.74 {^{+0.12}_{-0.10}}
{^{+0.21}_{-0.15}} \, .
\end{eqnarray}
However, all the experiments performed at
LEP~\cite{Akers:1994fz,Abreu:1994hj,Barate:1998cq,Acciarri:1999jx,Abbiendi:2000zv,Buskulic:1995mt}
used the inclusive or semi-exclusive $B$ decays, which made the
separation of the states of different spin-parity quantum numbers
impossible~\cite{Colangelo:2012xi} (see the following $B_1(5721)^0$
and $B^*_2(5747)^0$).

\begin{figure}[htb]
\begin{center}
\includegraphics*[width=0.6\textwidth]{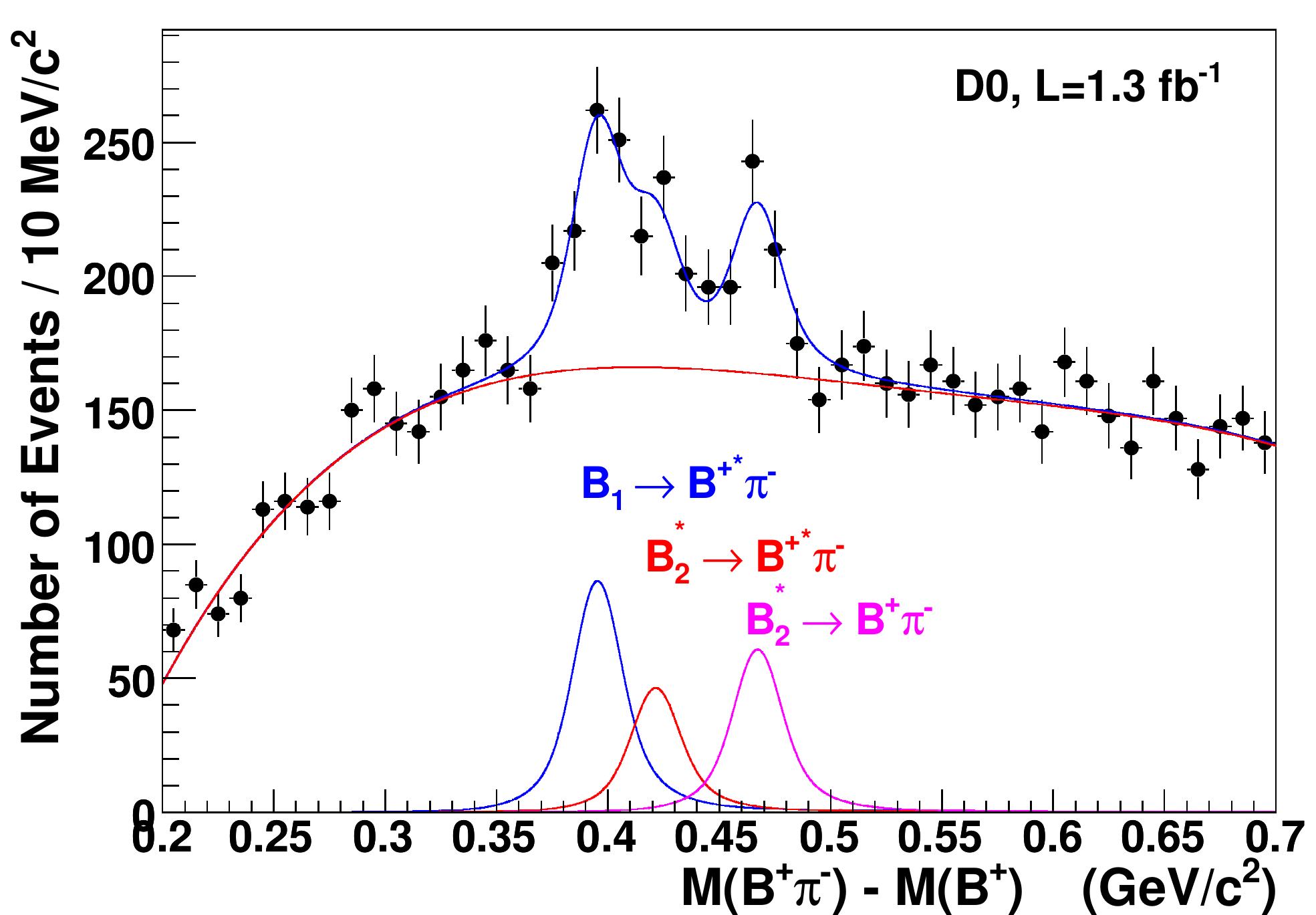}
\end{center}
\caption{(Color online) The mass difference distribution, $\Delta M = M(B^+
\pi^-) - M(B^+)$, for exclusive $B$ decays.
The three peaks are produced by the three decays $B_1 \to B^{+*} \pi^-$, $B^*_2 \to B^{+*} \pi^-$, and $B^*_2 \to B^+ \pi^-$.
Taken from D\O\,~\cite{Abazov:2007vq}. } \label{sec213:D0CDF}
\end{figure}

In 2007, the D\O\, Collaboration reported the observation of two
orbitally excited ($L = 1$) narrow bottom mesons in the $B^{(*)+}
\pi^-$ invariant mass distribution~\cite{Abazov:2007vq}, as shown in
Fig.~\ref{sec213:D0CDF}. They are the $B_1(5721)^0$ of $J^P = 1^+$
and the $B^*_2(5747)^0$ of $J^P = 2^+$. Using the mass of the
$B^+$~\cite{Yao:2006px}, their masses were determined to be $5720.6
\pm 2.4 \pm 1.4$ MeV and $5746.8 \pm 2.4 \pm 1.7$ MeV, respectively.

This observation was confirmed in the following CDF and LHCb
experiments~\cite{Aaltonen:2008aa,Aaltonen:2013atp,Aaij:2015qla} in
the same process. All the experimental information is listed in
Table~\ref{sec23:Bmeson}. The LHCb experiment also measured the
relative branching fractions for the $B^*_2(5747)^{0,+}$
decays~\cite{Aaij:2015qla}
\begin{eqnarray}
{\mathcal{B}(B^*_2(5747)^0 \to B^{*+} \pi^-) \over
\mathcal{B}(B^*_2(5747)^0 \to B^+ \pi^-)} &=& 0.71 \pm 0.14 \pm 0.30
\, ,
\\ \nonumber
{\mathcal{B}(B^*_2(5747)^+ \to B^{*0} \pi^+) \over
\mathcal{B}(B^*_2(5747)^+ \to B^0 \pi^+)} &=& 1.0 \pm 0.5 \pm 0.8 \,
.
\end{eqnarray}

\subsubsection{$B(5970)$, $B_J(5960)$, and $B_J(5840)$.}

\begin{figure}[htb]
\begin{center}
\includegraphics*[width=0.6\textwidth]{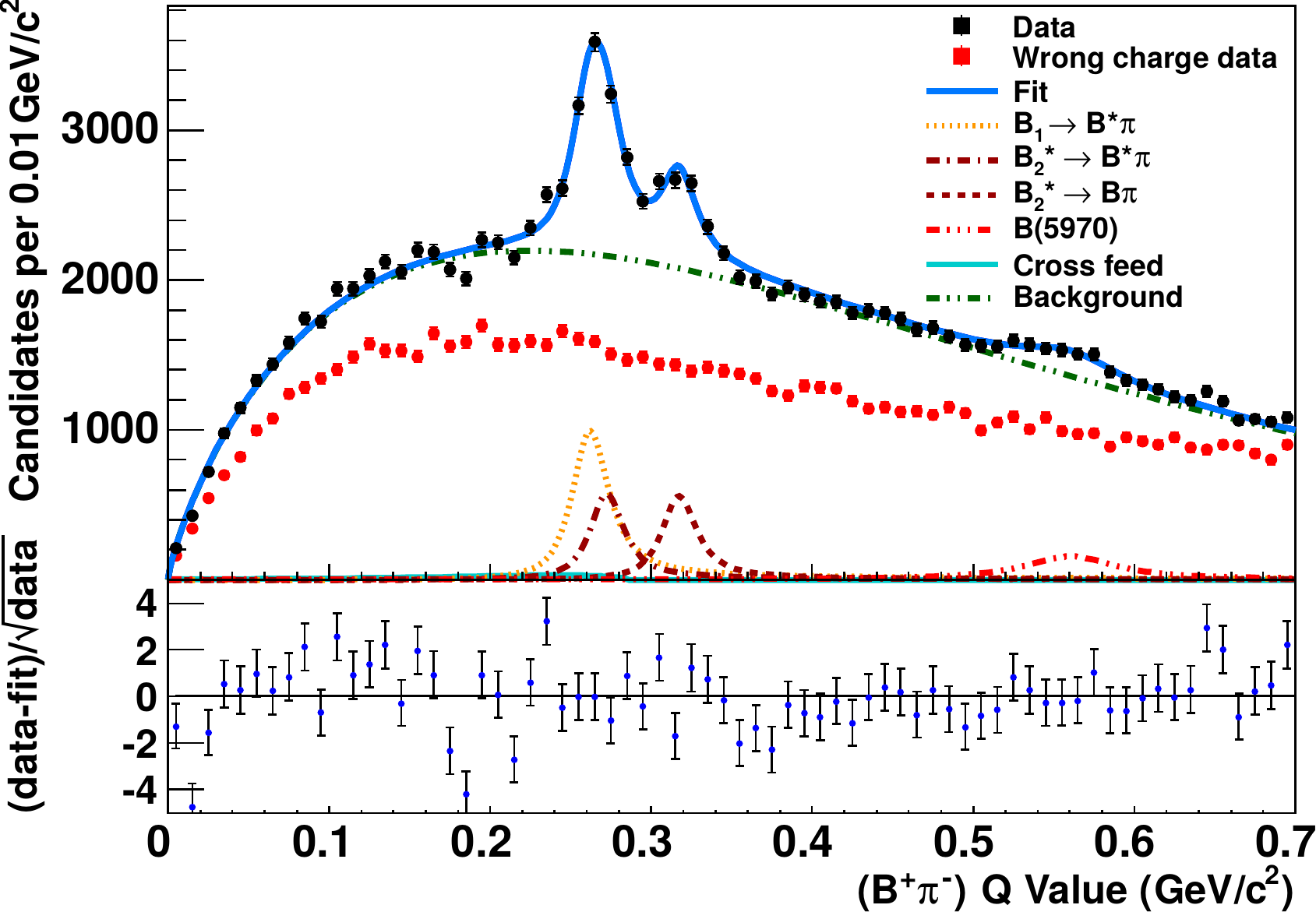}
\end{center}
\caption{(Color online) Distribution of $Q = m(B \pi) - m(B) -
m_{\pi}$ value of the $B^{**0}$ candidates.
Black points are experimental data, which are fitted with several channels as detailed in the legend.
Taken from CDF~\cite{Aaltonen:2013atp}. }
\label{sec213:CDFLHCb}
\end{figure}

In 2013, an excited bottom meson, the $B(5970)$, was observed by the
CDF Collaboration simultaneously in the $B^+ \pi^-$ and $B^0 \pi^+$
mass distributions~\cite{Aaltonen:2013atp}, as shown in
Fig.~\ref{sec213:CDFLHCb}. The masses of the $B(5970)$ resonances were
determined to be $5978 \pm 5 \pm 12$ MeV for the neutral state and
$5961 \pm 5 \pm 12$ MeV for the charged state. They used independent
parameters for the $B(5970)^0$ and $B(5970)^+$ signals and found
individual significance of $4.2\sigma$ and $3.7\sigma$ for the
neutral and charged states, respectively.

Later in 2015, another excited bottom meson, the $B_J(5960)$, was observed by the LHCb Collaboration in the $B \pi$ mass distributions~\cite{Aaij:2015qla}. 
The results were fitted using both the empirical model and the quark
model and the results obtained using the former model are
\begin{eqnarray}
\nonumber m_{B_J(5960)^0} &=& 5969.2 \pm 2.9 \pm 5.1 \pm 0.2 {\rm~MeV} \, ,
\\
\Gamma_{B_J(5960)^0} &=& 82.3 \pm 7.7 \pm 9.4 {\rm~MeV} \, ,
\\ \nonumber
m_{B_J(5960)^+} &=& 5964.9 \pm 4.1 \pm 2.5 \pm 0.2 {\rm~MeV} \, ,
\\ \nonumber
\Gamma_{B_J(5960)^+} &=& 63.0 \pm 14.5 \pm 17.2 {\rm~MeV} \, .
\end{eqnarray}
The properties of the $B_J(5960)$ states~\cite{Aaij:2015qla} are
consistent with those of the $B(5970)$~\cite{Aaltonen:2013atp}
observed by the CDF Collaboration when assuming their decays into
$B\pi$, so they may be the same state.

Besides the $B_J(5960)$, the LHCb Collaboration also reported the
observation of another excited bottom meson, the $B_J(5840)$, in the
$B \pi$ mass distributions~\cite{Aaij:2015qla}.
The results obtained using the empirical model are
\begin{eqnarray}
\nonumber m_{B_J(5840)^0} &=& 5862.9 \pm 5.0 \pm 6.7 \pm 0.2 {\rm~MeV} \, ,
\\
\Gamma_{B_J(5840)^0} &=& 127.4 \pm 16.7 \pm 34.2 {\rm~MeV} \, ,
\\ \nonumber
m_{B_J(5840)^+} &=& 5850.3 \pm 12.7 \pm 13.7 \pm 0.2 {\rm~MeV} \, ,
\\ \nonumber
\Gamma_{B_J(5840)^+} &=& 224.4 \pm 23.9 \pm 79.8 {\rm~MeV} \, .
\end{eqnarray}

\subsection{The bottom-strange mesons}
\label{sec2.4}

\renewcommand{\arraystretch}{1.4}
\begin{table*}[htb]
\scriptsize
\caption{Experimental information of the observed bottom-strange
mesons. Only the $1S$ bottom-strange states ($B_s$ and $B_s^*$) are
well established, completing one $S$-wave doublet $(0^-, 1^-)$.
Accordingly, we only list their averaged masses and widths from
PDG~\cite{Olive:2016xmw} together with the experiments first
observing them. However, the higher states starting from the
$B^*_{sJ}(5850)$ are not well established, so we list all the
relevant experiments together with their observed masses, widths, and
decay modes therein. \label{sec24:Bsmeson} } \centering
\begin{tabular}{cccccc}
\toprule[1pt] State & $J^P$ & Mass (MeV) & Width (MeV) & Experiments
& Observed Modes
\\ \midrule[1pt]
$B_s$ & $0^-$ & $5366.82 \pm 0.22$ & $(1510 \pm 5) \times 10^{-15}$
s & CUSB-II~\cite{LeeFranzini:1990gy} & --
\\
$B_s^{*}$ & $1^-$ &  $5415.4^{+1.8}_{-1.5}$ & -- &
CUSB-II~\cite{LeeFranzini:1990gy} & $B_s \gamma$
\\ \midrule[1pt]
$B^*_{sJ}(5850)$ & $?^?$ & $5853 \pm 15$  & $47 \pm 22$ &
OPAL~\cite{Akers:1994fz} & $B^{(*)} K$
\\ \hline
\multirow{3}{*}{$B_{s1}(5830)$} & \multirow{3}{*}{$1^+$} & $5829.4
\pm 0.7$ & -- & CDF~\cite{Aaltonen:2007ah} & $B^{*} K$
\\
&& $5828.40 \pm 0.04 \pm 0.04 \pm 0.41$ & -- &
LHCb~\cite{Aaij:2012uva} & $B^{*} K$
\\
&& $5828.3 \pm 0.1 \pm 0.2 \pm 0.4$ & $0.5 \pm 0.3 \pm 0.3$ &
CDF~\cite{Aaltonen:2013atp} & $B^{*} K$
\\ \hdashline[2pt/2pt]
\multirow{4}{*}{$B^*_{s2}(5840)$} & \multirow{4}{*}{$1^+$} &
$5839.7 \pm 0.7$ & -- & CDF~\cite{Aaltonen:2007ah} & $B^{(*)} K$
\\
                                                           && $5839.6 \pm 1.1 \pm 0.7$ & -- & D\O\,~\cite{Abazov:2007af} & $B^{(*)} K$
\\
&& $ 5839.99 \pm 0.05 \pm 0.11 \pm 0.17$ & $1.56 \pm 0.13 \pm 0.47$
& LHCb~\cite{Aaij:2012uva} & $B^{(*)} K$
\\
&& $5839.7 \pm 0.1 \pm 0.1 \pm 0.2$ & $1.4 \pm 0.4 \pm 0.2$ &
CDF~\cite{Aaltonen:2013atp} & $B^{(*)} K$
\\ \bottomrule[1pt]
\end{tabular}
\end{table*}

In this subsection we review the bottom-strange mesons. Their
experimental information is listed in Table~\ref{sec24:Bsmeson}.
Similar to the bottom mesons, only the $1S$ bottom-strange states
($B_s$ and $B_s^*$) are well established, completing one $S$-wave
doublet $(0^-, 1^-)$. Accordingly, we only list their averaged
masses and widths from PDG~\cite{Olive:2016xmw} together with the
experiments first observing them. However, the higher states
starting from the $B^*_{sJ}(5850)$ are not well established, so we
list all the relevant experiments together with their observed
masses, widths, and decay modes therein.

\subsubsection{$B_s$ and $B_s^*$.}


The lowest-lying bottom-strange meson, $B_s$ of $J^P = 0^-$, was observed in 1990 by the CUSB-II Collaboration in $\Upsilon(5S)$ decays~\cite{LeeFranzini:1990gy}. 
The lowest-lying vector bottom-strange mesons, $B_s^{*}$ of $J^P =
1^-$, was observed in the same experiment~\cite{LeeFranzini:1990gy}.
The properties of the $B_s$ are known very
well~\cite{Olive:2016xmw}: the $B_s^0$ meson has a mass $5366.82 \pm 0.22$ MeV and a mean life $(1510 \pm 5) \times 10^{-15}$ s; many
of its decay modes have been observed in experiments, where the
Cabibbo-allowed process $b \rightarrow c W^-$ is preferred. The
existence of the $B_s^{*}$ has been confirmed by many following
experiments: it has the mass $5415.4^{+1.8}_{-1.5}$ MeV, and mainly
decays into $B_s \gamma$.

\subsubsection{$B^*_{sJ}(5850)$, $B_{s1}(5830)^0$, and $B^*_{s2}(5840)^0$.}

\begin{figure}[htb]
\begin{center}
\includegraphics*[width=0.48\textwidth]{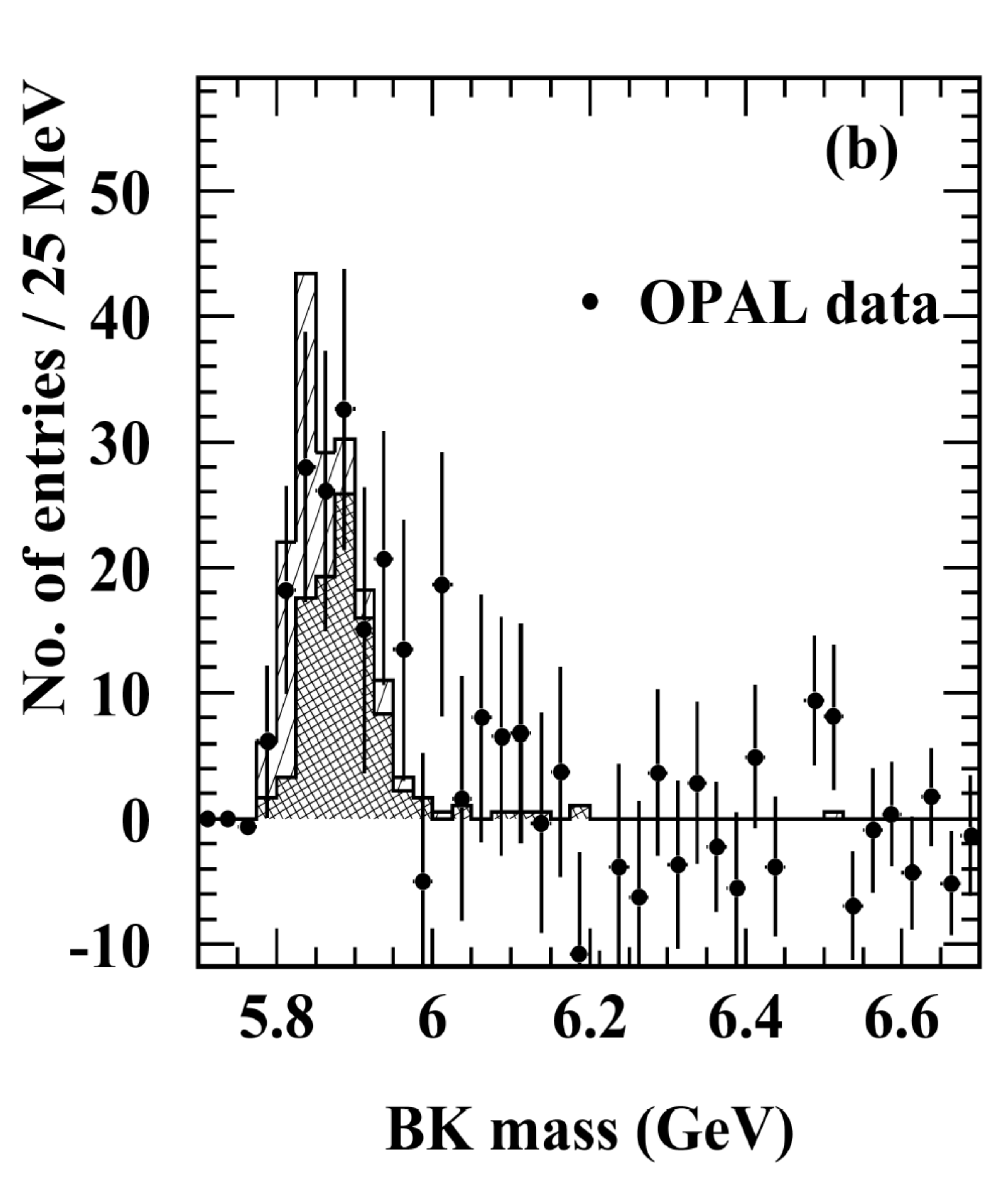}
\end{center}
\caption{The $B^+K^-$ invariant mass distribution, where the signal corresponds to the
$B^*_{sJ}(5850)$.
Taken from OPAL~\cite{Akers:1994fz}. } \label{sec214:OPAL}
\end{figure}

In 1994, the first excited bottom-strange meson, the
$B^*_{sJ}(5850)$, was observed by the OPAL detector at LEP in the
$B^{(*)+} K^-$ invariant mass distribution~\cite{Akers:1994fz}, as
shown in Fig.~\ref{sec214:OPAL}. Its mass and width were measured to
be $5853 \pm 15$ MeV and $47 \pm 22$ MeV respectively. Similar to
the $B^*_J(5732)$, this experiment performed at
LEP~\cite{Akers:1994fz} used inclusive (or semi-exclusive) $B$
decays, which made impossible the separation of the states of
different spin-parity quantum numbers~\cite{Colangelo:2012xi} (see
the following $B_{s1}(5830)^0$ and $B^*_{s2}(5840)^0$).

\begin{figure}[htb]
\begin{center}
\includegraphics*[width=0.48\textwidth]{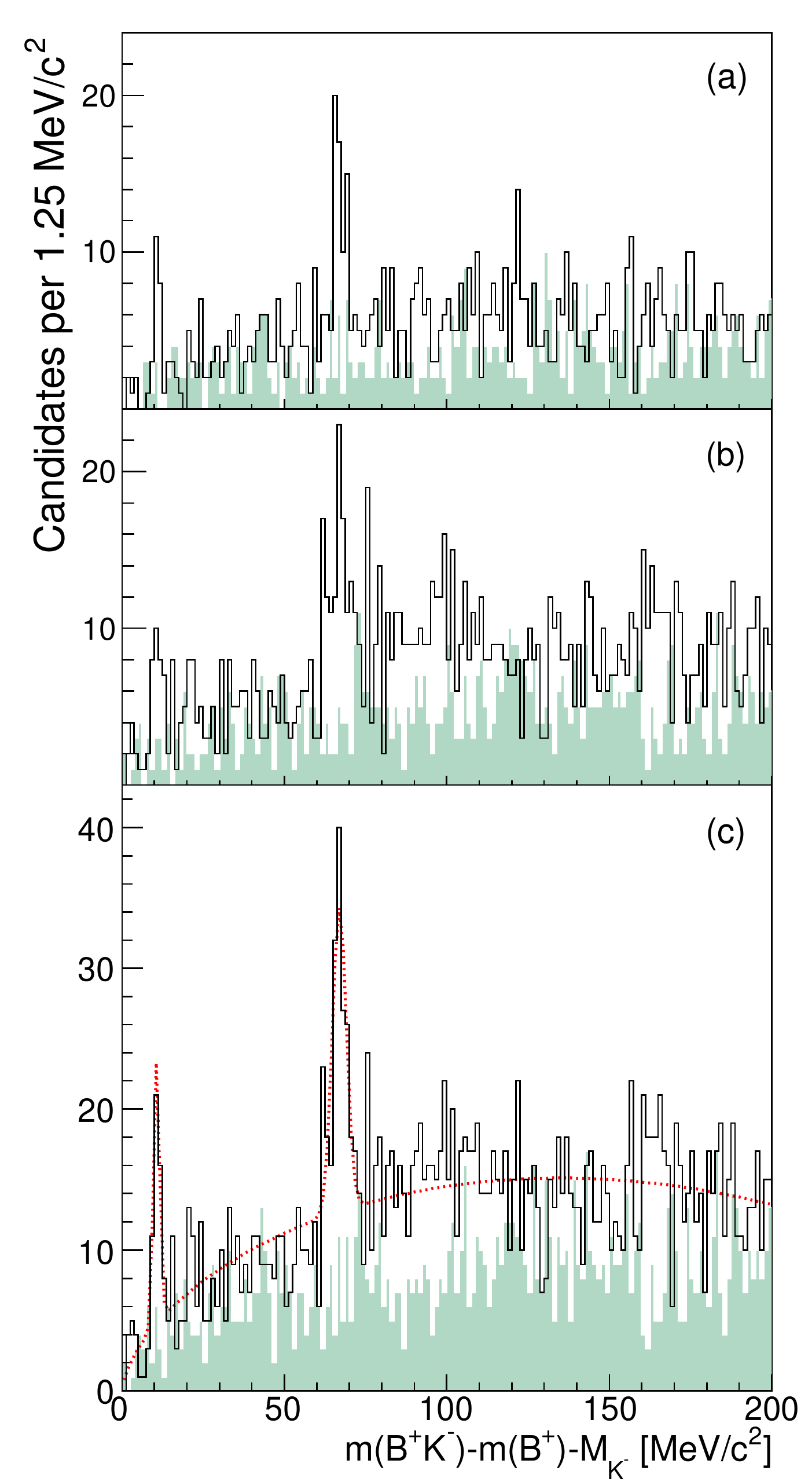}
\end{center}
\caption{(Color online) The mass difference distribution, $Q = m(B^+K^-) - m(B^+) -
M_{K^-}$, for the $B^{**}_s$ candidates with (a) $B^+ \to J/\psi
K^+$, (b) $B^+ \to D^0 \pi^+$, and (c) both channels combined.
The red dotted line is fitted with the $B_{s1}(5830)^0$ and the $B^*_{s2}(5840)^0$.
Taken from CDF~\cite{Aaltonen:2007ah}. } \label{sec214:CDF}
\end{figure}

In 2008, the CDF Collaboration reported the observation of two
orbitally excited ($L = 1$) narrow bottom-strange mesons in the
$B^{(*)+} K^-$ invariant mass distribution~\cite{Aaltonen:2007ah},
as shown in Fig.~\ref{sec214:CDF}. They are the $B_{s1}(5830)^0$ of
$J^P = 1^+$ and the $B^*_{s2}(5840)^0$ of $J^P = 2^+$. Using the
masses of the $B$, $B^*$, and $K$~\cite{Yao:2006px}, their masses were
determined to be $5829.4 \pm 0.7$ MeV and $5839.6 \pm 0.7$ MeV,
respectively. The $B^*_{s2}(5840)^0$ was also observed by the D\O\,
Collaboration in the same process~\cite{Abazov:2007af} and its mass
was measured to be $5839.6 \pm 1.1 \pm 0.7$ MeV. The collaboration also measured
its relative production rate with respect to the $B^+$ meson to be
\begin{equation}
{\mathcal{B}(b \to B^*_{s2} \to B^+ K^-) \over \mathcal{B}(b \to
B^+)} = (1.15 \pm 0.23 \pm 0.13) \% \, .
\end{equation}


The $B_{s1}(5830)^0$ and $B^*_{s2}(5840)^0$ were confirmed in the following LHCb experiment~\cite{Aaij:2012uva} in the $B^{(*)+} K^-$ invariant mass distribution. 
The collaboration reported the observation of the $B^*_{s2}(5840)^0$ meson
decaying to $B^{*+}K^-$ final states and first measured the width
of the $B^*_{s2}(5840)^0$ to be $1.56 \pm 0.13 \pm 0.47$ MeV. The
CDF experiment also confirmed their existence in
2013~\cite{Aaltonen:2013atp}.

%% file: section2.2.tex
\subsection{The charmed baryons}
\label{sec2.5}

To date, there are 23 singly-charmed, 1 doubly-charmed, and 10 bottom
baryons collected in PDG \cite{Olive:2016xmw}. Most of these heavy
baryons were reported by the B-factories and Tevatron. In this
subsection, we shall review these observed heavy baryons. Interested
readers may also consult Refs.~\cite{Cheng:2015iom,Asner:2008nq}.

\begin{figure}[htb]
\begin{center}
\includegraphics[width=1.0\textwidth]{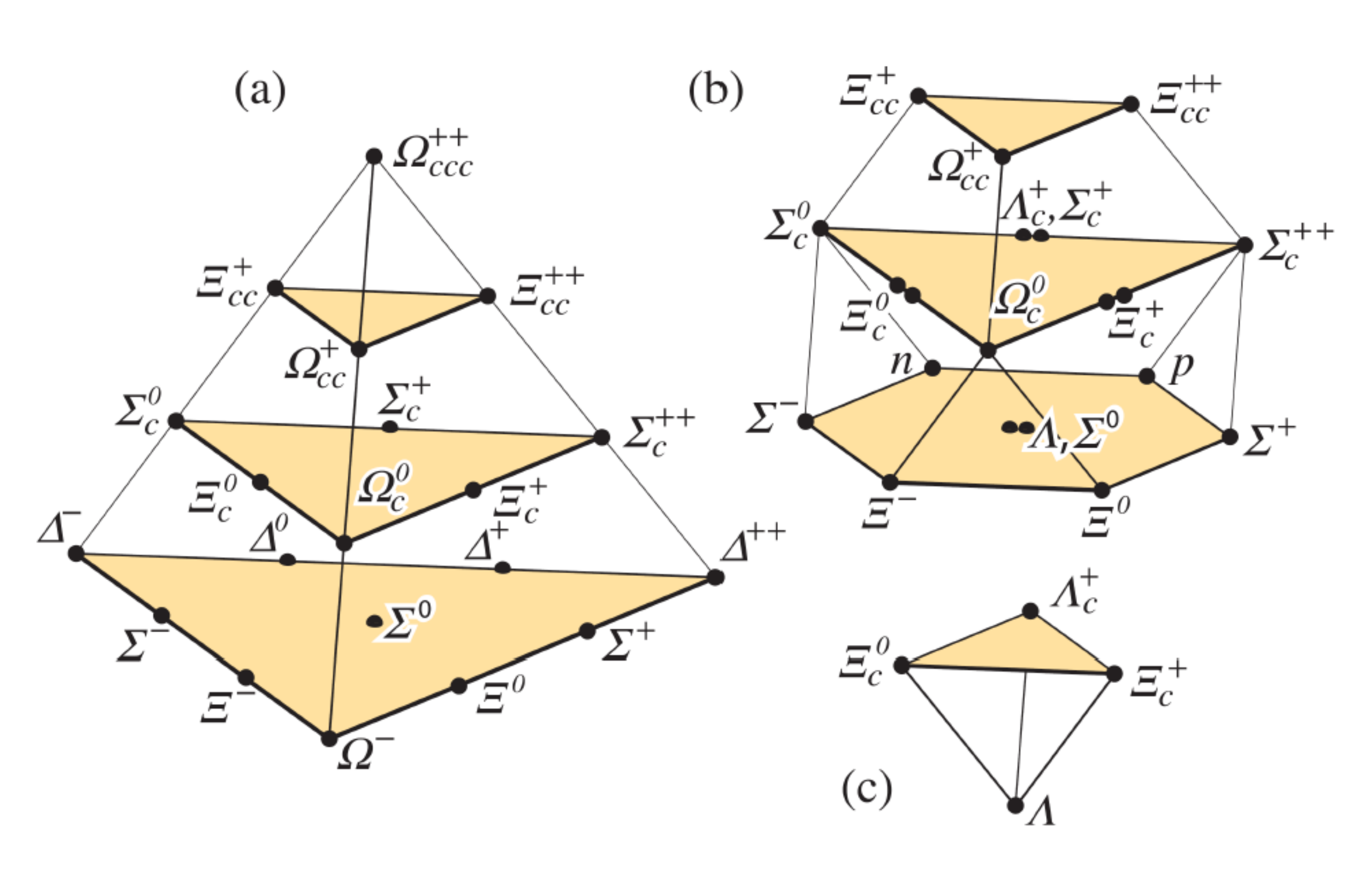}
\caption{(Color online) The SU(4) multiplets of baryons made of $u$, $d$, $s$, and
$c$ quarks: (a) the $\mathbf{20}$-plet, (b) the
$\mathbf{20}^\prime$-plet, and (c) the $\mathbf{\bar 4}$-plet. Taken
from PDG~\cite{Olive:2016xmw}. \label{Fig:2.2.Baryon}}
\end{center}
\end{figure}

First we use baryons containing charm quarks as an example to show
their naming scheme, and those containing bottom quarks can be
similarly named (see also discussions in PDG~\cite{Olive:2016xmw}).
Baryons made from $u$, $d$, $s$, and $c$ quarks belong to the SU(4)
multiplets:
\begin{eqnarray}
\mathbf{4} \otimes \mathbf{4} \otimes \mathbf{4} = \mathbf{20}
\oplus \mathbf{20}^\prime_1 \oplus \mathbf{20}^\prime_2 \oplus
\mathbf{\bar 4}.
\end{eqnarray}
We show the $\mathbf{20}$, $\mathbf{20}^\prime$ and $\mathbf{\bar
4}$ multiplets in Figs.~\ref{Fig:2.2.Baryon}(a),
\ref{Fig:2.2.Baryon}(b), and \ref{Fig:2.2.Baryon}(c), respectively.
Their bottom levels are SU(3) decuplet, octet, and singlet,
respectively. One level up from the bottom levels are the baryons
with one $c$ quark, and are SU(3) $\mathbf{6}$,
$\mathbf{6}\oplus\mathbf{\bar3}$, and $\mathbf{\bar 3}$ multiplets,
respectively.

The singly-charmed baryons contain a charm quark and two
relativistic light quarks which are often referred as a diquark.
Since the charm quark is much heavier than the light quarks, it is
almost static and provides the color source to the light quarks. The
whole system behaves as the QCD analogue of the familiar hydrogen in
QED. The charmed baryon system provides an ideal place to study
diquark correlation and the dynamics of the light quarks in the
environment of a heavy quark, using the heavy quark symmetry and the
heavy quark effective theory (HQET).

We list the observed singly-charmed baryons in Table
\ref{Table:2.2.singlycharmedbaryons}, including their masses,
widths, dominant decay modes, and the observed channels in
experiments. In the Baryon Summary Table in PDG
\cite{Olive:2016xmw}, only the 3- and 4-star status baryons are
included because the 1- and 2-star states are not established.
However, we collect all of them in Table
\ref{Table:2.2.singlycharmedbaryons} and will introduce their
experimental details in the following.

\renewcommand{\arraystretch}{1.4}
\begin{table*}[htb]
\tiny
\caption{The experimental information of the singly-charmed baryons.
The masses, widths and decay modes were taken from PDG
\cite{Olive:2016xmw}. The experiment column lists the discovery
experiments for these states. \label{Table:2.2.singlycharmedbaryons}}
\begin{center}
\begin{tabular}{ccccccc} \toprule[1pt]
~~State~~ &~~ Status ~~&~~$I(J^P)$~~ &
Mass (MeV) & Width (MeV) &~~Experiment~~&~Decay modes~\\
\midrule[1pt]
 $\Lambda_c^+$ & $\ast\ast\ast\ast$ & $0({1\over 2}^+)$  & $2286.46\pm0.14$ & $(200\pm6)\times 10^{-15} $ s &Fermilab \cite{Knapp:1976qw} & weak  \\
 $\Lambda_c(2595)^+$ &$\ast\ast\ast$ & $0({1\over 2}^-)$ & $2592.25\pm0.28$ &
 $2.59\pm0.56$ & CLEO \cite{Edwards:1994ar} & $\Lambda_c\pi\pi,\Sigma_c\pi$ \\
 $\Lambda_c(2625)^+$ & $\ast\ast\ast$ & $0({3\over 2}^-)$ & $2628.11\pm0.19$ &
 $<0.97$ & ARGUS \cite{Albrecht:1993pt} & $\Lambda_c\pi\pi,\Sigma_c\pi$ \\
 $\Lambda_c(2765)^+$ & $\ast$ & $?(?^?)$ & $2766.6\pm2.4$ & $50$ & CLEO \cite{Artuso:2000xy} & $\Sigma_c\pi,\Lambda_c\pi\pi$ \\
 $\Lambda_c(2880)^+$ & $\ast\ast\ast$ & $0({5\over 2}^+)$ & $2881.53\pm0.35$ & $5.8\pm1.1$
 & CLEO \cite{Artuso:2000xy} & $\Sigma_c^{(*)}\pi,\Lambda_c\pi\pi,D^0p$ \\
 $\Lambda_c(2940)^+$ &$\ast\ast\ast$ & $0(?^?)$ & $2939.3^{+1.4}_{-1.5}$ & $17^{+8}_{-6}$ & BaBar \cite{Aubert:2006sp} &
 $\Sigma_c^{(*)}\pi,\Lambda_c\pi\pi,D^0p$ \\
 \midrule[1pt]
 $\Sigma_c(2455)^{++}$ & $\ast\ast\ast\ast$& $1({1\over 2}^+$) & $2453.97 \pm 0.14$ &
 $1.89^{+0.09}_{-0.18}$ & BNL \cite{Cazzoli:1975et} & $\Lambda_c\pi$ \\
 $\Sigma_c(2455)^{+}$ & $\ast\ast\ast\ast$& $1({1\over 2}^+)$ & $2452.9\pm0.4$ &
 $<4.6$ & TST \cite{Calicchio:1980sc} & $\Lambda_c\pi$\\
 $\Sigma_c(2455)^{0}$ & $\ast\ast\ast\ast$& $1({1\over 2}^+)$ & $2453.75 \pm 0.14$
 & $1.83^{+0.11}_{-0.19}$ & BNL \cite{Cazzoli:1975et} & $\Lambda_c\pi$ \\
 $\Sigma_c(2520)^{++}$ & $\ast\ast\ast$& $1({3\over 2}^+)$ & $2518.41^{+0.21}_{-0.19}$
 & $14.78^{+0.30}_{-0.40}$ & SKAT \cite{Ammosov:1993pi} & $\Lambda_c\pi$\\
 $\Sigma_c(2520)^{+}$ & $\ast\ast\ast$& $1({3\over 2}^+)$ & $2517.5\pm2.3$
 & $<17$ & CLEO \cite{Ammar:2000uh} & $\Lambda_c\pi$ \\
 $\Sigma_c(2520)^{0}$ & $\ast\ast\ast$& $1({3\over 2}^+)$ & $2518.48\pm0.20$
 & $15.3^{+0.4}_{-0.5}$ & CLEO \cite{Brandenburg:1996jc} & $\Lambda_c\pi$ \\
 $\Sigma_c(2800)^{++}$ & $\ast\ast\ast$& $1(?^?)$ & $2801^{+4}_{-6}$ & $75^{+22}_{-17}$ & Belle \cite{Mizuk:2004yu} &
 $\Lambda_c\pi,\Sigma_c^{(*)}\pi,\Lambda_c\pi\pi$ \\
 $\Sigma_c(2800)^{+}$ & $\ast\ast\ast$& $1(?^?)$ & $2792^{+14}_{-~5}$ & $62^{+60}_{-40}$ & Belle \cite{Mizuk:2004yu} &
 $\Lambda_c\pi,\Sigma_c^{(*)}\pi,\Lambda_c\pi\pi$ \\
 $\Sigma_c(2800)^{0}$ & $\ast\ast\ast$& $1(?^?)$ & $2806^{+5}_{-7}$ & $72^{+22}_{-15}$ & Belle \cite{Mizuk:2004yu} &
 $\Lambda_c\pi,\Sigma_c^{(*)}\pi,\Lambda_c\pi\pi$\\
\midrule[1pt]
 $\Xi_c^+$ &$\ast\ast\ast$ & $\frac{1}{2}(\frac{1}{2}^+)$ & $2467.93^{+0.28}_{-0.40}$ &$(442\pm26)\times 10^{-15} $ s & CERN  \cite{Biagi:1983en} & weak \\
 $\Xi_c^0$ & $\ast\ast\ast$& $\frac{1}{2}(\frac{1}{2}^+)$ & $2470.85^{+0.28}_{-0.40}$ &$(112^{+13}_{-10})\times 10^{-15} $ s & CLEO \cite{Avery:1988uh} & weak \\
 $\Xi'^+_c$ & $\ast\ast\ast$& $\frac{1}{2}(\frac{1}{2}^+)$ & $2575.7\pm3.0$ & -- & CLEO \cite{Jessop:1998wt}  & $\Xi_c\gamma$ \\
 $\Xi'^0_c$ & $\ast\ast\ast$& $\frac{1}{2}(\frac{1}{2}^+)$ & $2577.9\pm2.9$ & -- &CLEO \cite{Jessop:1998wt}  & $\Xi_c\gamma$ \\
 $\Xi_c(2645)^+$ & $\ast\ast\ast$& $\frac{1}{2}(\frac{3}{2}^+)$ & $2645.9 \pm 0.5$ & $2.6\pm0.5$ & CLEO \cite{Gibbons:1996yv} & $\Xi_c\pi$ \\
 $\Xi_c(2645)^0$ & $\ast\ast\ast$& $\frac{1}{2}(\frac{3}{2}^+)$ & $2645.9\pm0.5$ & $<5.5$ & CLEO \cite{Avery:1995ps} & $\Xi_c\pi$ \\
 $\Xi_c(2790)^+$ & $\ast\ast\ast$& $\frac{1}{2}(\frac{1}{2}^-)$ & $2789.1\pm3.2$ & $<15$ & CLEO \cite{Csorna:2000hw} & $\Xi'_c\pi$\\
 $\Xi_c(2790)^0$ & $\ast\ast\ast$& $\frac{1}{2}(\frac{1}{2}^-)$ & $2791.9\pm3.3$ & $<12$ & CLEO \cite{Csorna:2000hw} & $\Xi'_c\pi$ \\
 $\Xi_c(2815)^+$ & $\ast\ast\ast$& $\frac{1}{2}(\frac{3}{2}^-)$ & $2816.6\pm0.9$ & $<3.5$ & CLEO \cite{Alexander:1999ud}  & $\Xi^*_c\pi,\Xi_c\pi\pi,\Xi_c'\pi$ \\
 $\Xi_c(2815)^0$ & $\ast\ast\ast$& $\frac{1}{2}(\frac{3}{2}^-)$ & $2819.6\pm1.2$ & $<6.5$ & CLEO \cite{Alexander:1999ud}  & $\Xi^*_c\pi,\Xi_c\pi\pi,\Xi_c'\pi$ \\
$\Xi_c(2930)^0$ & $\ast$& $?(?^?)$ & $2931\pm6$ & $36\pm13$
 & BaBar \cite{Aubert:2007eb} & $\Lambda_c \ov K$ \\
 $\Xi_c(2980)^+$ & $\ast\ast\ast$& $\frac{1}{2}(?^?)$ & $2970.7\pm2.2$ & $17.9\pm3.5$
 & Belle \cite{Chistov:2006zj} & $\Sigma_c \ov K,\Lambda_c \ov K\pi,\Xi_c\pi\pi$  \\
 $\Xi_c(2980)^0$ & $\ast\ast\ast$& $\frac{1}{2}(?^?)$ & $2968.0\pm2.6$ & $20\pm7$
 &Belle \cite{Chistov:2006zj} & $\Sigma_c \ov K,\Lambda_c \ov K\pi,\Xi_c\pi\pi$ \\
 $\Xi_c(3055)^+$ & $\ast\ast\ast$& $?(?^?)$ & $3055.1\pm1.7$ & $11\pm4$ & BaBar \cite{Aubert:2007dt} &
 $\Sigma_c \ov K,\Lambda_c \ov K\pi,D\Lambda$  \\
 $\Xi_c(3080)^+$ & $\ast\ast\ast$& $\frac{1}{2}(?^?)$ & $3076.94\pm0.28$ & $4.3\pm1.5$ & Belle \cite{Chistov:2006zj} &
 $\Sigma_c \ov K,\Lambda_c \ov K\pi,D\Lambda$  \\
 $\Xi_c(3080)^0$ & $\ast\ast\ast$& $\frac{1}{2}(?^?)$ & $3079.9\pm1.4$ & $5.6\pm2.2$
 & Belle \cite{Chistov:2006zj} & $\Sigma_c \ov K,\Lambda_c \ov K\pi,D\Lambda$ \\
$\Xi_c(3123)^+$ & $\ast$& $?(?^?)$ & $3122.9\pm1.3$ & $4.4\pm3.8$
 &BaBar \cite{Aubert:2007dt} & $\Sigma_c^* \ov K,\Lambda_c \ov K\pi$ \\
\midrule[1pt]
 $\Omega_c^0$ & $\ast\ast\ast$& $0({1\over 2}^+)$ & $2695.2\pm1.7$ &$(69\pm12)\times 10^{-15} $ s &WA62 \cite{Biagi:1984mu} & weak \\
 $\Omega_c(2770)^0$ &$\ast\ast\ast$ & $0({3\over 2}^+)$ & $2765.9\pm2.0$ & -- &Belle \cite{Solovieva:2008fw} & $\Omega_c\gamma$ \\
\bottomrule[1pt]
\end{tabular}
\end{center}
\end{table*}

\subsubsection{$\Lambda_c^+$.}

The lowest-lying charmed baryon is the $\Lambda_c^+$ ground state.
It was first reported by Fermilab in 1976 \cite{Knapp:1976qw}. To
date, there are numerous measurements for the branching fractions of
the $\Lambda_c^+$ \cite{Olive:2016xmw,Ablikim:2016tze}. Based on
the $\Lambda_c^+\to \Lambda K_S^0K^+$ and $\Lambda_c^+\to
\Sigma^0K_S^0K^+$ decay modes, BaBar reported the most precise
measurement of the $\Lambda_c^+$ mass in 2005:
\begin{equation}
m=(2286.46\pm 0.14)\, \mbox{MeV}\, ,
\end{equation}
which is now the most precise mass measurement of an open-charm
hadron and adopted by PDG for the $\Lambda_c^+$ mass without the
other measurement for averages.

\subsubsection{$\Lambda_c(2595)^+$ and $\Lambda_c(2625)^+$.}

The $\Lambda_c(2625)^+$ was the first $\Lambda_c^+$ orbital
excitation observed in the $\Lambda_c^+\pi^+\pi^-$ final states by
the ARGUS Collaboration at DESY in 1993 \cite{Albrecht:1993pt},
which was confirmed in the same channel in 1995 by the CLEO
Collaboration \cite{Edwards:1994ar}. In addition, the CLEO
Collaboration also reported another orbital excitation
$\Lambda_c(2595)^+$ in the same paper \cite{Edwards:1994ar}, which
was soon confirmed by the E687 Collaboration
\cite{Frabetti:1993hg,Frabetti:1995sb} and the ARGUS Collaboration
\cite{Albrecht:1997qa}.

The $\Lambda_c(2595)^+$ and $\Lambda_c(2625)^+$ lie above the
$\Sigma_c\pi$ threshold, which is the predominant S-wave decay mode
of the $\Lambda_c(2595)^+$. However, the $\Lambda_c(2625)^+$ can
only decay into $\Sigma_c\pi$ via the D-wave decay. Since both
the $\Lambda_c(2595)^+$ and $\Lambda_c(2625)^+$ have not been seen in
the $\Lambda_c^+\pi^0$ channel, they can not be the excited
$\Sigma_c^+$ states.

\subsubsection{$\Lambda_c(2765)^+$, $\Lambda_c(2880)^+$, and \bf$\Lambda_c(2940)^+$.}

\begin{figure}[htb]
\begin{center}
\includegraphics[width=0.48\textwidth]{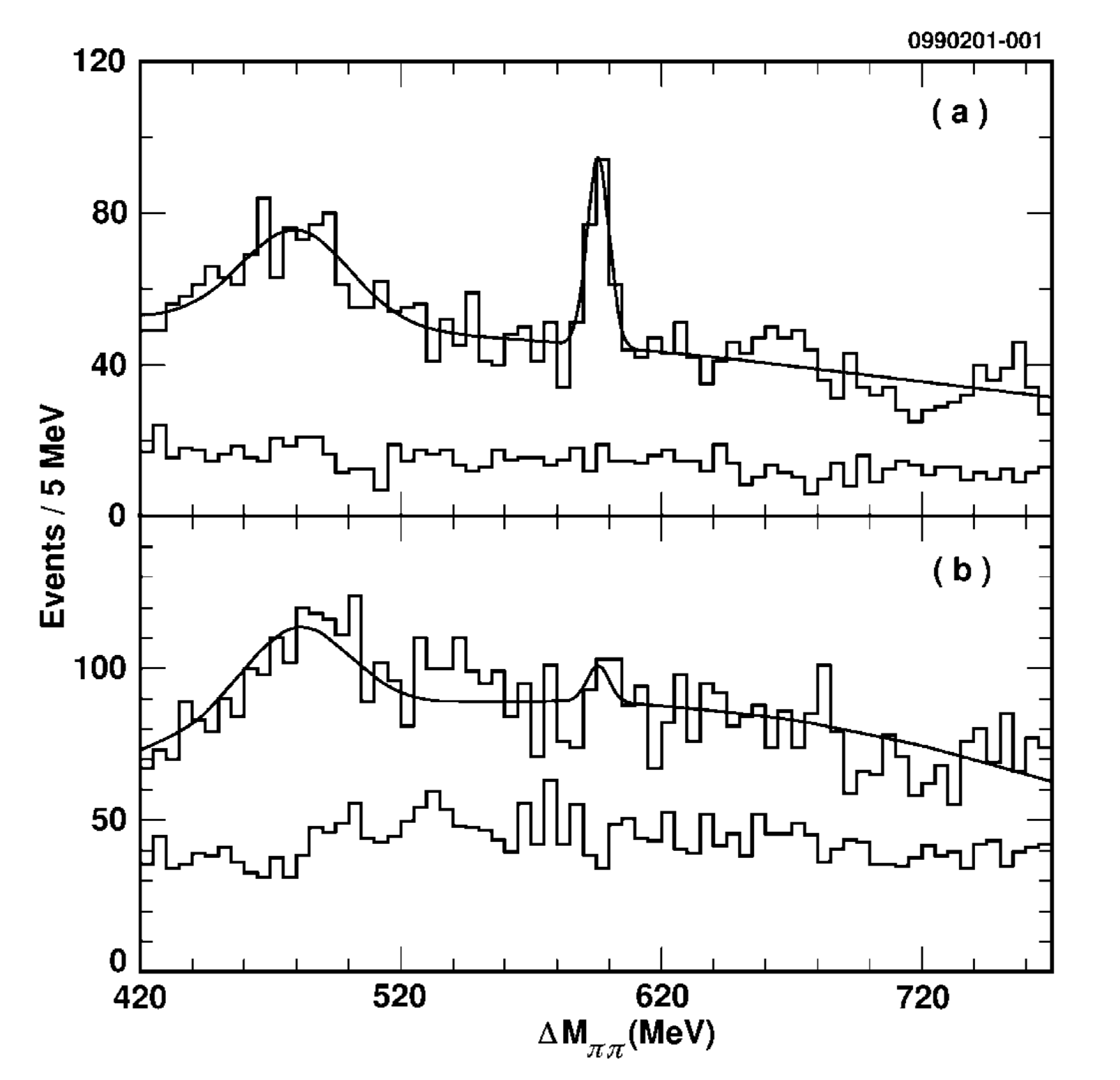}
\caption{The signal for the $\Lambda_c(2765)^+$ shown in $\Delta
M_{\pi\pi}= M(\Lambda_c^+\pi^+\pi^-)-M(\Lambda_c^+)$
with cuts that (a) $\Delta M_{\pi}= M(\Lambda_c^+\pi)-M(\Lambda_c^+)$ is
consistent with that expected for a $\Sigma_c$, and (b) $\Delta M_{\pi}= M(\Lambda_c^+\pi)-M(\Lambda_c^+)$ is consistent with that
expected for a $\Sigma^*_c$.
Taken from CLEO~\cite{Artuso:2000xy}.} \label{Fig:2.2.L2765}
\end{center}
\end{figure}

The CLEO Collaboration reported the
$\Lambda_c(2765)^+$ and $\Lambda_c(2880)^+$ in the
$\Lambda_c^+\pi^+\pi^-$ channel in 2001 \cite{Artuso:2000xy}. Since
the $\Lambda_c^+\pi^+\pi^-$ final states are accessible to both the
isoscalar and isovector channels, the $\Lambda_c(2765)^+$ can be
either a $\Lambda_c^+$ or a $\Sigma_c^+$ state. As shown in Fig.
\ref{Fig:2.2.L2765}, the $\Lambda_c(2765)^+$ was a rather broad
structure, which could be due to multiple overlapping states. The
$\Lambda_c(2765)^+$ was also observed by the Belle Collaboration
\cite{Abe:2006rz}.

\begin{figure}[htb]
\begin{center}
\includegraphics[width=0.48\textwidth]{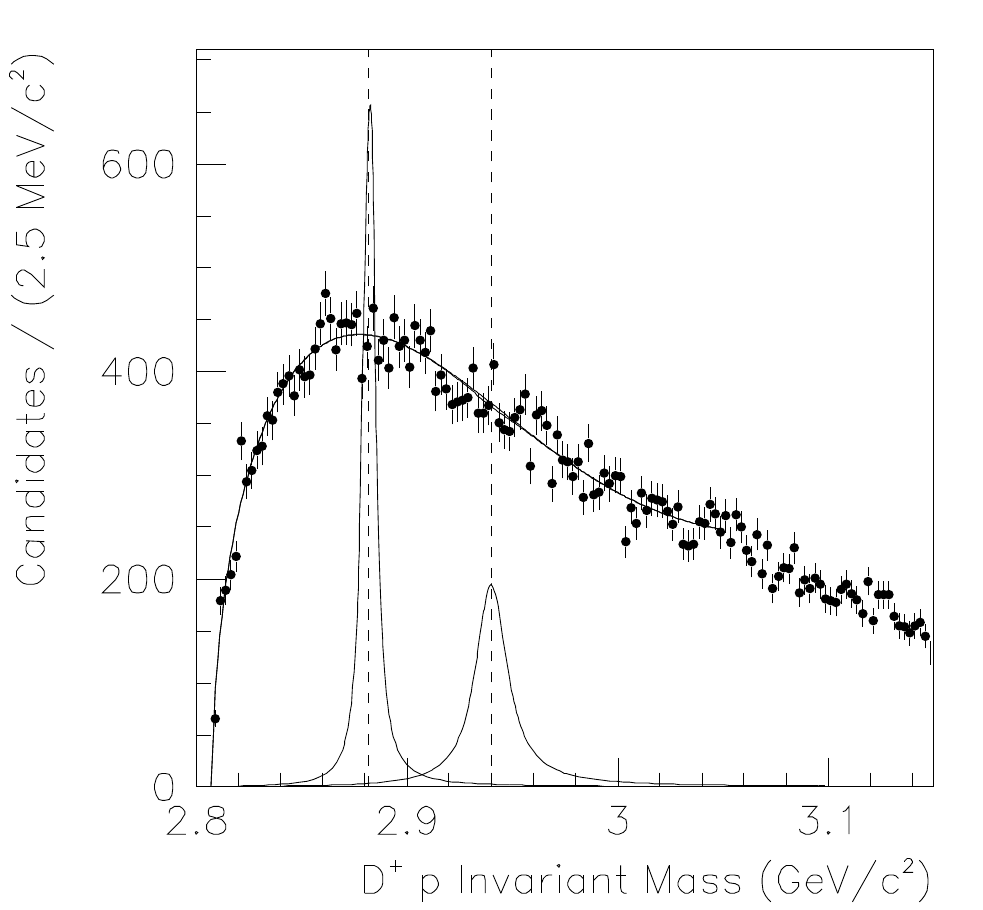}
\caption{The signals for the $\Lambda_c(2880)^+$ and
$\Lambda_c(2940)^+$ in the $D^0p$ invariant mass distribution.
Taken from BaBar~\cite{Aubert:2006sp}.} \label{Fig:2.2.L2880Barbar}
\end{center}
\end{figure}



For the $\Lambda_c(2880)^+$, CLEO did not determine its quantum
numbers \cite{Artuso:2000xy}. In 2007, the BaBar Collaboration
observed two narrow charmed baryons $\Lambda_c(2880)^+$ and
$\Lambda_c(2940)^+$ in the invariant $D^0p$ mass distribution
\cite{Aubert:2006sp}, as shown in Fig. \ref{Fig:2.2.L2880Barbar}.
The Belle Collaboration studied the $\Lambda_c\pi^+\pi^-$ channel
and also observed these two states \cite{Abe:2006rz}.
Belle studied the decay angular distribution of the
$\Lambda_c(2880)^+\to\Sigma_c^{0,++}\pi^\pm$ decay corresponding to
different spin hypothesis of the $\Lambda_c(2880)^+$. The result of
the fit favors the spin 5/2. To determine the parity of the
$\Lambda_c(2880)^+$, Belle also measured the ratio
\begin{equation}
R \equiv
\frac{\Gamma(\Sigma_c(2520)\pi)}{\Gamma(\Sigma_c(2455)\pi)}=0.225\pm
0.062\pm 0.025\, ,
\end{equation}
which is consistent with the prediction of the heavy quark effective
theory for the $5/2^+$ state in Refs.
\cite{Cheng:2006dk,Isgur:1991wq,Valcarce:2008dr}. However, there still exists
controversy on its parity, which needs to be determined in future
experiments.

\subsubsection{$\Sigma_c(2455)$ and $\Sigma_c(2520)$.}

\begin{figure}[htb]
\begin{center}
\includegraphics[width=0.48\textwidth]{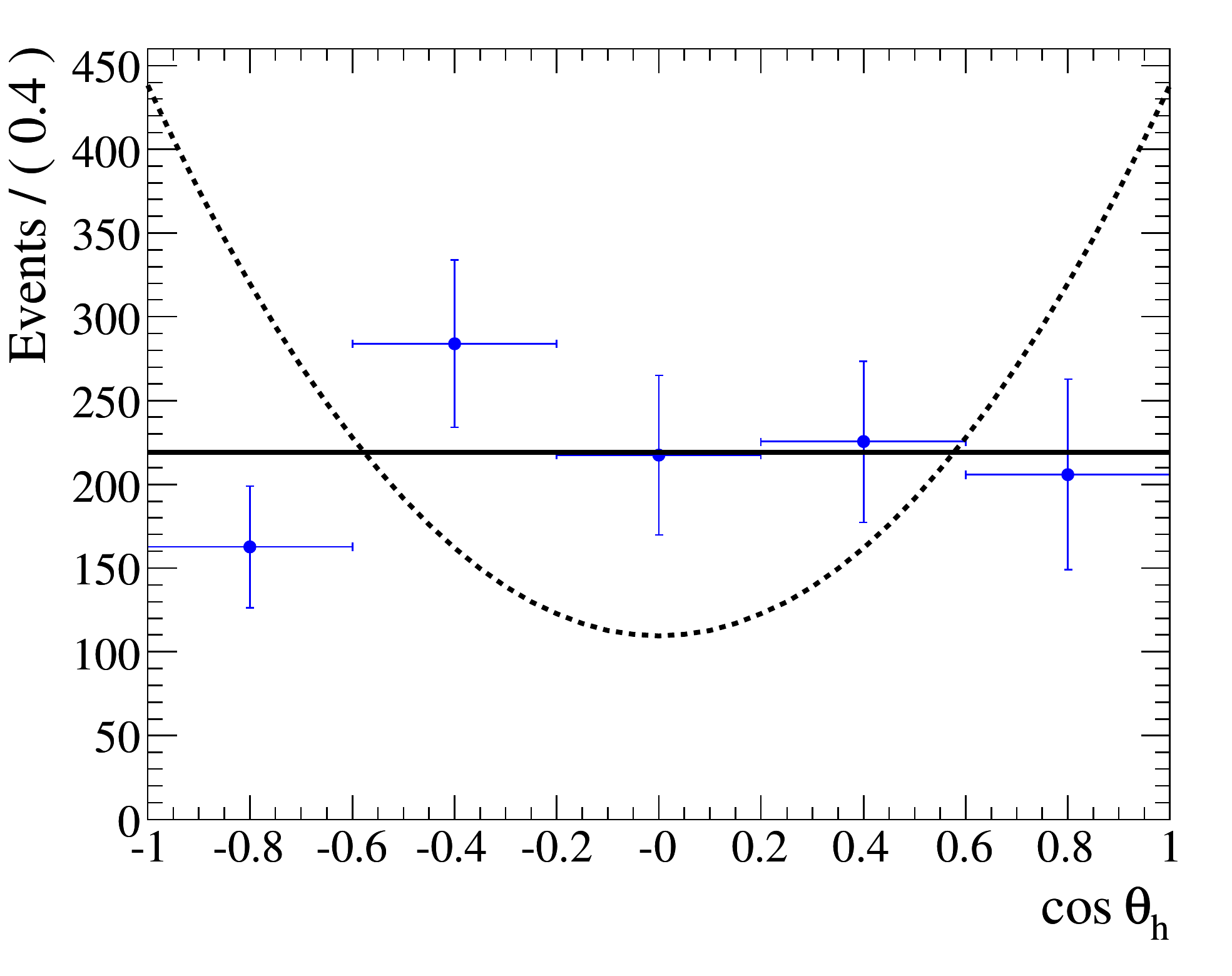}
\caption{(Color online) The efficiency-corrected helicity angle distribution for
$\Sigma_c(2455)^0$ candidate (points). The solid and dashed lines correspond to the
spin-1/2 and 3/2 hypothesis respectively.
Taken from BaBar~\cite{Aubert:2008ax}.
}
\label{Fig:2.2.S2455Angle}
\end{center}
\end{figure}

The $\Sigma_c(2455)$ and $\Sigma_c(2520)$ are the two lowest lying
$\Sigma_c$ ground states, which are well established and confirmed
by several experiments \cite{Olive:2016xmw}. The
$\Sigma_c(2455)^{++}$ and $\Sigma_c(2455)^0$ were observed by BNL
many years ago \cite{Cazzoli:1975et} while their isospin partner
$\Sigma_c(2455)^+$ is much more difficult to be detected. It was
first reported by the BEBC TST Neutrino Collaboration in 1980
\cite{Calicchio:1980sc} and confirmed in 1993 by CLEO
\cite{Crawford:1993pv}. The $\Sigma_c(2520)^{++}$ and
$\Sigma_c(2520)^{0}$ were reported in Refs.
\cite{Ammosov:1993pi,Brandenburg:1996jc} while the
$\Sigma_c(2520)^{+}$ state was observed later in Ref.
\cite{Ammar:2000uh} based on its decay into $\Lambda_c^+\pi^0$. In
Ref. \cite{Athar:2004ni}, CLEO measured the masses and decay widths
for $\Sigma_c(2520)^{++}$ and $\Sigma_c(2455)^{0}$, which were much
improved by the CDF Collaboration \cite{Aaltonen:2011sf}. The
angular momentum of $\Sigma_c(2455)$ has been measured by the BaBar
Collaboration by reconstructing $B^-\to\Lambda_c^+\bar p\pi^-$ decay
proceeding via $\Sigma_c(2455)^0$ \cite{Aubert:2008ax}, as shown in
Fig. \ref{Fig:2.2.S2455Angle}. The angular distribution prefers the
spin-1/2 hypothesis for $\Sigma_c(2455)^0$.

\subsubsection{$\Sigma_c(2800)$.}

\begin{figure}[htb]
\begin{center}
\includegraphics[width=1.0\textwidth]{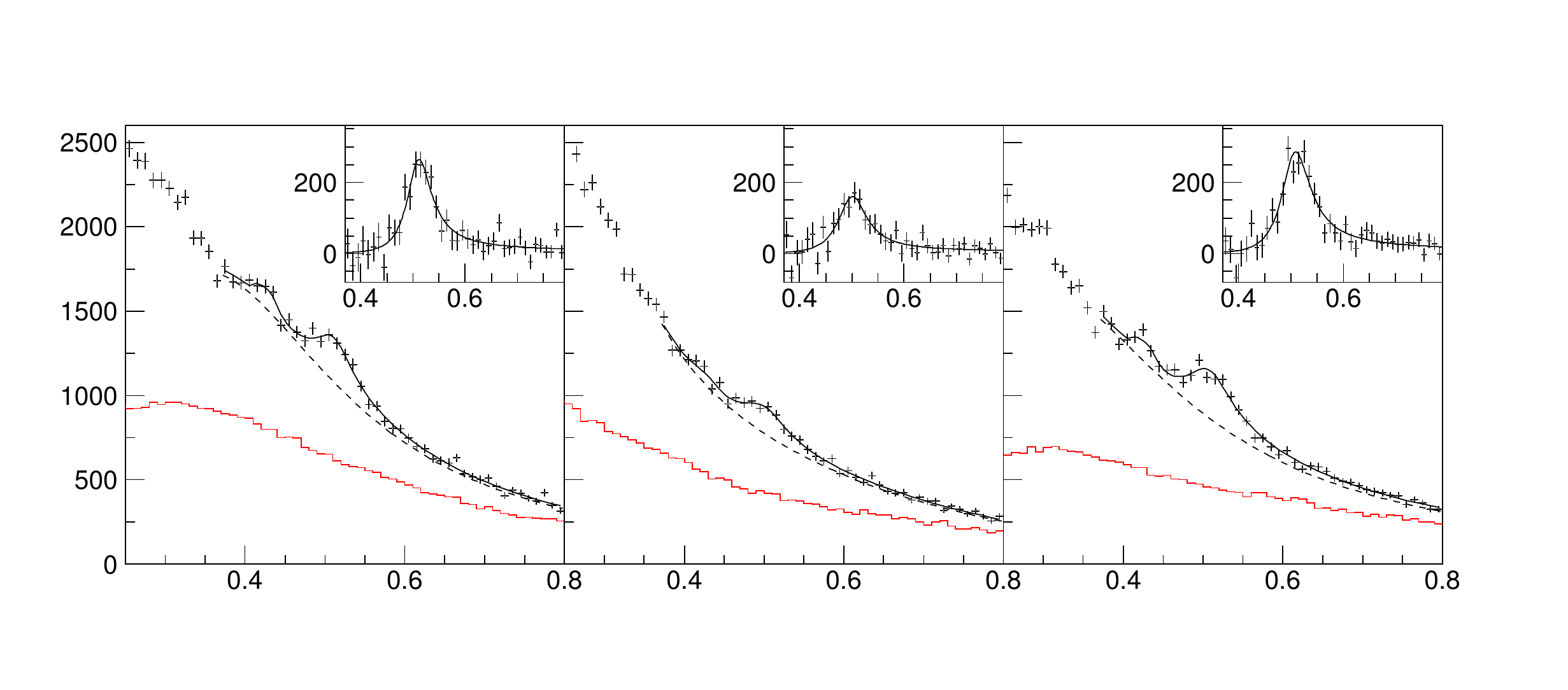}
\caption{(Color online) $M(\Lambda_c^+\pi)-M(\Lambda_c^+)$ distributions of the
selected $\Lambda_c^+\pi^-$ (left), $\Lambda_c^+\pi^0$ (middle), and
$\Lambda_c^+\pi^+$ (right) combinations.
The signals correspond to the $\Sigma_c(2800)$.
Taken from Belle~\cite{Mizuk:2004yu}.} \label{Fig:2.2.S2800}
\end{center}
\end{figure}

The $\Sigma_c(2800)$ resonance was observed in the $\Lambda_c^+\pi$
invariant mass distribution by the Belle Collaboration in 2005
\cite{Mizuk:2004yu}, as shown in Fig. \ref{Fig:2.2.S2800}. As
mentioned above, BaBar also reported a neutral charmed baryon state
in the decay $B^-\to\Lambda_c^+\pi^-\bar p$ \cite{Aubert:2008ax}.
Although the widths are consistent in both experiments, the fitted
mass is $(2846\pm8\pm10)\, \mbox{MeV}$ by BaBar
\cite{Aubert:2008ax}, about $3\sigma$ away from that of
$\Sigma_c(2800)^0$ reported by Belle \cite{Mizuk:2004yu}. If this
mass discrepancy is genuine and the states are distinct, the neutral
state seen by BaBar would be one of the missing $\Sigma_c$
resonances.

\subsubsection{$\Xi_c$, $\Xi_c^{\prime}$, and $\Xi_c(2645)$.}

The $\Xi_c^+$, $\Xi_c^0$ and $\Xi_c^{\prime +}$, $\Xi_c^{\prime 0}$
resonances form two isospin doublets, respectively. They all contain
three quarks with different flavors: $csu$ for the charged states
while $csd$ for the neutral states. The $\Xi_c^+$ ground state was
first observed in the reaction $\Sigma^-+Be\to (\Lambda
K^-\pi^+\pi^+)+X$ in an experiment at the CERN SPS hyperon beam
\cite{Biagi:1983en}. Its isospin partner $\Xi_c^0$ was discovered
later in the $\Xi^-\pi^+$ final states by the CLEO Collaboration in
1988 \cite{Avery:1988uh}. The mass splitting was measured to be
$m_{\Xi_c^0}-m_{\Xi_c^+}=(2.9\pm0.5)$ MeV. As shown in Table
\ref{Table:2.2.singlycharmedbaryons}, PDG identified their quantum
numbers to be $I(J^P)=1/2(1/2^+)$, where the
$J^P=1/2^+$ is the quark-model prediction~\cite{Olive:2016xmw}.

In 1998, CLEO discovered the second isospin doublet, $\Xi_c^{\prime
+}$ and $\Xi_c^{\prime 0}$, in their decays into $\Xi_c^+\gamma$ and
$\Xi_c^0\gamma$ \cite{Jessop:1998wt}, respectively. These two
resonances were explained as the symmetric partners of the
well-established antisymmetric $\Xi_c^+$ and $\Xi_c^0$. Their mass
differences were measured to be
\begin{eqnarray}
m_{\Xi_c^{\prime +}}-m_{\Xi_c^+}=(107.8\pm1.7\pm2.5)\, \mbox{MeV}\,
,
\\ \nonumber
m_{\Xi_c^{\prime 0}}-m_{\Xi_c^0}=(107.0\pm1.4\pm2.5)\, \mbox{MeV}\,
,
\end{eqnarray}
which are too small to allow the hadronic transitions
$\Xi_c^{\prime}\to \Xi_c\pi$. The only allowed decay modes between
them are the radiative decays, which were the observed channels.

In 1995, the $\Xi_c(2645)^0$ was reported by the CLEO Collaboration
in the $\Xi_c^+\pi^-$ final states with
$M(\Xi_c^+\pi^-)-M(\Xi_c^+)=178.2\pm0.5\pm1.0$ MeV and a width of
$\Gamma<5.5$ MeV \cite{Avery:1995ps}. Its charged partner
$\Xi_c(2645)^+$ was also reported later by CLEO in its decay mode
$\Xi_c^0\pi^+$ with $M(\Xi_c^0\pi^+)-M(\Xi_c^0)=174.3\pm0.5\pm1.0$ MeV
and a width of $\Gamma<3.1$ MeV \cite{Gibbons:1996yv}. The Belle
Collaboration confirmed these resonances with more precise mass
measurements in Ref. \cite{Lesiak:2008wz}. Although its spin-parity
has not been measured, the $\Xi_c(2645)$ was identified to be a
$J^P=3/2^+$ state in PDG~\cite{Olive:2016xmw}.

\subsubsection{$\Xi_c(2790)$ and $\Xi_c(2815)$.}

All these two excited $\Xi_c$ states were first observed by the CLEO
Collaboration. The $\Xi_c(2815)^+$ and $\Xi_c(2815)^0$ were first
observed by CLEO in the decays into $\Xi_c^{+(0)}\pi^+\pi^-$ via the
intermediate states $\Xi_c(2645)^+$ and $\Xi_c(2645)^0$
\cite{Alexander:1999ud}, respectively. Belle confirmed their
existence~\cite{Lesiak:2008wz}. The $\Xi_c(2815)$ states
were interpreted as the charmed-strange analogues of the
$\Lambda_c(2625)$, with $J^P=3/2^-$. Finally, the
$\Xi_c(2790)$ was observed in the decay $\Xi_c^\prime\pi$ by CLEO
\cite{Csorna:2000hw} and confirmed by Belle~\cite{Lesiak:2008wz},
which were explained as the charmed-strange
partners of the $\Lambda_c(2595)$ with $J^P=1/2^-$~\cite{Olive:2016xmw}.

\subsubsection{$\Xi_c(2930)^0$, $\Xi_c(2980)$, $\Xi_c(3055)$, $\Xi_c(3080)$, and $\Xi_c(3123)$.}

\begin{figure}[htb]
\begin{center}
\includegraphics[width=0.48\textwidth]{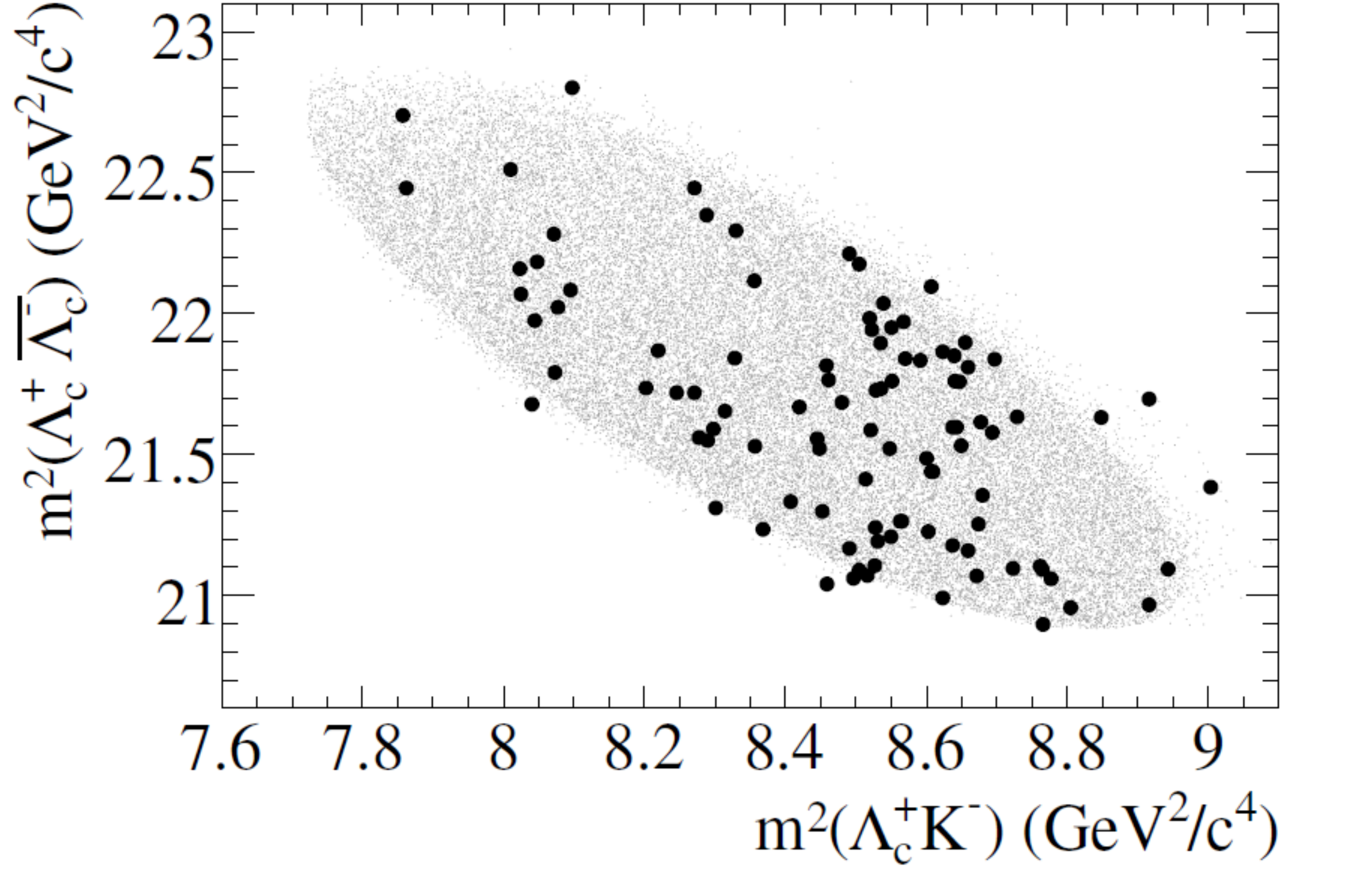}
\includegraphics[width=0.48\textwidth]{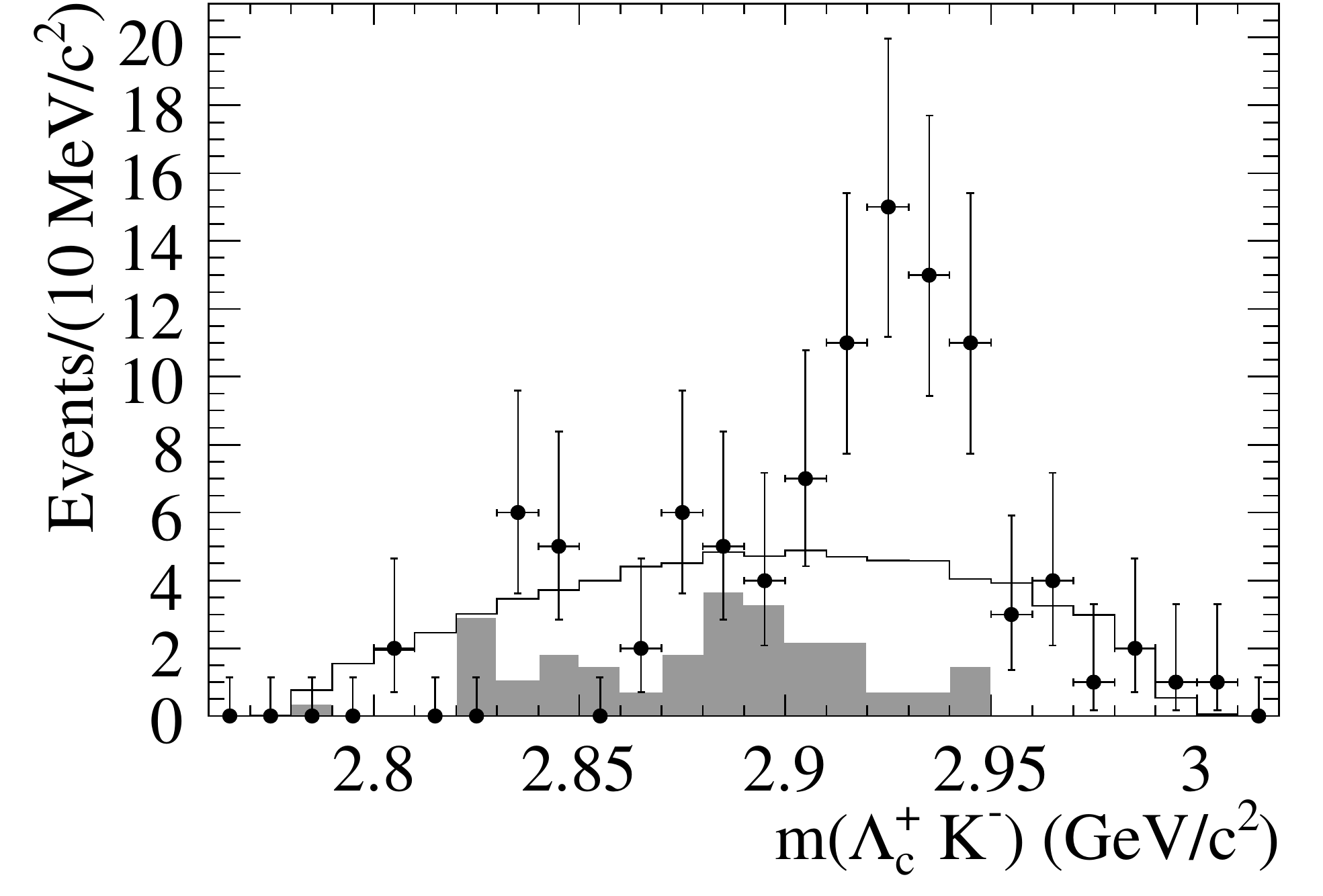}
\caption{Reconstructed $B^-\to \Lambda_c^+\bar \Lambda_c^-K^-$
candidate in the Dalitz plot (left) and the $\Lambda_c^+K^-$
invariant mass distribution (right).
The peak in the right panel corresponds to the $\Xi_c(2930)^0$.
Taken from BaBar~\cite{Aubert:2007eb}.}
\label{Fig:2.2X2930}
\end{center}
\end{figure}

The $\Xi_c(2930)^0$ state was seen in the $\Lambda_c^+K^-$ invariant
mass spectrum of the $B^-\to \Lambda_c^+\bar \Lambda_c^-K^-$ decay by
the BaBar Collaboration \cite{Aubert:2007eb}.
As shown in Fig.~\ref{Fig:2.2X2930}, both the Dalitz plot and the $\Lambda_c^+K^-$
projection supported the existence of a single resonance. However, a
more complicated explanation, such as two narrow resonances in close
proximity, cannot be excluded. PDG denotes the $\Xi_c(2930)^0$ to be
a 1-star resonance \cite{Olive:2016xmw}.

\begin{figure}[htb]
\begin{center}
\includegraphics[width=0.48\textwidth]{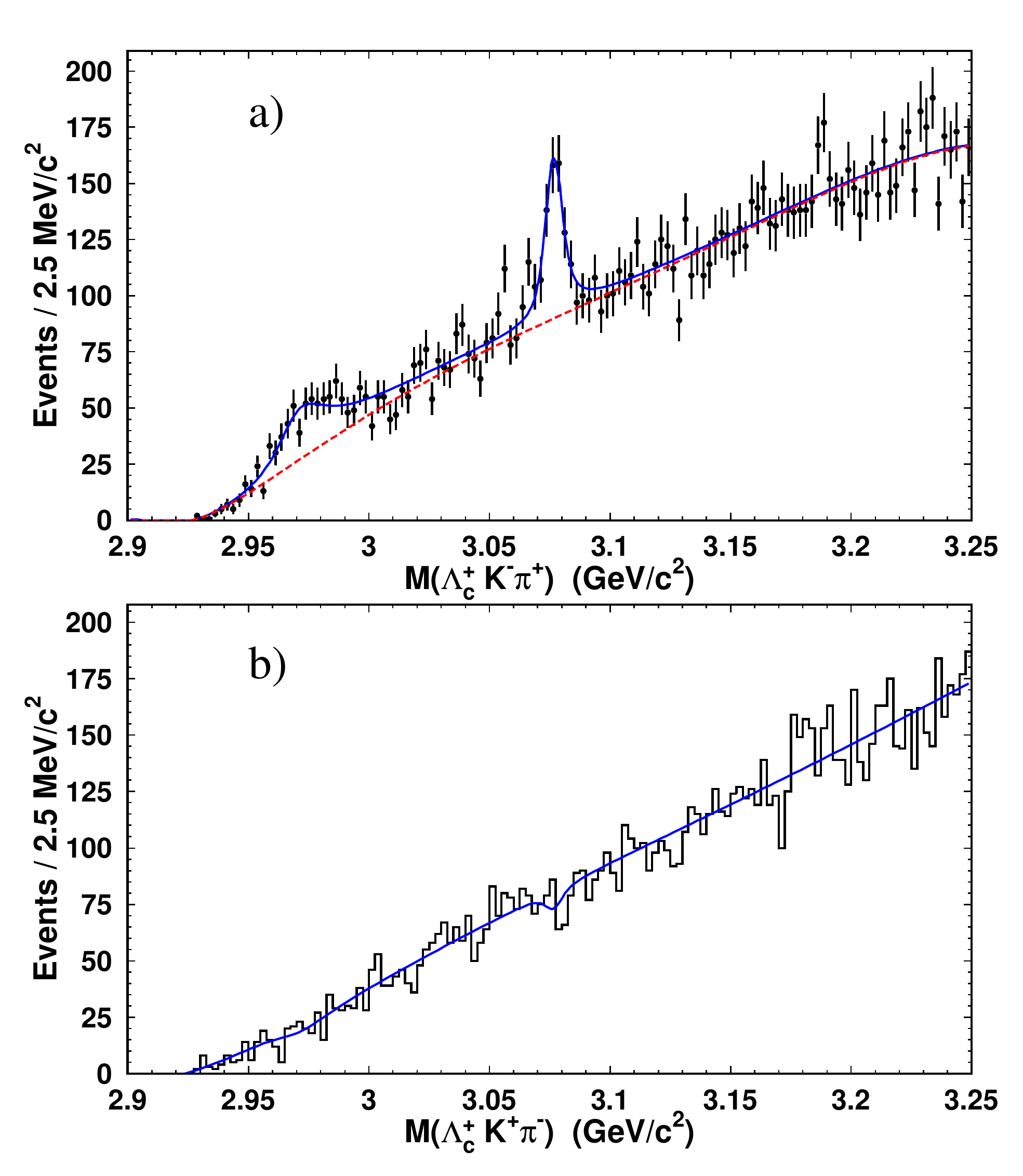}
\includegraphics[width=0.48\textwidth]{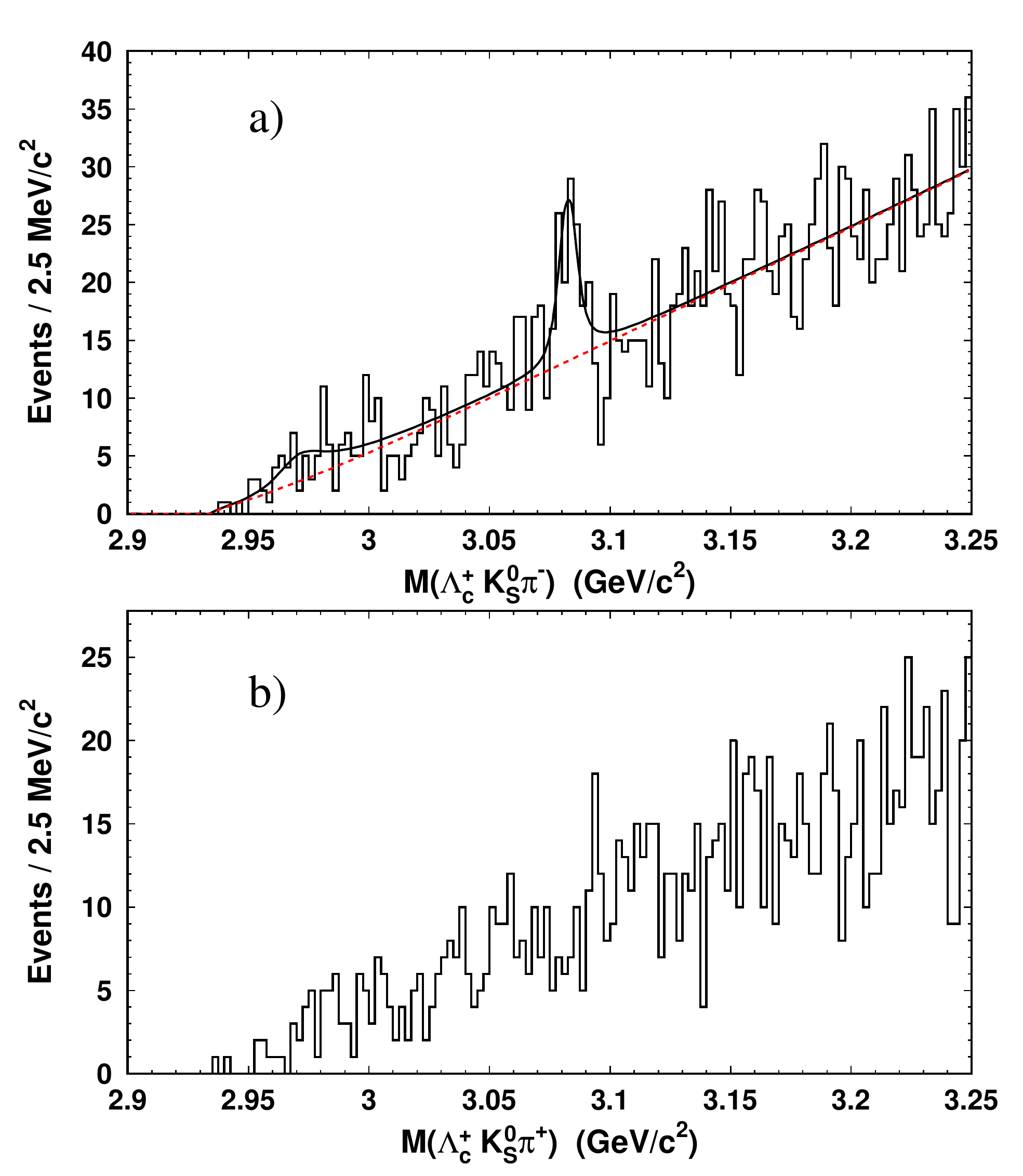}
\caption{(Color online) $M(\Lambda_c^+K^-\pi^+)$ and $M(\Lambda_c^+K^0_S\pi^-)$
distributions together with the overlaid fitting curves.
The curves are fitted with the $\Xi_c(2980)$ and $\Xi_c(3080)$.
Taken from Belle \cite{Chistov:2006zj}.} \label{Fig:2.2X2980}
\end{center}
\end{figure}

The remaining four resonances, $\Xi_c(2980)$, $\Xi_c(3055)$, $\Xi_c(3080)$, and $\Xi_c(3123)$, were all seen in the $\Lambda_c^+\bar
K\pi^+$ final states. In 2006, the Belle Collaboration reported two
new charmed-strange baryons, $\Xi_c(2980)$ and $\Xi_c(3080)$,
decaying into $\Lambda_c^+K^-\pi^+$ and $\Lambda_c^+K^0_S\pi^-$
\cite{Chistov:2006zj}, as shown in Fig. \ref{Fig:2.2X2980}. The
$\Xi_c(2980)$ was confirmed later by Belle in its decay into
$\Xi_c(2645)\pi$ \cite{Lesiak:2008wz}.

\begin{figure}[htb]
\begin{center}
\includegraphics[width=0.48\textwidth]{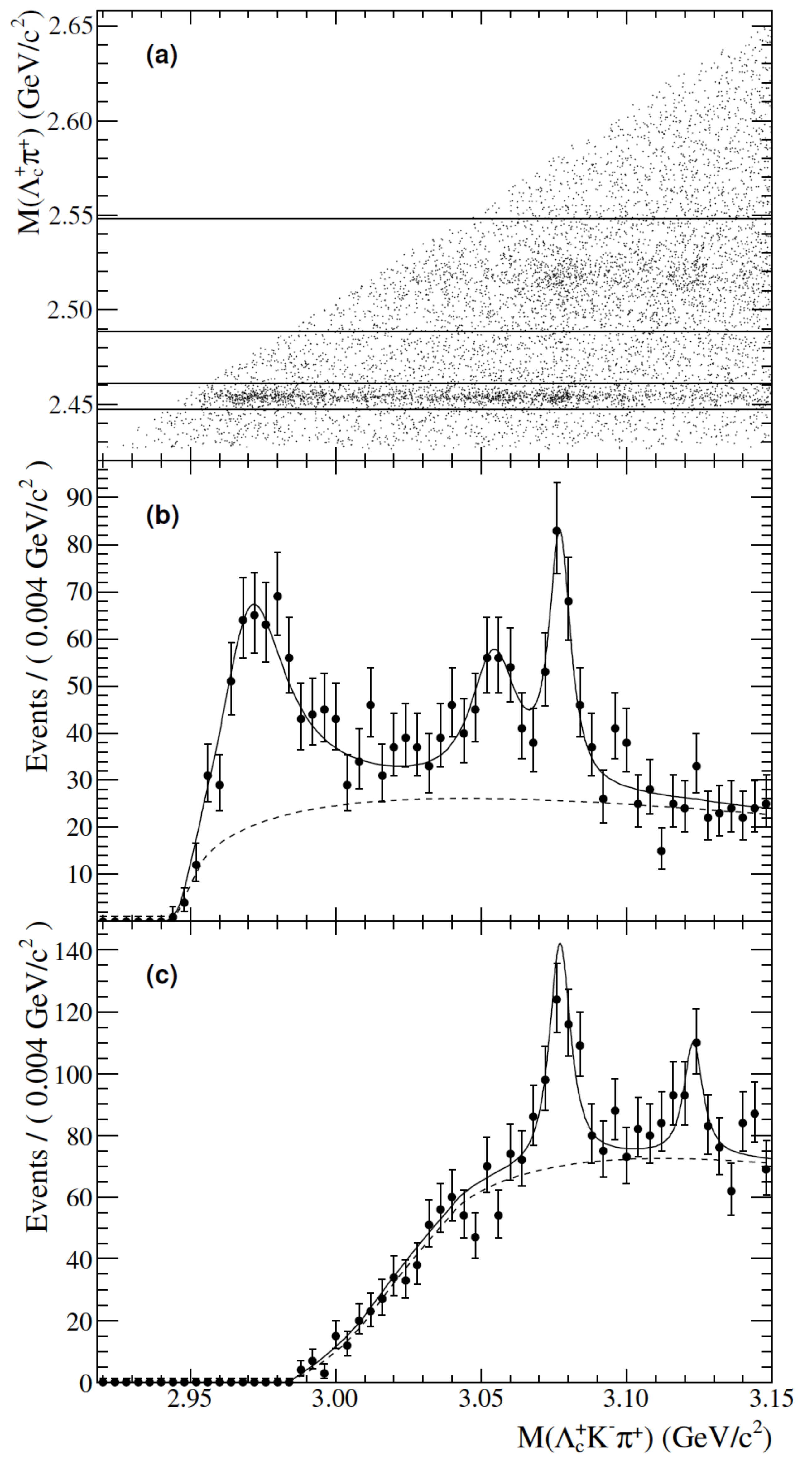}
\caption{The $M(\Lambda_c^+K^-\pi^+)$ distributions in the
$M(\Lambda_c^+\pi^+)$ ranges (b) within 3.0 natural widths of the
$\Sigma_c(2455)^{++}$ mass and (c) within 2.0 natural widths of the
$\Sigma_c(2520)^{++}$ mass.
The curves are fitted with the $\Xi_c(2980)$, $\Xi_c(3055)$, $\Xi_c(3080)$, and $\Xi_c(3123)$.
Taken from BaBar \cite{Aubert:2007dt}.}
\label{Fig:2.2X3123}
\end{center}
\end{figure}

These two states were also confirmed by the BaBar Collaboration~\cite{Aubert:2007dt},
in which two additional charmed-strange baryons, $\Xi_c(3055)$ and
$\Xi_c(3123)$, were reported. In Fig.~\ref{Fig:2.2X3123}, the
$M(\Lambda_c^+K^-\pi^+)$ distributions were shown in two ranges of
$M(\Lambda_c^+\pi^+)$, in which the $\Xi_c(2980)$ and $\Xi_c(3080)$
were clearly visible. Moreover, two new signals, $\Xi_c(3055)$ and
$\Xi_c(3123)$, were also observed with the resonance parameters
$m_{\Xi_c(3055)}=(3054.2\pm1.2\pm0.5)$ MeV,
$\Gamma_{\Xi_c(3055)}=(17\pm6\pm11)$ MeV,
$m_{\Xi_c(3123)}=(3122.9\pm1.3\pm0.3)$ MeV, and
$\Gamma_{\Xi_c(3123)}=(4.4\pm3.4\pm1.7)$ MeV. The
$\Xi_c(3123)$ had a limited statistical significance $3.6\sigma$,
which was identified as a 1-star state in PDG \cite{Olive:2016xmw}.
A recent experimental study on the $\Xi_c(3055)$ and $\Xi_c(3080)$
can be found in Ref.~\cite{Kato:2016hca} where the Belle Collaboration gave the following three branching ratios:
\begin{eqnarray}
 \nonumber {\mathcal{B}(\Xi_c(3055)^+ \to \Lambda D^+) \over
\mathcal{B}(\Xi_c(3055)^+ \to \Sigma_c^{++} K^-)} &=& 5.09 \pm 1.01
\pm 0.76 \, ,
\\ {\mathcal{B}(\Xi_c(3080)^+ \to \Lambda D^+) \over \mathcal{B}(\Xi_c(3080)^+ \to \Sigma_c^{++} K^-)} &=& 1.29 \pm 0.30 \pm 0.15 \, ,
\\ \nonumber {\mathcal{B}(\Xi_c(3080)^+ \to \Sigma_c^{*++} K^-) \over \mathcal{B}(\Xi_c(3080)^+ \to \Sigma_c^{++} K^-)} &=& 1.07 \pm 0.27 \pm 0.01 \, .
\end{eqnarray}
As listed in Table \ref{Table:2.2.singlycharmedbaryons}, the
quantum numbers for all these excited $\Xi_c$ states have not been
determined yet. More experimental information is required to
constrain the allowed possibilities.

Very recently, the Belle Collaboration reported the excited $\Xi_c$ states
decaying into $\Xi_c^0$ or $\Xi_c^+$ ground state, via the emission
of photons and/or charged pions~\cite{Yelton:2016fqw}. They
presented new measurements of the masses and decay widths of the
$\Xi^\prime$, $\Xi_c(2645), \Xi_c(2790), \Xi_c(2815)$, and
$\Xi_c(2980)$ isodoublets. These new mass measurements constitute
a considerable improvement in precision compared with previous
measurements~\cite{Olive:2016xmw}.

\subsubsection{$\Omega_c^0$ and $\Omega_c(2770)^0$.}

\begin{figure}[htb]
\begin{center}
\includegraphics[width=0.48\textwidth]{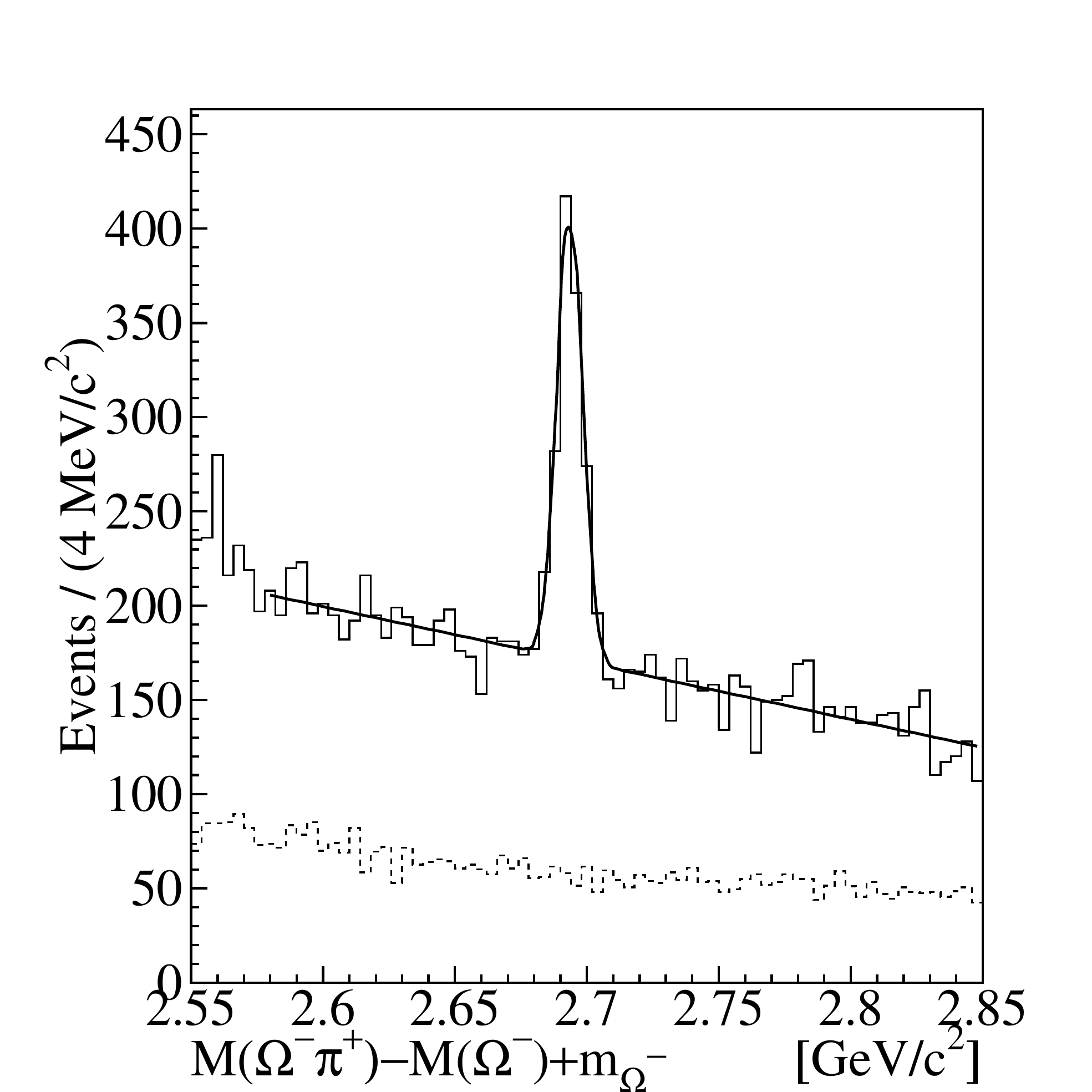}
\includegraphics[width=0.48\textwidth]{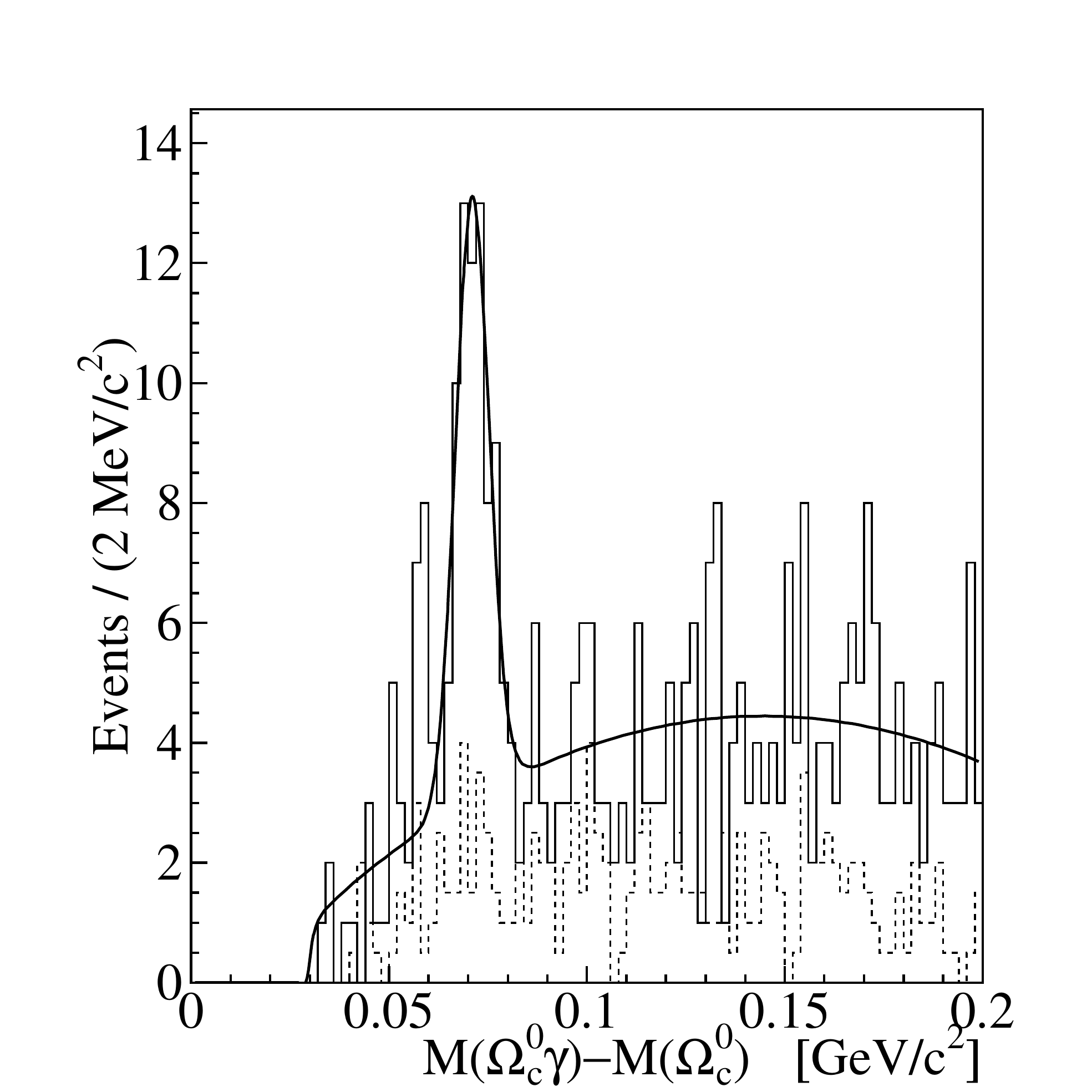}
\caption{Left: $M(\Omega^-\pi^+)-M(\Omega^-)+m_{\Omega^-}$ spectrum
in the $\Omega_c^0\to\Omega^-\pi^+$ process. Right:
$M(\Omega^0\gamma)-M(\Omega^0)$ spectrum in the $\Omega_c^{\ast
0}\to\Omega_c^0\gamma$ process.
The signal in the left panel corresponds to the $\Omega_c^0$ while the one in the right panel corresponds to the $\Omega_c(2770)^0$.
Taken from Belle
\cite{Solovieva:2008fw}.} \label{Fig:2.2O2695}
\end{center}
\end{figure}

The $\Omega_c^0$ ground state was first reported in 1985 by the
experiment WA62, which searched for the charmed-strange baryons in
the $\Sigma^-$ interaction in the SPS charged hyperon beam at CERN
\cite{Biagi:1984mu}. Although this signal was seen in several other
experiments
\cite{Frabetti:1994dp,Adamovich:1995pf,CroninHennessy:2000bz}, the
statistical significance was still limited due to very few events
(order of 10) before the B factory. In 2009, the Belle
Collaboration~\cite{Solovieva:2008fw} provided a more precise mass measurement of
the $\Omega_c^0$ to be $(2693.6\pm0.3^{+1.8}_{-1.5})$ MeV, which is
close to the PDG value shown in Table \ref{Table:2.2.singlycharmedbaryons}.
In the left panel of Fig.~\ref{Fig:2.2O2695}, the
$\Omega_c^0$ signal was clearly visible in the
$M(\Omega^-\pi^+)-M(\Omega^-)+m_{\Omega^-}$ spectrum of the
$\Omega_c^0\to\Omega^-\pi^+$ process.

\begin{figure}[htb]
\begin{center}
\includegraphics[width=0.48\textwidth]{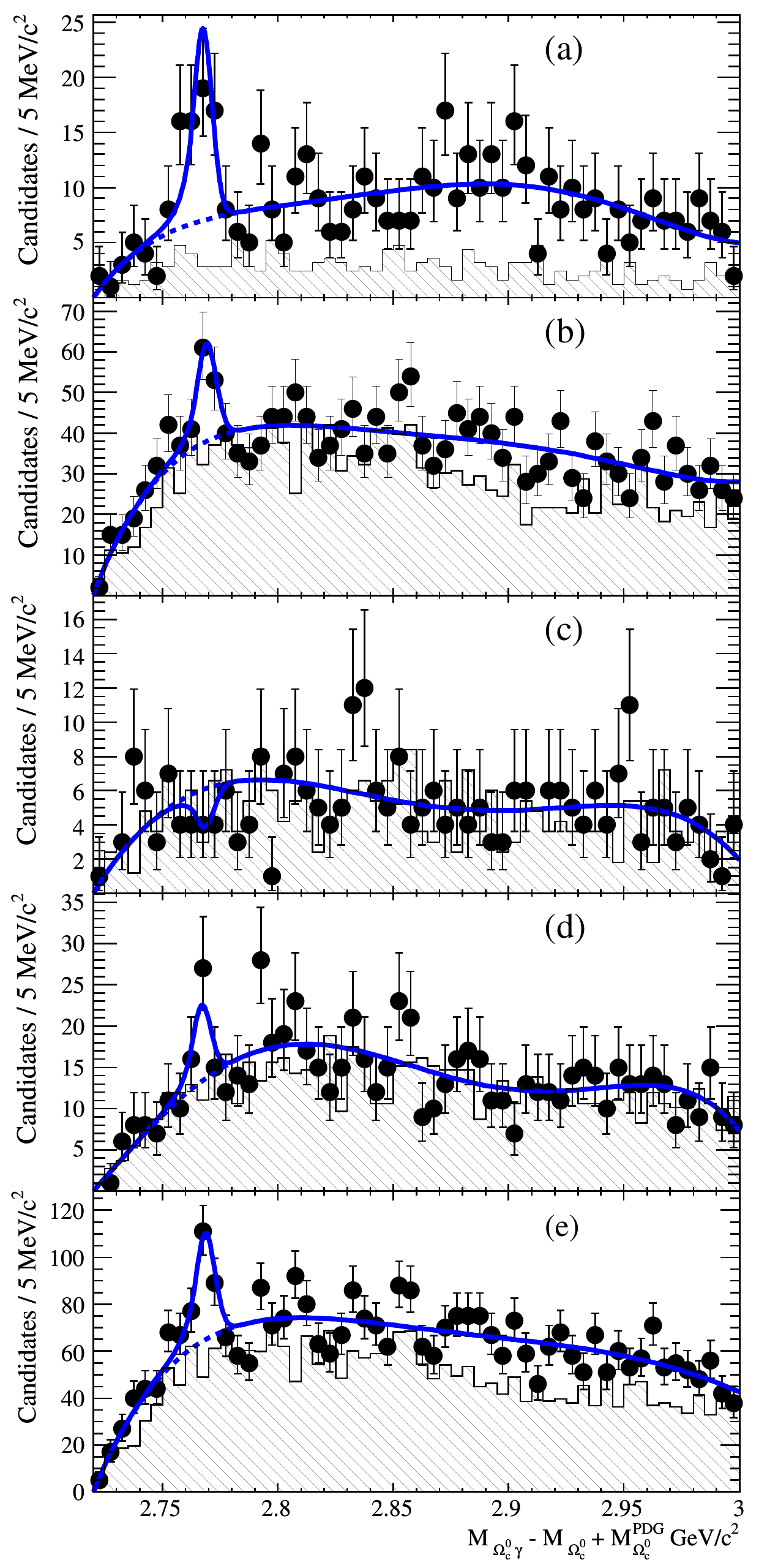}
\caption{(Color online) The invariant mass distributions of $\Omega_c^{\ast
0}\to\Omega_c^0\gamma$ candidates, with $\Omega_c^0$ reconstructed
in the decay modes (a) $\Omega^-\pi^+$, (b) $\Omega^-\pi^+\pi^0$,
(c) $\Omega^-\pi^+\pi^-\pi^+$, (d) $\Xi^-K^-\pi^+\pi^-$, and (e) for
the combined decay modes.
The signals correspond to the $\Omega_c(2770)^0$.
Taken from BaBar
\cite{Aubert:2006je}.} \label{Fig:2.2O2766}
\end{center}
\end{figure}

In the same experiment of Belle \cite{Solovieva:2008fw}, the excited
state $\Omega_c(2770)^0$ was also reconstructed in the
$\Omega^0\gamma$ mode, as shown in the right panel of Fig.
\ref{Fig:2.2O2695}. This resonance $\Omega_c(2770)^0$ was originally
discovered by BaBar in the same channel \cite{Aubert:2006je}, as
shown in Fig. \ref{Fig:2.2O2766}. Both BaBar and Belle measured the
mass difference of $m_{\Omega_c^{\ast 0}}-m_{\Omega_c^0}$ and their
results were both consistent with the PDG average value
$(70.7^{+0.8}_{-0.9})$ MeV \cite{Olive:2016xmw}. Such a mass
difference is too small for any hadronic strong decay to occur.
Although its $J^P$ has not been measured, the $\Omega_c(2770)^0$ was
predicted to be the $J^P=3/2^+$ partner of the $\Sigma_c(2520)$
and $\Xi_c(2645)$ \cite{Olive:2016xmw}. To date, no other radially
or orbitally excited $\Omega_c$ resonances have been discovered.


\subsection{The bottom baryons}
\label{sec2.6}

In this subsection we review the bottom baryons. The number of such baryons collected in PDG~\cite{Olive:2016xmw} is 10.
Their experimental information is listed in Table~\ref{sec22:Bbaryon}. All the $1S$
bottom baryons have been observed, except the $\Omega_b^*$ of $J^P =
3/2^+$. Hence, we only list their averaged masses and widths from
PDG~\cite{Olive:2016xmw} together with the experiments first
observing them, but we note that not all of them are well known.
There are only two excited bottom baryons observed in experiments,
the $\Lambda_b(5912)^0$ of $1/2^-$ and the $\Lambda_b(5920)^0$ of
$3/2^-$, and we list all the relevant experiments together with
their observed masses, widths, and decay modes therein. We shall
separately review them in the following.

\renewcommand{\arraystretch}{1.4}
\begin{table*}[htb]
\tiny
\caption{Experimental information of the observed bottom baryons.
All the $1S$ bottom baryons have been observed, except the
$\Omega_b^*$ of $J^P = 3/2^+$. Hence, we only list their averaged
masses and widths from PDG~\cite{Olive:2016xmw} together with the
experiments first observing them, but we note that not all of them
are well known. There are only two excited bottom baryons observed
in experiments, the $\Lambda_b(5912)^0$ of $1/2^-$ and the
$\Lambda_b(5920)^0$ of $3/2^-$, and we list all the relevant
experiments together with their observed masses, widths, and decay
modes therein. \label{sec22:Bbaryon} } \centering
\begin{tabular}{ccccccc}
\toprule[1pt] State & Status & $J^P$ & Mass (MeV) & Width (MeV) &
Experiments & Decay Modes
\\ \midrule[1pt]
$\Lambda_b^0$ & $\ast\ast\ast$ & $1/2^+$ & $5619.51 \pm 0.23$ &
$(1466 \pm 10) \times 10^{-15}$ s & CERN R415~\cite{Basile:1981wr} &
$p K^- \pi^+ \pi^-$
\\ \hline
$\Lambda_b(5912)^0$ & $\ast\ast\ast$ & $1/2^-$ & $5911.97 \pm 0.12
\pm 0.02 \pm 0.66$ & $< 0.66$ & LHCb~\cite{Aaij:2012da} &
$\Lambda_b^0 \pi^+ \pi^-$
\\ \hdashline[2pt/2pt]
\multirow{2}{*}{$\Lambda_b(5920)^0$} &
\multirow{2}{*}{$\ast\ast\ast$} & \multirow{2}{*}{$3/2^-$} &
$5919.77 \pm 0.08 \pm 0.02 \pm 0.66$ & $< 0.63$ &
LHCb~\cite{Aaij:2012da} & $\Lambda_b^0 \pi^+ \pi^-$
\\
                                                       &&& $5919.22 \pm 0.35 \pm 0.30 \pm 0.60$  & -- & CDF~\cite{Aaltonen:2013tta} & $\Lambda_b^0 \pi^+ \pi^-$
\\ \midrule[1pt]
$\Sigma_b^+$ & \multirow{2}{*}{$\ast\ast\ast$} & $1/2^+$ & $5811.3
^{+0.9}_{-0.8} \pm 1.7$ & $9.7 {^{+3.8}_{-2.8}} {^{+1.2}_{-1.1}}$ &
CDF~\cite{Aaltonen:2007ar} & $\Lambda_b \pi$
\\
$\Sigma_b^-$ && $1/2^+$ & $5815.5 ^{+0.6}_{-0.5} \pm 1.7$ & $4.9
^{+3.1}_{-2.1} \pm 1.1$ & CDF~\cite{Aaltonen:2007ar} & $\Lambda_b
\pi$
\\ \hdashline[2pt/2pt]
$\Sigma_b^{*+}$ & \multirow{2}{*}{$\ast\ast\ast$} & $3/2^+$ &
$5832.1 \pm 0.7 ^{+1.7}_{-1.8}$ & $11.5 {^{+2.7}_{-2.2}}
{^{+1.0}_{-1.5}}$ & CDF~\cite{Aaltonen:2007ar} & $\Lambda_b \pi$
\\
$\Sigma_b^{*-}$ && $3/2^+$ & $5835.1 \pm 0.6 ^{+1.7}_{-1.8}$ & $7.5
{^{+2.2}_{-1.8}} {^{+0.9}_{-1.4}}$ & CDF~\cite{Aaltonen:2007ar} &
$\Lambda_b \pi$
\\ \midrule[1pt]
$\Xi_b^0$ & \multirow{2}{*}{$\ast\ast\ast$} &
\multirow{2}{*}{$1/2^+$} & $5791.9 \pm 0.5$ & $(1464 \pm 31) \times 10^{-15}$ s &  \multirow{2}{*}{DELPHI~\cite{Abreu:1995kt}} &
semileptonic
\\
$\Xi_b^-$ & & & $5794.5 \pm 1.4$ & $(1560 \pm 40) \times
10^{-15}$ s
&& decays
\\ \hdashline[2pt/2pt]
$\Xi^\prime_b(5935)^-$ & $\ast\ast\ast$ & $1/2^+$ & $5935.02 \pm
0.02 \pm 0.05$ & $<0.08$ & LHCb~\cite{Aaij:2014yka} &
$\Xi^0_b \pi^-$
\\ \hline
$\Xi_b(5945)^0$ & $\ast\ast\ast$ & $3/2^+$ & $5948.9 \pm 0.8 \pm 1.4$ & $2.1 \pm 1.7$ & CMS~\cite{Chatrchyan:2012ni} & $\Xi^-_b
\pi^+$
\\
$\Xi^*_b(5955)^-$ & $\ast\ast\ast$ & $3/2^+$ & $5955.33 \pm 0.12 \pm
0.05$ & $1.65 \pm 0.31 \pm 0.10$ & LHCb~\cite{Aaij:2014yka}
& $\Xi^0_b \pi^-$
\\ \midrule[1pt]
$\Omega_b^-$ & $\ast\ast\ast$ & $1/2^+$ & $6046.4 \pm 1.9$ & $(1570
^{+230}_{-200}) \times 10^{-15}$ s & D\O\,~\cite{Abazov:2008qm} &
$J/\psi \Omega^-$
\\ \bottomrule[1pt]
\end{tabular}
\end{table*}

\subsubsection{$\Lambda_b^0$.}


The lowest-lying bottom baryon, the $\Lambda_b^0$ of $J^P = 1/2^+$,
was first reported by the CERN R415 Collaboration in
1981~\cite{Basile:1981wr}. 
It has
a mass $5619.51 \pm 0.23$ MeV and the mean life $(1466 \pm 10)
\times 10^{-15}$ s. Many of its decay modes have been observed in
experiments, where the Cabibbo-allowed process $b \rightarrow c W^-$
is preferred~\cite{Olive:2016xmw}.

\subsubsection{$\Lambda_b(5912)^0$ and $\Lambda_b(5920)^0$.}

\begin{figure}[htb]
\begin{center}
\includegraphics*[width=0.6\textwidth]{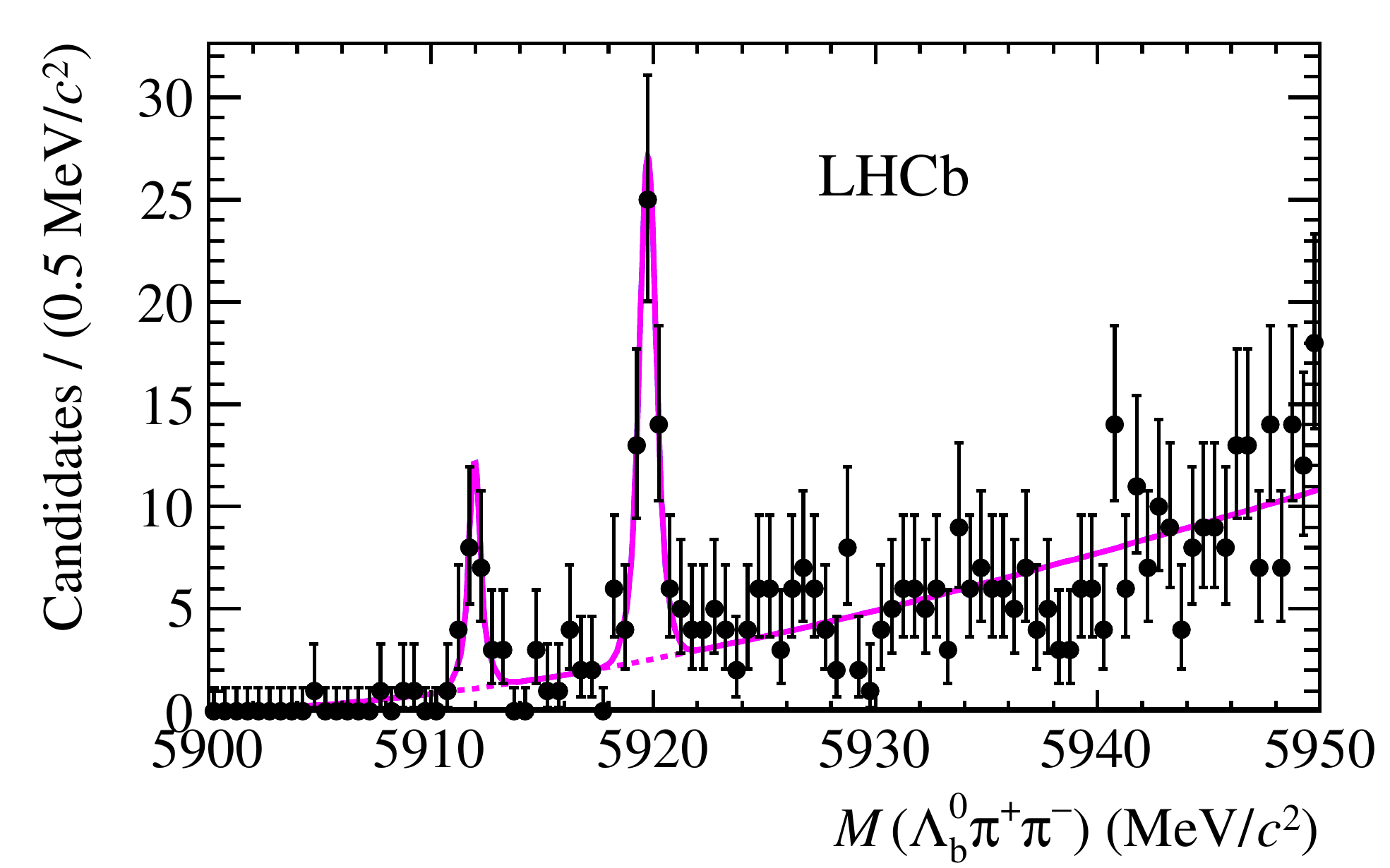}
\end{center}
\caption{(Color online) The $\Lambda_b^0 \pi^+ \pi^-$ invariant mass spectrum,
where the signals correspond to the $\Lambda_b(5912)^0$
and $\Lambda_b(5920)^0$.
Taken from LHCb~\cite{Aaij:2012da}. } \label{sec22:LHCb}
\end{figure}

The two excited bottom baryons, the $\Lambda_b(5912)^0$ of $1/2^-$
and the $\Lambda_b(5920)^0$ of $3/2^-$, were first observed in the
$\Lambda_b^0 \pi^+ \pi^-$ invariant mass spectrum by the LHCb
Collaboration in 2012~\cite{Aaij:2012da}, as shown in
Fig.~\ref{sec22:LHCb}. Their masses were measured to be
\begin{eqnarray}
M_{\Lambda^{*0}_b(5912)} &=& 5911.97 \pm 0.12 \pm 0.02 \pm
0.66 \mbox{ MeV} \, ,
\\ \nonumber M_{\Lambda^{*0}_b(5920)} &=& 5919.77 \pm 0.08 \pm 0.02 \pm 0.66 \mbox{ MeV} \, ,
\end{eqnarray}
and the upper limits of their widths were determined to be
\begin{eqnarray}
\Gamma_{\Lambda^{*0}_b(5912)} &<& 0.66 \mbox{ MeV} \, ,
\\ \nonumber \Gamma_{\Lambda^{*0}_b(5920)} &<& 0.63 \mbox{ MeV} \, ,
\end{eqnarray}
at the 90\% C.L.

Later in 2013, the $\Lambda_b(5920)^0$ was confirmed by the CDF
Collaboration~\cite{Aaltonen:2013tta}. Its mass was measured to be
$5919.22 \pm 0.35 \pm 0.30 \pm 0.60$ MeV, consistent with the LHCb
experiment~\cite{Aaij:2012da}.

\subsubsection{$\Sigma_b$ and $\Sigma_b^*$.}

\begin{figure}[htb]
\begin{center}
\includegraphics*[width=0.48\textwidth]{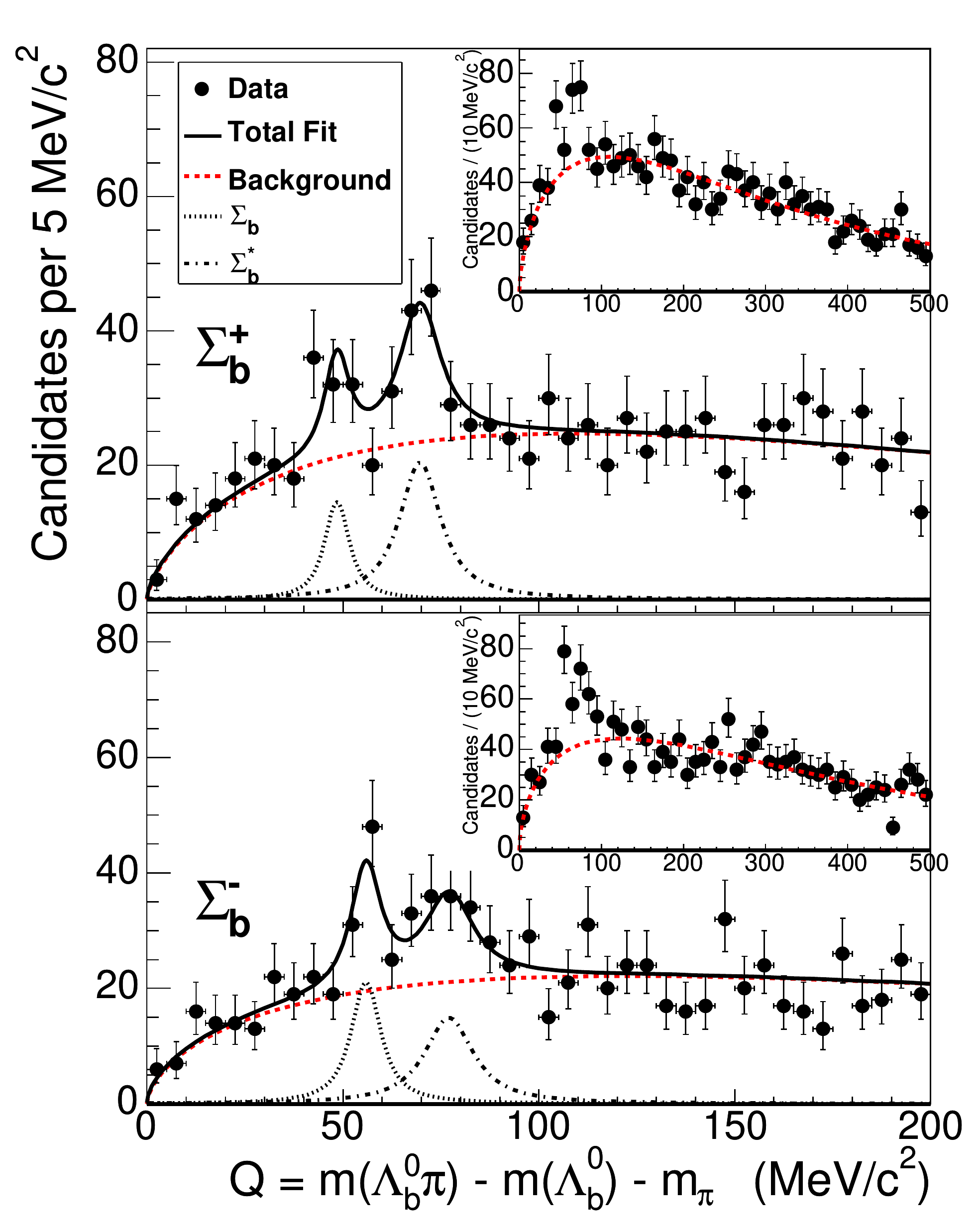}
\end{center}
\caption{(Color online) The $\Sigma^{(*)}_b$ fit to the $\Lambda^0_b \pi^+$ (top)
and $\Lambda^0_b \pi^-$ (bottom) subsamples,
where the signals correspond to the $\Sigma_b$ and $\Sigma_b^*$. Taken from
CDF~\cite{Aaltonen:2007ar}. } \label{sec22:CDF}
\end{figure}

The two ground state $\Sigma_b$ baryons, the $\Sigma_b$ of $J^P =
1/2^+$ and the $\Sigma_b^*$ of $J^P = 3/2^+$, were both first
observed in the $\Lambda_b^0 \pi$ invariant mass spectrum by the CDF
Collaboration~\cite{Aaltonen:2007ar}, as shown in
Fig.~\ref{sec22:CDF}. Their masses were measured to be
\begin{eqnarray}
\nonumber M_{\Sigma^+_b} &=& 5807.8 ^{+2.0}_{-2.2} \pm 1.7 \mbox{
MeV} \, ,
\\ M_{\Sigma^-_b} &=& 5815.2 \pm 1.0 \pm 1.7 \mbox{ MeV} \, ,
\\ \nonumber M_{\Sigma^{*+}_b} &=& 5829.0 {^{+1.6}_{-1.8}}{^{+1.7}_{-1.8}}  \mbox{ MeV} \, ,
\\ \nonumber M_{\Sigma^{*-}_b} &=& 5836.4 \pm 2.0 ^{+1.8}_{-1.7}  \mbox{ MeV} \, .
\end{eqnarray}
Five years later, the CDF Collaboration confirmed their previous
results, measured their masses and widths to be
\begin{eqnarray}
\nonumber M_{\Sigma^+_b} &=& 5811.3 ^{+0.9}_{-0.8} \pm 1.7  \mbox{
MeV} \, ,
\\ \nonumber \Gamma_{\Sigma^+_b} &=& 9.7 {^{+3.8}_{-2.8}} {^{+1.2}_{-1.1}}  \mbox{ MeV} \, ,
\\ \nonumber M_{\Sigma^-_b} &=& 5815.5 ^{+0.6}_{-0.5} \pm 1.7 \mbox{ MeV} \, ,
\\ \Gamma_{\Sigma^-_b} &=& 4.9 ^{+3.1}_{-2.1} \pm 1.1 \mbox{ MeV} \, ,
\\ \nonumber M_{\Sigma^{*+}_b} &=& 5832.1 \pm 0.7 ^{+1.7}_{-1.8} \mbox{ MeV} \, ,
\\ \nonumber \Gamma_{\Sigma^{*+}_b} &=& 11.5 {^{+2.7}_{-2.2}} {^{+1.0}_{-1.5}} \mbox{ MeV} \, ,
\\ \nonumber M_{\Sigma^{*-}_b} &=& 5835.1 \pm 0.6 ^{+1.7}_{-1.8} \mbox{ MeV} \, ,
\\ \nonumber \Gamma_{\Sigma^{*-}_b} &=& 7.5 {^{+2.2}_{-1.8}} {^{+0.9}_{-1.4}} \mbox{ MeV} \, .
\end{eqnarray}

\subsubsection{$\Xi_b$, $\Xi_b^\prime$ and $\Xi_b^*$.}

The $\Xi_b$ ground state of $J^P = 1/2^+$ was first observed by the
DELPHI Collaboration in 1995 in its semileptonic decay
process~\cite{Abreu:1995kt}. It has been confirmed in many other
experiments, and its properties are known very
well~\cite{Olive:2016xmw}: the $\Xi_b^-$ has a mass $5794.5 \pm
1.4$ MeV and a mean life $(1560 \pm 40) \times 10^{-15}$ s; the
$\Xi_b^0$ has a mass $5791.9 \pm 0.5$ MeV and a mean life $(1464 \pm
31) \times 10^{-15}$ s; several of their decay modes were observed
in experiments.

Many years later in 2014, the other $\Xi_b$ state of $J^P = 1/2^+$,
the $\Xi^\prime_b(5935)^-$, was observed in the $\Xi^0_b \pi^-$ mass
spectrum by the LHCb Collaboration~\cite{Aaij:2014yka}.
Its mass was measured to be $5935.02 \pm 0.02 \pm 0.01 \pm 0.50$
MeV, and the upper limit of its decay width was determined to be
0.08 MeV at 95\% C.L.

\begin{figure}[htb]
\begin{center}
\includegraphics*[width=0.48\textwidth]{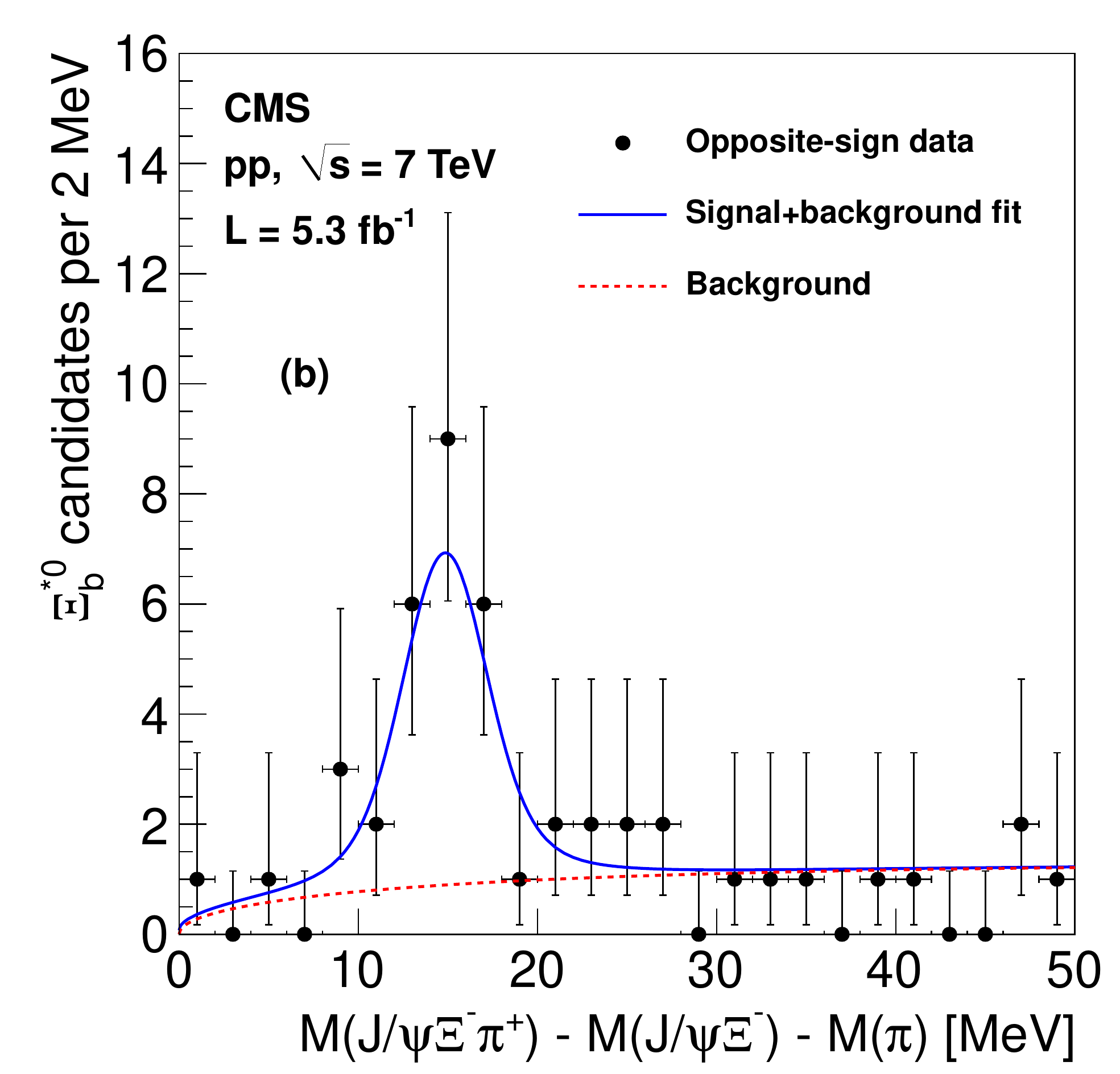}
\end{center}
\caption{(Color online) Opposite-sign $Q = M(J/\psi \Xi^- \pi^+) - M(J/\psi \Xi^-)
- M(\pi)$ distribution in the $0 < Q < 50$ MeV range.
The blue solid curve is fitted with the $\Xi_b(5945)^0$.
Taken from CMS~\cite{Chatrchyan:2012ni}. } \label{sec22:CMS}
\end{figure}

In 2012 the $\Xi_b(5945)^0$ was observed by the CMS
Collaboration~\cite{Chatrchyan:2012ni}. It was observed in the
distribution of the difference between the mass of the $\Xi^-_b
\pi^+$ system and the sum of the masses of the $\Xi^-_b$ and
$\pi^+$, as shown in Fig.~\ref{sec22:CMS}. Its mass was measured to
be $5945.0 \pm 0.7 \pm 0.3 \pm 2.7$ MeV, and its Breit-Wigner width
was fitted to be $2.1 \pm 1.7$ MeV. Given its measured mass and
decay mode, this state was suggested to be the $\Xi^{*0}_b$, the
$J^P = 3/2^+$ companion of the $\Xi_b^{(\prime)}$.

Another $\Xi_b$ state of $J^P = 3/2^+$, the $\Xi^*_b(5955)^-$ was
observed by the LHCb Collaboration~\cite{Aaij:2014yka} together with
the $\Xi^\prime_b(5935)^-$. Its mass and width were measured to be
$5955.33 \pm 0.12 \pm 0.06 \pm 0.50$ MeV and $1.65 \pm 0.31 \pm
0.10$ MeV, and can also be explained as the $J^P = 3/2^+$ companion
of the $\Xi_b^{(\prime)}$. However, the mass difference between the
$\Xi_b(5945)^0$ and the $\Xi^*_b(5955)^-$ seems too large, which
needs to be clarified in future experiments. Besides the up and down
mass difference, the Coulomb interaction among the three quarks
should contribute around 5 MeV to the rather large mass splitting.

\subsubsection{$\Omega_b^-$.}

\begin{figure}[htb]
\begin{center}
\includegraphics*[width=0.48\textwidth]{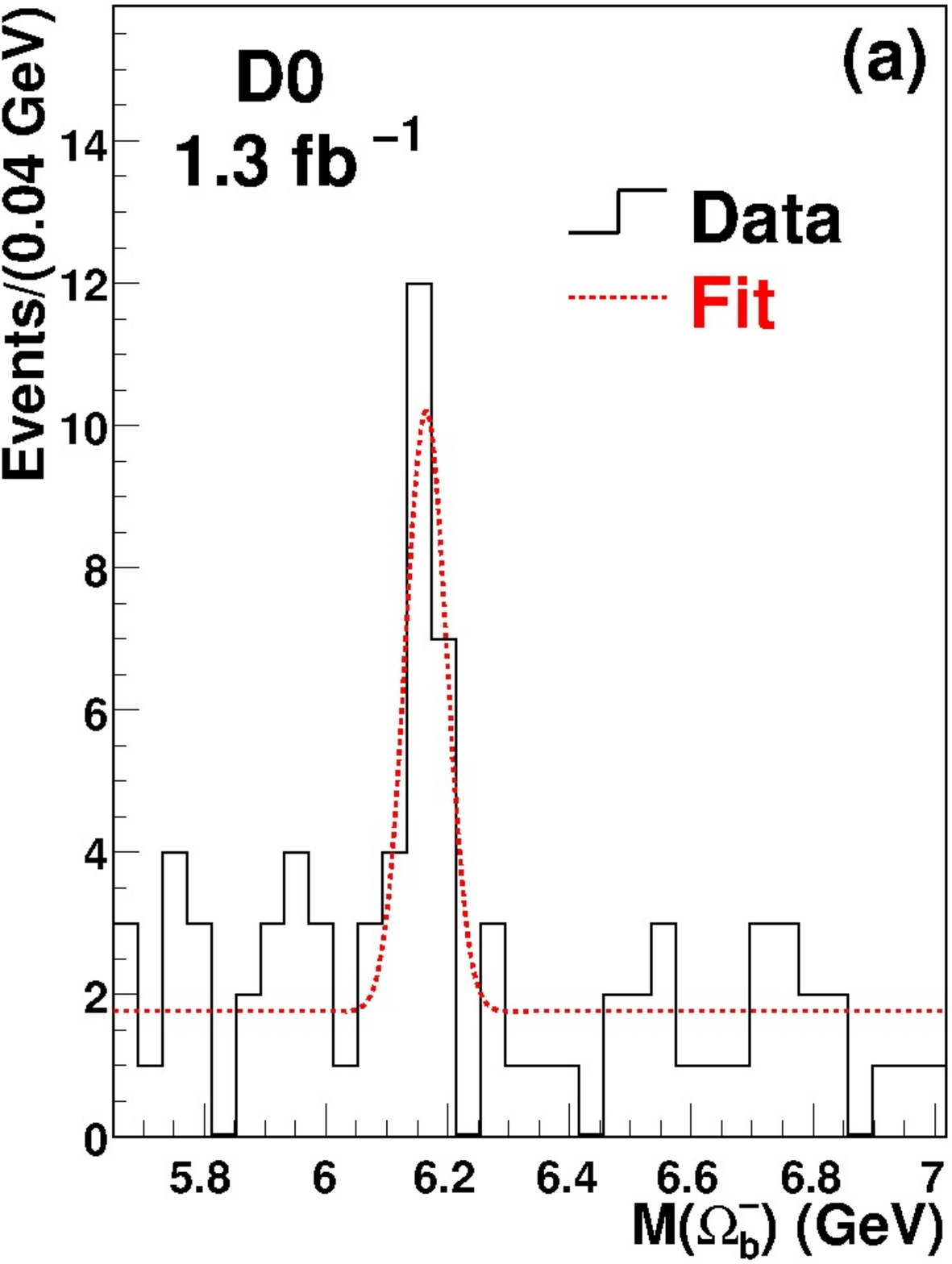}
\end{center}
\caption{(Color online) The mass distribution of the $\Omega^-_b$
candidates. Taken from
D\O\,~\cite{Abazov:2008qm}. } \label{sec22:D0}
\end{figure}

Only the ground state $\Omega_b$ baryon, the $\Omega_b^-$ of $J^P =
1/2^+$, was observed by the D\O\, Collaboration~\cite{Abazov:2008qm}, as shown in
Fig.~\ref{sec22:D0}. D\O\, reported the doubly strange $\Omega_b^-$
state in the decay channel $\Omega_b^-\to J/\psi \Omega^-$ with
$J/\psi\to\mu^+\mu^-$ and $\Omega^-\to\Lambda K^-\to(p\pi^-)K^-$ in
$p\bar p^-$ collisions at $\sqrt{s}=1.96$ TeV. It has been confirmed in many other
experiments~\cite{Olive:2016xmw}, and its mass and mean life were
determined to be $6046.4 \pm 1.9$ MeV and $(1570 ^{+230}_{-200})
\times 10^{-15}$ s, respectively.

\subsection{The doubly-charmed baryons}
\label{sec2.7}

The lightest doubly-charmed baryon has the quark content $ccu$ or
$ccd$. In 2002, the SELEX Collaboration at Fermilab discovered the
first doubly-charmed baryon $\Xi_{cc}^+$ in the charged decay mode
$\Xi_{cc}^+\to\Lambda_c^+ K^-\pi^+$ with a statistical significance
of $6.3\sigma$ \cite{Mattson:2002vu}. This structure was soon
confirmed by SELEX in its decay mode $\Xi_{cc}^+\to p D^+K^-$ with a
signal significance $4.8\sigma$ \cite{Ocherashvili:2004hi}. The
average mass value provided by PDG is \cite{Olive:2016xmw}
\begin{equation}
m=3518.9\pm0.9\, \mbox{MeV}\, .
\end{equation}
SELEX also measured the lifetime of $\Xi_{cc}^+$ with $\tau<33$ fs
\cite{Mattson:2002vu}. This result is much smaller than the
theoretical calculations \cite{Olive:2016xmw,Wei:2015gsa}.
See also lattice QCD studies in Refs.~\cite{Vijande:2015faa,Garcilazo:2016piq}.

To reproduce the structure of $\Xi_{cc}^+$ in SELEX, BaBar
\cite{Aubert:2006qw}, Belle \cite{Chistov:2006zj}, and the FOCUS
photoproduction experiment \cite{Ratti:2003ez} studied the
$\Lambda_c^+ K^-\pi^+$ decay mode along with various other final
states. None of them found any signal of the $\Xi_{cc}^+$. However, all
these experiments used the $\pi^-$-induced reactions while SELEX
used a hyperon beam. They have very different production mechanisms.
Thus it cannot be excluded that the SELEX had a higher double-charm
baryon cross-section than other experiments.

%% file: section2.3.tex
\subsection{The $X(5568)$}
\label{sec2.8}

\begin{figure}[htb]
\begin{center}
\includegraphics*[width=0.6\textwidth]{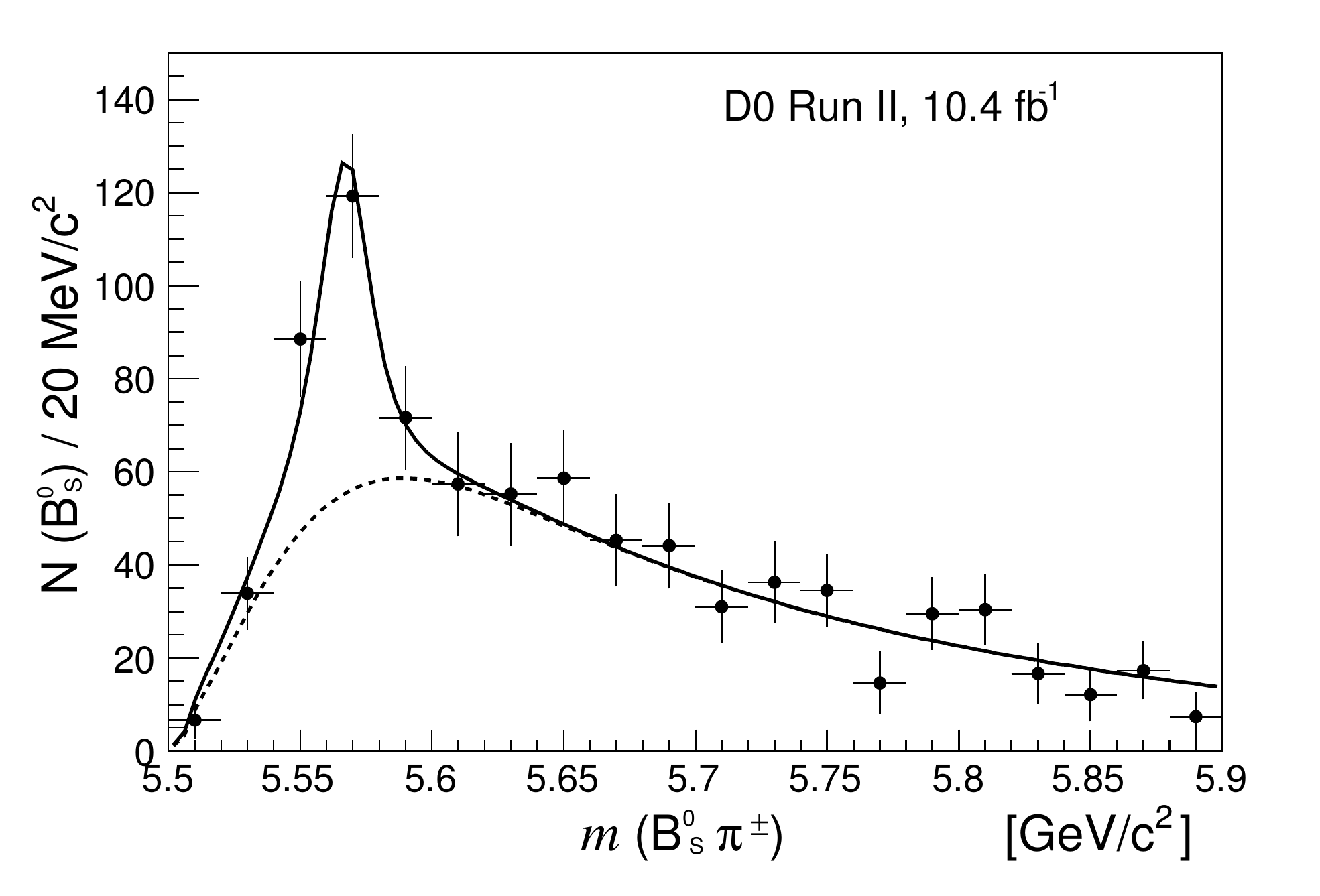}
\end{center}
\caption{The $B_s^0\pi^\pm$ invariant mass distribution, where the signal corresponds to the $X(5568)$.
Taken from D\O\,~\cite{D0:2016mwd}. } \label{sec214:D0X5568}
\end{figure}

Very recently, the D\O\, Collaboration reported evidence for a narrow
structure $X(5568)$ in the $B_s^0\pi^\pm$ invariant mass spectrum
with 5.1$\sigma$ significance~\cite{D0:2016mwd}, as shown in
Fig.~\ref{sec214:D0X5568}. The measured mass and width of the
$X(5568)$ are
\begin{eqnarray}
m_{X(5568)} &=& 5567.8\pm2.9({\rm stat})^{+0.9}_{-1.9}({\rm syst})
{\rm~MeV} \, ,
\\ \nonumber
\Gamma_{X(5568)} &=& 21.9\pm6.4({\rm stat})^{+5.0}_{-2.5}({\rm
syst}){\rm~MeV} \, .
\end{eqnarray}
Due to the $B_s^0\pi^\pm$ decay final states, the $X(5568)$ will be
the first evidence for a hadronic state with valence quarks of four
different flavors $su\bar{b}\bar{d}$ (or $sd\bar{b}\bar{u}$). Hence,
the reported $X(5568)$ state, if it exists, is a good candidate for
exotic tetraquark state.

Later, the LHCb Collaboration presented preliminary results for
their analysis of the $pp$ collision data at energies 7 TeV and 8
TeV~\cite{Aaij:2016iev}. They didn't find any resonance structure in
the $B_s^0\pi^\pm$ invariant mass distribution. The CMS Collaboration
also can not confirm the peaking structure of $X(5568)\to B_s\pi^+$
\cite{CMS:2016X5568}.

The D0 Collaboration also saw an enhancement in $m(B_s^0\pi^\pm)$ with
$B_s^0\to D_s\mu\nu$ at the same mass and at the expected width
and rate \cite{D0:2016X5568}. This observation is a confirmation of
the $X(5568)$ state in a new channel.
However, the
production mechanisms are different at Tevatron and LHC. It is
possible that the $X(5568)$ cross-section in the $p\bar p$
collisions may be higher than that in the $pp$ collisions. In other
words, the existence of the exotic state $X(5568)$ needs further
experimental confirmation.

%% file: section3.1.tex
\section{Candidates of the conventional excited heavy mesons}
\label{sec3}

The heavy mesons can be categorized into the charmed mesons,
charmed-strange mesons, bottom mesons and bottom-strange mesons, all
of which are composed of one heavy quark (charm or bottom, usually
denoted as $Q$) and one light quark (up, down or strange, usually
denoted as $q$). In the heavy mesons, the light degrees of freedom
circle around the nearly static heavy quark. The whole system
behaves as the QCD analogue of the familiar hydrogen.

The ground-state heavy mesons have no orbital excitations ($L=0$).
Its total angular momentum ($J$) is the same as its spin angular
momentum ($S$), which is the sum of the heavy quark spin ($s_Q$) and
the light quark spin ($s_l$):
\begin{equation}
J = S = s_Q \otimes s_l = 1/2 \otimes 1/2 = 0 \oplus 1 \, .
\end{equation}
Hence, there are two ground-state heavy mesons, $1^1S_0$ and
$1^3S_1$ (the symbol $n^{2S+1}L_J$ is used here, where $n$ is the
principal quantum number). These two states compose a spin doublet
($J^P = 0^-$ and $1^-$). In the heavy quark limit ($m_Q \rightarrow
\infty$), their masses are degenerate. Since the heavy quark
symmetry is explicitly broken, there exists a mass splitting between
them.

The $P$-wave heavy mesons ($L=1$) are a bit more complicated. We
denote the total angular momentum of its light degrees of freedom as
$j_l$, which is the sum of the orbital angular momentum ($L$) and
the light quark spin ($s_l$):
\begin{equation}
j_l = L \otimes s_l = 1 \otimes 1/2 = 1/2 \oplus 3/2 \, .
\end{equation}
Then its total angular momentum ($J$) is:
\begin{equation}
J = j_l \otimes s_Q = 0 \oplus 1 \oplus 1^\prime \oplus 2 \, .
\end{equation}
Hence, there are four $P$-wave heavy mesons, $1^3P_0$, $1^1P_1$,
$1^3P_1$ and $1^3P_2$. The two spin-1 states, $1^1P_1$ and $1^3P_1$,
can mix with each other to form the two physical states $1P_1$ and
$1P_1^\prime$. In the heavy quark limit, these four states further
compose two spin doublets $(0^+, 1^+)$ and $(1^+, 2^+)$. The former
doublet has $j_l = 1/2$ while the latter has $j_l = 3/2$. Again, the
masses of the two states belonging to the same doublet are
degenerate.

Similarly, we can categorize the excited heavy mesons with higher
orbital and nonzero radial excitations into:
\begin{eqnarray}
\nonumber S\mbox{-wave} &:& (n^1S_0, n^3S_1)~{\rm or}~(0^-,
1^-)~{\rm with}~j_l = 1/2 \, ,
\\ \nonumber P\mbox{-wave} &:&
(n^3P_0, n^1P_1, n^3P_1, n^3P_2)~{\rm or}~ \left\{ \begin{array}{c}
(0^+, 1^+)~{\rm with}~j_l = 1/2 \, ,
\\
(1^+, 2^+)~{\rm with}~j_l = 3/2 \, ,
\end{array} \right.
\\ \nonumber D\mbox{-wave} &:&
(n^3D_1, n^1D_2, n^3D_2, n^3D_3)~{\rm or}~ \left\{ \begin{array}{c}
(1^-, 2^-)~{\rm with}~j_l = 3/2 \, ,
\\
(2^-, 3^-)~{\rm with}~j_l = 5/2 \, ,
\end{array} \right.
\\ \nonumber F\mbox{-wave} &:&
(n^3F_2, n^1F_3, n^3F_3, n^3F_4)~{\rm or}~ \left\{ \begin{array}{c}
(2^+, 3^+)~{\rm with}~j_l = 5/2 \, ,
\\
(3^+, 4^+)~{\rm with}~j_l = 7/2 \, ,
\end{array} \right.
\\ && \nonumber \cdots
\end{eqnarray}
Because the heavy meson system is similar to the hydrogen, various
quark potential models have been applied to evaluate their mass
spectra (see reviews in Sec.~\ref{sec1}), and the results can be
used to explain the heavy meson signals observed in particle
experiments. Moreover, their productions and decay properties are
also important to understand their inner structure, which have been
studied using various models and methods.

In the following subsections, we shall review the theoretical
progress on the charmed, charmed-strange, bottom and bottom-strange
mesons.

\subsection{The charmed mesons}
\label{sec3.1}

\begin{figure*}
\begin{center}
\includegraphics[width=1.0\textwidth]{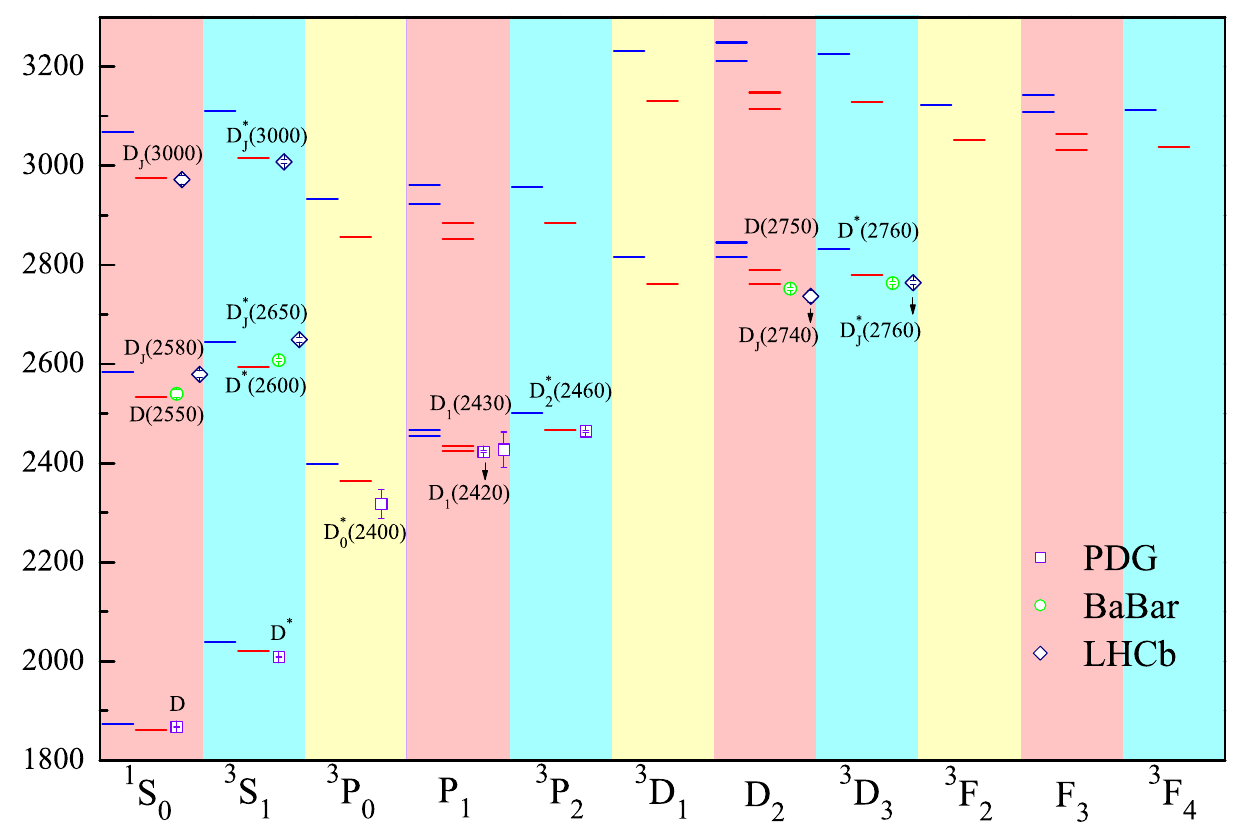}
\end{center}
\caption{(Color online) Mass spectrum of the charmed mesons, in
units of MeV. The blue lines are obtained by the GI
model~\cite{Godfrey:1985xj}, while the red lines are obtained by the
modified GI model where the screening effect is taken into account.
The purple squares, green circles and blue lozenges denote the data
from PDG~\cite{Beringer:1900zz} and the BaBar and LHCb
experiments~\cite{delAmoSanchez:2010vq,Aaij:2013sza}, respectively.
The symbol $^{2S+1}L_J$ is listed on the abscissa to describe quantum
numbers, and another notation $L_L$ is used when there exists a mixture
between the $n^1L_L$ and $n^3L_L$ states. Taken from
Ref.~\cite{Song:2015fha}. } \label{sec3:GIscreen1}
\end{figure*}

The mass spectrum of the charmed mesons has been calculated by many
theoretical groups using various models. In this review we list
three investigations:
\begin{enumerate}

\item The first one is the original GI model~\cite{Godfrey:1985xj} updated by Godfrey and Moats~\cite{Godfrey:2015dva}.
We have detailly reviewed this model in Sec.~\ref{sec1.1.1}. Its
potential is given in Eq.~(\ref{Sec11:EQGI}), containing two main
ingredients: the short-distance one-gluon-exchange interaction and
the long-distance linear confining interaction. We refer interested
readers to read their old reference~\cite{Godfrey:1985xj} for more
information.

\item The second one is calculated within the framework of the QCD-motivated relativistic quark model based on the quasipotential approach~\cite{Ebert:2009ua}.
Again, this model has been reviewed in Sec.~\ref{sec1.1.1}, whose
quasipotential is given in Eq.~(\ref{sec11:EQRQM}). We refer
interested readers to read Refs.~\cite{Ebert:2009ua,Ebert:1997nk}
for more information.

\item The third one is calculated still by the GI model but taking into account the screening effect~\cite{Song:2015fha}.
We have also reviewed this modified model in Sec.~\ref{sec1.1.2},
whose potential is just the GI one with its linear confining
interaction modified by Eq.~(\ref{sec11:EQscreen}). We refer
interested readers to read Ref.~\cite{Song:2015nia} for more
information.

\end{enumerate}
We summarize the results obtained using these three methods in
Table~\ref{sec3:charm}, and note that the results obtained by using
the second and third methods are consistent with each other. The
results obtained by using the third method are also shown in
Fig.~\ref{sec3:GIscreen1}. See also studies using the constituent
quark model~\cite{Vijande:2003ki,Vijande:2004he,Vijande:2006hj} as
well as lattice QCD studies in
Refs.~\cite{AliKhan:1995ub,Kawanai:2011jt,Mohler:2011ke,Atoui:2013ksa,Kawanai:2015tga,Nochi:2016wqg,Cheung:2016bym}
and QCD sum rule studies in
Refs.~\cite{Azizi:2013aua,Torres:2013saa,Gelhausen:2014jea,Alhendi:2015rka}.
More discussions using other models and methods can be found in
Refs.~\cite{Falk:1995th,Chliapnikov:2001tm,Matsuki:2007zza,Becattini:2008tx,vanBeveren:2010jz,Wang:2011zzw,Badalian:2011tb,Colangelo:2012xi,Segovia:2013sxa,Liu:2013maa,Liu:2015lka,Klein:2015doa,Batra:2015cua,Geng:2016pyr,Zhao:2016mxc,Liu:2016efm,Xiao:2016kak,Zhang:2016dom,Matsuki:2016hzk,Albaladejo:2016lbb,Kalashnikova:2016bta}.

\renewcommand{\arraystretch}{1.6}
\begin{table*}[htb]
\tiny
\caption{ Comparison of the experimental data and theoretical
results of the charmed mesons obtained using the original GI model
updated by Godfrey and Moats (GI-Original)~\cite{Godfrey:2015dva},
the QCD-motivated relativistic quark model based on the
quasipotential approach (R.~Q.~M.)~\cite{Ebert:2009ua}, and the
modified GI model taking into account the screening effect
(GI-Screen)~\cite{Song:2015fha}. The notation $L_L$ is introduced to
express mixing states of $^1L_L$ and $^3L_L$. The masses are in
units of MeV.} \centering
\begin{tabular}{ c c c c c c }\toprule[1pt]
& $n \ ^{2S+1}L_J$ & Experimental values~\cite{Olive:2016xmw} &
GI-Original~\cite{Godfrey:2015dva} & R.~Q.~M.~\cite{Ebert:2009ua} &
GI-Screen~\cite{Song:2015fha}
\\ \midrule[1pt]
$D^0$                   & $1 \ ^1S_0$      & $1864.83 \pm 0.05$                                & 1877  & 1871  & 1861 \\
$D^{\ast0}$             & $1 \ ^3S_1$      & $2006.85 \pm 0.05$                                & 2041  & 2010  & 2020 \\
\hline
$D_{0}^\ast(2400)^0$    & $1 \ ^3P_0$      & $2318 \pm 29$                                     & 2399  & 2406  & 2365 \\
$D_{1}(2430)^0$         & $1 \ P_1$        & $2427\pm26\pm25$                                  & 2456  & 2426  & 2424 \\
$D_{1}(2420)^0$         & $1 \ P_1^\prime$ & $2420.8 \pm 0.5$                                  & 2467  & 2469  & 2434 \\
$D_{2}^\ast(2460)^0$    & $1 \ ^3P_2$      & $2460.57 \pm 0.15$                                & 2502  & 2460  & 2468 \\
\hline
$D_1^*(2760)^0$         & $1 \ ^3D_1$      & $2781 \pm 18 \pm 11 \pm 6$~\cite{Aaij:2015vea}    & 2817  & 2788  & 2762 \\
\hdashline[2pt/2pt] \multirow{2}{*}{$D(2750)^0/D_J(2740)^0$}   & $1
\ D_2$        &
\multirow{2}{*}{$2752.4\pm1.7\pm2.7$~\cite{delAmoSanchez:2010vq}}
                                                                                               & 2816  & 2806  & --   \\
                        & $1 \ D_2^\prime$ &                                                   & 2845  & 2850  & 2789 \\
\hdashline[2pt/2pt]
$D_3^*(2760)^0$         & $1 \ ^3D_3$      & $2775.5\pm4.5\pm4.5\pm4.7$~\cite{Aaij:2016fma}    & 2833  & 2863  & 2779 \\
\hline
--                      & $1 \ ^3F_2$      & --                                                & 3132  & 3090  & 3053 \\
--                      & $1 \ F_3$        & --                                                & 3108  & 3129  & -- \\
--                      & $1 \ F_3^\prime$ & --                                                & 3143  & 3145  & -- \\
--                      & $1 \ ^3F_4$      & --                                                & 3113  & 3187  & 3037 \\
\hline
$D(2550)^0/D_J(2580)^0$ & $2 \ ^1S_0$      & $2539.4\pm4.5\pm6.8$~\cite{delAmoSanchez:2010vq}  & 2581  & 2581  & 2534 \\
$D_1^*(2600)^0$         & $2 \ ^3S_1$      & $2608.7\pm2.4\pm2.5$~\cite{delAmoSanchez:2010vq}  & 2643  & 2632  & 2593 \\
\hline
--                      & $2 \ ^3P_0$      & --                                                & 2931  & 2919  & 2856 \\
--                      & $2 \ P_1$        & --                                                & 2924  & 2932  & -- \\
--                      & $2 \ P_1^\prime$ & --                                                & 2961  & 3021  & -- \\
--                      & $2 \ ^3P_2$      & --                                                & 2957  & 3012  & 2884 \\
\hline
--                      & $2 \ ^3D_1$      & --                                                & 3231  & 3228  & 3131 \\
--                      & $2 \ D_2$        & --                                                & 3212  & 3259  & -- \\
--                      & $2 \ D_2^\prime$ & --                                                & 3248  & 3307  & -- \\
--                      & $2 \ ^3D_3$      & --                                                & 3226  & 3335  & 3129 \\
\hline
$D_J(3000)^0$           & $3 \ ^1S_0$      & $2971.8\pm8.7$~\cite{Aaij:2013sza}                & 3068  & 3062  & 2976 \\
$D_J^*(3000)^0$         & $3 \ ^3S_1$      & $3008.1\pm4.0$~\cite{Aaij:2013sza}                & 3110  & 3096  & 3015 \\
\hline
--                      & $3 \ ^3P_0$      & --                                                & 3343  & 3346  & --   \\
--                      & $3 \ P_1$        & --                                                & 3328  & 3365  & --   \\
--                      & $3 \ P_1^\prime$ & --                                                & 3360  & 3461  & --   \\
$D^*_2(3000)$           & $3 \ ^3P_2$      & $ 3214 \pm 29 \pm 33 \pm 36$~\cite{Aaij:2016fma}  & 3353  & 3407  & --   \\
\hline
--                      & $4 \ ^1S_0$      & --                                                & 3468  & 3452  & --   \\
--                      & $4 \ ^3S_1$      & --
& 3497  & 3482  & --
\\ \bottomrule[1pt]
\end {tabular}
\label{sec3:charm}
\end{table*}

\begin{figure*}
\begin{center}
\includegraphics[width=0.6\textwidth]{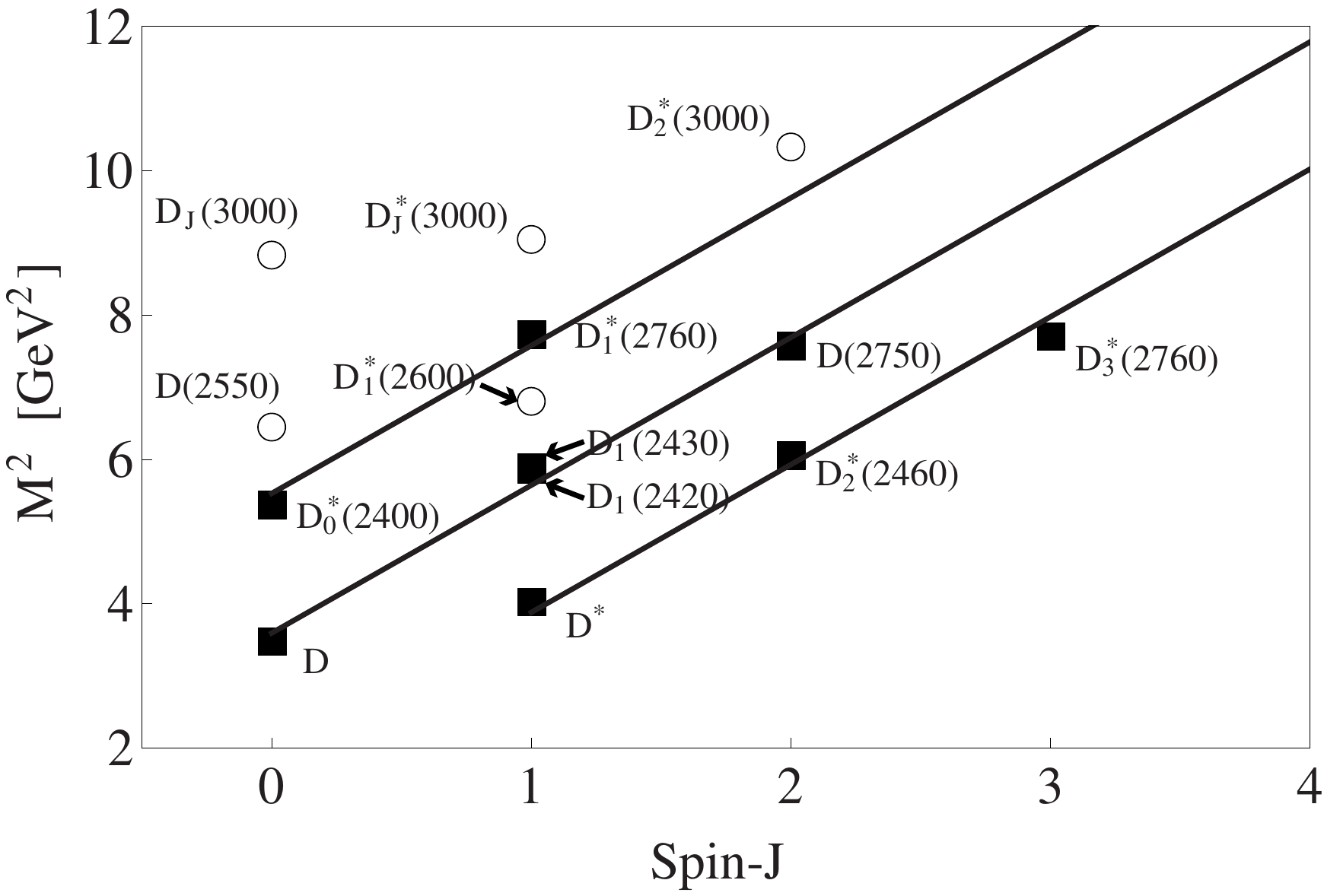}
\end{center}
\caption{ Regge trajectories in the $(J,M^2)$ plane for all the
charmed mesons observed in experiments, where experimental data are
given by solid squares ($1S$ and $1P$ states and $1D$ candidates)
and hollow circles (other excited states) with particle names. }
\label{Fig:3.1.Regge1}
\end{figure*}

In Table~\ref{sec3:charm} and Fig.~\ref{sec3:GIscreen1}, we further
make a comparison between the experimental data and the above
theoretical values, and conclude from the mass spectrum analysis,
i.e.,
\begin{enumerate}

\item Two $1S$ states ($D$ and $D^*$) and four $1P$ states ($D_{0}^\ast(2400)$, $D_{1}(2430)$, $D_{1}(2420)$ and $D_{2}^\ast(2460)$) in the charmed meson family are reproduced quite well.

\item The $D(2750)$ and $D_J(2740)$ are probably the same state, and can be a candidate of $D(1^3D_2)$.
      The $D^*(2760)$ and $D^*_J(2760)$ are separated into the $D^*_1(2760)$ and $D_3^*(2760)$
      by the LHCb experiments~\cite{Aaij:2015vea,Aaij:2015sqa}, which may correspond to $D(1^3D_1)$ and $D(1^3D_3)$, respectively.

\item The $D(2550)$ and $D_J(2580)$ are probably the same state, and usually considered as a candidate of $D(2^1S_0)$.
      The $D^*(2600)$ and $D^*_J(2650)$ are probably the same state, and can be a candidate of $D(2^3S_1)$.

\item The $D_J(3000)$ and $D_J^*(3000)$ can be candidates of $D(3^1S_0)$ and $D(3^3S_1)$, respectively.
      In addition, they can also be candidates of $D(1F)$ or $D(2P)$ etc.
      The recently observed $D^*_2(3000)$ may be a candidate of $D(3^3P_2)$.

\end{enumerate}
We also use the charmed mesons to construct Regge trajectories (see
discussions in Sec.~\ref{sec1.5}), as shown in
Fig.~\ref{Fig:3.1.Regge1} in the $(J,M^2)$ plane. The results
similarly suggest that the $D$, $D^*$, $D_{0}^\ast(2400)$,
$D_{1}(2430)$, $D_{1}(2420)$ and $D_{2}^\ast(2460)$ can be well
interpreted as the $1S$ and $1P$ charmed mesons, the $D^*_1(2760)$,
$D(2750)/D_J(2740)$ and $D_3^*(2760)$ may be interpreted as the $1D$
charmed mesons, but the $D(2550)/D_J(2580)$,
$D^*(2600)/D^*_J(2650)$, $D_J(3000)$, $D_J^*(3000)$ and
$D^*_2(3000)$ can not be simply explained.

In the following paragraphs we start to review the theoretical
progress on the excited charmed mesons.

\subsubsection{$D(2550)$ and $D_J(2580)$.}

The $D(2550)$ and $D_J(2580)$ are probably the same state, whose
mass is consistent with the theoretical prediction of
$D(2^1S_0)$~\cite{Godfrey:1985xj}. In addition, the decay width of
$D(2^1S_0)$ was calculated by the QPC model in
Ref.~\cite{Close:2005se}, and the result is also close to the lower
limit of the experimental width of the $D(2550)/D_J(2580)$.

Besides these studies, the $D(2550)$ was investigated using the
Regge trajectory phenomenology~\cite{Chen:2011rr}, the relativistic
quark model~\cite{Lu:2014zua}, and the improved Bethe-Salpeter
method~\cite{Wang:2012wk}. Their results also suggested it to be the
$D(2^1S_0)$ state, the first radial excitation of the $D$ meson.

However, there exist opposite opinions: the theoretical total width
of $D(2^1S_0)$ was evaluated using the chiral quark
model~\cite{DiPierro:2001dwf} and the constituent quark
model~\cite{Zhong:2010vq,Li:2010vx}. Its mass and decay properties
were also studied in Ref.~\cite{Sun:2010pg}. These studies indicated
that the total width of $D(2^1S_0)$ is far below the experimental
value of the decay width of the $D(2550)$.

In Ref.~\cite{Song:2015fha} Song {\it et al.} systematically studied
the charmed meson family and investigated their decay properties
using the QPC model. They considered the $D(2550)$ as a $D(2^1S_0)$
state, and found that the main decay channels of the $D(2550)$ are
$D^*\pi$ and $D^*_0(2400)\pi$. This can explain why BaBar and LHCb
first observed the $D(2550)$/$D_J(2580)$ in the $D^*\pi$ channel.
Its total width was obtained as 71.65 MeV comparable with the lower
bound of the BaBar data~\cite{delAmoSanchez:2010vq} but smaller than
the LHCb value~\cite{Aaij:2013sza}. Considering this situation, a
more precise measurement of the resonance parameters of the
$D(2550)$/$D_J(2580)$ will be helpful.

\subsubsection{$D^*(2600)$ and $D^*_J(2650)$.}

The $D^*(2600)$ and $D^*_J(2650)$ are probably the same state, which
can be a candidate of $D(2^3S_1)$. In 1994 the $D(2^3S_1)$ state was
studied via the constituent quark model, and its mass was predicted
to be 2620 MeV~\cite{Zeng:1994vj}, already in good agreement with
the experimental value of the
$D^*(2600)$~\cite{delAmoSanchez:2010vq}. Later in 1998, the ratio
$\Gamma(D(2^3S_1)^0\rightarrow D^+\pi^-) /
\Gamma(D(2^3S_1)^0\rightarrow D^{*+}\pi^-)=0.47$ was predicted via
the relativistic chiral quark model \cite{Goity:1998jr}, which is close to
the upper bound determined by the BaBar
experiment~\cite{delAmoSanchez:2010vq}:
\begin{eqnarray}
\frac{\mathcal{B}(D^{*0}(2600)\rightarrow
D^+\pi^-)}{\mathcal{B}(D^{*0}(2600)\rightarrow
D^{*+}\pi^-)}=0.32\pm0.02\pm0.09\, . \label{sec3:eq2600}
\end{eqnarray}

In Ref.~\cite{DiPierro:2001dwf}, Pierro and Eichten calculated the
mass spectrum of the charmed mesons via a relativistic quark model
and estimated their hadronic decay widths. They evaluated the mass
of $D(2^3S_1)$ to be 2692 MeV, heavier than the $D^*(2600)$.
However, their predicted total width of $D(2^3S_1)$ is consistent
with the experimental value for the $D^*(2600)$.

The $2^3S_1$ assignment of the $D^*(2600)$ is also supported by the
investigations using the constituent quark
model~\cite{Zhong:2010vq,Li:2010vx}, the relativistic quark
model~\cite{Lu:2014zua}, the Regge trajectory
phenomenology~\cite{Chen:2011rr}, the QCD sum
rule~\cite{Wang:2010ydc}, and by investigating its mass and decay
properties~\cite{Sun:2010pg}, etc.

\begin{figure}[htb]
\begin{center}
\includegraphics[width=0.6\textwidth]{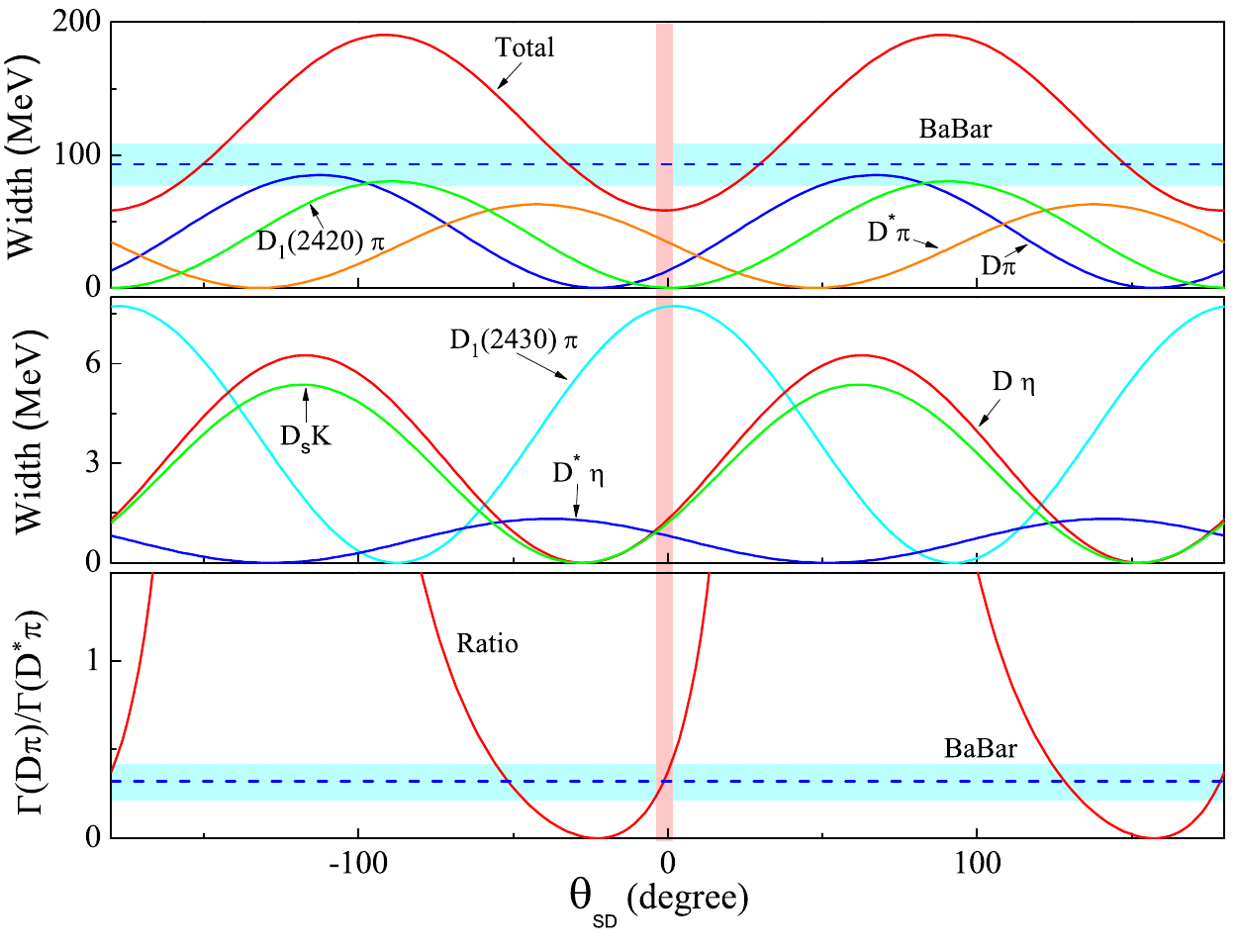}
\end{center}
\caption{(Color online) The $\theta_{SD}$ dependence of the total
width (the red curve in the top panel), partial decay widths (the
other seven colored curves in the top and middle panels), and the
ratio $\Gamma(D\pi)/\Gamma(D^*\pi)$ (the red curve in the bottom
panel) of the $D^*(2600)$, calculated using the QPC
model~\cite{Song:2015fha}. The two dashed curves in the top and
bottom panels correspond to the data from the BaBar
experiment~\cite{delAmoSanchez:2010vq} that $\Gamma(D^*(2600)) =
93\pm6\pm13$ MeV and $\Gamma(D^+\pi^-)/\Gamma(D^{*+}\pi^-) =
0.32\pm0.02\pm0.09$. Taken from Ref.~\cite{Song:2015fha}.}
\label{sec3:D2600}
\end{figure}

More generally, the $D^*(2600)/D^*_J(2650)$ can be interpreted as a
mixture of the $2^3S_1$ and $1^3D_1$ states~\cite{Song:2015fha}:
\begin{equation}
 \left(
  \begin{array}{c}
   |D^*(2600)\rangle\\
   |D^{*\prime}(1^-)\rangle\\
  \end{array}
\right )= \left(
  \begin{array}{cc}
    \cos\theta_{SD} & \sin\theta_{SD} \\
   -\sin\theta_{SD} & \cos\theta_{SD}\\
  \end{array}
\right) \left(
  \begin{array}{c}
    |2^3S_1  \rangle \\
   |1^3D_1 \rangle\\
  \end{array}
\right).
\end{equation}
The relevant mixing angle $\theta_{SD}$ was evaluated and discussed
in Ref.~\cite{Song:2015fha}. The $\theta_{SD}$ dependence of the
total width, partial decay widths, and the ratio $\mathcal{B}(
D^+\pi)/\mathcal{B}( D^{*}\pi)$ of the $D^\ast(2600)$ is shown in
Fig.~\ref{sec3:D2600}. When taking the range
$-3.6^\circ<\theta_{SD}<1.8^\circ$, the obtained theoretical ratio
is consistent with the BaBar measurement of Eq.~(\ref{sec3:eq2600}),
and the total width was estimated to be about 60 MeV also comparable
to the experimental data $\Gamma=93\pm6\pm13$
MeV~\cite{delAmoSanchez:2010vq}.

\subsubsection{$D(2750)$ and $D_J(2740)$.}

The $D(2750)$ and $D_J(2740)$ are two states with unnatural
spin-parity. They are probably the same state due to their
similarities. Their spin-parity quantum number may be $J^P = 2^-$.
They may belong to either the $(1^-,2^-)$ or $(2^-,3^-)$ doublet.

The D-wave charmed $c\bar q$ meson with $J^P = 2^-$ was studied in
Ref. \cite{Chen:2011qu} in the framework of the QCD sum rule
approach. The authors studied the following tensor interpolating
current with $J^{P(C)}=2^{-(-)}$
\begin{eqnarray}
  J_{\mu\nu}=\bar Q_1(x)(\gamma_{\mu}\gamma_5\stackrel{\longleftrightarrow}{D_{\nu}}
  +\gamma_{\nu}\gamma_5\stackrel{\longleftrightarrow}{D_{\nu}}-\frac{2}{3}\eta_{\mu\nu}\gamma_5
  \stackrel{\longleftrightarrow}{D\!\!\!\slash})Q_2(x),
\end{eqnarray}
where $\eta_{\mu\nu}=q_{\mu}q_{\nu}/q^2-g_{\mu\nu}$ and the
covariant derivative $\stackrel{\longleftrightarrow}{D_{\nu}}$ is
defined as
\begin{eqnarray}
  \stackrel{\longleftrightarrow}{D_{\nu}}&=\overrightarrow{D_{\mu}}-\overleftarrow{D_{\mu}},
  \\ \nonumber
  \overrightarrow{D_{\mu}}&=\overrightarrow{\partial_{\mu}}+ig\frac{\lambda^a}{2}A^a_{\mu},
  \overleftarrow{D_{\mu}}=\overleftarrow{\partial_{\mu}}-ig\frac{\lambda^a}{2}A^a_{\mu}\, .
\end{eqnarray}
By exploring the correlation functions induced by the above current,
they studied the $\bar qq, \bar qs, \bar ss, \bar qc, \bar sc, \bar
cc \bar qb, \bar sb$, $\bar cb$ and $\bar bb$ systems and obtained
their masses. For the $\bar qc$ system, they gave the hadron mass
$m=2.86\pm0.14$ GeV, which is consistent with the masses of the
$D(2750)$ and $D_J(2740)$ and supports them to be a $\bar qc$ meson
with $J^P = 2^-$.

In Ref.~\cite{Colangelo:2012xi} Colangelo {\it et al.} calculated
the ratio ${\Gamma(D^*(2760)^0\rightarrow D^+\pi^-) \over
\Gamma(D(2750)^0\rightarrow D^{*+}\pi^-)}$ with the effective
Lagrangian approach, and their result suggested the $D(2750)$ as the
$2^-$ state in the $(2^-,3^-)$ doublet. This assignment is supported
by many other
studies~\cite{Chen:2011rr,Zhong:2010vq,Li:2010vx,Wang:2010ydc}.

\begin{figure}[htb]
\begin{center}
\includegraphics[width=0.6\textwidth]{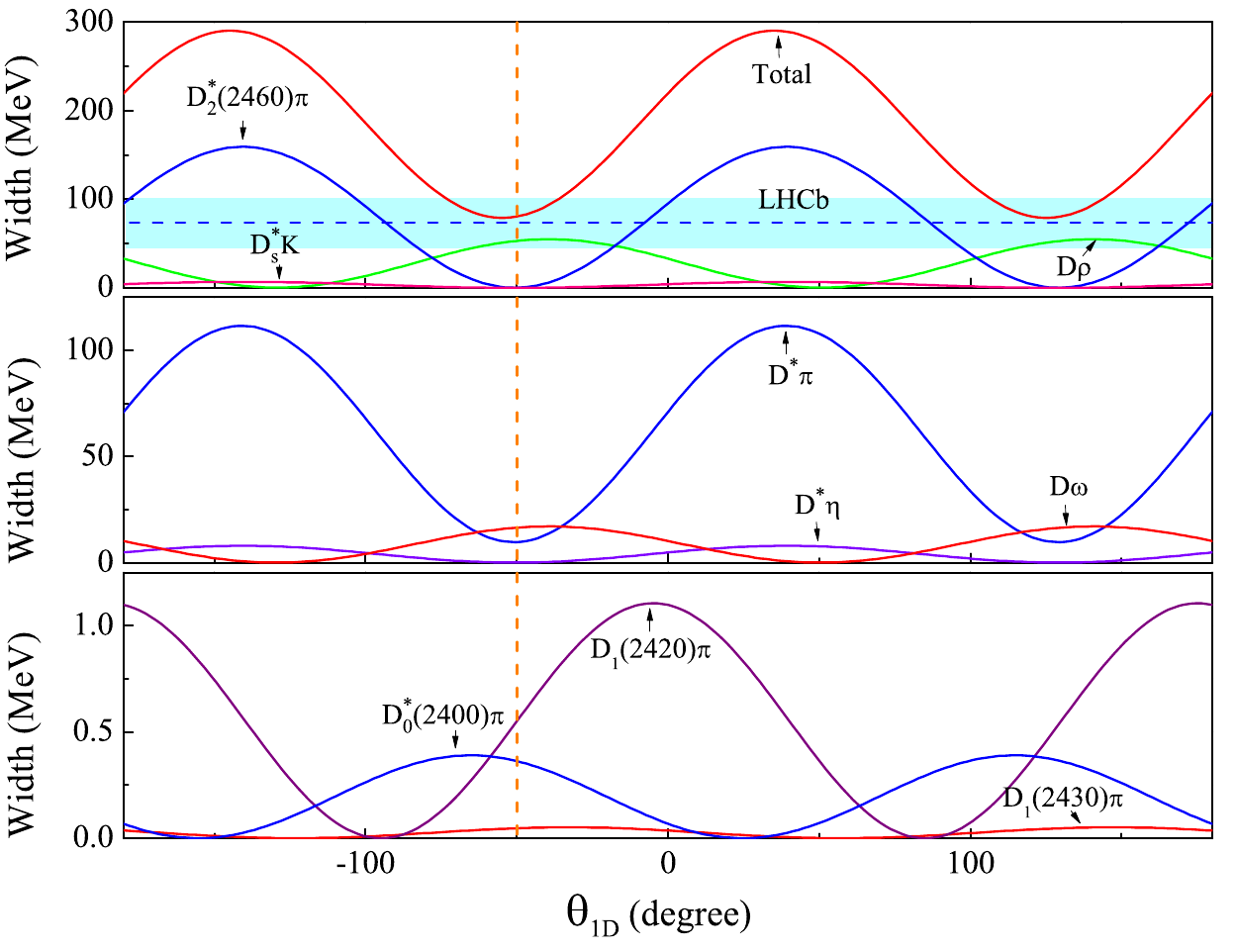}
\end{center}
\caption{ (Color online) The $\theta_{1D}$ dependence of the total
width (the red curve in the top panel) and partial decay widths (the
other nine colored curves in the top, middle and bottom panels) of
the $D(2750)$/$D_J(2740)$, calculated using the QPC
model~\cite{Song:2015fha}. The blue dashed curve in the top panel
corresponds to the data from the LHCb experiment~\cite{Aaij:2013sza}
that $\Gamma(D_J(2740)) = 73.2\pm13.4\pm25.0$. Taken from
Ref.~\cite{Song:2015fha}.} \label{sec3:D2750}
\end{figure}

More generally, the $D(2750)/D_J(2740)$ can be interpreted as a
mixture of the $1^1D_2$ and $1^3D_2$ states~\cite{Song:2015fha}:
 \begin{equation}
 \left(
  \begin{array}{c}
   |1D(2^-)\rangle\\
   |D(2750)\rangle\\
  \end{array}
\right )= \left(
  \begin{array}{cc}
    \cos\theta_{1D} & \sin\theta_{1D} \\
   -\sin\theta_{1D} & \cos\theta_{1D}\\
  \end{array}
\right) \left(
  \begin{array}{c}
    |1^1D_2  \rangle \\
   |1^3D_2 \rangle\\
  \end{array}
\right).
\end{equation}
The relevant mixing angle $\theta_{1D}$ was evaluated and discussed
in Refs.~\cite{Song:2015fha}, and the $\theta_{1D}$ dependence of
the corresponding partial and total decay widths is given in
Fig.~\ref{sec3:D2750}. The range of a mixing angle was obtained as
$-73.8^\circ<\theta_{1D}<-35.7^\circ$ so that the calculated total
width is consistent with the experimental data.

\subsubsection{$D^*(2760)$, $D^*_J(2760)$, $D^*_1(2760)^0$ and $D_3^*(2760)^-$.}

The $D^*(2760)$ and $D^*_J(2760)$ are two natural states. They may
be the same state. There are many possible assignments in this
energy region, such as the $D(1^3D_1)$, $D(1^3D_3)$ and $D(2^3S_1)$
states, etc. Comparing the prediction of the relativistic quark
model~\cite{DiPierro:2001dwf} and the experimental data of the
$D^*(2760)$, one notes that the $D^*(2760)$ can be interpreted as
either the $D(1^3D_1)$ or $D(1^3D_3)$ states. The total widths of
these two assignments were calculated in Ref.~\cite{Close:2005se},
which are far larger than the experimental
value~\cite{Beringer:1900zz}.

In Ref.~\cite{Sun:2010pg} the $D^*(2760)$ was suggested to be a
mixture of the $2^3S_1$ and $1^3D_1$ states by studying its mass and
decay properties. This is supported by the study done within the
Regge trajectory phenomenology~\cite{Chen:2011rr}. However, the
$D(1^3D_3)$ assignment for the $D^*(2760)$ is still possible and
supported by studies using various
models~\cite{Colangelo:2012xi,Zhong:2010vq,Li:2010vx,Sun:2010pg,Wang:2010ydc}.

Later in the LHCb experiments~\cite{Aaij:2015vea,Aaij:2015sqa}, the
$D^*(2760)$ was further separated into two states, the
$D^*_1(2760)^0$ of $J^P = 1^-$ and $D_3^*(2760)^-$ of $3^-$. These
two states were studied in Ref.~\cite{Chen:2015lpa} by performing a
combined study of the $2S$ and $1D$ open-charm mesons with natural
spin-parity, and the obtained results suggested that the
$D^*_1(2760)$ is predominantly the $1^3D_1$ charmed meson, while the
$D_3^*(2760)$ can be regarded as the $1^3D_3$ charmed meson.

\subsubsection{$D_J(3000)$, $D^*_J(3000)$ and $D^*_2(3000)$.}

Many theoretical groups have studied the $D_J(3000)$ and
$D^*_J(3000)$ using various methods, but their nature are still
unclear. In Ref.~\cite{Sun:2013qca}, Sun, Liu and Matsuki studied
the $D_J(3000)$ by analyzing its mass and decay behaviors. Their
results suggested that $D_J(3000)$ and $D^*_J(3000)$ can be
explained as the $2P$ states in the $D$ meson family. This was
partly supported by studies using the chiral quark
model~\cite{Xiao:2014ura}, the QPC model~\cite{Yu:2014dda}, and the
heavy meson effective theory~\cite{Wang:2013tka}, etc. Different
assignments to the $D_J(3000)/D^*_J(3000)$ are also possible, such
as the $D(3^1S_0)$ state~\cite{Lu:2014zua} and the $D(3^+)$
state~\cite{Yu:2014dda}, etc. The semi-leptonic production of the
$D_J(3000)$ in $B_s$ and $B$ decays was recently studied in
Ref.~\cite{Li:2016tbj} by Li {\it et al.}, and their results using
the improved Bethe-Salpeter method indicated that these decays have
considerable branching ratios.

Recently, the $D^*_2(3000)$ was observed by the LHCb
experiment~\cite{Aaij:2016fma}, but still the situation in this
energy region is not very clear because there are too many
possibilities. In Ref.~\cite{Wang:2016krl}, the authors studied the
decay behaviors of the $3P$ and $2F$ charmed mesons using the QPC
model. Their results are summarized in Table~\ref{sec3:D23000},
suggesting that the most possible assignment for the $D^*_2(3000)$
is the $3^3P_2$, while the assignment of the $2^3F_2$ can not be
fully excluded. The decay properties of the $D^*_2(3000)$ were also
studied in Refs.~\cite{Wang:2016ewb,Yu:2016mez} by Wang with the
heavy meson effective theory. His result suggested that the
$D^*_2(3000)$ can be tentatively assigned as the $1F$ $2^+$ state.

\begin{table*}[htb]
\footnotesize
\caption{Masses and decay behaviors of the $3P$ and $2F$
charmed mesons. The corresponding branching ratios for different
assignments are also given in the brackets. Taken from
Ref.~\cite{Wang:2016krl}. \label{sec3:D23000}}
\begin{tabular}{ccccc}
\toprule[1pt]\toprule[1pt]
&  Mass (MeV) & Width (MeV) & Main channels \\
\midrule[1pt]
Experiment & $3214 \pm 29 \pm 33 \pm 36$  & $186 \pm 38 \pm 34 \pm 63 $ & $D\pi$ \\
$3^3P_2$  & 3234 & 102.4  & $D^\ast \pi$, $D \rho$, $D^\ast \rho$, $D_2^\ast(2460)\pi$, $D\pi(4.18\%)$  \\
$2^3F_2$  & 3364 & 302.2 & $D(1D_2\pi)$, $D_1(2420) \pi$, $D^\ast \rho$,  \\
&&&$D\pi(1300)$, $D\pi(4.76\%)$, $D^\ast a_0(980)$ \\
\midrule[1pt]
$3^3P_0$  & 3219 &251.1   & $D\pi(1300)$, $D_1(2420) \pi$, $Db_1(1235)$, \\
&&&$D(1D_2) \pi$, $D \pi$, $D(2550) \pi$\\
$3P(1^+)$ & 3200      &144     &$Da_2(1320)$, $D^\ast \pi$ \\
$3P^\prime(1^+)$&3245 &185    &$D_2^\ast(2460) \pi$, $D(1^3D_3) \pi$             \\
$2F(3^+)$ & 3335       &165    &$D_2^\ast(2460)\rho$, $D^\ast \rho$, $D \rho$, $D_2^\ast(2460) \pi$ \\
$2F^\prime(3^+)$  &3377  &248 &$D(1^3D_3) \pi$, $D_2^*(2460) \pi$, $D^* \pi$                \\
$2^3F_4$ &3345 &155        &$D_2^\ast(2460) \rho$, $D^* \rho$, $D^* a_2(1320)$, $D^* f_2(1270)$ \\
\bottomrule[1pt]\bottomrule[1pt]
\end{tabular}
\end{table*}

\subsection{The charmed-strange mesons}
\label{sec3.2}

\begin{figure*}
\begin{center}
\includegraphics[width=1.0\textwidth]{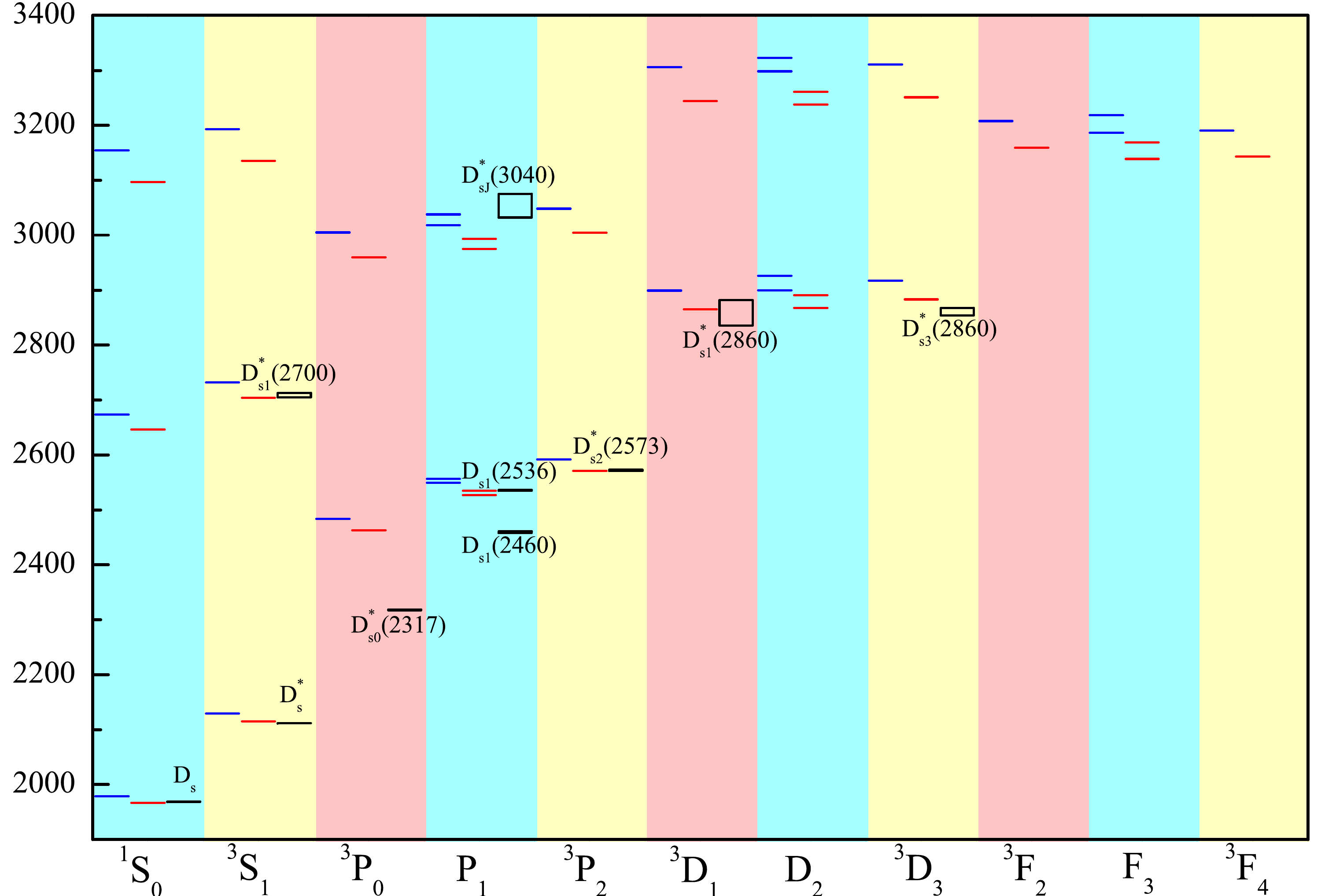}
\end{center}
\caption{ (Color online) Mass spectrum of the charmed-strange
mesons, in units of MeV. The blue lines are obtained by the GI
model~\cite{Godfrey:1985xj}, while the red lines are obtained by the
modified GI model where the screening effect is taken into account.
The blue lozenges denote the experimental data from
PDG~\cite{Beringer:1900zz}. The symbol $^{2S+1}L_J$ is listed on the
abscissa to describe quantum numbers, and another notation $L_L$ is
used when there exists a mixture between the $n^1L_L$ and $n^3L_L$
states. Taken from Ref.~\cite{Song:2015nia}.} \label{sec3:GIscreen2}
\end{figure*}

\renewcommand{\arraystretch}{1.3}
\begin{table*}[htbp]
\scriptsize
\caption{ Comparison of the experimental data and theoretical
results of the charmed-strange mesons obtained using the original GI
model updated by Godfrey and Moats
(GI-Original)~\cite{Godfrey:2015dva}, the QCD-motivated relativistic
quark model based on the quasipotential approach
(R.~Q.~M.)~\cite{Ebert:2009ua}, and the modified GI model taking
into account the screening effect (GI-Screen)~\cite{Song:2015nia}.
The notation $L_L$ is introduced to express mixing states of $^1L_L$
and $^3L_L$. The masses are in units of MeV. } \centering
\begin{tabular}{ c c c c c c }\toprule[1pt]
&$n \ ^{2S+1}L_J$ &Experimental values~\cite{Olive:2016xmw} &
GI-Original~\cite{Godfrey:2015dva} & R.~Q.~M.~\cite{Ebert:2009ua} &
GI-Screen~\cite{Song:2015nia}
\\ \midrule[1pt]
$D_s$               & $1 \ ^1S_0$      & $1968.27 \pm 0.10$                                         & 1979  & 1969  & 1967 \\
$D_s^{\ast}$        & $1 \ ^3S_1$      & $2112.1 \pm 0.4$                                           & 2129  & 2111  & 2115 \\
\hline
$D_{s0}^\ast(2317)$ & $1 \ ^3P_0$      & $2317.7 \pm 0.6$                                           & 2484  & 2509  & 2463 \\
$D_{s1}(2460)$      & $1 \ P_1$        & $2459.5 \pm 0.6$                                           & 2549  & 2536  & 2529 \\
$D_{s1}(2536)$      & $1 \ P_1^\prime$ & $2535.10 \pm 0.06$                                         & 2556  & 2574  & 2534 \\
$D_{s2}^\ast(2573)$ & $1 \ ^3P_2$      & $2569.1 \pm 0.8$                                           & 2592  & 2571  & 2571 \\
\hline
$D_{s1}^\ast(2860)$ & $1 \ ^3D_1$      & $2859\pm12\pm6\pm23$~\cite{Aaij:2014xza,Aaij:2014baa}      & 2899  & 2913  & 2865 \\
--                  & $1 \ D_2$        & --                                                         & 2900  & 2931  & -- \\
--                  & $1 \ D_2^\prime$ & --                                                         & 2926  & 2961  & -- \\
$D_{s3}^\ast(2860)$ & $1 \ ^3D_3$      & $2860.5\pm2.6\pm2.5\pm6.0$~\cite{Aaij:2014xza,Aaij:2014baa}& 2917  & 2971  & 2883 \\
\hline
--                  & $1 \ ^3F_2$      & --                                                         & 3208  & 3230  & 3159 \\
--                  & $1 \ F_3$        & --                                                         & 3186  & 3254  & -- \\
--                  & $1 \ F_3^\prime$ & --                                                         & 3218  & 3266  & -- \\
--                  & $1 \ ^3F_4$      & --                                                         & 3190  & 3300  & 3143 \\
\hline
--                  & $2 \ ^1S_0$      & --                                                         & 2673  & 2688  & 2646 \\
$D_{s1}^\ast(2700)$ & $2 \ ^3S_1$      & $2708.3^{+4.0}_{-3.4}$                                     & 2732  & 2731  & 2704 \\
\hline
--                  & $2 \ ^3P_0$      & --                                                         & 3005  & 3054  & 2960 \\
\hdashline[2pt/2pt] \multirow{2}{*}{$D_{sJ}(3040)$}        & $2 \
P_1$        &
\multirow{2}{*}{$3044\pm8^{+30}_{-5}$~\cite{Aubert:2009ah}}
                                                                                                    & 3018  & 3067  & -- \\
                    & $2 \ P_1^\prime$ &                                                            & 3038  & 3154  & 2992 \\
\hdashline[2pt/2pt]
--                  & $2 \ ^3P_2$      & --                                                         & 3048  & 3142  & 3004 \\
\hline
--                  & $2 \ ^3D_1$      & --                                                         & 3306  & 3383  & 3244 \\
--                  & $2 \ D_2$        & --                                                         & 3298  & 3403  & -- \\
--                  & $2 \ D_2^\prime$ & --                                                         & 3323  & 3456  & -- \\
--                  & $2 \ ^3D_3$      & --                                                         & 3311  & 3469  & 3251 \\
\hline
--                  & $3 \ ^1S_0$      & --                                                         & 3154  & 3219  & -- \\
--                  & $3 \ ^3S_1$      & --                                                         & 3193  & 3242  & -- \\
\hline
--                  & $3 \ ^3P_0$      & --                                                         & 3412  & 3513  & -- \\
--                  & $3 \ P_1$        & --                                                         & 3416  & 3519  & -- \\
--                  & $3 \ P_1^\prime$ & --                                                         & 3433  & 3618  & -- \\
--                  & $3 \ ^3P_2$      & --                                                         & 3439  & 3580  & -- \\
\hline
--                  & $4 \ ^1S_0$      & --                                                         & 3547  & 3652  & -- \\
--                  & $4 \ ^3S_1$      & --
& 3575  & 3669  & --
\\ \bottomrule[1pt]
\end {tabular}
\label{sec3:cs}
\end{table*}

In the following, we present the numerical results of the mass
spectrum of the charmed-strange meson family. We summarize in
Table~\ref{sec3:cs} three
investigations~\cite{Godfrey:2015dva,Ebert:2009ua,Song:2015nia},
which use the same methods as those listed in Sec.~\ref{sec3.1} for
the charmed mesons. The results obtained by using the GI model
taking into account the screening effect are also shown in
Fig.~\ref{sec3:GIscreen2}.

\begin{figure*}
\begin{center}
\includegraphics[width=0.6\textwidth]{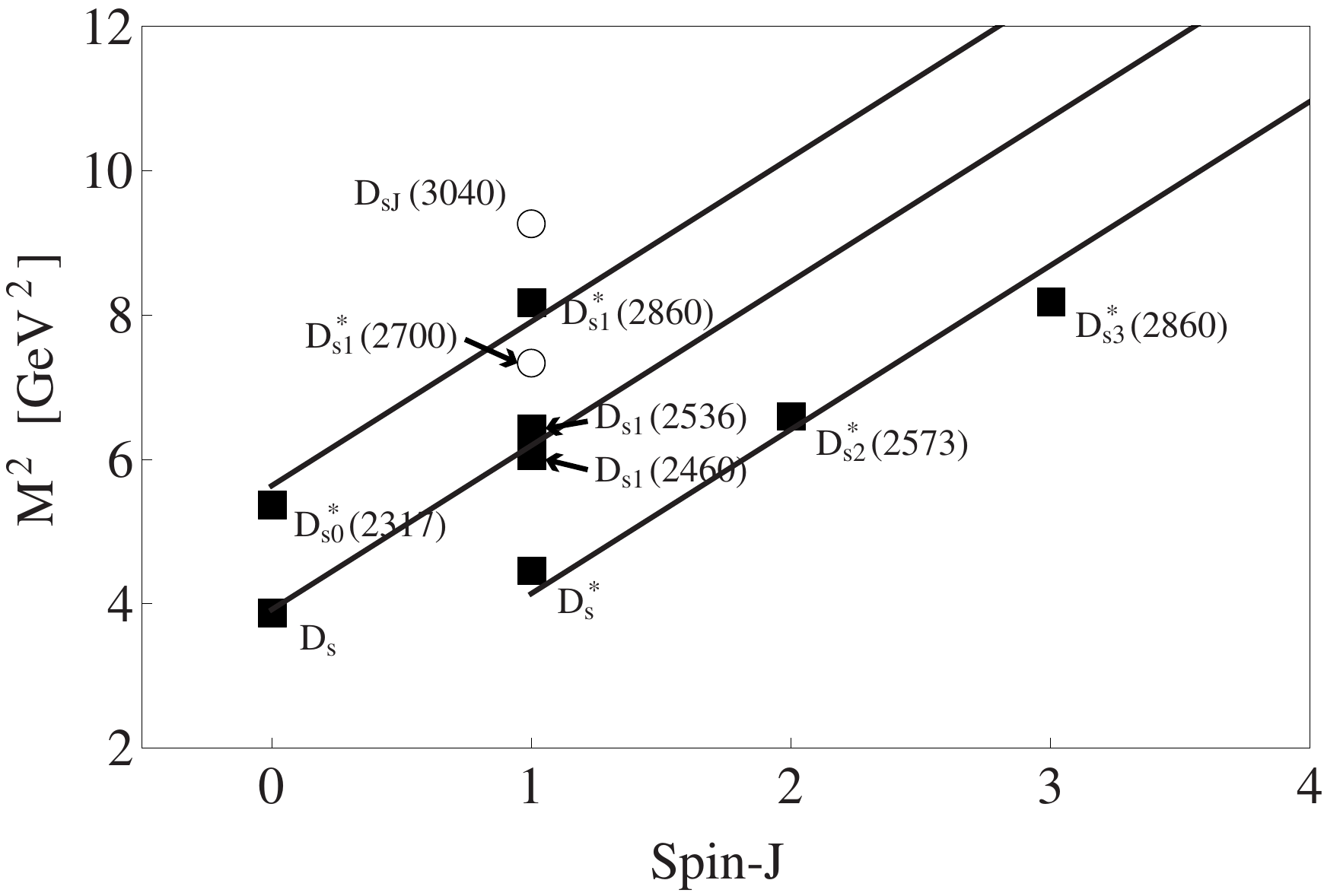}
\end{center}
\caption{ Regge trajectories in the $(J,M^2)$ plane for all the
charmed-strange mesons observed in experiments, where experimental
data are given by solid squares ($1S$ and $1P$ states and $1D$
candidates) and hollow circles (other excited states) with particle
names. } \label{Fig:3.1.Regge2}
\end{figure*}

In Table~\ref{sec3:cs} and Fig.~\ref{sec3:GIscreen2}, the above
theoretical results are also compared with the experimental data,
where we conclude from the mass spectrum analysis, i.e.,
\begin{enumerate}

\item The experimental masses of the $D_{s0}^\ast(2317)$ and
$D_{s1}(2460)$ can not be reproduced by simply assuming they are the
charmed-strange mesons. We shall review the relevant theoretical
studies in Sec.~\ref{sec4}.

\item The two $1S$ states ($D_s$ and $D_s^*$) and the other two $1P$ states ($D_{s1}(2536)$ and $D_{s2}^\ast(2573)$) in the charmed-strange
meson family can be reproduced well.

\item The $D_{s1}^\ast(2860)$ and $D_{s3}^\ast(2860)$ are good
candidates for $D_s(1^3D_1)$ and $D_s(1^3D_3)$, respectively.

\item The $D_{s1}^\ast(2700)$ is a good candidate for $D_s(2^3S_1)$.

\item The $D_{sJ}(3040)$ may be a candidate for $D_s(2P_1)$. In addition,
it can also be interpreted as $1^3D_2$ or $1^1D_2$ states, etc.

\end{enumerate}
Again, we use the charmed-strange mesons to construct Regge
trajectories, as shown in Fig.~\ref{Fig:3.1.Regge2} in the $(J,M^2)$
plane. The results similarly suggest that the $D_s$, $D_s^*$,
$D_{s0}^\ast(2317)$, $D_{s1}(2460)$, $D_{s1}(2536)$ and
$D_{s2}^\ast(2573)$ can be interpreted as the $1S$ and $1P$
charmed-strange mesons, the $D_{s1}^\ast(2860)$ and
$D_{s3}^\ast(2860)$ may be interpreted as the $1D$ charmed-strange
mesons. However, the $D_{s1}^\ast(2700)$ and $D_{sJ}(3040)$ can not be
simply explained.

In the following paragraphs we start to review the theoretical
progress on the excited charmed-strange mesons. See also
Refs.~\cite{vanBeveren:2006st,Li:2007px,Vijande:2008zn,Li:2010bb,Guo:2011dd,Gan:2014jxa,Ke:2014ega,Ge:2015fxa,Wang:2016hkf}
for more information.

\subsubsection{$D_{s1}^\ast(2700)$.}
\label{sec3.2.ds2700}

The $D_{s1}^\ast(2700)$ is a vector charmed-strange state. It was
observed in the $D^{(*)}K$ invariant mass spectrum by many
experiments, including the BaBar, Belle and LHCb
ones~\cite{Aubert:2006mh,Brodzicka:2007aa,Aaij:2012pc}. Its measured
mass is close to the prediction of the $2^3S_1$ charmed-strange
meson~\cite{Godfrey:1985xj}. Based on this assignment, its strong
decay behavior was investigated using the QPC model in
Ref.~\cite{Zhang:2006yj}. The $D_s(2^3S_1)$ assignment is also
supported by the constituent quark model~\cite{Segovia:2015dia},
where the mass and decay width of the $D_{s1}^\ast(2700)$ were
evaluated and are both consistent with the experimental values. In
Ref.~\cite{Wang:2009as}, Wang, Zhang, and Wang studied the
production of the $D_{s1}^\ast(2700)$ from the $B$ meson decay
through a naive factorization method based on the Bethe-Salpeter
method. They calculated the branching ratio of $B^+ \to \bar D^0
D^+_{sJ}(2S) \to \bar D^0 D^0 K^+$, again suggesting that the
$D_{s1}^\ast(2700)$ could be explained as the first radial
excitation of the $D_{s}^\ast(2112)$.

In Ref.~\cite{Colangelo:2007ds} Colangelo {\it et al.} studied the
decay modes of the $D_{s1}^\ast(2700)$ using an effective lagrangian
approach with heavy quark and chiral symmetries. They evaluated the
ratio $\mathcal{B}(D_{s1}^\ast(2700) \to D^\ast
K)/\mathcal{B}(D_{s1}^\ast(2700) \to DK)$, but their result favors
the $D_s(2^1S_0)$ assignment.

In Ref.~\cite{Close:2006gr} Close, Thomas, Lakhina, and Swanson
suggested the $D_{s1}^\ast(2700)$ to be a mixture of the $2^3S_1$
and $1^3D_1$ $c\bar{s}$ states:
\begin{equation}
\label{sec3:mix2700}
 \left(
  \begin{array}{c}
   |D_{s1}^*(2700)\rangle\\
   |D_{s1}^{*}(2860)\rangle\\
  \end{array}
\right )= \left(
  \begin{array}{cc}
    \cos\theta_{SD} & \sin\theta_{SD} \\
   -\sin\theta_{SD} & \cos\theta_{SD}\\
  \end{array}
\right) \left(
  \begin{array}{c}
    |2^3S_1  \rangle \\
   |1^3D_1 \rangle\\
  \end{array}
\right).
\end{equation}
This assignment was supported by Ref.~\cite{Li:2009qu}, where the
authors studied strong decays of the $D_{s1}^\ast(2700)$ using the
QPC model. They used the experimental measurement from
BaBar~\cite{Aubert:2009ah} (see discussions in Sec.~\ref{sec2.2}):
\begin{eqnarray}
\frac{\mathcal{B}(D_{s1}^\ast(2700) \to D^\ast
K)}{\mathcal{B}(D_{s1}^\ast(2700) \to DK)} = 0.91 \pm 0.13 \pm 0.12
\, ,
\end{eqnarray}
to determined the mixing angle $\theta_{SD}$ to be in the range
$-1.38$ rad $\leq \theta_{SD} \leq -1.12$ rad, which was further
used to study the $D_{s1}^{*}(2860)$. Besides these studies, some
other investigations including the Regge
Phenomenology~\cite{Li:2007px,Chen:2011rr} and the constituent quark
model~\cite{Zhong:2009sk} also support the assignment of the
$D_{s1}^\ast(2710)$ as a mixing of the $2^3S_1$ and $1^3D_1$
charmed-strange states.

\begin{figure}[htb]
\begin{center}
\includegraphics[width=0.6\textwidth]{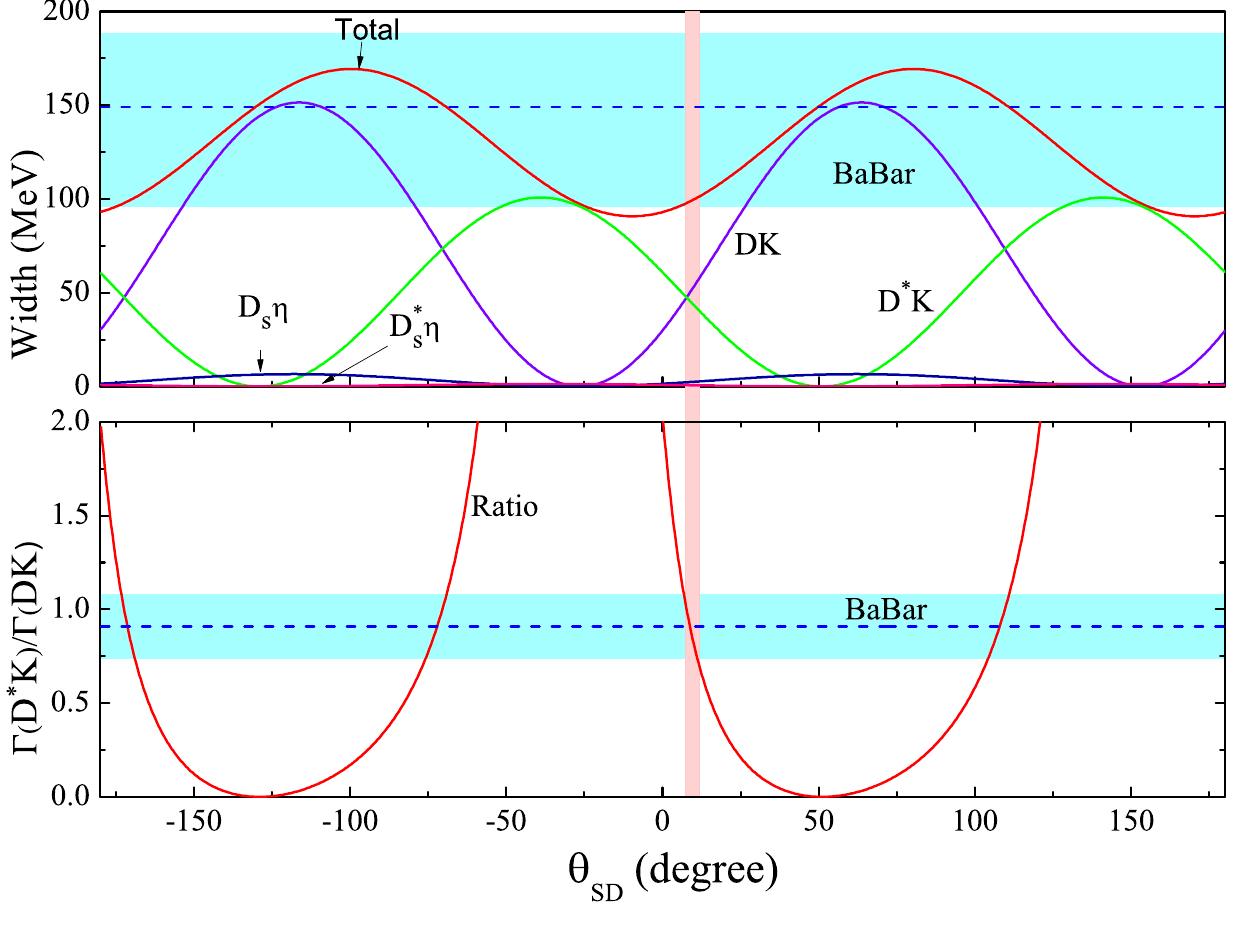}
\end{center}
\caption{ (Color online) The $\theta_{SD}$ dependence of the total
width (the red curve in the top panel), partial decay widths (the
other three colored curves in the top panel) and the ratio
$\Gamma(D^*K)/\Gamma(DK)$ (the red curve in the bottom panel) of the
$D_{s1}^\ast(2700)$, calculated using the QPC
model~\cite{Song:2015nia}. The two blue dashed curves in the top and
bottom panels correspond to the data from the BaBar
experiment~\cite{Aubert:2009ah} that $\Gamma(D_{s1}^\ast(2700)) =
149\pm7^{+39}_{-52}$ and $\Gamma(D^\ast K)/\Gamma(DK) = 0.91 \pm
0.13 \pm 0.12$. Taken from Ref.~\cite{Song:2015nia}.}
\label{sec3:ds2700}
\end{figure}

Especially, the mixing angle $\theta_{SD}$ was evaluated and
discussed in Refs.~\cite{Song:2015nia,Chen:2015lpa}. The decay
properties of the $D_{s1}^\ast(2700)$ do depend on this angle, as
shown in Fig.~\ref{sec3:ds2700} for the $\theta_{SD}$ dependence of
its total decay widths and the ratio $\Gamma(D^*K)/\Gamma(DK)$.
There exists the $\theta_{SD}$ range, $6.8^\circ$-$11.2^\circ$, in
which both of the calculated width and this ratio overlap with the
BaBar experiment~\cite{Aubert:2009ah}. This small $\theta_{SD}$
value is consistent with the estimation in
Ref.~\cite{Godfrey:1985xj}.

Besides the interpretation of the $D_{s1}^\ast(2700)$ as a
charmed-strange state, a $DK^*$ molecule explanation was proposed in
Ref.~\cite{Vinodkumar:2008zd} within the framework of
phenomenological potential models.

\subsubsection{$D_{sJ}^\ast(2860)$, $D_{s1}^\ast(2860)$ and $D_{s3}^\ast(2860)$.}

The $D_{sJ}^\ast(2860)$ was observed in the $D^{(*)}K$ invariant
mass spectrum by the BaBar and LHCb
experiments~\cite{Aubert:2006mh,Aaij:2012pc}. In the later LHCb
experiments~\cite{Aaij:2014xza,Aaij:2014baa}, it was further
separated into two states, the $D_{s1}^\ast(2860)$ and
$D_{s3}^\ast(2860)$. We shall review the relevant theoretical
studies both before and after the LHCb
experiments~\cite{Aaij:2014xza,Aaij:2014baa}.

Before the measurement by the LHCb
Collaboration~\cite{Aaij:2014xza,Aaij:2014baa}, the properties of
the $D_{sJ}^\ast(2860)$ have been widely discussed. Experimentally,
its interpretation as the first radial excitation of the
$D_{s0}^\ast(2317)$ was ruled out due to its decay into $D^\ast
K$~\cite{Aubert:2009ah}, and we shall not discuss this possibility
any more.

Theoretically, its assignment as a $1^3D_3$ charmed strange meson is
supported by various models, including the QPC
model~\cite{Zhang:2006yj}, the Regge phenomenology~\cite{Li:2007px},
the chiral quark model~\cite{Zhong:2008kd}, and the flux tube
model~\cite{Chen:2009zt}. The ratio of
$\mathcal{B}(D_{sJ}^\ast(2860) \to D^\ast K)
/\mathcal{B}(D_{sJ}^\ast(2860) \to DK)$ was evaluated based on the
$1^3D_3$ assignment, which is 0.59 in the QPC
model~\cite{Zhang:2006yj}, 0.43 in the chiral quark
model~\cite{Zhong:2008kd} and 0.81 in the flux tube
model~\cite{Li:2009qu}. Note that these values are smaller than the
experiment data measured by BaBar~\cite{Aubert:2009ah} (see
discussions in Sec.~\ref{sec2.2}):
\begin{eqnarray}
\frac{\mathcal{B}(D_{sJ}^\ast(2860)^+ \to D^\ast K
)}{\mathcal{B}(D_{sJ}^\ast(2860)^+ \to D K )} = 1.10 \pm 0.15 \pm
0.19 \, . \label{sec3:ds2860ratio}
\end{eqnarray}

Besides the simple $1^3D_3$ assignment, the $2S$-$1D$ mixing was
proposed to explain the $D_{sJ}^\ast(2860)$ in
Ref.~\cite{Li:2009qu}, where the $D_{sJ}^\ast(2860)$ and
$D_{s1}^\ast(2710)$ were treated as a mixture of $2^3S_1$ and
$1^3D_1$ charmed-strange mesons, as shown in
Eq.~(\ref{sec3:mix2700}). With a proper mixing angle, the ratios of
$\mathcal{B}(D_{sJ}^\ast(2860) \to D^\ast K)
/\mathcal{B}(D_{sJ}^\ast(2860) \to DK)$ and
$\mathcal{B}(D_{sJ}^\ast(2700) \to D^\ast K)
/\mathcal{B}(D_{sJ}^\ast(2700)  \to DK)$ can be well explained at
the same time.

In Ref.~\cite{Zhong:2009sk} Zhong and Zhao proposed a two-state
scenario for the $D_{sJ}^\ast(2860)$: one is likely to be the
$1^3D_3$ and the other to be the higher mixing state of $1^1D_2$ and
$1^1D_2$. In Ref.~\cite{vanBeveren:2009jq}, Beveren and Rupp also
indicated that there exist two resonances around $2.86$ GeV but with
quantum numbers $J^P=0^+$ and $2^+$. The structure in the $DK$
invariant mass spectrum near 2.86 GeV contains both of these two
resonances, but the structure in the $D^\ast K$ invariant mass
spectrum contains only one resonance of $J^P=2^+$.

In 2014, the LHCb experiments observed two separated states,
$D_{s1}^\ast(2860)$ and $D_{s3}^\ast(2860)$, in the $DK$ invariant
mass spectrum near 2.86 GeV~\cite{Aaij:2014xza,Aaij:2014baa}. Since
there are actually two states, the ratio in
Eq.~(\ref{sec3:ds2860ratio}) observed by the BaBar Collaboration can
be changed according to which state is assigned as the
$D_{sJ}^\ast(2860)$ in both the denominator and numerator. Thus, we
suggest new measurement of this ratio when considering the LHCb
results~\cite{Aaij:2014xza,Aaij:2014baa}.

After the LHCb experiments~\cite{Aaij:2014xza,Aaij:2014baa}, the
decay behaviors of the $D_{s1}^\ast(2860)$ and $D_{s3}^\ast(2860)$
were evaluated by the QPC model in
Refs.~\cite{Song:2014mha,Godfrey:2014fga}, and the result suggested
that these two states can be good $D_s(1D)$ candidates. Based on
these interpretations, their decay behaviors were studied using the
effective Lagrangian approach~\cite{Wang:2014jua} and the
constituent quark model~\cite{Segovia:2015dia}.

By using the QCD sum rule
method~\cite{Shifman:1978bx,Reinders:1984sr} based on the heavy
quark effective
theory~\cite{Grinstein:1990mj,Eichten:1989zv,Falk:1990yz}, the
masses of $1D$ charmed-strange mesons were calculated in
Ref.~\cite{Zhou:2014ytp}, also supporting their interpretations as
the $1D$ charmed-strange mesons. This method has also been developed
in Ref.~\cite{Zhou:2015ywa} to study the $F$ heavy meson doublets
$(2^+,3^+)$ and $(3^+,4^+)$.

Here is a natural picture for the $D_{s1}^\ast(2700)$,
$D_{s1}^\ast(2860)$ and $D_{s3}^\ast(2860)$. The $D_{s1}^\ast(2700)$
and $D_{s1}^\ast(2860)$ can be interpreted as a mixture of $D_s(2
^3S_1)$ and $D_s(1^3D_1)$, as defined in Eq.~(\ref{sec3:mix2700}),
and the $D_{s3}^\ast(2860)$ is a good candidate of $D_s(1^3D_3)$.
This picture was used in Ref.~\cite{Song:2015nia} (see also
discussions in Ref.~\cite{Zhong:2009sk}), where the following ratio
is obtained:
\begin{eqnarray}
\frac{\mathcal{B}(D_s(1^3D_3)\rightarrow
D^{*0}K)}{\mathcal{B}(D_s(1^3D_3)\rightarrow D^0K)}=0.802 \, .
\end{eqnarray}
Later in Ref.~\cite{Chen:2015lpa}, a combined study of $2S$ and $1D$
open-charm mesons with natural spin-parity was performed, where the
$2S$-$1D$ mixing effect was investigated. Their results indicate
that the $D_{s1}^\ast(2700)$ and $D_{s1}^\ast(2860)$ are
predominantly the $2^3S_1$ and $1^3D_1$ charmed-strange mesons,
respectively, while the $D_{s3}^\ast(2860)$ can be regarded as the
$1^3D_3$ charmed-strange meson.

\begin{figure}[htb]
\begin{center}
\includegraphics[width=0.6\textwidth]{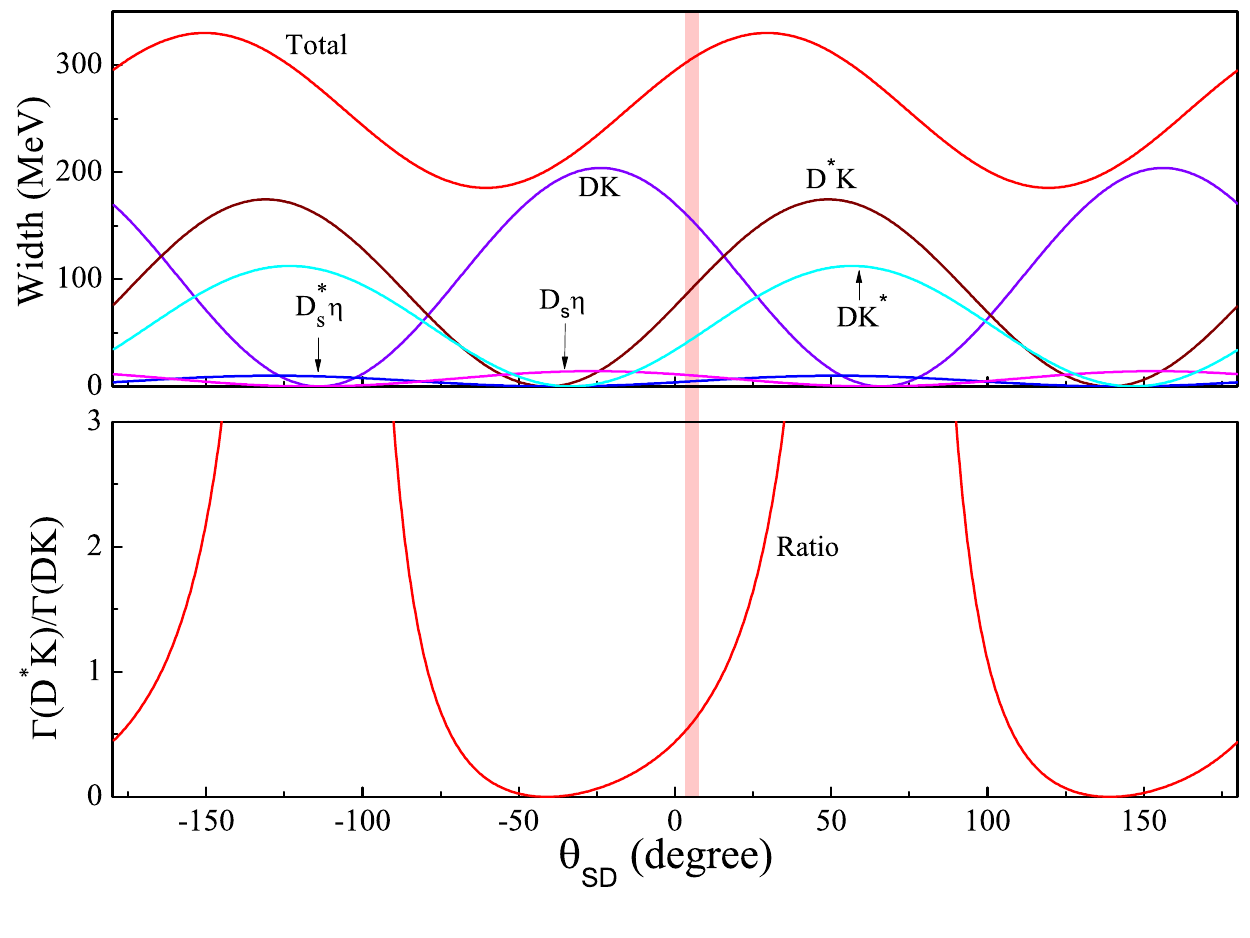}
\end{center}
\caption{ (Color online) The $\theta_{SD}$ dependence of the total
width (the red curve in the top panel), partial decay widths (the
other five colored curves in the top panel) and the ratio
$\Gamma(D^*K)/\Gamma(DK)$ (the red curve in the bottom panel) of the
$D_{s1}^*(2860)$, calculated using the QPC
model~\cite{Song:2015nia}. The red vertical band corresponds to the
common range of $\theta_{SD}$ used for the $D_{s1}^\ast(2700)$ and
shown in Fig.~\ref{sec3:ds2700}. Taken from
Ref.~\cite{Song:2015nia}. } \label{sec3:ds2860}
\end{figure}

Again, the decay properties of the $D_{s1}^*(2860)$ depend on the
mixing angle $\theta_{SD}$, as shown in
Fig.~\ref{sec3:ds2860}~\cite{Song:2015nia}. If taking
$6.8^\circ$-$11.2^\circ$ for the range of $\theta_{SD}$ obtained in
the study of the $D_{s1}^\ast(2700)$, the total decay width
of the $D_{s1}^*(2860)$ would reach up to $\sim 300$ MeV comparable
with the LHCb data \cite{Aaij:2014xza,Aaij:2014baa}, and the ratio
is $\mathcal{B}(D_{s1}^*(2860)\rightarrow D^*K) /
\mathcal{B}(D_{s1}^*(2860)\rightarrow DK)=0.6\sim0.8$ which can be
tested in future experiments.

Recently in Ref.~\cite{Wang:2016enc}, the OZI allowed two-body
strong decays of $3^-$ heavy-light mesons were systematically
studied, and the total strong decay width of the $D^*_{s3}(2860)$
was evaluated to be 47.6 MeV, which is consistent with the
experiments~\cite{Aaij:2014xza,Aaij:2014baa}.

\subsubsection{$D_{sJ}(3040)$.}

The observed mass of the $D_{sJ}(3040)$ and its unnatural parity are
consistent with the quark model prediction for the $2^3P_1$ charmed
strange meson~\cite{Godfrey:1985xj}. Hence, it can be interpreted as
the $D_s(2^3P_1)$ state, the first radial excitation of the
$D_{s1}(2460)$.

The calculations in the QPC model~\cite{Sun:2009tg} also support the
$D_{sJ}(3040)$ as the $1^+$ state in the $(0^+,1^+)$ spin doublet.
In addition, studies using the flux tube model~\cite{Chen:2009zt},
the constitute quark model~\cite{Xiao:2014ura,Zhong:2009sk} and the
effective approach~\cite{Colangelo:2010te} all indicated the
possible interpretation of the $D_{sJ}(3040)$ as a $1^+$
charmed-strange meson. Moreover, in
Refs.~\cite{Segovia:2013wma,Segovia:2012cd} Segovia {\it et al.}
calculated the decay widths of the $D_{sJ}(3040)$ as a
$n(J^P)=3(1^+)$ or $4(1^+)$ state, and their results are compatible
with the experimental data~\cite{Beringer:1900zz}.

Besides the above $J^P=1^+$ assignment, the $D_{sJ}(3040)$ was
interpreted as a mixture of the $1^3D_2$ and $1^1D_2$
charmed-strange meson in Ref.~\cite{Colangelo:2010te} with an
effective Lagrangian approach.

%% file: section3.2.tex
\subsection{The bottom mesons}
\label{sec3.3}

\begin{figure*}
\begin{center}
\includegraphics[width=0.6\textwidth]{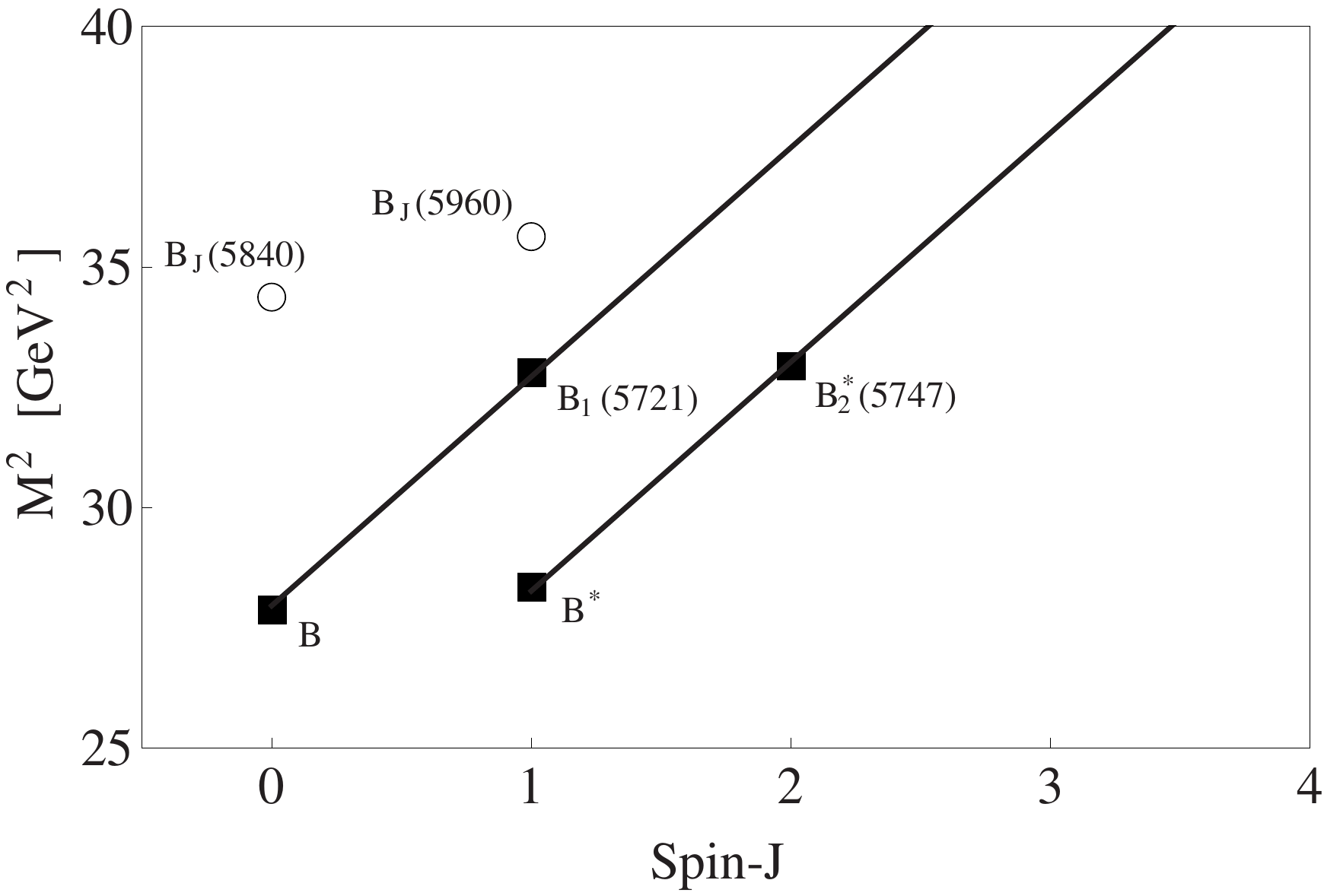}
\end{center}
\caption{
Regge trajectories in the $(J,M^2)$ plane for all the bottom mesons observed in experiments, where experimental data are given by solid squares ($1S$ and $1P$ states) and hollow circles (other excited states) with particle names.
} \label{Fig:3.1.Regge3}
\end{figure*}

In the following, we discuss the mass spectrum of the bottom meson
family. We investigate the following two methods:
\begin{enumerate}
\item The alternate relativized model and the original GI model updated by Godfrey, Moats and Swanson~\cite{Godfrey:2016nwn},
\item The QCD-motivated relativistic quark model based on the quasipotential approach~\cite{Ebert:2009ua}.
\end{enumerate}
These results are summarized in Table~\ref{sec3:bottom} and compared with the experimental data. We conclude
from the mass spectrum analysis, i.e.,
\begin{enumerate}

\item The two $1S$ states ($B$ and $B^*$) in the bottom meson family can be reproduced well.

\item The $B_1(5721)$ can be regarded as the mixture of $B(1^1P_1)$
and $B(1^3P_1)$ states, and the $B^*_2(5747)$ can be regarded as the $B(1^3P_2)$ state.

\item The $B_J(5840)$ may be a candidate for $B(2^1S_0)$. The $B(5970)$ and $B_J(5960)$ are probably the same state, and may be a candidate for $B(2^3S_1)$.

\end{enumerate}
Regge trajectories constructed using the bottom mesons are shown in Fig.~\ref{Fig:3.1.Regge3} in the $(J,M^2)$ plane, and similar conclusions can be obtained that the $B$, $B^*$, $B_1(5721)$ and $B^*_2(5747)$ can be interpreted as the $1S$ and $1P$ bottom mesons, but the $B_J(5840)$ and $B(5970)/B_J(5960)$ can not be simply explained.

In the following paragraphs we start to review the theoretical progress on the excited bottom mesons. See also Refs.~\cite{Vijande:2007ke,Collins:1996kb,Wurtz:2015mqa,Soler:2015hna,Grinstein:2015aua,Bondar:2016pox} for more information.

\renewcommand{\arraystretch}{1.3}
\begin{table*}[htbp]
\footnotesize
\caption{Comparison of the experimental data and theoretical results
of the bottom mesons obtained using the original GI model
updated by Godfrey, Moats and Swanson (GI-Original)~\cite{Godfrey:2016nwn} and the relativistic quark model
(R.~Q.~M.)~\cite{Ebert:2009ua}. The notation $L_L$ is
introduced to express the mixing states of $^1L_L$ and $^3L_L$.
The masses are in units of MeV. }
\centering
\begin{tabular}{ c c c c c }\toprule[1pt]
&$n \ ^{2S+1}L_J$ & Experimental values~\cite{Olive:2016xmw} &
GI-Original~\cite{Godfrey:2016nwn} & R.~Q.~M.~\cite{Ebert:2009ua}
 \\ \midrule[1pt]
$B^0$           & $1 \ ^1S_0$      & $5279.62 \pm 0.15$                                    & 5312  & 5280 \\
$B^{*}$         & $1 \ ^3S_1$      & $5324.65 \pm 0.25$                                    & 5371  & 5326 \\
\hline
--              & $1 \ ^3P_0$      & --                                                    & 5756  & 5749 \\
\hdashline[2pt/2pt]
\multirow{2}{*}{$B_1(5721)^0$}     & $1 \ P_1$        & \multirow{2}{*}{$5727.7\pm0.7\pm1.4\pm0.17\pm0.4$~\cite{Aaij:2015qla}}
                                                                                           & 5777  & 5723 \\
                & $1 \ P_1^\prime$ &                                                       & 5784  & 5774 \\
\hdashline[2pt/2pt]
$B^*_2(5747)^0$ & $1 \ ^3P_2$      & $5739.44\pm0.37\pm0.33\pm0.17$~\cite{Aaij:2015qla}    & 5797  & 5741 \\
\hline
--              & $1 \ ^3D_1$      & --                                                    & 6110  & 6119 \\
--              & $1 \ D_2$        & --                                                    & 6095  & 6103 \\
--              & $1 \ D_2^\prime$ & --                                                    & 6124  & 6121 \\
--              & $1^3D_3$         & --                                                    & 6106  & 6091 \\
\hline
--              & $1 \ ^3F_2$      & --                                                    & 6387  & 6412 \\
--              & $1 \ F_3$        & --                                                    & 6358  & 6391 \\
--              & $1 \ F_3^\prime$ & --                                                    & 6396  & 6420 \\
--              & $1 \ ^3F_4$      & --                                                    & 6364  & 6380 \\
\hline
$B_J(5840)^0$   & $2 \ ^1S_0$      & $5862.9\pm5.0\pm6.7\pm0.2$~\cite{Aaij:2015qla}        & 5904  & 5890 \\
$B_J(5960)^0$   & $2 \ ^3S_1$      & $5969.2 \pm 2.9 \pm 5.1 \pm 0.2$~\cite{Aaij:2015qla}  & 5933  & 5906 \\
\hline
--              & $2 \ ^3P_0$      & --                                                    & 6213  & 6221 \\
--              & $2 \ P_1$        & --                                                    & 6197  & 6209 \\
--              & $2 \ P_1^\prime$ & --                                                    & 6228  & 6281 \\
--              & $2 \ ^3P_2$      & --                                                    & 6213  & 6260 \\
\hline
--              & $2 \ ^3D_1$      & --                                                    & 6475  & 6534 \\
--              & $2 \ D_2$        & --                                                    & 6450  & 6528 \\
--              & $2 \ D_2^\prime$ & --                                                    & 6486  & 6554 \\
--              & $2 \ ^3D_3$      & --                                                    & 6460  & 6542 \\
\hline
--              & $3 \ ^1S_0$      & --                                                    & 6335  & 6379 \\
--              & $3 \ ^3S_1$      & --                                                    & 6355  & 6387 \\
\hline
--              & $3 \ ^3P_0$      & --                                                    & 6576  & 6629 \\
--              & $3 \ P_1$        & --                                                    & 6557  & 6650 \\
--              & $3 \ P_1^\prime$ & --                                                    & 6585  & 6685 \\
--              & $3 \ ^3P_2$      & --                                                    & 6570  & 6678 \\
\hline
--              & $4 \ ^1S_0$      & --                                                    & 6689  & 6781 \\
--              & $4 \ ^3S_1$      & --                                                    & 6703  & 6786
\\ \bottomrule[1pt]
\end {tabular}
\label{sec3:bottom}
\end{table*}

\subsubsection{$B^*_J(5732)$, $B_1(5721)^0$ and $B^*_2(5747)^0$.}

The excited bottom meson $B^*_J(5732)$, first observed by the OPAL
detector at LEP~\cite{Akers:1994fz}, was later separated into two
states, the $B_1(5721)^0$ and $B^*_2(5747)^0$, by the D0
Collaboration~\cite{Abazov:2007vq}. The $B_1(5721)^0$ can be
regarded as either the $1^1P_1$ or $1^3P_1$ state or their mixture,
while the $B^*_2(5747)^0$ can be regarded as the $1^3P_2$ state.

In Ref.~\cite{Zhong:2008kd} Zhong and Zhao studied the strong decays
of the heavy-light mesons in a chiral quark model. By assigning the
$B^*_2(5747)^0$ as a $1^3P_2$ state, they obtained its total width
as a sum of $\Gamma(B\pi)$ and $\Gamma(B^*\pi)$ to be about 47 MeV,
consistent with the CDF measurement $\Gamma(B^*_2(5747)) \approx
22^{+7}_{-6}$ MeV~\cite{Aaltonen:2008aa}. They also obtained the
ratio
\begin{eqnarray}
R \equiv { \Gamma(B^* \pi) \over \Gamma(B^* \pi) + \Gamma(B\pi)} =
0.47  \, ,
\end{eqnarray}
which is also in good agreement with the D0 measurement, i.e., $R =
0.475 \pm 0.095 \pm 0.069$~\cite{Abazov:2007vq}.

Since the heavy-light mesons are not charge conjugation eigenstates,
the state mixing between spin $S = 0$ and $S = 1$ states with the
same $J^P$ can occur:
\begin{equation}
\left(
  \begin{array}{c}
   |1P_1\rangle\\
   |1P^\prime_1\rangle\\
  \end{array}
\right )= \left(
  \begin{array}{cc}
    \cos\theta_{1P} & \sin\theta_{1P} \\
   -\sin\theta_{1P} & \cos\theta_{1P} \\
  \end{array}
\right) \left(
  \begin{array}{c}
    |1^1P_1  \rangle \\
   |1^3P_1 \rangle\\
  \end{array}
\right),
\end{equation}
After choosing $\theta_{1P} = -(55 \pm 5)^{\rm o}$, the authors of
Ref.~\cite{Zhong:2008kd} found that the $B_1(5721)^0$ can be
interpreted as the $|1P^\prime_1\rangle$ state (note that the
notations here are different from those used in
Ref.~\cite{Zhong:2008kd}), and its $B^*\pi$ partial width was
evaluated to be about 30 MeV. This mixing angle is similar to the
one in the heavy quark limit, $\theta_{1P} = -54.7^{\rm o}$. With
the above strong decay widths for the $B^*_2(5747)$, the authors of
Ref.~\cite{Zhong:2008kd} obtained the following ratio
\begin{eqnarray}
R \equiv { \Gamma(B_1(5721)) \over \Gamma(B_1(5721)) +
\Gamma(B^*_2(5747))} = 0.34 \, ,
\end{eqnarray}
which is also consistent with the D0 experiment, i.e., $R = 0.477
\pm 0.069 \pm 0.062$~\cite{Abazov:2007vq}.

These two assignments, that the $B_1(5721)^0$ as a mixture of
$1^1P_1$ and $1^3P_1$ states and the $B^*_2(5747)^0$ as a $1^3P_2$
state, are supported by studies using the nonrelativistic quark
model~\cite{Lu:2016bbk}. Based on these assignments, their decay
properties were studied in Ref.~\cite{Wang:2014cta} within the heavy
meson effective theory.

\begin{figure}[htb]
\centering%
\begin{tabular}{c}
\includegraphics[width=0.6\textwidth]{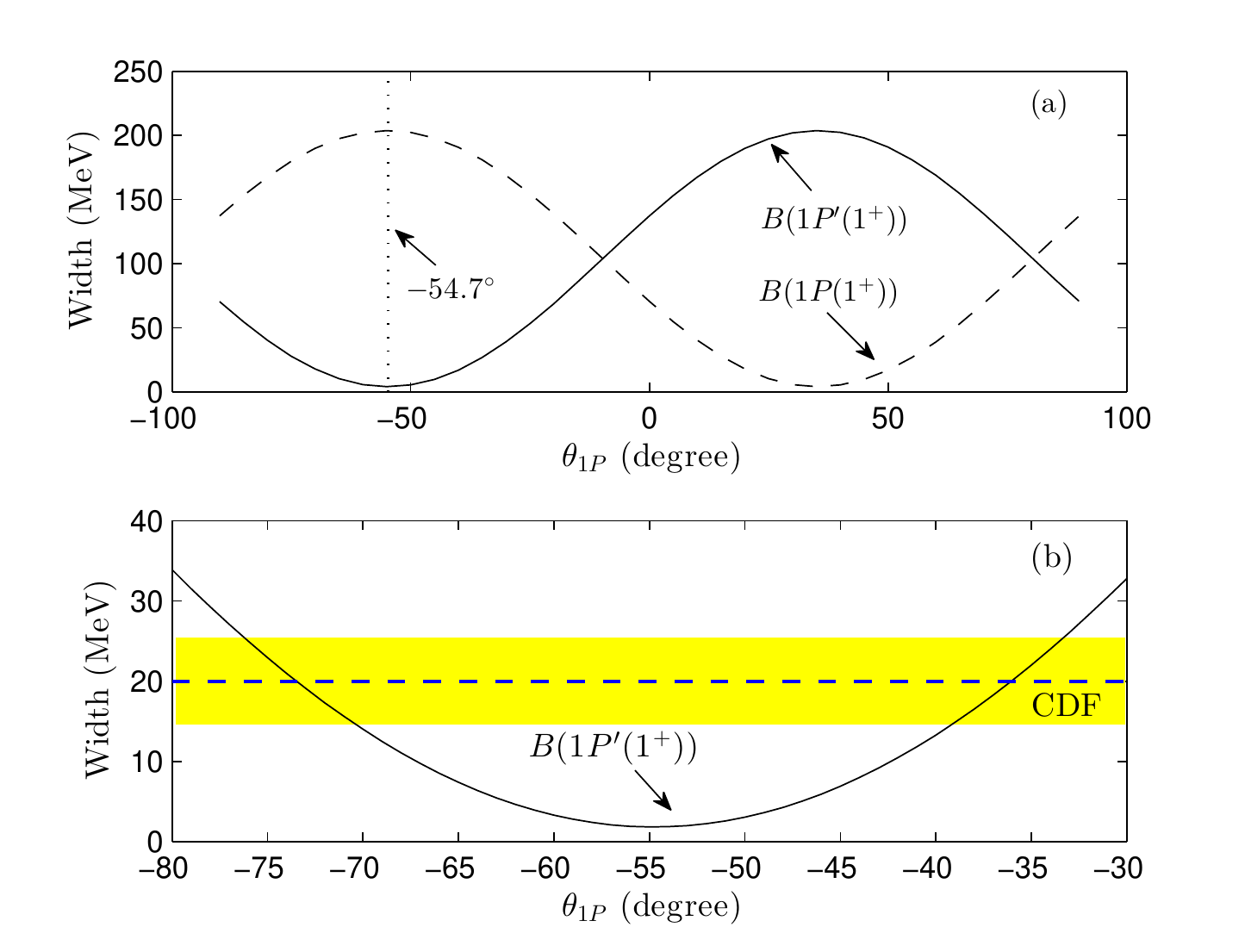}
\end{tabular}
\caption{(Color online) The $\theta_{1P}$ dependence of the total decay
widths of the $B(1P_1) = B(1P(1^+))$ (the dashed curve in the top panel) and
the $B(1P^\prime_1) = B(1P^\prime(1^+))$ (the two solid curves in the top and bottom panels).
The vertical dashed line in the top panel corresponds to the ideal mixing angle $\theta_{1P}=-54.7^\circ$ from the heavy quark limit,
and the blue dashed curve in the bottom panel corresponds to the CDF data that $\Gamma(B_1(5721)) = 20 \pm 2 \pm 5$ MeV~\cite{Aaltonen:2013atp}.
Taken from Ref.~\cite{Sun:2014wea}.}
\label{sec3:b5721}
\end{figure}

In Ref.~\cite{Sun:2014wea} Sun {\it et al.} systematically studied
the mass spectrum and strong decay patterns of the excited bottom
and bottom-strange mesons using the QPC model. They evaluated and
discussed the mixing angle $\theta_{1P}$. The $\theta_{1P}$
dependence of the total decay widths of the $B(1P_1)$ and
$B(1P^\prime_1)$ is shown in Fig.~\ref{sec3:b5721}. Sun {\it et al.}
found that the $B_1(5721)$ is a good candidate for the $1P^\prime_1$
bottom state since it has a narrow width. Moreover, the calculated
width of the $B(1P^\prime_1)$ overlaps with the experimental width
of the $B_1(5721)$ when $\theta_{1P}$ is in the range of
$-77^\circ\sim-70^\circ$ or $-40^\circ\sim -33^\circ$. This mixing
angle $\theta_{1P}$ was also studied in Ref.~\cite{Chen:2012zk} by
Chen, Yuan and Zhang, which systematically investigated the strong
decays of the $P$-wave heavy-light mesons within the
Eichten-Hill-Quigg (EHQ) formalism.

\subsubsection{$B(5970)$, $B_J(5960)$ and $B_J(5840)$.}

The $B(5970)$ and $B_J(5960)$ are probably the same state, whose
mass is close to the estimated masses of the $2^1S_0$ and $2^3S_1$
states of the $B$ meson family. Since the $B(5970)$ decays into
$B\pi$, we can exclude the $B(2^1S_0)$ assignment.

In Ref.~\cite{Sun:2014wea} Sun {\it et al.} evaluated the total
width of $B(2^3S_1)$ using the QPC model to be 47 MeV, which is in
agreement with the experimental width of the $B(5970)$. They also
calculated several partial decay widths, and their result indicated
that the $B(5970)$ is very probably the $B(2^3S_1)$ state. They also
suggested the experimental search for the $B(5970)$ via its $\pi
B^*$ decay.

Later in Ref.~\cite{Wang:2014cta}, Wang studied the two-body strong
decays of the $B(5970)$ within the heavy meson effective theory by
assuming it to be the $2S$ $1^-$, $1D$ $1^-$ and $1D$ $3^-$ states.
Its decay behavior as the $2^3S_1$ state was also investigated in
Ref.~\cite{Xu:2014mka} using the effective Lagrangian approach.

There exist other possible interpretations. In
Ref.~\cite{Xiao:2014ura}, Xiao and Zhong investigated the strong
decay properties of the $B(5970)$ using a chiral quark model, and
their result suggested that the $B(5970)$ resonance is most likely
to be the $1^3D_3$ with $J^P = 3^-$. Later in
Ref.~\cite{Lu:2016bbk}, the authors studied the excited bottom
mesons in the nonrelativistic quark model, and their results
suggested that the $B(5970)/B_J(5960)$ can be interpreted as either
$B(2^3S_1)$ or $B(1^3D_3)$ states.

In Ref.~\cite{Lu:2016bbk}, the authors studied the $B_J(5840)$ in
the nonrelativistic quark model, and their results suggested that
the $B_J(5840)$ can be interpreted as the $B(2^1S_0)$ state, which
is also suggested by LHCb Collaboration~\cite{Aaij:2015qla}.

\subsection{The bottom-strange mesons}
\label{sec3.4}

\begin{figure*}
\begin{center}
\includegraphics[width=0.6\textwidth]{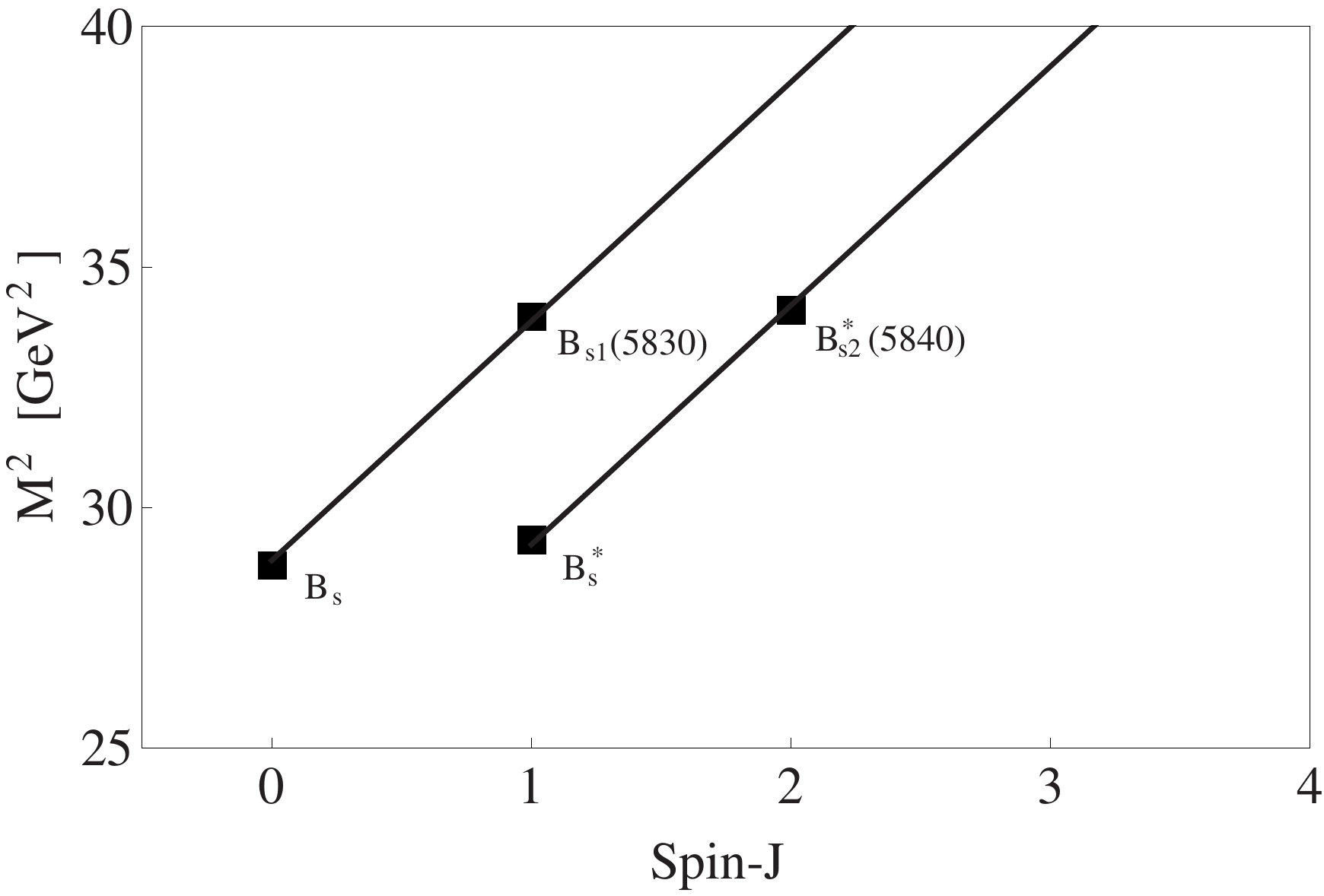}
\end{center}
\caption{
Regge trajectories in the $(J,M^2)$ plane for all the bottom-strange mesons observed in experiments, where experimental data are given by solid squares ($1S$ and $1P$ states) with particle names.
} \label{Fig:3.1.Regge4}
\end{figure*}

In the following, we discuss the mass spectrum of the bottom-strange meson family.
We summarize in Table~\ref{sec3:bs} the same investigations~\cite{Godfrey:2016nwn,Ebert:2009ua} as those listed in Sec.~\ref{sec3.3} for the bottom mesons.
Comparing these theoretical values with the experimental data, we conclude
\begin{enumerate}

\item The two $1S$ states ($B_s$ and $B_s^*$) in the bottom-strange meson family are reproduced well.

\item The $B_{s1}(5830)$ can be regarded as the mixture
of the $B_s(1^1P_1)$ and $B_s(1^3P_1)$ states, and the
$B^*_{s2}(5840)$ can be regarded as the $B_s(1^3P_2)$ state.

\end{enumerate}
More discussions can be found in Refs.~\cite{Koponen:2007nr,Luo:2009wu,Zhang:2009ewa,Zhang:2010gp,Gan:2010hw,Albaladejo:2016ztm}.
Again, we use the bottom-strange mesons to construct Regge trajectories, as shown in Fig.~\ref{Fig:3.1.Regge4} in the $(J,M^2)$ plane. The results similarly suggest that the $B_s$, $B_s^*$, $B_{s1}(5830)$ and $B^*_{s2}(5840)$ can be interpreted as the $1S$ and $1P$ bottom-strange mesons.

In the following paragraphs we start to review the theoretical progress on the excited bottom-strange mesons.

\renewcommand{\arraystretch}{1.3}
\begin{table*}[htbp]
\footnotesize
\caption{Comparison of the experimental data and theoretical results
of the bottom-strange mesons obtained using the original GI model
updated by Godfrey, Moats and Swanson (GI-Original)~\cite{Godfrey:2016nwn} and the
relativistic quark model (R.~Q.~M.)~\cite{Ebert:2009ua}.
The notation $L_L$ is introduced to express mixing states of $^1L_L$
and $^3L_L$.
The masses are in units of MeV. } \centering
\begin{tabular}{ c c c c c c c }\toprule[1pt]
&$n \ ^{2S+1}L_J$ & Experimental values~\cite{Olive:2016xmw} &
GI-Original~\cite{Godfrey:2016nwn} & R.~Q.~M.~\cite{Ebert:2009ua}
 \\ \midrule[1pt]
$B_s$           & $1 \ ^1S_0$      & $5366.82 \pm 0.22$                                    & 5394  & 5372 \\
$B_s^{*}$       & $1 \ ^3S_1$      & $5415.4^{+1.8}_{-1.5}$                                & 5450  & 5414 \\
\hline
--              & $1 \ ^3P_0$      & --                                                    & 5831  & 5833 \\
\hdashline[2pt/2pt]
\multirow{2}{*}{$B_{s1}(5830)$}    & $1 \ P_1$        & \multirow{2}{*}{$5828.3\pm0.1\pm0.2\pm0.4$~\cite{Aaltonen:2013atp}}
                                                                                           & 5857  & 5831 \\
                & $1 \ P_1^\prime$ &                                                       & 5861  & 5865 \\
\hdashline[2pt/2pt]
$B^*_{s2}(5840)$& $1 \ ^3P_2$      & $5839.7\pm0.1\pm0.1\pm0.2$~\cite{Aaltonen:2013atp}    & 5876  & 5842 \\
\hline
--              & $1 \ ^3D_1$      & --                                                    & 6182  & 6209 \\
--              & $1 \ D_2$        & --                                                    & 6169  & 6189 \\
--              & $1 \ D_2^\prime$ & --                                                    & 6196  & 6218 \\
--              & $1^3D_3$         & --                                                    & 6179  & 6191 \\
\hline
--              & $1 \ ^3F_2$      & --                                                    & 6454  & 6501 \\
--              & $1 \ F_3$        & --                                                    & 6425  & 6468 \\
--              & $1 \ F_3^\prime$ & --                                                    & 6462  & 6515 \\
--              & $1 \ ^3F_4$      & --                                                    & 6432  & 6475 \\
\hline
--              & $2 \ ^1S_0$      & --                                                    & 5984  & 5976 \\
--              & $2 \ ^3S_1$      & --                                                    & 6012  & 5992 \\
\hline
--              & $2 \ ^3P_0$      & --                                                    & 6279  & 6318 \\
--              & $2 \ P_1$        & --                                                    & 6279  & 6321 \\
--              & $2 \ P_1^\prime$ & --                                                    & 6296  & 6345 \\
--              & $2 \ ^3P_2$      & --                                                    & 6295  & 6359 \\
\hline
--              & $2 \ ^3D_1$      & --                                                    & 6542  & 6629 \\
--              & $2 \ D_2$        & --                                                    & 6526  & 6625 \\
--              & $2 \ D_2^\prime$ & --                                                    & 6553  & 6651 \\
--              & $2 \ ^3D_3$      & --                                                    & 6535  & 6637 \\
\hline
--              & $3 \ ^1S_0$      & --                                                    & 6410  & 6467 \\
--              & $3 \ ^3S_1$      & --                                                    & 6429  & 6475 \\
\hline
--              & $3 \ ^3P_0$      & --                                                    & 6639  & 6731 \\
--              & $3 \ P_1$        & --                                                    & 6635  & 6761 \\
--              & $3 \ P_1^\prime$ & --                                                    & 6650  & 6768 \\
--              & $3 \ ^3P_2$      & --                                                    & 6648  & 6780 \\
\hline
--              & $4 \ ^1S_0$      & --                                                    & 6759  & 6874 \\
--              & $4 \ ^3S_1$      & --                                                    & 6773  & 6879
\\ \bottomrule[1pt]
\end {tabular}
\label{sec3:bs}
\end{table*}

\subsubsection{$B^*_{sJ}(5850)$, $B_{s1}(5830)^0$ and $B^*_{s2}(5840)^0$.}

The properties of $B^*_{sJ}(5850)$, $B_{s1}(5830)^0$ and
$B^*_{s2}(5840)^0$ are quite similar to those of the $B^*_J(5732)$,
$B_1(5721)^0$ and $B^*_2(5747)^0$. The excited bottom-strange meson
$B^*_{sJ}(5850)$, first observed by the OPAL detector at
LEP~\cite{Akers:1994fz}, was later separated into two states, the
$B_{s1}(5830)^0$ and $B^*_{s2}(5840)^0$, by the CDF
Collaboration~\cite{Aaltonen:2007ah}. The $B_{s1}(5830)^0$ can be
regarded as the mixture of the $1^1P_1$ and $1^3P_1$ states, while
the $B^*_{s2}(5840)^0$ can be regarded as the $1^3P_2$ state.

Again in Ref.~\cite{Zhong:2008kd}, Zhong and Zhao studied the strong
decays of the $B_{s1}(5830)^0$ and $B^*_{s2}(5840)^0$ in a chiral
quark model. By assigning the $B^*_{s2}(5840)^0$ as a $1^3P_2$
state, they obtained its total width as a sum of $\Gamma(BK)$ and
$\Gamma(B^*K)$ to be about 2 MeV, consistent with the CDF
measurement $\Gamma(B^*_{s2}(5840)) \approx 1$
MeV~\cite{Aaltonen:2008aa}. They also obtained the ratio
\begin{eqnarray}
R \equiv { \Gamma(B^* K) \over \Gamma(B K)} \approx 6\%  \, .
\end{eqnarray}

The $B_s(1^1P_1)$ and $B_s(1^3P_1)$ states mix with each other
\begin{equation}
\left(
  \begin{array}{c}
   |1P_1\rangle\\
   |1P^\prime_1\rangle\\
  \end{array}
\right )= \left(
  \begin{array}{cc}
    \cos\theta_{1P} & \sin\theta_{1P} \\
   -\sin\theta_{1P} & \cos\theta_{1P} \\
  \end{array}
\right) \left(
  \begin{array}{c}
    |1^1P_1  \rangle \\
   |1^3P_1 \rangle\\
  \end{array}
\right),
\end{equation}
With $\theta_{1P} = -(55 \pm 5)^{\rm o}$, the authors of
Ref.~\cite{Zhong:2008kd} found that the $B_{s1}(5830)^0$ can be
interpreted as the $|1P^\prime_1\rangle$ state. They also evaluated
the decay width of the $B_1(5721)^0$ to be $0.4\sim1$ MeV, and
obtained the following ratio
\begin{eqnarray}
R \equiv { \Gamma(B_{s1}(5830)) \over \Gamma(B_{s1}(5830)) +
\Gamma(B^*_{s2}(5840))} = 0.02 \sim 0.6 \, .
\end{eqnarray}

\begin{figure}[htbp]
\centering%
\begin{tabular}{c}
\includegraphics[width=0.6\textwidth]{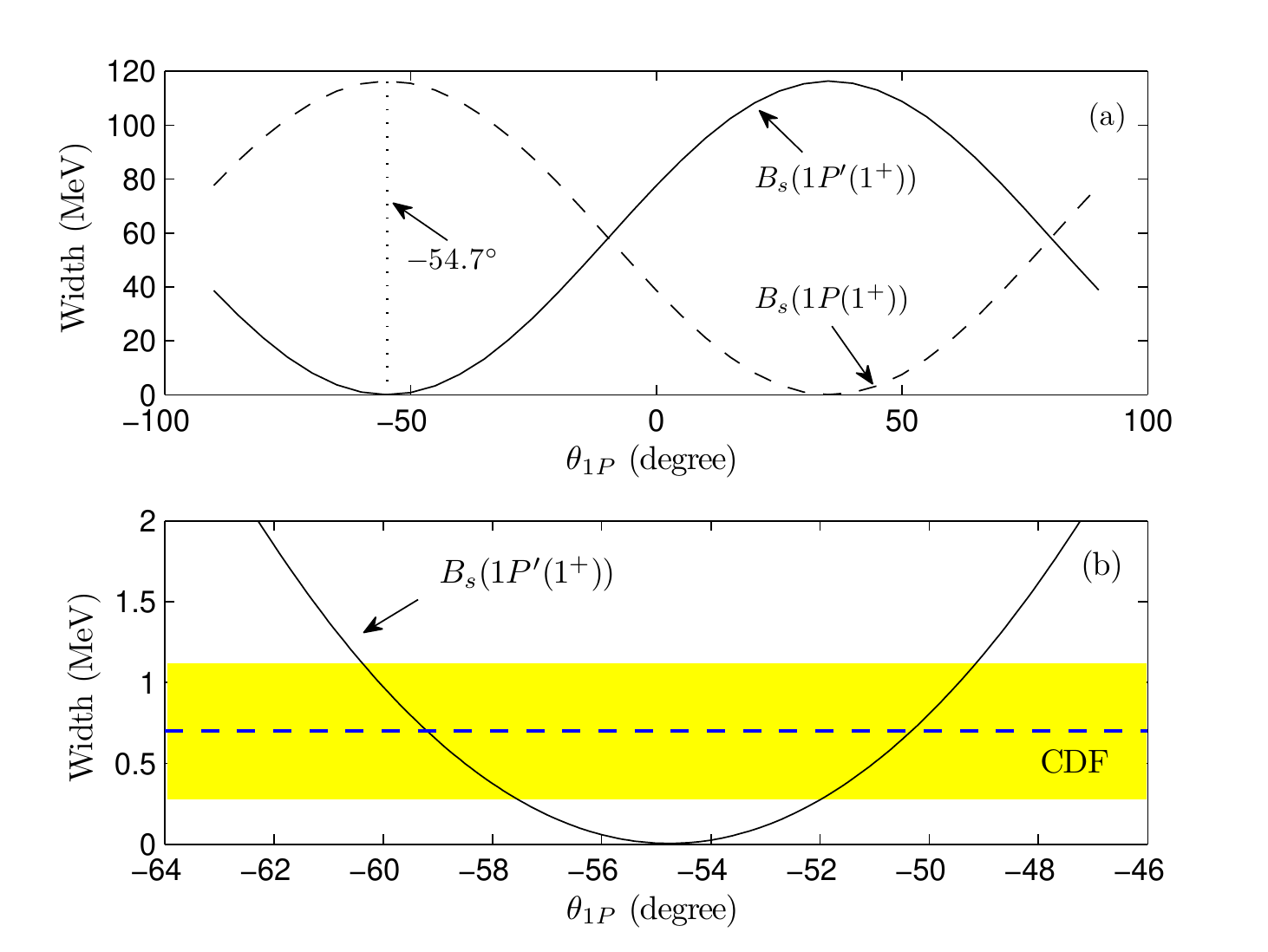}
\end{tabular}
\caption{
(Color online) The $\theta_{1P}$ dependence of the total decay
widths of the $B_s(1P_1) = B_s(1P(1^+))$ (the dashed curve in the top panel) and
the $B_s(1P^\prime_1) = B_s(1P^\prime(1^+))$ (the two solid curves in the top and bottom panels).
The vertical dashed line in the top panel corresponds to the ideal mixing angle $\theta_{1P}=-54.7^\circ$ from the heavy quark limit,
and the blue dashed curve in the bottom panel corresponds to the CDF data that $\Gamma(B_{s1}(5830)) = 0.7 \pm 0.3 \pm 0.3$ MeV~\cite{Aaltonen:2013atp}.
Taken from Ref.~\cite{Sun:2014wea}.}
\label{sec3:bs5721}
\end{figure}

These assignments for the $B_{s1}(5830)^0$ and $B^*_{s2}(5840)^0$
are supported by studies using the nonrelativistic quark
model~\cite{Lu:2016bbk} and the QPC model~\cite{Sun:2014wea}. Here
we show the mixing angle $\theta_{1P}$ dependence of the total decay
widths of the $B_s(1P_1)$ and $B_s(1P^\prime_1)$ in
Fig.~\ref{sec3:bs5721}, where the calculated width of the
$B_s(1P^\prime_1)$ overlaps with the experimental width of the
$B_{s1}(5830)^0$ when $\theta_{1P}$ is in the range of
$-60.5^\circ\sim-57.5^\circ$ or $-52.0^\circ\sim
-49.0^\circ$~\cite{Sun:2014wea}. Their decay properties
were also studied in Ref.~\cite{Wang:2012pf} by the improved Bethe-Salpeter method.

Lattice QCD was also applied to study the spectrum of $1P$ bottom-strange states in Ref.~\cite{Lang:2015hza}.
Their results for the $B_{s1}(5830)^0$ and $B^*_{s2}(5840)^0$ mesons are in good agreement with the experimental values.
They also predict other two states: one is a $J^P=0^+$ bound state $B^*_{s0}$ with mass $m_{B^*_{s0}}=5.711\pm13\pm19$ GeV,
and the other is $J^P=1^+$ bound state $B_{s1}$ with mass $m_{B_{s1}}=5.750\pm17\pm19$ GeV.

%% file: section6.1.tex
\section{Candidates for the singly heavy baryons}
\label{sec6}

The heavy baryons can be categorized into the singly heavy baryons
($Qqq$), doubly heavy baryons ($QQq$), and triply heavy baryons
($QQQ$), where $Q$ denotes the heavy (charm and bottom) quark, and
$q$ denotes the light (up, down and strange) quark. The singly heavy
baryons ($Qqq$) can be further categorized into the singly charmed
baryons ($cqq$) and singly bottom baryons ($bqq$), which will be
separately reviewed in this section. The doubly heavy baryons
($QQq$) and triply heavy baryons ($QQQ$) will be reviewed in the
next section, Sec.~\ref{sec7}. We note that we shall omit the
notation ``singly'' in this section for simplicity.

\begin{figure*}[htb]
\begin{center}
\includegraphics[width=0.48\textwidth]{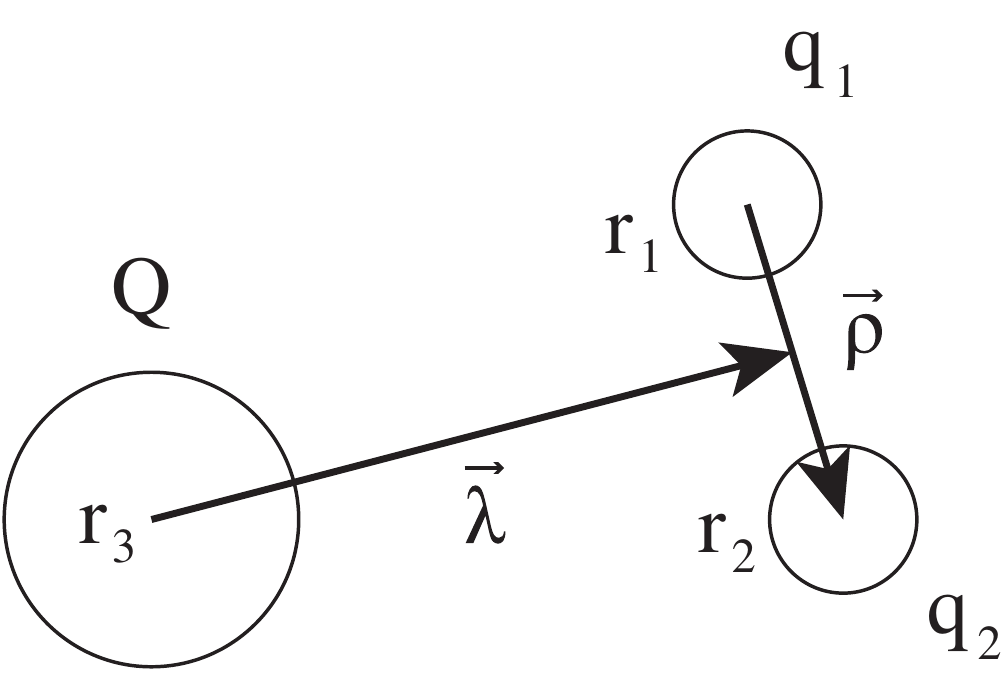}
\end{center}
\caption{ Jacobi coordinates $\vec \rho$ and $\vec \lambda$ for the
three-body system. } \label{Fig:6.1.Jacobi}
\end{figure*}

\begin{figure*}[htb]
\begin{center}
\includegraphics[width=0.6\textwidth]{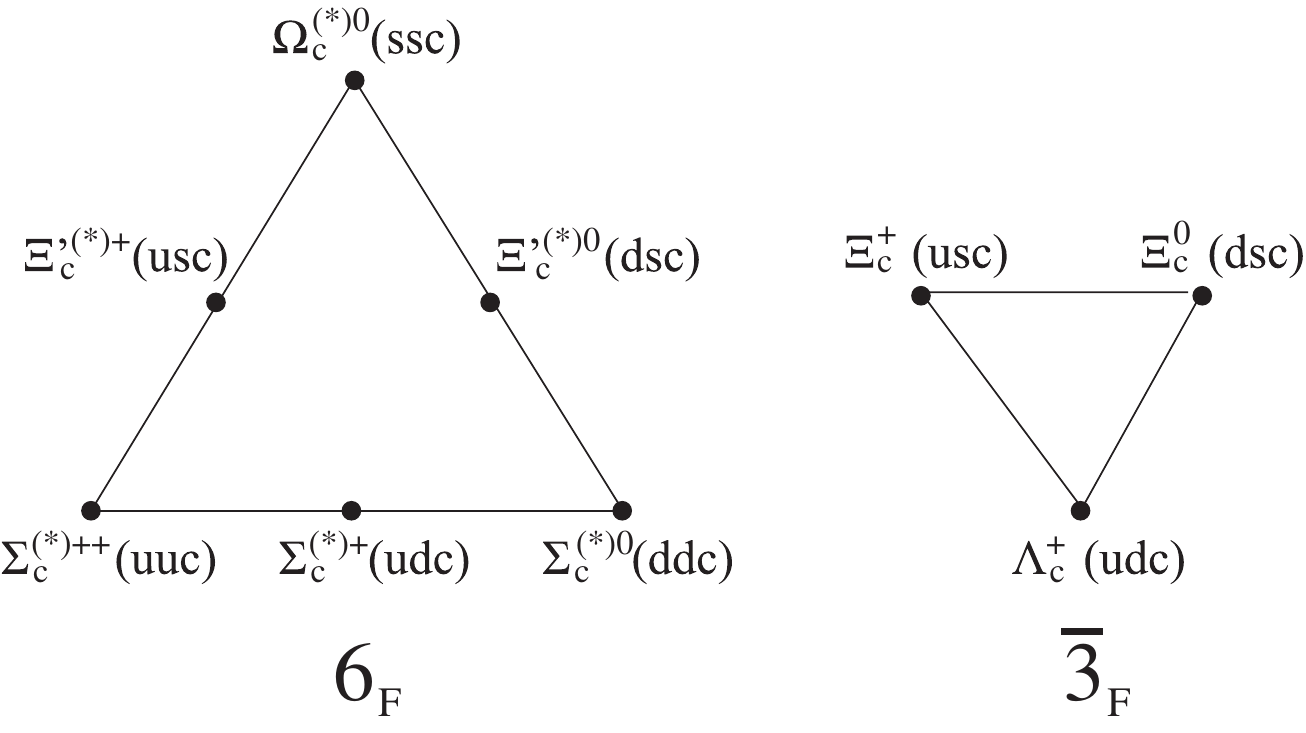}
\end{center}
\caption{ SU(3) flavor multiplets of charmed baryons. }
\label{Fig:6.1.diagrambaryon}
\end{figure*}

The heavy baryons are composed of one heavy quark and two light
quarks. Similar to the heavy mesons, the light diquark circles
around the nearly static heavy quark. However, their internal
structure is much more complicated than that of the heavy mesons. To
study this three-body system, the Jacobi coordinates are sometimes
used as shown in Fig.~\ref{Fig:6.1.Jacobi}, where
\begin{eqnarray}
\vec \rho = \vec r_2 - \vec r_1 \, ~{\rm and}~\, \vec \lambda =
(\vec r_2 + \vec r_1)/2 - \vec r_3 \, .
\end{eqnarray}
Accordingly, we use $l_\rho$ to denote the orbital angular momentum
between the two light quarks and $l_\lambda$ to denote the orbital
angular momentum between the heavy quark and the light diquark
system. Then the total orbital angular momentum is $L = l_\rho
\otimes l_\lambda$.

The heavy baryons contain two light quarks, which compose a light
diquark obeying the Pauli principle. The structure of the light
diquark is simple. The two light quarks have the antisymmetric color
structure $\mathbf{\bar 3}_C$. They can have either the symmetric
flavor structure $\mathbf{6}_F$ or the antisymmetric flavor
structure $\mathbf{\bar 3}_F$ (see
Fig.~\ref{Fig:6.1.diagrambaryon}). They can have either the
symmetric spin angular momentum ($s_l \equiv s_{qq} = 1$) or the
antisymmetric spin angular momentum ($s_l=0$). Together with the
internal orbital angular momentum ($l_\rho$), we arrive at the
$S$-wave scalar (``good'') and axial-vector (``bad'')
diquarks~\cite{Jaffe:2004ph} as well as the excited diquarks with
orbital excitations:
\begin{eqnarray*}
\nonumber S{\mbox{-wave diquark}}~(l_\rho = 0,~{\bf S}) \left\{
\begin{array}{l} s_l = 0~({\bf A}) \, , \, \mathbf{\bar 3}_F~({\bf
A}) \, , \, j_{qq} = 0 \, , (\mbox{``good''})
\\
s_l = 1~({\bf S}) \, , \, \mathbf{6}_F~({\bf S}) \, , \, j_{qq} = 1
\, , (\mbox{``bad''})
\end{array}\right.
\\ P{\mbox{-wave diquark}}~(l_\rho = 1,~{\bf A})
\left\{ \begin{array}{l} s_l = 0~({\bf A}) \, , \,
\mathbf{6}_F~({\bf S}) \, , \, j_{qq} = 1 \, ,
\\
s_l = 1~({\bf S}) \, , \, \mathbf{\bar 3}_F~({\bf A}) \, , \, j_{qq}
= 0/1/2 \, ,
\end{array}\right.
\\ \nonumber D{\mbox{-wave diquark}}~(l_\rho = 2,~{\bf S})
\left\{ \begin{array}{l} s_l = 0~({\bf A}) \, , \, \mathbf{\bar
3}_F~({\bf A}) \, , \, j_{qq} = 2 \, ,
\\
s_l = 1~({\bf S}) \, , \, \mathbf{6}_F~({\bf S}) \, , \, j_{qq} =
1/2/3 \, ,
\end{array}\right.
\\ \nonumber \cdots
\end{eqnarray*}
where we have denoted the total angular momentum of the light
diquark as $j_{qq}$.

These light diquarks and the heavy quark form the heavy baryons. The
total angular momentum of the light degrees of freedom of the heavy
baryons ($j_l$) is
\begin{eqnarray}
j_l = j_{qq} \otimes l_\lambda = s_l \otimes l_\rho \otimes
l_\lambda \, ,
\end{eqnarray}
and their total angular momentum ($J$) is
\begin{eqnarray}
J = s_Q \otimes j_l = s_Q \otimes j_{qq} \otimes l_\lambda = s_Q
\otimes s_l \otimes l_\rho \otimes l_\lambda \, .
\end{eqnarray}
As an example, the $S$-wave heavy baryons with $L =  l_\rho =
l_\lambda = 0~({\bf S})$ and $\mathbf{\bar 3}_C~({\bf A})$ can be
categorized into
\begin{eqnarray*}
\left\{ \begin{array}{l} s_l = 0~({\bf A}) \, , \, \mathbf{\bar
3}_F~({\bf A}) \, , \, j_l = 0 \, : \, \Lambda_c({1\over2}^+) \, ,
\Xi_c({1\over2}^+) \, ,
\\
s_l = 1~({\bf S}) \, , \, \mathbf{6}_F~({\bf S}) \, , \, j_l = 1 \,
: \, \Sigma_c({1\over2}^+, {3\over2}^+) \, ,
\Xi^\prime_c({1\over2}^+, {3\over2}^+) \, , \Omega_c({1\over2}^+,
{3\over2}^+) \, .
\end{array}\right.
\end{eqnarray*}
Hence, the ground-state heavy baryons contain one flavor
$\mathbf{\bar 3}_F$ multiplet of $J^P = 1/2^+$, one flavor
$\mathbf{6}_F$ multiplet of $J^P = 1/2^+$, and one flavor
$\mathbf{6}_F$ multiplet of $J^P = 3/2^+$. The flavor $\mathbf{\bar
3}_F$ multiplet of $J^P = 1/2^+$ composes a heavy baryon multiplet
where the light diquark spin is $j_l=0$, while the two flavor
$\mathbf{6}_F$ multiplets of $J^P = 1/2^+$ and $3/2^+$ compose
another heavy baryon multiplet where the diquark spin is $j_l=1$.
All the $S$-wave charmed and bottom baryons have been observed,
except the $\Omega_b^*$ of $J^P = 3/2^+$~\cite{Olive:2016xmw}. This
is a great success of the quark model in the classification of heavy
hadrons.

\begin{figure*}[htb]
\begin{center}
\includegraphics[width=1.0\textwidth]{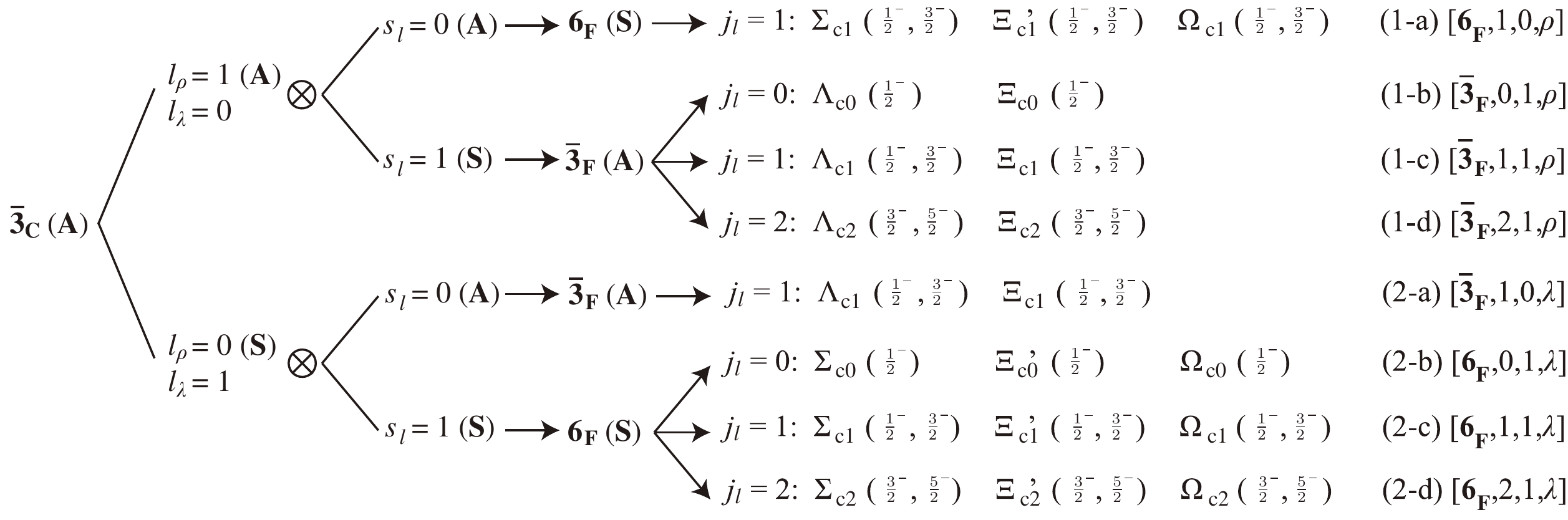}
\end{center}
\caption{The $P$-wave charmed baryons containing one orbital
excitation ($L = l_\lambda \otimes l_\rho = 1$). The two light
quarks compose a light diquark with: a) the antisymmetric color
structure $\mathbf{\bar 3}_C$; b) either the symmetric flavor
structure $\mathbf{6}_F$ or the antisymmetric flavor structure
$\mathbf{\bar 3}_F$; c) either the symmetric spin angular momentum
($s_l = 1$) or the antisymmetric spin angular momentum ($s_l=0$).
The two light quarks obey the Pauli principle, taking into account
these structures and the orbital angular momentum between them
($l_\rho$). The total angular momentum of the light degrees of
freedom is $j_l = l_\rho \otimes l_\lambda \otimes s_l$. Taken from
Ref.~\cite{Chen:2015kpa}. \label{fig:pwave}}
\end{figure*}

\begin{figure*}[htb]
\begin{center}
\includegraphics[width=1.0\textwidth]{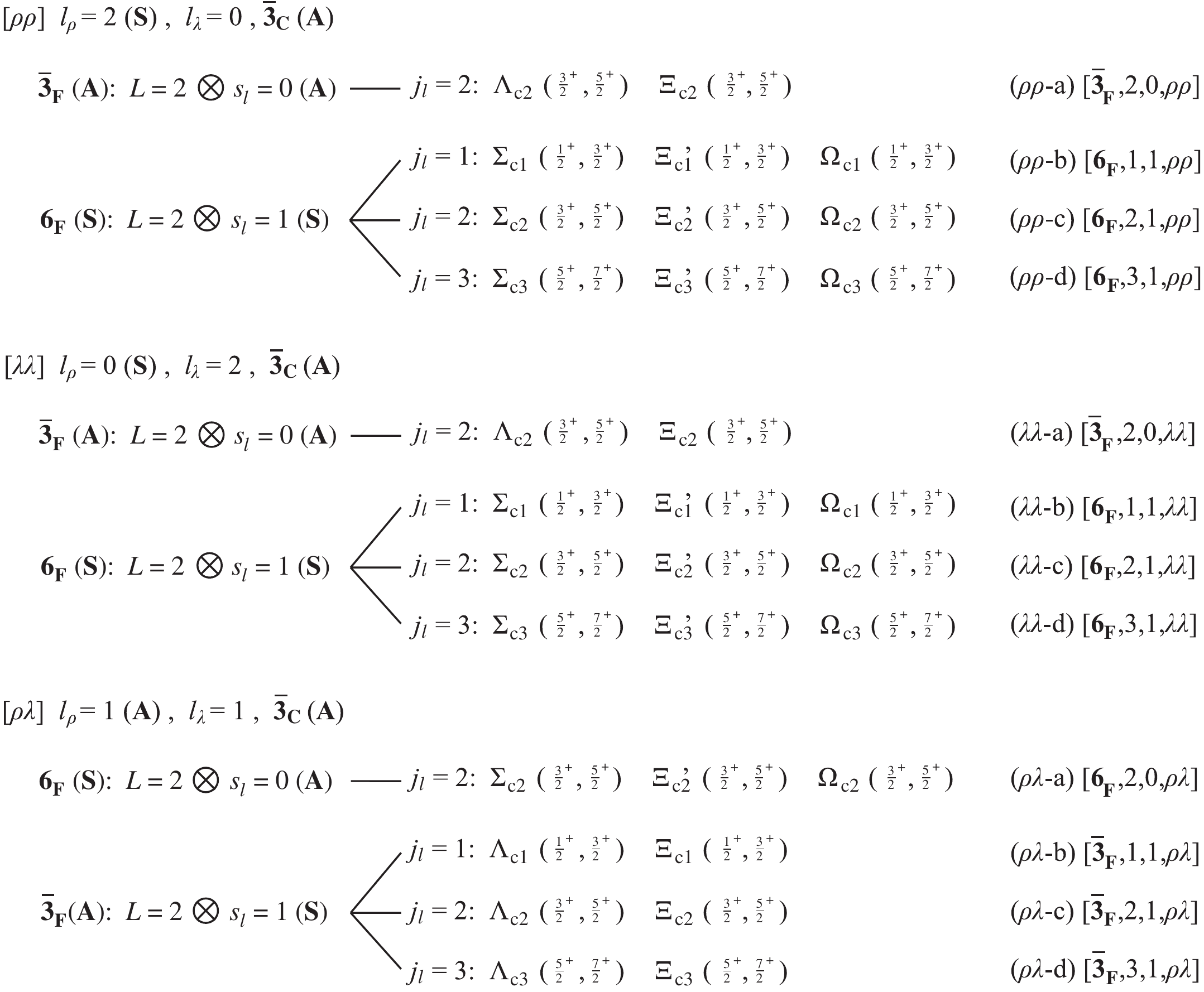}
\end{center}
\caption{The $D$-wave charmed baryons containing two orbital
excitations ($L = l_\lambda \otimes l_\rho = 2$). Taken from
Ref.~\cite{Chen:2016phw}. \label{fig:dwave}}
\end{figure*}

The excited heavy baryons can be similarly categorized. In
Figs.~\ref{fig:pwave} and \ref{fig:dwave} we show the results for
the $P$ and $D$-wave charmed baryons, whose internal structure is
very complicated. Moreover, the present experimental data are far
from complete. Similar to the heavy mesons, various methods and
models have been applied to study masses and decay properties of the
heavy baryons. In the following two subsections, we shall review the
theoretical progress on these heavy baryons, separately for the
charmed and bottom baryons.

\subsection{The charmed baryons}
\label{sec6.1}

The mass spectrum of the charmed baryons has been calculated by many
theoretical groups using various models. In this review we list five
investigations:
\begin{enumerate}

\item The first one is calculated within the framework of the QCD-motivated relativistic quark model based on the quasipotential approach~\cite{Ebert:2011kk}. Its potential is generalized from the $q \bar q$ quasipotential~\cite{Ebert:2009ua,Ebert:1997nk}, which has been reviewed in Sec.~\ref{sec1.1.1}. See Refs.~\cite{Ebert:2011kk,Ebert:2007nw,Ebert:2005xj} for more information.

\item The second one is a non-relativistic quark model~\cite{Roberts:2007ni}, whose Hamiltonian contains a spin independent confining potential, a spin independent confining potential and a simplified spin-orbit potential. See Refs.~\cite{Copley:1979wj,Roberts:2007ni,Yoshida:2015tia} for more information.

\item The third one is the method of QCD sum rule in the framework of the heavy quark effective theory (HQET), which has been reviewed in Sec.~\ref{sec1.4}. We refer interested readers to Refs.~\cite{Liu:2007fg,Chen:2015kpa,Mao:2015gya,Chen:2016phw} for detailed discussions.

\item The fourth one is a relativistic quark model for the three-quark system, proposed in Ref.~\cite{Capstick:1986bm} by Capstick and Isgur. Its potential is the immediate and essentially unique generalization of the GI model~\cite{Godfrey:1985xj} from $q \bar q$ to $qqq$ (see discussions in Sec.~\ref{sec1.1.1}), which contains two main ingredients: the short-distance one-gluon-exchange interaction and the long-distance linear confining interaction. See Ref.~\cite{Capstick:1986bm} for more information.

\item The fifth one is a constituent quark model incorporating the basic properties of QCD~\cite{Garcilazo:2007eh}, which takes into account the QCD nonperturbative effects (chiral symmetry breaking and confinement) as well as QCD perturbative effects (a flavor dependent one-gluon exchange potential). See Refs.~\cite{Garcilazo:2007eh,Valcarce:2008dr} for more information.

\end{enumerate}
More discussions using various models and methods can be found in
Refs.~\cite{Hasenfratz:1980ka,Stanley:1980fe,Izatt:1981pt,Richard:1983tc,Hwang:1986ee,Itoh:1989qi,Korner:1994nh,Savage:1995dw,Chernyshev:1995gj,Lu:1995te,Hussain:1995xs,Ito:1996mr,Falk:1996qm,Jenkins:1996rr,SilvestreBrac:1996bg,Jenkins:1996de,Coester:1997ki,Jin:2001jx,Lewis:2001iz,Narodetskii:2001bq,Wang:2003it,Albertus:2003sx,Pervin:2005ve,Albertus:2005zy,Faessler:2006ft,Migura:2006ep,Garcilazo:2007eh,Dhir:2009ax,Huang:2009is,Majethiya:2008fe,Aliev:2010ev,Vijande:2014uma}.
See also lattice QCD studies in
Refs.~\cite{Bowler:1996ws,AliKhan:1999yb,Flynn:2003vz,Liu:2009jc,Detmold:2011bp,Lin:2011ti,Alexandrou:2012xk,Briceno:2012wt,Detmold:2012ge,Brown:2014ena}.

\begin{figure*}[htb]
\begin{center}
\includegraphics[width=0.6\textwidth]{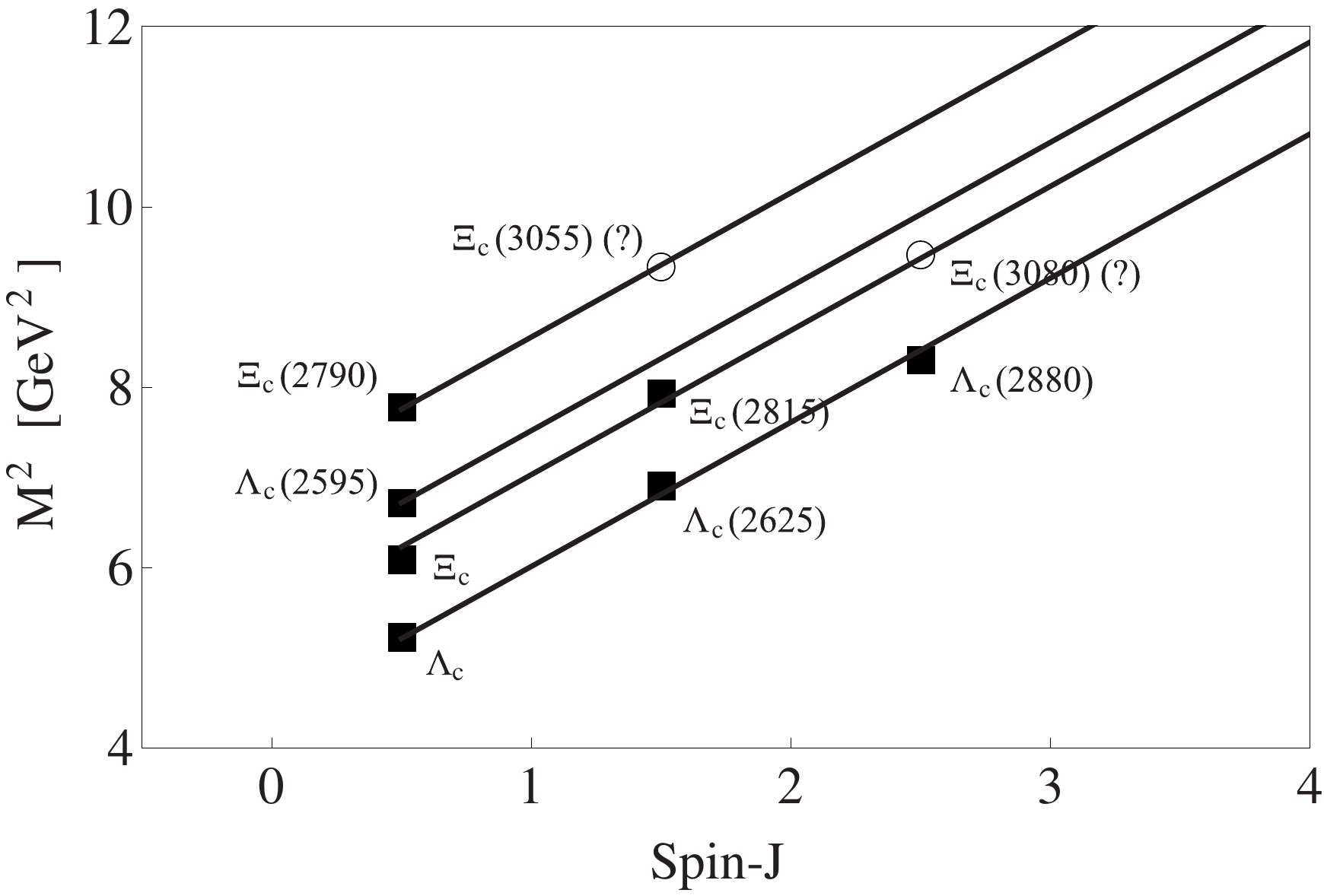}
\end{center}
\caption{ Regge trajectories in the $(J,M^2)$ plane for some
selected charmed baryons, where experimental data are given by solid
squares and hollow circles with particle names. }
\label{Fig:6.1.Regge5}
\end{figure*}

The results of these five investigations are summarized in
Tables~\ref{sec6:charm1} and \ref{sec6:charm2}, where we further
make a comparison between the theoretical and experimental results
(see also Tables~\ref{sec6:excitedcharm} and \ref{sec6:excitedXi}),
and conclude from the mass spectrum analysis, i.e.,
\begin{enumerate}

\item All the $1S$ charmed baryons are reproduced quite well, which complete one flavor $\mathbf{\bar 3}_F$ multiplet of $J^P = 1/2^+$ and two flavor $\mathbf{6}_F$ multiplets of $J^P = 1/2^+$ and $3/2^+$.

\item The $\Lambda_c(2595)$, $\Lambda_c(2625)$, $\Xi_c(2790)$ and $\Xi_c(2815)$ are good candidates for the $P$-wave charmed baryons, which complete two flavor $\mathbf{\bar3}_F$ multiplets of $J^P = 1/2^-$ and $3/2^-$.

\item The mass spectrum of these higher excited states is quite complicated. There are many possible interpretations of the $\Lambda_c(2765)$, $\Lambda_c(2880)$, $\Lambda_c(2940)$, $\Sigma_c(2800)$, $\Xi_c(2815)$, $\Xi_c(2930)$, $\Xi_c(2980)$, $\Xi_c(3055)$, $\Xi_c(3080)$ and $\Xi_c(3123)$. Among them, the $\Lambda_c(2880)$, $\Xi_c(3055)$ and $\Xi_c(3080)$, together with a missing $\Lambda_c(3/2^+)$ state, may be the $D$-wave charmed baryons completing two flavor $\mathbf{\bar3}_F$ multiplets of $J^P = 3/2^+$ and $5/2^+$.

\end{enumerate}
We select some of the charmed baryons to construct Regge
trajectories (see discussions in Sec.~\ref{sec1.5}). The result is
shown in Fig.~\ref{Fig:6.1.Regge5} in the $(J,M^2)$ plane, which
suggests that the $\Lambda_c$, $\Xi_c$, $\Lambda_c(2595)$,
$\Lambda_c(2625)$, $\Xi_c(2790)$ and $\Xi_c(2815)$ may be
interpreted as the $1S$ and $1P$ charmed baryons, while the
$\Lambda_c(2880)$, $\Xi_c(3055)$ and $\Xi_c(3080)$ may be
interpreted as the $1D$ charmed baryons of $J^P = 3/2^+$ and
$5/2^+$.

In the following paragraphs we review the theoretical progress on
the excited charmed baryons.

\renewcommand{\arraystretch}{1.6}
\begin{table*}[htb]
\tiny
\caption{Comparison of the experimental data and theoretical
results of the $\Lambda_c$, $\Sigma_c$ and $\Omega_c$ baryons,
obtained using the QCD-motivated relativistic quark model based on
the quasipotential approach (R.~Q.~M.)~\cite{Ebert:2011kk}, the
non-relativistic quark model (Non-RQM)~\cite{Roberts:2007ni}, the
QCD sum rules within HQET
(QSR)~\cite{Liu:2007fg,Chen:2015kpa,Mao:2015gya,Chen:2016phw}, and
the relativistic quark model generalized from the GI model
(C.~I.)~\cite{Capstick:1986bm}. See also
Table~\ref{sec6:excitedcharm}. The masses are in units of MeV.}
\centering
\begin{tabular}{ c c c c c c c }\toprule[1pt]
& $J^P~(nL)$ & Experimental values~\cite{Olive:2016xmw} &
R.~Q.~M.~\cite{Ebert:2011kk} & Non-RQM~\cite{Roberts:2007ni} &
QSR~\cite{Liu:2007fg,Chen:2015kpa,Chen:2016phw} &
C.~I.~\cite{Capstick:1986bm}
\\ \midrule[1pt]
$\Lambda_c$             & $1/2^+~(1S)$     & $2286.46\pm0.14$                                  & 2286  & 2268  & $2271^{+67}_{-49}$   & 2265  \\
$\Sigma_c^+$            & $1/2^+~(1S)$     & $2452.9\pm0.4$                                    & 2443  & 2455  & $2411^{+93}_{-81}$   & 2440  \\
$\Sigma_c^{*+}$         & $3/2^+~(1S)$     & $2517.5\pm2.3$                                    & 2519  & 2519  & $2534^{+96}_{-81}$   & 2495  \\
$\Omega_c^{0}$          & $1/2^+~(1S)$     & $2695.2\pm1.7$                                    & 2698  & 2718  & $2657^{+102}_{-99}$  & --    \\
$\Omega_c^{*0}$         & $3/2^+~(1S)$     & $2765.9\pm2.0$                                    & 2768  & 2776  & $2790^{+109}_{-105}$ & --    \\
\hline
$\Lambda_c$             & $1/2^-~(1P)$     & $\Lambda_c(2595)=2592.25\pm0.28$                  & 2598  & 2625  & $2.60\pm0.14$        & 2630  \\
$\Lambda_c$             & $3/2^-~(1P)$     & $\Lambda_c(2625)=2628.11\pm0.19$                  & 2627  & 2636  & $2.65\pm0.14$        & 2640  \\
\hdashline[2pt/2pt]
$\Sigma_c$              & $1/2^-~(1P)$ & \multirow{5}{*}{$\Sigma_c(2800)=2792^{+14}_{-~5}$~(?)}& 2713  & 2748  & $2.73\pm0.18$        & 2765  \\
$\Sigma_c$              & $1/2^-~(1P)$     &                                                   & 2799  & 2768  & --                   & 2770  \\
$\Sigma_c$              & $3/2^-~(1P)$     &                                                   & 2773  & 2763  & $2.75\pm0.18$        & 2770  \\
$\Sigma_c$              & $3/2^-~(1P)$     &                                                   & 2798  & 2776  & $2.80\pm0.15$        & 2805  \\
$\Sigma_c$              & $5/2^-~(1P)$     &                                                   & 2789  & 2790  & $2.89\pm0.15$        & 2815  \\
\hdashline[2pt/2pt]
$\Omega_c$              & $1/2^-~(1P)$     & --                                                & 2966  & 2977  & $3.25\pm0.20$        & --    \\
$\Omega_c$              & $1/2^-~(1P)$     & --                                                & 3055  & 2990  & --                   & --    \\
$\Omega_c$              & $3/2^-~(1P)$     & --                                                & 3029  & 2986  & $3.26\pm0.19$        & --    \\
$\Omega_c$              & $3/2^-~(1P)$     & --                                                & 3054  & 2994  & $3.27\pm0.17$        & --    \\
$\Omega_c$              & $5/2^-~(1P)$     & --                                                & 3051  & 3014  & $3.32\pm0.17$        & --    \\
\hline
$\Lambda_c$             & $1/2^+~(2S)$     & $\Lambda_c(2765)=2766.6\pm2.4$~(?)                & 2769  & 2791  & --   & 2775  \\
$\Sigma_c$              & $1/2^+~(2S)$     & --                                                & 2901  & 2958  & --   & 2890  \\
$\Sigma_c$              & $3/2^+~(2S)$     & --                                                & 2936  & 2995  & --   & 2985  \\
$\Omega_c$              & $1/2^+~(2S)$     & --                                                & 3088  & 3152  & --   & --    \\
$\Omega_c$              & $3/2^+~(2S)$     & --                                                & 3123  & 3190  & --   & --    \\
\hline
$\Lambda_c$             & $3/2^+~(1D)$     & --                                                & 2874  & 2887  & $2.81^{+0.33}_{-0.18}$   & 2910  \\
$\Lambda_c$             & $5/2^+~(1D)$     & $\Lambda_c(2880)=2881.53\pm0.35$~(?)              & 2880  & 2887  & $2.84^{+0.37}_{-0.20}$   & 2910  \\
$\Sigma_c$              & $1/2^+~(1D)$     & --                                                & 3041  & --    & --   & 3005  \\
$\Sigma_c$              & $3/2^+~(1D)$     & --                                                & 3040  & --    & --   & 3060  \\
$\Sigma_c$              & $3/2^+~(1D)$     & --                                                & 3043  & --    & --   & 3065  \\
$\Sigma_c$              & $5/2^+~(1D)$     & --                                                & 3023  & 3003  & --   & 3065  \\
$\Sigma_c$              & $5/2^+~(1D)$     & --                                                & 3038  & 3010  & --   & 3080  \\
$\Sigma_c$              & $7/2^+~(1D)$     & --                                                & 3013  & 3015  & --   & 3090  \\
$\Omega_c$              & $1/2^+~(1D)$     & --                                                & 3287  & --    & --   & --    \\
$\Omega_c$              & $3/2^+~(1D)$     & --                                                & 3282  & --    & --   & --    \\
$\Omega_c$              & $3/2^+~(1D)$     & --                                                & 3298  & --    & --   & --    \\
$\Omega_c$              & $5/2^+~(1D)$     & --                                                & 3286  & 3196  & --   & --    \\
$\Omega_c$              & $5/2^+~(1D)$     & --                                                & 3297  & 3203  & --   & --    \\
$\Omega_c$              & $7/2^+~(1D)$     & --
& 3283  & 3206  & --   & --
\\ \bottomrule[1pt]
\end {tabular}
\label{sec6:charm1}
\end{table*}

\renewcommand{\arraystretch}{1.6}
\begin{table*}[htb]
\tiny \caption{Comparison of the experimental data and theoretical
results of the $\Xi_c$ baryons, obtained using the QCD-motivated
relativistic quark model based on the quasipotential approach
(R.~Q.~M.)~\cite{Ebert:2011kk}, the non-relativistic quark model
(Non-RQM)~\cite{Roberts:2007ni}, the QCD sum rules within HQET
(QSR)~\cite{Liu:2007fg,Chen:2015kpa,Mao:2015gya,Chen:2016phw}, and
the constituent quark model (C.~Q.~M.)~\cite{Garcilazo:2007eh}. See
also Table~\ref{sec6:excitedXi}. The masses are in units of MeV.
Here we use $\Xi_c$ and $\Xi_c^\prime$ to denote the $\Xi_c$ baryons
belonging to the flavor $\mathbf{\bar 3}_F$ and $\mathbf{6}_F$
respectively, but note that the superscript $^\prime$ is often
omitted. Actually, the $\Xi_c$ and $\Xi_c^\prime$ can mix with each
other, which effect was taken into account in
Ref.~\cite{Roberts:2007ni}. } \centering
\begin{tabular}{ c c c c c c c }\toprule[1pt]
& $J^P~(nL)$ & Experimental values~\cite{Olive:2016xmw} &
R.~Q.~M.~\cite{Ebert:2011kk} & Non-RQM~~\cite{Roberts:2007ni} &
QSR~\cite{Liu:2007fg,Chen:2015kpa,Chen:2016phw} &
C.~Q.~M.~\cite{Garcilazo:2007eh}
\\ \midrule[1pt]
$\Xi_c^+$               & $1/2^+~(1S)$     & $2467.93^{+0.28}_{-0.40}$                         & 2476  & 2466  & $2432^{+79}_{-68}$  & 2496  \\
$\Xi_c^{\prime+}$       & $1/2^+~(1S)$     & $2575.7\pm3.0$                                    & 2579  & 2594  & $2508^{+97}_{-91}$  & 2574  \\
$\Xi_c^{\prime+}$       & $3/2^+~(1S)$     & $\Xi_c^{*+}=2645.9\pm0.5$                         & 2649  & 2649  & $2634^{+102}_{-94}$ & 2633  \\
\hline
$\Xi_c^+$               & $1/2^-~(1P)$     & $\Xi_c(2790)=2789.1\pm3.2$                        & 2792  & 2773  & $2.79\pm0.15$       & 2749  \\
$\Xi_c^+$               & $3/2^-~(1P)$     & $\Xi_c(2815)=2816.6\pm0.9$                        & 2819  & 2783  & $2.83\pm0.15$       & 2749  \\
\hdashline[2pt/2pt]
$\Xi^\prime_c$          & $1/2^-~(1P)$     & \multirow{5}{*}{$\Xi_c(2930)=2931\pm6$~(?)}       & 2854  & 2855  & $2.96\pm0.15$       & 2829  \\
$\Xi^\prime_c$          & $1/2^-~(1P)$     &                                                   & 2936  & --    & --                  & --    \\
$\Xi^\prime_c$          & $3/2^-~(1P)$     &                                                   & 2912  & 2866  & $2.98\pm0.15$       & 2829  \\
$\Xi^\prime_c$          & $3/2^-~(1P)$     &                                                   & 2935  & --    & $2.98\pm0.21$       & --    \\
$\Xi^\prime_c$          & $5/2^-~(1P)$     &                                                   & 2929  & 2895  & $3.05\pm0.21$       & --    \\
\hline
$\Xi_c^+$               & $1/2^+~(2S)$     & --                                                & 2959  & --    & --    & --    \\
$\Xi_c^{\prime+}$       & $1/2^+~(2S)$     & --                                                & 2983  & --    & --    & --    \\
$\Xi_c^{\prime+}$       & $3/2^+~(2S)$     & --                                                & 3026  & --    & --    & --    \\
\hline
$\Xi_c$                 & $3/2^+~(1D)$     & $\Xi_c(3055)=3055.1\pm1.7$~(?)                    & 3059  & 3012  & $3.04\pm0.15$           & 2951  \\
$\Xi_c$                 & $5/2^+~(1D)$     & $\Xi_c(3080)=3076.94\pm0.28$~(?)                  & 3076  & 3004  & $3.05^{+0.15}_{-0.16}$  & --    \\
$\Xi^\prime_c$          & $1/2^+~(1D)$     & --                                                & 3163  & --    & --    & --    \\
$\Xi^\prime_c$          & $3/2^+~(1D)$     & --                                                & 3160  & --    & --    & --    \\
$\Xi^\prime_c$          & $3/2^+~(1D)$     & --                                                & 3167  & --    & --    & --    \\
$\Xi^\prime_c$          & $5/2^+~(1D)$     & --                                                & 3153  & 3080  & --    & --    \\
$\Xi^\prime_c$          & $5/2^+~(1D)$     & --                                                & 3166  & --    & --    & --    \\
$\Xi^\prime_c$          & $7/2^+~(1D)$     & --
& 3147  & 3094  & --    & --
\\ \bottomrule[1pt]
\end {tabular}
\label{sec6:charm2}
\end{table*}

\subsubsection{$\Lambda_c(2595)$, $\Lambda_c(2625)$, $\Xi_c(2790)$ and $\Xi_c(2815)$.}

The $\Lambda_c(2595)$, $\Lambda_c(2625)$, $\Xi_c(2790)$ and
$\Xi_c(2815)$ can be well interpreted in the conventional quark
model as the $P$-wave charmed baryons with the quark content $cqq$.
They complete two flavor $\mathbf{\bar3}_F$ multiplets of $J^P =
1/2^-$ and $3/2^-$. See their mass spectrum analysis at the
beginning of this subsection and
Refs.~\cite{Chow:1994sz,Schechter:1994ip,Schechter:1995vr,Adamov:1997yk,Tawfiq:1998nk,Ivanov:1999bk,Tawfiq:1999cf,Hussain:1999sp,Lee:2000wb,Lee:2002ka,Shah:2016mig}
for more information. The QCD sum rule formalism was also applied in
Refs.~\cite{Huang:2000tn,Wang:2003zp} to evaluate the masses of the
$P$-wave excited heavy baryons up to the $1/m_Q$ order in the heavy
quark expansion. The extracted splitting between the spin $1/2$ and
$3/2$ doublets is consistent with the experiment measurement.

The productions and decay properties of the heavy baryons are also
important to understand their nature, which were investigated in
Refs.~\cite{Cho:1994vg,Ivanov:1995pk,Elwood:1995hm,Rosner:1995yu,Chow:1995nw,Pirjol:1997nh,Chiladze:1997ev,AzizaBaccouche:2001jc,AzizaBaccouche:2001pu}.
Particularly, the $\Lambda_c(2595)$ is very close to the $\pi
\Sigma_c$ threshold. Its strong decays are sensitive to the finite
width effects, which were studied in Ref.~\cite{Blechman:2003mq}.

In Ref.~\cite{Ivanov:1998qe} the authors studied one-pion
transitions between the charmed baryon states in the framework of a
relativistic three-quark model. They evaluated two coupling
constants, $f_{\Lambda_c(2595) \Sigma_c \pi}=0.52$ ($S$-wave
transition) and $f_{\Lambda_c(2625) \Sigma_c \pi} = 21.5$ GeV$^2$
($D$-wave transition), and their derived partial decay widths are
consistent with the experiments~\cite{Caso:1998tx}. The same method
was used in Ref.~\cite{Ivanov:1998wj} to study the one-photon
transitions between heavy baryon states. They evaluated the
one-photon transition rates of some specific excited states, and
obtained $\Gamma(\Lambda_c(2595) \to \Lambda_c \gamma) = 104.3 \pm
1.3$ keV.

The methods of QCD sum rules and light-cone sum rules were also
applied to study the productions and decay properties of the
$P$-wave heavy baryons. In Ref.~\cite{Zhu:2000py}, Zhu calculated
the pionic and electromagnetic coupling constants of the
lowest-lying $P$-wave heavy baryon doublet in the leading order of
the heavy quark expansion. He obtained $\Gamma(\Lambda_c(2595) \to
\Sigma_c \pi, \Sigma_c \gamma, \Sigma^*_c \gamma) = 2.7 \, , \,
0.011 \, , \, 0.001$ MeV and $\Gamma(\Lambda_c(2625) \to \Sigma_c
\pi, \Sigma_c \gamma, \Sigma^*_c \gamma, \Lambda_c(2595)\gamma) = 33
\, , \, 5 \, , \, 6 \, , \, 0.014$ keV, respectively. In
Ref.~\cite{Huang:2000xw} the authors studied the semileptonic
transitions $\Lambda_b \to \Lambda_c(2595) l \bar \nu$  and
$\Lambda_b \to \Lambda_c(2625) l \bar \nu$ using the method of QCD
sum rules in the framework of heavy quark effective theory, and
evaluated the branching ratios $\mathcal{B}(\Lambda_b \to
\Lambda_c(2595) e \bar \nu_e)$ and $\mathcal{B}(\Lambda_b \to
\Lambda_c(2625) e \bar \nu_e)$ to be around 0.21-0.28\%.

Decays of the charmed baryons were recently systematically
investigated in Ref.~\cite{Nagahiro:2016nsx} based on the quark
model together with the heavy quark symmetry. Their results indicated
that the low-lying $\Lambda_c(2595)$ and $\Lambda_c(2625)$ can be
well described as the $P$-wave charmed baryons with one
$\lambda$-mode orbital excitation ($l_\lambda = 1$ and $l_\rho =
0$).

There exist some other interpretations for the excited charmed
baryon picture. The Skyrme model was applied to study the heavy
baryons in Refs.~\cite{Oh:1995ey,Oh:1997tp} where the soliton moves
around the fixed heavy meson. Bound states were obtained, which
could be naively compared with the $\Lambda_c(2595)$ and
$\Lambda_c(2625)$. A model of the $DN$ interaction is proposed in
Ref.~\cite{Haidenbauer:2010ch} where the main ingredients of the
interaction are provided by the vector meson exchange and
higher-order box diagrams involving $D^*N$, $D\Delta$, and
$D^*\Delta$ intermediate states. Their results suggested that the
$\Lambda_c(2595)$ resonance is dynamically generated as a $DN$
quasi-bound state. Recently in Ref.~\cite{Guo:2016wpy} the $\pi
\Sigma_c$ scattering around its threshold was studied to investigate
the nature of the $\Lambda_c(2595)$. They developed a general
framework to properly handle the CDD pole accompanied by the nearby
thresholds, and their results suggested that the $\pi^0 \Sigma_c^+$
component is subdominant inside the $\Lambda_c(2595)$.

The non-linear chiral $SU(3)$ Lagrangians were used in
Ref.~\cite{Lutz:2003jw} to study the charmed baryons of $J^P =
1/2^-$. Through the scattering of the ground-state pseudoscalar
mesons and heavy baryons, the $\Lambda_c(2595)$, $\Lambda_c(2880)$
and $\Xi_c(2790)$ were dynamically generated. A similar method was
used in
Refs.~\cite{Gamermann:2010zz,Romanets:2012hm,Garcia-Recio:2013gaa},
where the charmed and strange baryon resonances were dynamically
generated with a unitary baryon-meson coupled-channel model
incorporating heavy-quark spin symmetry. Their model produced
resonances with negative parity from the $S$-wave interaction of the
ground-state pseudoscalar and vector mesons with baryons. The
authors identified the $\Xi_c(2790)$ and $\Xi_c(2815)$ as possible
candidates forming a heavy-quark spin doublet. As a dynamically
generated state, the radiative decays of the $\Lambda_c(2595)$ were
evaluated in Ref.~\cite{Gamermann:2010ga} to be
$\Gamma(\Lambda_c(2595) \to \Lambda_c \gamma)$ = 278 keV and
$\Gamma(\Lambda_c(2595) \to \Sigma^+_c \gamma)$ = 2 keV.

\subsubsection{$\Lambda_c(2765)$, $\Lambda_c(2880)$, $\Lambda_c(2940)$ and $\Sigma_c(2800)$.}

\renewcommand{\arraystretch}{1.4}
\begin{table*}[htb]
\footnotesize
\caption{Possible interpretations for the $\Lambda_c(2765)$,
$\Lambda_c(2880)$, $\Lambda_c(2940)$ and $\Sigma_c(2800)$. Many
studies use the conventional excited charmed baryon picture ($cqq$)
to study their mass spectrum and decay properties, for such studies
we show the possible spin-parity quantum numbers. See also
Ref.~\cite{Cheng:2015iom} for more information. } \centering
\begin{tabular}{ c c c c c }\toprule[1pt]
References & $\Lambda_c(2765)$ & $\Lambda_c(2880)$ &
$\Lambda_c(2940)$ & $\Sigma_c(2800)$
\\ \midrule[1pt]
Ref.~\cite{Garcilazo:2007eh} & ${1\over2}^+$ & ${1\over2}^- \big/
{3\over2}^-$ & ${3\over2}^+$ & ${1\over2}^- \big/ {3\over2}^-$
\\
Ref.~\cite{Valcarce:2008dr}  & ${1\over2}^+$ & ${1\over2}^- \big/
{5\over2}^+$ & ${3\over2}^+$ & ${3\over2}^-$
\\
Ref.~\cite{Yoshida:2015tia}  &  --  &  ${5\over2}^+$  &  --  &   --
\\
Ref.~\cite{Chen:2016iyi}   &  ${1\over2}^+~(2S)$  &  --   &  --  &
${3\over2}^- \big/ {5\over2}^-$
\\ \hdashline[2pt/2pt]
Ref.~\cite{Gerasyuta:2007un} & ${5\over2}^-$ & ${1\over2}^-$ & -- &
${5\over2}^-$
\\
Ref.~\cite{Ebert:2011kk}   &  ${1\over2}^+~(2S)$  &
${5\over2}^+~(1D)$  & ${1\over2}^-~(2P)$  &  ${1\over2}^- \big/
{3\over2}^-~(1P)$
\\
Ref.~\cite{Lu:2016ctt}     &  ${1\over2}^+~(2S)$  &
${3\over2}^+~(1D)$  & ${1\over2}^-~(2P)$  &  --
\\ \hdashline[2pt/2pt]
Ref.~\cite{Selem:2006nd}   &  --   &  ${5\over2}^+$  &  --  &  --
\\
Ref.~\cite{Chen:2009tm}    &  ${3\over2}^+$  & --              &
${5\over2}^-$  & --
\\
Ref.~\cite{Chen:2014nyo}   &  ${1\over2}^+~(2S)$  &
${5\over2}^+(1D)$   &  ${1\over2}^- \big/ {3\over2}^-(2P)$  & --
\\ \hdashline[2pt/2pt]
Refs.~\cite{Chen:2015kpa,Chen:2016phw}  &  --     &
${5\over2}^+~(1D)$  & --                  &  ${1\over2}^- \big/
{3\over2}^-~(1P)$
\\ \hline
Refs.~\cite{Cheng:2006dk}  & --              & ${5\over2}^+$   &
${5\over2}^- \big/ {3\over2}^+$  & --
\\
Ref.~\cite{Cheng:2015naa}  & --              & ${5\over2}^+$   &  --
&   ${3\over2}^-$
\\ \hdashline[2pt/2pt]
Ref.~\cite{Zhong:2007gp}   & ${1\over2}^-$   & ${3\over2}^+$   &
${5\over2}^+$                    & ${1\over2}^- \big/ {5\over2}^-$
\\
Ref.~\cite{Nagahiro:2016nsx}  & --           & ${5\over2}^+$   &
${7\over2}^+$  &  --
\\
Ref.~\cite{Chen:2007xf}    & --              & ${5\over2}^+$   & --
& ${3\over2}^- \big/ {5\over2}^-$
\\ \midrule[1pt]
Ref.~\cite{Lutz:2003jw}    & --  & dynamically generated  & --  & --
\\
Refs.~\cite{JimenezTejero:2009vq,JimenezTejero:2011fc}     & --  &
--  & --  & dynamically generated
\\ \hdashline[2pt/2pt]
Ref.~\cite{He:2006is}     & --  & --  &  $D^{*0}p$ molecule  &  --
\\
Ref.~\cite{Ortega:2012cx}     & --  & --  &  $D^{*}N$ molecule  &
--
\\
Ref.~\cite{Zhao:2016zhf}  & --  &  --  &  $D^{*}N$ molecule  &  $DN$
molecule
\\ \bottomrule[1pt]
\end {tabular}
\label{sec6:excitedcharm}
\end{table*}

There are many possible interpretations for the $\Lambda_c(2765)$,
$\Lambda_c(2880)$, $\Lambda_c(2940)$ and $\Sigma_c(2800)$, which are
summarized in Table~\ref{sec6:excitedcharm}. We shall briefly review
the theoretical interpretations of these states. See also
Refs.~\cite{He:2011jp,Cheng:2012fq,Lin:2014jza,Mu:2014iaa} for more
discussions.

In Ref.~\cite{Garcilazo:2007eh} the authors employed a constituent
quark model to study heavy baryon spectroscopy by solving exactly
the three-quark problem using the Faddeev method in momentum space.
Their results suggested that the $\Lambda_c(2765)$ may be an excited
$\Lambda_c$ state of $J^P = 1/2^+$ or an excited $\Sigma_c$ state of
$J^P = 1/2^-$ or $3/2^-$, the $\Lambda_c(2880)$ is an excited
$\Lambda_c$ state of $J^P = 1/2^-$ or $3/2^-$, the $\Lambda_c(2940)$
is an excited $\Lambda_c$ state of $J^P = 3/2^+$, and the
$\Sigma_c(2800)$ is an excited $\Sigma_c$ state of $J^P = 1/2^-$ or
$3/2^-$. In Ref.~\cite{Valcarce:2008dr} the authors also adopted the
Faddeev method and noticed that the $\Lambda_c(2765)$ can be
interpreted as an excited $\Lambda_c$ state of $J^P = 1/2^+$ or an
excited $\Sigma_c$ state of $J^P = 1/2^-$; the $\Lambda_c(2880)$ can
be interpreted as an excited $\Lambda_c$ state of $J^P = 1/2^-$ or
$J^P = 5/2^+$; the $\Lambda_c(2940)$ and $\Sigma_c(2800)$ can be
interpreted as excited $\Sigma_c$ states of $J^P = 3/2^+$ and
$3/2^-$, respectively. Later in Ref.~\cite{Yoshida:2015tia} the
authors also used the constituent quark model to study the heavy
baryon spectrum, and found that the $\Lambda_c(2880)$ is a good
$\Lambda_c(1D)$ candidate with $J^P = 5/2^+$. Recently in
Ref.~\cite{Chen:2016iyi} the authors systematically studied the mass
spectra and strong decays of $1P$ and $2S$ charmed baryons using the
nonrelativistic constituent quark model, and pointed out that the
$\Lambda_c(2765)$ could be explained as the $\Lambda_c(2S)$ state of
$J^P = 1/2^+$ or the excited $\Sigma_c$ state of $J^P = 1/2^-$, and
the $\Sigma_c(2800)$ can be assigned as an excited $\Sigma_c$ state
of $J^P = 3/2^-$ or $5/2^-$.

Besides the constituent quark model, the relativistic quark model
was also applied to study the heavy baryon spectrum. In
Ref.~\cite{Gerasyuta:2007un} the authors calculated the masses of
the negative parity charmed baryons in the relativistic quark model,
and their results suggested that the $\Sigma_c(2800)$ is an excited
$\Sigma_c$ state of $J^P = 5/2^-$, and the $\Lambda_c(2765)$ and
$\Lambda_c(2880)$ are excited $\Lambda_c$ states of $J^P = 5/2^-$
and $1/2^-$, respectively. Later in Ref.~\cite{Ebert:2011kk} the
authors calculated the mass spectra of heavy baryons in the
heavy-quark-light-diquark picture in the framework of the
QCD-motivated relativistic quark model. Some of their results have
been reviewed in Tables~\ref{sec6:charm1} and \ref{sec6:charm2},
suggesting that the $\Lambda_c(2765)$, $\Lambda_c(2880)$ and
$\Lambda_c(2940)$ can be interpreted as the $1/2^+~(2S)$,
$5/2^+~(1D)$ and $1/2^-~(2P)$ excited $\Lambda_c$ states,
respectively. The $\Lambda_c(2765)$ and $\Lambda_c(2940)$ may also
be interpreted as the $3/2^-~(1P)$ and $3/2^+~(2S)$ excited
$\Sigma_c$ states, respectively. The $\Sigma_c(2800)$ can be
interpreted as the $1/2^-~(1P)$ or $3/2^-~(1P)$ excited $\Sigma_c$
state. Recently in Ref.~\cite{Lu:2016ctt} the authors adopted the
interactions proposed by the relativized GI model to calculate
baryon masses. Their results suggested that the $\Lambda_c(2880)$
can be interpreted as a $D$-wave $\Lambda_c$ state with $J^P =
3/2^+$, and the $\Lambda_c(2765)$ and $\Lambda_c(2940)$ can be
interpreted as the $1/2^+~(2S)$ and $1/2^-~(2P)$ excited $\Lambda_c$
states, respectively.

A semi-classical model of the color flux tubes was proposed in
Ref.~\cite{Selem:2006nd} where the $\Lambda_c(2880)$ can be well
interpreted as an excited $\Lambda_c$ state of $J^P = 5/2^+$. Later
in Ref.~\cite{Chen:2009tm} the authors employed the ``good'' diquark
to study the $\Lambda_c$ baryons within a mass loaded flux tube
model. Their results suggested that the $\Lambda_c(2765)$ can be an
excited $\Lambda_c$ state of $J^P = 3/2^+$, and the
$\Lambda_c(2940)$ is possibly an orbitally excited $\Lambda_c$ state
of $J^P = 5/2^-$. A mass formula derived analytically from the
relativistic flux tube model was used in Ref.~\cite{Chen:2014nyo} to
investigate the mass spectra of the charmed baryons, and the results
suggested that the $\Lambda_c(2765)$ can be assigned as the first
radial excitation of the $\Lambda_c$ with $J^P = 1/2^+$, the
$\Lambda_c(2880)$ is a good $\Lambda_c(1D)$ candidate with $J^P =
5/2^+$, and the $\Lambda_c(2940)$ might be the $2P$ excitation of
the $\Lambda_c$.

The method of QCD sum rules in the framework of the heavy quark
effective theory was also applied in
Refs.~\cite{Chen:2015kpa,Chen:2016phw} to systematically investigate
the $P$ and $D$-wave charmed baryons, where the $\Lambda_c(2880)$
was interpreted as a $D$-wave $\Lambda_c$ state of $J^P = 5/2^+$,
and the $\Sigma_c(2800)$ was interpreted as a $P$-wave $\Sigma_c$
state of $J^P = 1/2^-$ or $3/2^-$.

Besides the mass spectrum analysis, there are also many studies
investigating the decay properties of the excited charmed baryons in
order to understand their nature. In
Refs.~\cite{Cheng:2006dk,Cheng:1997rp,Cheng:1997xba} the authors
studied strong decays of the charmed baryons in the framework of
heavy hadron chiral perturbation theory which synthesizes both the
heavy quark symmetry and chiral symmetry. Their results suggested
that the $\Lambda_c(2880)$ is a $D$-wave excited $\Lambda_c$ state
of $J^P = 5/2^+$, while the possible spin-parity quantum numbers of
the $\Lambda_c(2940)$ are $5/2^-$ and $3/2^+$. Later in
Ref.~\cite{Cheng:2015naa}, the $\Lambda_c(2880)$ was still
interpreted as a $D$-wave $\Lambda_c$ state of $J^P = 5/2^+$ and the
$\Sigma_c(2800)$ was interpreted as a $P$-wave $\Sigma_c$ state of
$J^P = 3/2^-$.

Strong decays of the charmed baryons were also systematically
investigated using other methods. In Ref.~\cite{Zhong:2007gp} the
authors used a chiral quark model, and their results suggested that
the $\Lambda_c(2765)$ may be a $\rho$-mode ($l_\rho = 1$) $P$-wave
$\Lambda_c$ state of $J^P = 1/2^-$; the $\Lambda_c(2880)$ and
$\Lambda_c(2940)$ could be the $D$-wave excited $\Lambda_c$ states
of $J^P = 3/2^+$ and $5/2^+$, respectively; the $\Sigma_c(2800)$ can
be interpreted as a $\lambda$-mode ($l_\lambda = 1$) $P$-wave
$\Sigma_c$ state of $J^P = 1/2^-$ or $5/2^-$. In
Ref.~\cite{Chen:2007xf} the authors used the $^3P_0$ model, and
obtained that the $\Lambda_c(2880)$ can be interpreted as a $D$-wave
excited $\Lambda_c$ state of $J^P = 5/2^+$, and the $\Sigma_c(2800)$
can be interpreted as a $P$-wave excited $\Sigma_c$ state of $J^P =
3/2^-$ or $5/2^-$. In Ref.~\cite{Nagahiro:2016nsx}, the authors used
the quark model together with the heavy quark symmetry, and their
results suggested that the $\Lambda_c(2880)$ can be interpreted as a
$D$-wave excited $\Lambda_c$ state of $J^P = 5/2^+$, and the
$\Lambda_c(2940)$ can be its partner belonging to the same heavy
quark spin doublet having $J^P = 7/2^+$.

Besides the conventional quark model $cqq$ picture, there also exist
other interpretations of the $\Lambda_c(2765)$, $\Lambda_c(2880)$,
$\Lambda_c(2940)$ and $\Sigma_c(2800)$. In Ref.~\cite{Lutz:2003jw}
the authors studied the scattering of the pseudoscalar mesons and
ground state charmed baryons in terms of the non-linear chiral SU(3)
Lagrangian and found that the $\Lambda_c(2880)$ can be dynamically
generated in the $J^P = 1/2^-$ channel. Later in
Ref.~\cite{Lutz:2005ip} they proposed that the $\Sigma_c(2800)$ can
be interpreted as a chiral molecule. See also
Refs.~\cite{Mizutani:2006vq,Tolos:2009nn,Kumar:2011ff} which discuss
the $D$ and $D^*$ mesons in the nuclear and hyperonic medium. The
interaction of the pseudoscalar mesons and ground state baryons was
also studied in
Refs.~\cite{JimenezTejero:2009vq,JimenezTejero:2011fc} within a
coupled channel approach, where the authors used a $t$-channel
vector-exchange driving force and concluded that the
$\Sigma_c(2800)$ can be interpreted as a dynamically generated
resonance with a dominant $ND$ configuration and $J^P = 1/2^-$.

In Ref.~\cite{He:2006is} the authors proposed that the
$\Lambda_c(2940)^+$ is a $D^{*0}p$ molecular state, which can
naturally explain why its mass is a few MeV below the threshold and
why its observed channels are $D^0p$ and $\Lambda^+_c \pi^+ \pi^-$.
They also proposed the experimental search of the channels such as
$D^+n$, $D^0\pi^0 p$, $D^0 \gamma p$ and $\Lambda^+_c \pi^0 \pi^0$
to further test this molecular interpretation. In
Refs.~\cite{Ortega:2012cx,Ortega:2014eoa} the authors studied the
$\Lambda_c(2940)$ in a constituent quark model. They suggested that
the $\Lambda_c(2940)$ may be interpreted as a molecular state
composed of nucleons and $D^*$. They could obtain the right binding
energy in the $J^P = 3/2^-$ channel, and their calculated partial
widths for the $\Lambda_c(2940) \to ND$ and $\Lambda_c(2940) \to
\Sigma_c \pi$ decays are consistent with the experimental data.
Moreover, they predicted its bottom partner, a $\bar B^* N$
molecular state around 6248 MeV.

The quark delocalization color screening model was recently applied
in Ref.~\cite{Zhao:2016zhf} to study the $ND$ system, and their
results suggested that the $\Sigma_c(2800)$ could be explained as
the $DN$ bound state with $J^P=1/2^-$ but its coupling to the $ND^*$
channel should be taken into account, and the $\Lambda_c(2940)$
could be explained as the $D^*N$ bound state with $J^P=3/2^-$. They
further proposed a possible $\Delta D^*$ resonance state around
3210.1 MeV with $I = 1$ and $J^P = 5/2^-$. The method of QCD sum
rules was also applied in Ref.~\cite{Zhang:2012jk} to test these
interpretations. See discussions in
Refs.~\cite{Bugg:2008wu,He:2010zq,Chen:2014mwa} for more
information.

As a $D^*N$ hadronic molecule, the two-body strong decays of the
$\Lambda_c(2940)$ were studied in Ref.~\cite{Dong:2009tg}, where
they excluded the spin-parity $J^P = 1/2^-$ assignment, and
calculated the dominant decay channels $\Sigma^{++}_c \pi^-$ and
$\Sigma^0_c \pi^+$ for the choice of $J^P = 1/2^+$. Later in
Refs.~\cite{Dong:2010xv,Dong:2011ys}, the authors investigated the
radiative decay $\Lambda_c(2940)^+ \to \Lambda_c(2286)^+ \gamma$ and
the strong three-body decays $\Lambda_c(2940)^+ \to
\Lambda_c(2286)^+ \pi^+ \pi^-$ and $\Lambda_c(2940)^+ \to
\Lambda_c(2286)^+ \pi^0 \pi^0$, assuming the $\Lambda_c(2940)$ as a
$D^*N$ hadronic molecule of $J^P = 1/2^+$. They also studied the
annihilation process $p \bar p \to p D^0 \bar \Lambda_c(2286)$ in
Ref.~\cite{Dong:2014ksa}, and found that the contribution from the
intermediate state $\Lambda_c(2940)$ is sizeable near the threshold
of $p \bar p \to \Lambda_c(2286) \bar \Lambda_c(2286)$ and can be
observed at the PANDA experiment. Its role in the $\pi^- p \to D^-
D^0 p$ reaction near threshold was investigated in
Ref.~\cite{Xie:2015zga} within an effective Lagrangian approach. The
$\Sigma_c(2800)$ can also be interpreted as a $DN$ bound state,
under which assumption its strong two-body decay $\Sigma_c(2800) \to
\Lambda_c \pi$ was investigated in Ref.~\cite{Dong:2010gu}, and the
evaluated width with the $J^P = 1/2^+$ and $3/2^-$ assignments is
consistent with the experimental data.

As a $D^{*0} p$ molecular state, the photoproduction of the
$\Lambda_c(2940)$ in the $\gamma n \to D^- \Lambda_c(2940)^+$
process was investigated in Ref.~\cite{Wang:2015rda} with an
effective Lagrangian approach, where the authors estimated the total
cross section of this process, and proposed to search for the
$\Lambda_c(2940)$ at the COMPASS experiment. A similar method was
applied in Ref.~\cite{Huang:2016ygf} to study the production of the
$\Lambda_c(2940)$ by the kaon-induced reaction on a proton target,
through the $K^- p \to D^-_s \Lambda_c(2940) (\to D^0 p )$ reaction,
and their results suggested that the $\Lambda_c(2940)$ can be
studied in the experiment with high-energy kaon beam on a proton
target.

\subsubsection{$\Xi_c(2930)$, $\Xi_c(2980)$, $\Xi_c(3055)$, $\Xi_c(3080)$ and $\Xi_c(3123)$.}

\renewcommand{\arraystretch}{1.6}
\begin{table*}[htb]
\tiny \caption{Possible interpretations of the $\Xi_c(2930)$,
$\Xi_c(2980)$, $\Xi_c(3055)$, $\Xi_c(3080)$ and $\Xi_c(3123)$. We
show the possible spin-parity quantum numbers for the conventional
quark model picture. We use $\Xi_c$ and $\Xi_c^\prime$ to denote the
$\Xi_c$ baryons belonging to the flavor $\mathbf{\bar 3}_F$ and
$\mathbf{6}_F$ respectively, but note that the superscript $^\prime$
is often omitted. See also Ref.~\cite{Cheng:2015iom} for more
information. } \centering
\begin{tabular}{ c c c c c c }\toprule[1pt]
References & $\Xi_c(2930)$ & $\Xi_c(2980)$ & $\Xi_c(3055)$ &
$\Xi_c(3080)$ & $\Xi_c(3123)$
\\ \midrule[1pt]
Ref.~\cite{Valcarce:2008dr}  &  --  &  $\Xi_c\big({1\over2}^-\big)$
&  $\Xi_c\big({5\over2}^+\big)$  &  $\Xi_c\big({3\over2}^+\big)$  &
$\Xi_c\big({1\over2}^+\big) \big/ \Xi_c^\prime\big({5\over2}^+\big)$
\\
Ref.~\cite{Guo:2008he}   &  --  &  --  &
$\Xi_c\big({5\over2}^+\big)$  &  $\Xi_c\big({5\over2}^+\big)$  &
$\Xi_c^\prime\big({5\over2}^+\big)$
\\
Ref.~\cite{Ebert:2011kk} & ${1\over2}^- \big/ {3\over2}^- \big/
{5\over2}^-~\Xi_c^\prime(1P)$ & ${1\over2}^+~\Xi_c^\prime(2S)$ &
${3\over2}^+~\Xi_c(1D)$ & ${5\over2}^+~\Xi_c(1D)$ &
${7\over2}^+~\Xi_c^\prime(1D)$
\\
Refs.~\cite{Chen:2015kpa,Chen:2016phw}  &  --  &   ${1\over2}^-
\big/ {3\over2}^- ~\Xi_c^\prime(1P)$  &  ${3\over2}^+~\Xi_c(1D)$  &
${5\over2}^- ~\Xi_c^\prime(1P)  \big/ {5\over2}^+~\Xi_c(1D)$  &  --
\\
Ref.~\cite{Chen:2014nyo}  &  --  &   ${1\over2}^+~\Xi_c(2S)$  &
${3\over2}^+~\Xi_c(1D)$  &  ${5\over2}^+~\Xi_c(1D)$  &  ${1\over2}^-
\big/ {3\over2}^-~\Xi_c(2P)$
\\
Ref.~\cite{Chen:2016iyi}  &  ${3\over2}^- \big/ {5\over2}^-~\Xi_c^\prime(1P)$  &  ${1\over2}^+ ~\Xi_c(2S)$  &  --  &  --  &  --
\\ \hline
Refs.~\cite{Cheng:2006dk,Cheng:2015naa}  &  --  &
$\Xi_c^\prime\big({1\over2}^+\big)$   &  --  &
$\Xi_c\big({5\over2}^+\big)$  &  --
\\
Ref.~\cite{Liu:2012sj} &  ${1\over2}^-~\Xi_c^\prime(1P)$  &
${1\over2}^-\big/ {3\over2}^-~\Xi_c^\prime(1P)$  &
${3\over2}^+~\Xi_c(1D)$  &  ${1\over2}^+~\Xi_c(2S)$  &
${3\over2}^+\big/ {5\over2}^+$
\\
Ref.~\cite{Zhao:2016qmh}  &  --  &  --  &  ${5\over2}^+\big/
{7\over2}^+$  &  --  &  --
\\ \midrule[1pt]
Ref.~\cite{JimenezTejero:2009vq}  &  --  &  dynamically generated  &
--   &  --   &  --
\\ \bottomrule[1pt]
\end {tabular}
\label{sec6:excitedXi}
\end{table*}

There are also many possible interpretations for the $\Xi_c(2930)$,
$\Xi_c(2980)$, $\Xi_c(3055)$, $\Xi_c(3080)$ and $\Xi_c(3123)$.
Actually, these states are even more complicated than the excited
$\Lambda_c$ and $\Sigma_c$ baryons, because they can belong to both
the flavor $\mathbf{\bar 3}_F$ and $\mathbf{6}_F$ representations.
We summarize them in Table~\ref{sec6:excitedXi}, and briefly review
these efforts here. See also
Refs.~\cite{Liu:2007ge,Zhang:2008pm,Cheng:2009yz,Albuquerque:2009pr,Aliev:2009jt,Wang:2010it}
for more discussions.

In Ref.~\cite{Valcarce:2008dr} the authors studied the heavy baryon
spectroscopy within the Faddeev method, and found that the
$\Xi_c(2980)$ can be interpreted as an excited $\Xi_c$ state of $J^P
= 1/2^-$; the $\Xi_c(3055)$ and $\Xi_c(3080)$ can be interpreted as
excited $\Xi_c$ states of $J^P = 5/2^+$ and $3/2^+$, respectively;
the $\Xi_c(3123)$ can be interpreted as an excited $\Xi_c$ state of
$J^P = 1/2^+$ or an excited $\Xi_c^\prime$ state of $J^P = 5/2^+$.
In Ref.~\cite{Guo:2008he} the authors studied the Regge
phenomenology, and their results suggested that both the
$\Xi_c(3055)$ and $\Xi_c(3080)$ can be assigned as the excited
$\Xi_c$ state with $J^P = 5/2^+$, and the $\Xi_c(3123)$ may be the
$D$-wave excited state of the $\Xi^\prime_c$ with $J^P = 5/2^+$.

The QCD-motivated relativistic quark model was applied in
Ref.~\cite{Ebert:2011kk} to study the heavy baryon spectrum, and
their results suggested that the $\Xi_c(2930)$ can be interpreted as
the excited $\Xi_c^\prime$ state with $J^P = 1/2^-$, $3/2^-$ or
$5/2^-$; the $\Xi_c(2980)$ can be interpreted as the $1/2^+~(2S)$
excited $\Xi_c^\prime$ state; the $\Xi_c(3055)$ and $\Xi_c(3080)$
can be interpreted as the $3/2^+~(1D)$ and $5/2^+~(1D)$ excited
$\Xi_c$ states, respectively; the $\Xi_c(3123)$ can be interpreted
as the $7/2^+~(1D)$ excited $\Xi_c^\prime$ state. The method of QCD
sum rules within HQET was applied in
Refs.~\cite{Chen:2015kpa,Chen:2016phw} to systematically investigate
the heavy baryon spectrum, and their results suggested that the
$\Xi_c(2980)$ can be interpreted as the $P$-wave excited
$\Xi_c^\prime$ state of $J^P = 1/2^-$ or $3/2^-$, and the
$\Xi_c(3080)$ can be interpreted as the $P$-wave excited
$\Xi_c^\prime$ state of $J^P = 5/2^-$; while the $\Xi_c(3080)$ can
also be interpreted as the $D$-wave excited $\Xi_c$ state of $J^P =
5/2^+$, and the $\Xi_c(3055)$ can be its partner belonging to the
same heavy quark spin doublet with $J^P = 3/2^+$.

In Ref.~\cite{Chen:2014nyo} the authors investigated the mass
spectra of the charmed baryons using the relativistic flux tube
model, and obtained that the $\Xi_c(2980)$ can be interpreted as the
first radial excitation of the $\Xi_c$ with $J^P = 1/2^+$, and the
$\Xi_c(3123)$ might be the $2P$ excitation of the $\Xi_c$. They also
obtained that the $\Xi_c(3080)$ is a good $\Xi_c(1D)$ candidate with
$J^P = 5/2^+$, and the $\Xi_c(3055)$ could be its doublet partner
with $J^P = 3/2^+$. Recently in Ref.~\cite{Chen:2016iyi} the authors
systematically studied the excited charmed baryons using the
nonrelativistic constituent quark model. Their results suggested
that the $\Xi_c(2980)$ can be interpreted as the first radial
excited state of the $\Xi_c$ with $J^P = 1/2^+$, and the
$\Xi_c(2930)$ can be assigned to the $1P$ excited state of the
$\Xi_c^\prime$ with $J^P = 3/2^-$ or $5/2^-$.

Besides the mass spectrum, the decay properties of the
$\Xi_c(2930)$, $\Xi_c(2980)$, $\Xi_c(3055)$, $\Xi_c(3080)$ and
$\Xi_c(3123)$ are also interesting. In Ref.~\cite{Cheng:2006dk} the
authors studied the strong decays of the charmed baryons using the
heavy hadron chiral perturbation theory, and concluded that the
spin-parity quantum numbers of the $\Xi_c(2980)$ and $\Xi(3080)$ are
$1/2^+$ and $5/2^+$, respectively. Later in
Ref.~\cite{Cheng:2015naa}, the authors further identified the
$\Xi_c(2980)$ and $\Xi(3080)$ as the excited $\Xi_c^\prime$ state
$J^P = 1/2^+$ and the excited $\Xi_c$ state of $J^P = 5/2^+$,
respectively.

The strong decays of the charmed baryons were also systematically
investigated in Ref.~\cite{Liu:2012sj} using a chiral quark model.
Their results suggested that the $\Xi_c(2930)$ might be the first
$P$-wave excited $\Xi_c^\prime$ state with $J^P = 1/2^-$; the
$\Xi_c(2980)$ might be the $P$-wave excited $\Xi_c^\prime$ state
with $J^P = 1/2^-$ or $3/2^-$; the $\Xi_c(3055)$ is most likely to
be the first $D$-wave excitation of the $\Xi_c$ with $J^P = 3/2^+$,
and the $\Xi_c(3080)$ can be interpreted as the first radial
excitation of the $\Xi_c$ with $J^P = 1/2^+$; the $\Xi_c(3123)$
might be assigned as the $D$-wave excitation of the $\Xi_c^\prime$
with $J^P = 3/2^+$ or $5/2^+$ or the $D$-wave excitation of the
$\Xi_c$ with $J^P = 5/2^+$. Recently in Ref.~\cite{Zhao:2016qmh} the
authors systematically studied the decay properties of the
$\Xi_c(3055)$ and $\Xi_c(3080)$ using the $^3P_0$ model. Their
results suggested that the $\Xi_c(3055)$ can be interpreted as the
$D$-wave excite $\Xi_c$ state of $J^P = 5/2^+$ or $7/2^+$, and its
total decay width was evaluated to be around $10$ MeV. This state
can also be interpreted as the $D$-wave excite $\Xi_c^\prime$ state
of $J^P = 5/2^+$ or $7/2^+$, and its total decay width was evaluated
to be around $7$ MeV. Their results also suggested that the
$\Xi_c(3080)$ seems impossible to be identified as a $D$-wave
charmed strange baryon.

Besides the conventional excited charmed baryon picture ($cqq$),
there also exist other interpretations of the $\Xi_c(2930)$,
$\Xi_c(2980)$, $\Xi_c(3055)$, $\Xi_c(3080)$ and $\Xi_c(3123)$. In
Ref.~\cite{JimenezTejero:2009vq} the authors studied the interaction
of the pseudoscalar mesons and ground state baryons within a coupled
channel approach, and noticed that the $\Xi_c(2980)$ can be a
dynamically generated resonance with $J^P = 1/2^-$.

%% file: section6.2.tex
\subsection{The bottom baryons}
\label{sec6.2}

In this subsection we discuss the mass spectrum of the bottom baryon
family. Here we investigate five theoretical approaches as listed in
Sec.~\ref{sec6.1} and summarize the results in
Tables~\ref{sec6:bottom1} and \ref{sec6:bottom2}. More discussions
can be found in
Refs.~\cite{Bagan:1991sc,Duraes:2007te,Wang:2007sqa,Dhir:2013nka,Yamaguchi:2014era,Valcarce:2014fma,Aliev:2015qea,Thakkar:2016dna}.
Particularly, the dipion decays of the $P$-wave and $D$-wave excited
bottom baryons were systematically investigated in
Ref.~\cite{Mu:2014iaa} in the framework of the QPC model. See also
the lattice QCD studies in Refs.~\cite{Burch:2008qx,Burch:2015pka}.

\renewcommand{\arraystretch}{1.2}
\begin{table*}[htb]
\scriptsize
\caption{Comparison of the experimental data and theoretical
results of the $\Lambda_b$, $\Sigma_b$ and $\Omega_b$ baryons,
obtained using the QCD-motivated relativistic quark model based on
the quasipotential approach (R.~Q.~M.)~\cite{Ebert:2011kk}, the
non-relativistic quark model (Non-RQM)~\cite{Roberts:2007ni}, the
QCD sum rules within HQET
(QSR)~\cite{Liu:2007fg,Chen:2015kpa,Mao:2015gya,Chen:2016phw}, and
the relativistic quark model generalized from the GI model
(C.~I.)~\cite{Capstick:1986bm}. The masses are in unit of MeV.}
\centering
\begin{tabular}{ c c c c c c c }\toprule[1pt]
& $J^P~(nL)$ & Experimental values~\cite{Olive:2016xmw} &
R.~Q.~M.~\cite{Ebert:2011kk} & Non-RQM~\cite{Roberts:2007ni} &
QSR~\cite{Liu:2007fg,Mao:2015gya,Chen:2016phw} &
C.~I.~\cite{Capstick:1986bm}
\\ \midrule[1pt]
$\Lambda_b$             & $1/2^+~(1S)$     & $5619.51\pm0.23$                                  & 5620  & 5612  & $5637^{+68}_{-56}$  & 5585  \\
$\Sigma_b^+$            & $1/2^+~(1S)$     & $5811.3^{+0.9}_{-0.8}\pm1.7$                      & 5808  & 5833  & $5809^{+82}_{-76}$  & 5795  \\
$\Sigma_b^{*+}$         & $3/2^+~(1S)$     & $5832.1\pm0.7^{+1.7}_{-1.8}$                      & 5834  & 5858  & $5835^{+82}_{-77}$  & 5805  \\
$\Omega_b^{0}$          & $1/2^+~(1S)$     & $6046.4\pm1.9$                                    & 6064  & 6081  & $6036\pm81$         & --    \\
$\Omega_b^{*0}$         & $3/2^+~(1S)$     & --                                                & 6088  & 6102  & $6063^{+83}_{-82}$  & --    \\
\hline
$\Lambda_b$             & $1/2^-~(1P)$     & $\Lambda_b(5912)=5911.97\pm0.67$                  & 5930  & 5939  & $5.87\pm0.12$       & 5912  \\
$\Lambda_b$             & $3/2^-~(1P)$     & $\Lambda_b(5920)=5919.77\pm0.67$                  & 5942  & 5941  & $5.88\pm0.11$       & 5920  \\
$\Sigma_b$              & $1/2^-~(1P)$     & --                                                & 6095  & 6099  & $5.91\pm0.14$       & 6070  \\
$\Sigma_b$              & $1/2^-~(1P)$     & --                                                & 6101  & 6106  & $6.02\pm0.12$       & 6070  \\
$\Sigma_b$              & $3/2^-~(1P)$     & --                                                & 6087  & 6101  & $5.92\pm0.14$       & 6070  \\
$\Sigma_b$              & $3/2^-~(1P)$     & --                                                & 6096  & 6105  & $5.96\pm0.18$       & 6085  \\
$\Sigma_b$              & $5/2^-~(1P)$     & --                                                & 6084  & 6172  & $5.98\pm0.18$       & 6090  \\
$\Omega_b$              & $1/2^-~(1P)$     & --                                                & 6330  & 6301  & $6.34\pm0.13$       & --    \\
$\Omega_b$              & $1/2^-~(1P)$     & --                                                & 6339  & 6312  & $6.50\pm0.11$       & --    \\
$\Omega_b$              & $3/2^-~(1P)$     & --                                                & 6331  & 6304  & $6.34\pm0.13$       & --    \\
$\Omega_b$              & $3/2^-~(1P)$     & --                                                & 6340  & 6311  & $6.43\pm0.13$       & --    \\
$\Omega_b$              & $5/2^-~(1P)$     & --                                                & 6334  & 6311  & $6.43\pm0.13$       & --    \\
\hline
$\Lambda_b$             & $1/2^+~(2S)$     & --                                                & 6089  & 6107  & --  & 6045  \\
$\Sigma_b$              & $1/2^+~(2S)$     & --                                                & 6213  & 6294  & --  & 6200  \\
$\Sigma_b$              & $3/2^+~(2S)$     & --                                                & 6226  & 6308  & --  & 6250  \\
$\Omega_b$              & $1/2^+~(2S)$     & --                                                & 6450  & 6472  & --  & --    \\
$\Omega_b$              & $3/2^+~(2S)$     & --                                                & 6461  & 6478  & --  & --    \\
\hline
$\Lambda_b$             & $3/2^+~(1D)$     & --                                                & 6190  & 6181  & $6.01^{+0.20}_{-0.12}$  & 6145  \\
$\Lambda_b$             & $5/2^+~(1D)$     & --                                                & 6196  & 6183  & $6.01^{+0.20}_{-0.13}$  & 6165  \\
$\Sigma_b$              & $1/2^+~(1D)$     & --                                                & 6311  & --    & --  & 6200  \\
$\Sigma_b$              & $3/2^+~(1D)$     & --                                                & 6285  & --    & --  & 6250  \\
$\Sigma_b$              & $3/2^+~(1D)$     & --                                                & 6326  & --    & --  & 6320  \\
$\Sigma_b$              & $5/2^+~(1D)$     & --                                                & 6270  & 6325  & --  & 6325  \\
$\Sigma_b$              & $5/2^+~(1D)$     & --                                                & 6284  & 6328  & --  & 6335  \\
$\Sigma_b$              & $7/2^+~(1D)$     & --                                                & 6260  & 6333  & --  & 6340  \\
$\Omega_b$              & $1/2^+~(1D)$     & --                                                & 6540  & --    & --  & --    \\
$\Omega_b$              & $3/2^+~(1D)$     & --                                                & 6530  & --    & --  & --    \\
$\Omega_b$              & $3/2^+~(1D)$     & --                                                & 6549  & --    & --  & --    \\
$\Omega_b$              & $5/2^+~(1D)$     & --                                                & 6520  & 6492  & --  & --    \\
$\Omega_b$              & $5/2^+~(1D)$     & --                                                & 6529  & 6494  & --  & --    \\
$\Omega_b$              & $7/2^+~(1D)$     & --
& 6517  & 6497  & --  & --
\\ \bottomrule[1pt]
\end {tabular}
\label{sec6:bottom1}
\end{table*}

\renewcommand{\arraystretch}{1.6}
\begin{table*}[htb]
\tiny
\caption{Comparison of the experimental data and theoretical
results of the $\Xi_b$ baryons, obtained using the QCD-motivated
relativistic quark model based on the quasipotential approach
(R.~Q.~M.)~\cite{Ebert:2011kk}, the non-relativistic quark model
(Non-RQM)~\cite{Roberts:2007ni}, the QCD sum rules within HQET
(QSR)~\cite{Liu:2007fg,Chen:2015kpa,Mao:2015gya,Chen:2016phw}, and
the constituent quark model (C.~Q.~M.)~\cite{Garcilazo:2007eh}. The
masses are in unit of MeV. Here we use $\Xi_b$ and $\Xi_b^\prime$
to denote the $\Xi_b$ baryons belonging to the flavor $\mathbf{\bar
3}_F$ and $\mathbf{6}_F$ respectively, but note that the superscript
$^\prime$ is often omitted. Actually, the $\Xi_b$ and $\Xi_b^\prime$
can mix with each other, which effect was taken into account in
Ref.~\cite{Roberts:2007ni}. } \centering
\begin{tabular}{ c c c c c c c }\toprule[1pt]
& $J^P~(nL)$ & Experimental values~\cite{Olive:2016xmw} &
R.~Q.~M.~\cite{Ebert:2011kk} & Non-RQM~\cite{Roberts:2007ni} &
QSR~\cite{Liu:2007fg,Mao:2015gya,Chen:2016phw} &
C.~Q.~M.~\cite{Garcilazo:2007eh}
\\ \midrule[1pt]
$\Xi_b^-$               & $1/2^+~(1S)$     & $5794.5\pm1.4$                                    & 5803  & 5806  & $5780^{+73}_{-68}$  & 5825  \\
$\Xi_b^{\prime-}$       & $1/2^+~(1S)$     & $5935.02\pm0.02\pm0.05$                           & 5936  & 5970  & $5903^{+81}_{-79}$  & 5913  \\
$\Xi_b^{\prime-}$       & $3/2^+~(1S)$     & $\Xi_b^{*-}=5955.33\pm0.12\pm0.05$                & 5963  & 5980  & $5929^{+83}_{-79}$  & 5967  \\
\hline
$\Xi_b$                 & $1/2^-~(1P)$     & --                                                & 6120  & 6090  & $6.06\pm0.13$       & 6076  \\
$\Xi_b$                 & $3/2^-~(1P)$     & --                                                & 6130  & 6093  & $6.07\pm0.13$       & 6076  \\
$\Xi^\prime_b$          & $1/2^-~(1P)$     & --                                                & 6227  & 6188  & $6.11\pm0.13$       & 6157  \\
$\Xi^\prime_b$          & $1/2^-~(1P)$     & --                                                & 6233  & --    & $6.24\pm0.11$       & --    \\
$\Xi^\prime_b$          & $3/2^-~(1P)$     & --                                                & 6224  & 6190  & $6.11\pm0.13$       & 6157  \\
$\Xi^\prime_b$          & $3/2^-~(1P)$     & --                                                & 6234  & --    & $6.17\pm0.17$       & --    \\
$\Xi^\prime_b$          & $5/2^-~(1P)$     & --                                                & 6226  & 6201  & $6.18\pm0.16$       & --    \\
\hline
$\Xi_b^+$               & $1/2^+~(2S)$     & --                                                & 6266  & --    & --   & --   \\
$\Xi_b^{\prime+}$       & $1/2^+~(2S)$     & --                                                & 6329  & --    & --   & --   \\
$\Xi_b^{\prime+}$       & $3/2^+~(2S)$     & --                                                & 6342  & --    & --   & --   \\
\hline
$\Xi_b$                 & $3/2^+~(1D)$     & --                                                & 6366  & 6311  & $6.19^{+0.10}_{-0.12}$  & 6275  \\
$\Xi_b$                 & $5/2^+~(1D)$     & --                                                & 6373  & 6300  & $6.19^{+0.10}_{-0.12}$  & --    \\
$\Xi^\prime_b$          & $1/2^+~(1D)$     & --                                                & 6447  & --    & --   & --   \\
$\Xi^\prime_b$          & $3/2^+~(1D)$     & --                                                & 6431  & --    & --   & --   \\
$\Xi^\prime_b$          & $3/2^+~(1D)$     & --                                                & 6459  & --    & --   & --   \\
$\Xi^\prime_b$          & $5/2^+~(1D)$     & --                                                & 6420  & 6393  & --   & --   \\
$\Xi^\prime_b$          & $5/2^+~(1D)$     & --                                                & 6432  & --    & --   & --   \\
$\Xi^\prime_b$          & $7/2^+~(1D)$     & --
& 6414  & 6395  & --   & --
\\ \bottomrule[1pt]
\end {tabular}
\label{sec6:bottom2}
\end{table*}

In Tables~\ref{sec6:bottom1} and \ref{sec6:bottom2}, we also compare
the theoretical results with the experimental data, and conclude
that
\begin{enumerate}

\item The $\Omega_b^*$ of $J^P = 3/2^+$ has not been observed yet.
Various theoretical calculations of all the other $1S$ bottom baryon
massed agree with the data quite well. The states nearly complete
the flavor $\mathbf{\bar 3}_F$ multiplet of $J^P = 1/2^+$ and two
flavor $\mathbf{6}_F$ multiplets of $J^P = 1/2^+$ and $3/2^+$.

\item The $\Lambda_b(5912)$ and $\Lambda_b(5920)$ are good
candidates for the $P$-wave bottom baryons, which belong to the
flavor $\mathbf{\bar3}_F$ multiplets of $J^P = 1/2^-$ and $3/2^-$.

\item Many bottom baryons remain to be discovered experimentally,
probably at LHCb.

\end{enumerate}
In the following we review the theoretical progress on the excited
bottom baryons, i.e., the $\Lambda_b(5912)$ and $\Lambda_b(5920)$.

\subsubsection{$\Lambda_b(5912)$ and $\Lambda_b(5920)$.}

The $\Lambda_b(5912)$ and $\Lambda_b(5920)$ were both observed by
the LHCb Collaboration in 2012~\cite{Aaij:2012da}. In 1986, Capstick
and Isgur studied the $P$-wave bottom baryons using the relativistic
quark model~\cite{Capstick:1986bm}, and their predicted masses are
exactly the same as the experimental values (see
Table~\ref{sec6:bottom1}). This agreement is a big success of the
relativistic quark model. Besides this
work, many other models and methods were applied to study the
$\Lambda_b(5912)$ and $\Lambda_b(5920)$. See discussions at the
beginning of this subsection and
Refs.~\cite{Xiao:2013yca,Yasui:2013iga} for more information.

In Ref.~\cite{Karliner:2008sv} the authors used the color hyperfine
interaction to study the bottom baryons, and predicted the masses of
the baryons with $l_\lambda = 1$ to be $M(\Lambda_b(1/2^-)) = 5929
\pm 2$ MeV and $M(\Lambda_b(3/2^-)) = 5940 \pm 2$ MeV. They also
predicted that $M(\Xi_b(1/2^-)) = 6106 \pm 4$ MeV and
$M(\Xi_b(1/2^-)) = 6115 \pm 4$ MeV. See also studies in
Refs.~\cite{Karliner:2006ny}. Later in Ref.~\cite{Karliner:2015ema},
the authors further explored the $\Sigma_b$ baryons with $l_\lambda
= 1$. Their results suggested that the $\Sigma_b$ states of $J^P =
3/2^-$ and $5/2^-$ with $j_l = 2$ lie around 6100 MeV.

In Ref.~\cite{Chen:2014nyo} the authors applied a mass formula
derived from the relativistic flux tube model to investigate the
heavy baryons. With the heavy-quark-light-diquark picture, the
$\Lambda_b(5912)$ and $\Lambda_b(5920)$ were assigned as the $1P$
bottom baryons of $J^P = 1/2^-$ and $3/2^-$, respectively.

Different from the conventional $bqq$ picture, the $\Lambda_b(5912)$
and $\Lambda_b(5920)$ are explained as the dynamically generated
states in
Refs.~\cite{GarciaRecio:2012db,Lu:2014ina,Garcia-Recio:2015jsa,Liang:2014eba,Torres-Rincon:2014ffa}.
A unitarized meson-baryon coupled-channel dynamical model was used
to investigate the $\Lambda_b(5912)$ and $\Lambda_b(5920)$ in
Refs.~\cite{GarciaRecio:2012db,Garcia-Recio:2015jsa}, where these
two states were identified as the dynamically generated meson-baryon
molecular states. With the heavy quark spin symmetry, the authors
predicted two $bsq$ baryons, the $\Xi_b(6035.4)$ of $J^P = 1/2^-$
and the $\Xi_b(6043.3)$ of $J^P = 3/2^-$.

In Ref.~\cite{Liang:2014eba} the authors studied the bottom baryons
in the extended local hidden gauge approach. Under the assumption
that the heavy quarks act as spectators, they found two states with
nearly zero width. These two states couple mostly to $\bar B^* N$,
and were identified as the $\Lambda_b(5912)$ and $\Lambda_b(5920)$.
In Ref.~\cite{Torres-Rincon:2014ffa} the authors investigated the
interaction of the $\bar B$ mesons with $N$ and $\Delta$ within a
unitarized approach based on effective models compatible with chiral
and heavy-quark symmetries. They identified several $\Lambda_b$ and
$\Sigma_b$ doublets, two of which can be associated with the
$\Lambda_b(5912)$ and $\Lambda_b(5920)$. They also identified
another bottom baryon, the $\Sigma_b^*(5904)$ of $J = 3/2$, as the
bottom counterpart of the $\Sigma^*(1670)$ and $\Sigma^*_c(2549)$.

%% file: section7.1.tex
\section{The doubly and triply charmed baryons}
\label{sec7}

The only experimental evidence for baryons containing two or more
heavy quarks is the doubly charmed baryon $\Xi_{cc}^+$ reported by
the SELEX Collaboration in the $\Xi_{cc}^+\to\Lambda_c^+ K^-\pi^+$
process \cite{Mattson:2002vu}, as mentioned in Sec. \ref{sec2.7}. Up
to now, no other experiments confirmed the existence of the
$\Xi_{cc}^+$ \cite{Aubert:2006qw,Chistov:2006zj,Ratti:2003ez}.
However, the doubly charmed baryon systems with the quark contents
$ccu, ccd, ccs$ have been studied extensively using various
theoretical methods, such as the various quark models, the bag
model, QCD sum rules, heavy quark effective theory, lattice QCD
simulation, etc.~\cite{LlanesEstrada:2011kc,Wang:2011ae,Flynn:2011gf,Meinel:2012qz,Aliev:2012tt,Aliev:2014lxa,Padmanath:2013zfa}. In this section, we briefly review these
investigations for the doubly and triply charmed/bottom baryon
systems.

Forty years ago, De Rujula, Georgi and Glashow investigated the
doubly charmed baryons and estimated their masses in a
renormalizable gauge field theory \cite{DeRujula:1975qlm}. Jaffe and
Khakis calculated the mass spectra of the doubly and triply charmed
baryons in the bag model \cite{Jaffe:1975us}. The similar
investigations were extended to the bottom sector to study the
masses of the doubly and triply bottom baryons in Ref.
\cite{Ponce:1978gk}. Using the hyper-spherical formalism, Hasenfratz
{\it et al.} solved the Schr\"odinger equation and obtained the masses
of the triply charmed baryons \cite{Hasenfratz:1980ka}. Later, Fleck
and Richard calculated the mass spectra of the doubly charmed
baryons in potential models and several versions of the bag model
\cite{Fleck:1989mb}. Bagan {\it et al.} discussed the masses of the
$bcq, ccq, bbq$ baryons by combining the potential model and the QCD
sum rules \cite{Bagan:1994dy,Bagan:1992za}. The hadron spectroscopy
of the baryons containing two or more heavy quarks were also studied
in the heavy quark effective theory \cite{Korner:1994nh}, mass sum
rules \cite{Lichtenberg:1995kg}, quark models
\cite{Ebert:1996ec,Gerasyuta:1999pc,Itoh:2000um}, relativistic
quark-diquark model \cite{Ebert:2002ig} and some other approaches
\cite{Roncaglia:1995az,SilvestreBrac:1996wp,Kiselev:2001fw,Narodetskii:2002ib}.
Lewis, Mathur and Woloshyn calculated the masses of the doubly
charmed baryons and the mass differences between the spin $3/2$ and
spin $1/2$ baryons states in quenched Lattice QCD
\cite{Lewis:2001iz}. The masses of the $ccu$ and $ccd$ states with
$J^P=\frac{1}{2}^+$ were often predicted above 3.6 GeV, which are
higher than the mass of the $\Xi_{cc}^+(3520)$
\cite{Mattson:2002vu}.

After the announcement of the $\Xi_{cc}^+(3520)$ by SELEX
collaboration, there are more theoretical efforts to study the
spectroscopy of the doubly and triply charmed baryons. In Ref.
\cite{He:2004px}, the authors evaluated the mass spectra of baryons
consisting of two heavy and one light quarks in the MIT bag model.
They considered both the scalar and axial-vector diquark formalisms
for the two heavy quarks. Accordingly, the mass spectra of the
$QQ^\prime q$ ($Q, Q^\prime=b, c$ and  $q=u, d, s$) states with
$J^P=\frac{1}{2}^+, \frac{3}{2}^+$ were obtained. In the framework
of the potential models, Richard and Stancu revisited the doubly
charmed baryons and calculated the mass of the $ccd (1/2^+)$ state
to be around 3.6 GeV  \cite{Richard:2005jz}.

In Ref. \cite{Migura:2006ep}, the authors investigated the mass
spectra of the doubly and triply charmed baryons in the framework of
a relativistically covariant constituent quark model. They
considered the Bethe-Salpeter equation with the instantaneous
approximation and used a linearly rising three-body confinement
potential and a flavor dependent two-body force derived from QCD
instanton effects. A simple quark model was also applied to
calculate the spectrum of baryons containing two and three heavy
quarks $QQ^\prime q, QQQ^\prime$ in Ref. \cite{Roberts:2007ni}. The
phenomenological Hamiltonian was considered by including the kinetic
energy term, the spin independent confining potential, the
spin-dependent hyperfine potential and a simplified spin-orbit
potential, in which the spin independent confining potential
consists of the linear and Coulomb components. In a nonrelativistic
quark model \cite{Albertus:2006ya}, many static properties of the
doubly heavy baryons were evaluated including the masses, charge
radii and magnetic moments. They used five different quark-quark
potentials and solved the three-body problem with a simple
variational approach and Jastrow type orbital wave functions. All
the theoretical approaches predicted the mass of the $\Xi_{cc}$ with
$J^P=\frac{1}{2}^+$ to be around 3.6-3.7 GeV.

In Ref. \cite{Weng:2010rb}, the authors derived the Bethe-Salpeter
equations for the heavy diquarks and the doubly heavy baryons in
leading order in the $1/m_Q$ expansion. They solved the
Bethe-Salpeter equations numerically under the covariant
instantaneous approximation with the kernels containing the scalar
confinement and one-gluon-exchange terms and calculated the masses
and non-leptonic decay widths of the doubly heavy baryons.

In the chiral perturbation theory, Sun {\it et al.} constructed the
chiral effective Lagrangians describing the interactions of the
light mesons and doubly charmed baryons \cite{Sun:2014aya}. They
further made the non-relativistic reduction and obtained the chiral
Lagrangians up to $O(p^4)$ in the heavy baryon limit. They derived
the chiral corrections to the mass of the doubly heavy baryons up to
N$^3$LO and predicted the mass of the $\Xi_{cc}$ to be
$m_{\Xi_{cc}}=3.665^{+.093}_{-.097}$ GeV.

Karliner and Rosner estimated the masses of the doubly heavy $J=1/2$
and $3/2$ baryons using the hyperfine interaction
\cite{Karliner:2014gca}. For the $J=1/2$ $\Xi_{cc}$ state, they
predicted its mass to be about $3627\pm12$ GeV. They also discussed
the P-wave excitations, production mechanisms, decay modes,
lifetimes and prospects for the detection of the doubly heavy
baryons.

In Ref. \cite{Wei:2015gsa}, the authors studied the masses of the
doubly and triply charmed baryons in the Regge phenomenology. They
first expressed the mass of the ground state $\Omega_{cc}^{\ast +}$
as a function of the masses of the well established light baryons
and singly charmed baryons. Then they calculated the masses of the
ground state triply charmed baryon and the doubly charmed baryons
with the quadratic mass relations. The extracted mass of the
$\Xi_{cc}$ with $J^P=1/2^+$ was about 3.52 GeV, which was in good
agreement with the SELEX's value.

The QCD sum rules have also been used to study the mass spectra of
the doubly and triply heavy baryons
\cite{Zhang:2008rt,Zhang:2009re,Wang:2010it,Wang:2010vn,Wang:2010hs}.
In Ref. \cite{Zhang:2008rt}, Zhang and Huang proposed the doubly
heavy baryonic interpolating currents with $J^P={1/2}^+$ and
${3/2}^+$ in a tentative heavy-diquark-light-quark configuration.
They calculated the two-point correlation functions up to the
dimension six nonperturbative contributions in the operator product
expansion. They found the mass of the $ccq$ with $J^P={1/2}^+$ to be
about 4.3 GeV, which is much higher than the predictions from other
methods. Using the same interpolating current, Wang also performed
QCD sum rule analyses for the ${1/2}^+$ $QQq$ baryons
\cite{Wang:2010hs} and obtained the mass of $\Xi_{cc}$ to be around
3.57 GeV. Later, Wang studied the $J^P={3/2}^+$ \cite{Wang:2010vn}
and ${1/2}^-, {3/2}^-$ \cite{Wang:2010it} doubly heavy baryon
states. In Ref. \cite{Zhang:2009re}, Zhang and Huang also studied
the triply heavy baryons with one or two heavy quark flavors.

\renewcommand{\arraystretch}{1.4}
\begin{table*}[!htb]
\footnotesize
\caption{Masses of the doubly charmed baryon $\Xi_{cc}(ccq)$ with
$J^P=1/2^+$ in various models. \label{mass:doublycharm}}
\begin{center}
\begin{tabular}{lcc} \hline \hline
Method & Reference & Mass (MeV) \\ \hline
\multirow{6}{*}{Quark models} &\cite{DeRujula:1975qlm} & 3550--3760 \\
&\cite{Ebert:1996ec} & $3660 $  \\
&\cite{Ebert:2002ig} & $3620 $ \\
&\cite{Itoh:2000um} & $ 3646 \pm 12$ \\
&\cite{Roberts:2007ni} & $3678 $  \\
&\cite{Karliner:2014gca} & $3627 \pm 12$ \\
\hline
\multirow{4}{*}{Potential models} &\cite{Fleck:1989mb} & 3613 \\
&\cite{Bagan:1994dy} & 3630  \\
&\cite{Kiselev:2001fw} & $3480 \pm 50$ \\
&\cite{Richard:2005jz} & 3643  \\
\hline
\multirow{2}{*}{Bag models} &\cite{Fleck:1989mb} & 3516 \\
&\cite{He:2004px} & $3520 $ \\
\hline
Feynman-Hellmann theorem & \cite{Roncaglia:1995az} & $3660 \pm70$ \\
\hline
Heavy quark effective theory & \cite{Korner:1994nh} & 3610  \\
\hline
Chiral perturbation theory & \cite{Sun:2014aya} & $3665^{+93}_{-97}$ \\
\hline
Regge phenomenology & \cite{Wei:2015gsa} & $3520^{+41}_{-40}$ \\
\hline
Nonperturbative string & \cite{Narodetskii:2002ib} & $3690 $ \\
\hline
\multirow{2}{*}{Faddeev equations} &\cite{SilvestreBrac:1996wp} & $3607 $ \\
&\cite{Gerasyuta:1999pc}&3527 \\
\hline
\multirow{2}{*}{Bethe-Salpeter equations} &\cite{Migura:2006ep} & 3642 \\
&\cite{Weng:2010rb} & $ 3540 \pm 20 $ \\
\hline
\multirow{2}{*}{QCD sum rules} &\cite{Zhang:2008rt} & $4260 \pm 190$  \\
&\cite{Wang:2010hs} & $3570 \pm 140$ \\
\hline
\multirow{8}{*}{Lattice QCD} &\cite{Lewis:2001iz} & $3608(15)(^{13}_{35})$ \\
&\cite{Flynn:2003vz} & 3549(13)(19)(92) \\
&\cite{Liu:2009jc} & $3665 \pm 17 \pm 14^{+0}_{-78}$ \\
&\cite{Namekawa:2013vu} & $3603(15)(16)$  \\
&\cite{Briceno:2012wt} & 3595(39)(20)(6)  \\
&\cite{Alexandrou:2014sha} & 3568(14)(19)(1) \\
&\cite{Bali:2015lka} & 3610(90)(120) \\
&\cite{Brown:2014ena} & 3610(23)(22)
 \\ \hline \hline
\end{tabular}
\end{center}
\end{table*}

There are also various lattice simulations on the mass spectra for
the doubly and triply charmed baryons. In Ref. \cite{Flynn:2003vz},
the UKQCD Collaboration presented results for the masses of the
spin-1/2 and spin-3/2 doubly charmed baryons in quenched lattice QCD
with non-perturbatively improved clover action at $\beta=6.2$. The
mass of the $\Xi_{cc}$ was $m=3549(13)(19)(92)$ MeV. Liu {\it et
al.} computed the masses of the $J=1/2$ doubly charmed baryons in
full lattice QCD \cite{Liu:2009jc}. They used the low-lying
charmonium spectrum to tune the heavy-quark action and as a guide to
understand the discretization errors associated with the heavy
quark. Their result for the mass of the $\Xi_{cc}$ was a bit higher.
In Ref. \cite{Namekawa:2013vu}, the PACS-CS Collaboration
investigated the doubly and triply charmed baryon mass spectra using
the relativistic heavy quark action on $2+1$ flavor lattice QCD at
the physical point with the inverse lattice spacing
$a^{-1}=2.194(10)$ GeV. The mass of the $\Xi_{cc}$ was calculated to
be approximately 85 MeV higher than the SELEX's result. Alexandrou
{\it et al.} calculated the masses of the doubly and triply charmed
baryons with the pion mass in the range of about 260 MeV to 450 MeV.
They used three values of the lattice spacing to check the
dependence of the baryon masses on the lattice spacing and the charm
quark mass \cite{Alexandrou:2012xk}. Later, they also evaluated
these mass spectra using a total of ten ensembles of dynamical
twisted mass fermion gauge configurations \cite{Alexandrou:2014sha}.
Their results for the doubly charmed $\Xi_{cc}$ mass were in good
agreement with the SELEX's measurement.

In Ref. \cite{Briceno:2012wt}, the authors calculated the mass
spectra of the positive-parity doubly and triply charmed baryons
from lattice QCD with $N_f=2+1+1$ flavors of dynamical quarks. They
used a relativistic heavy-quark action for the valence charm quark,
clover-Wilson fermions for the valence light and strange quarks and
HISQ sea quarks. They used three lattice spacings $a=0.12$ fm, 0.09
fm, and 0.06 fm to extrapolate to the continuum with a lightest pion
mass around 220 MeV. For the doubly charmed $\Xi_{cc}$ with
$J^P=1/2^+$, they obtained the isospin-averaged value
$M=3595(39)(20)(6)$ MeV, which was in good agreement with the
SELEX's result. In Ref. \cite{Bali:2015lka}, the authors determined
the ground state and first excited state masses of the spin-1/2 and
spin-3/2 doubly charmed baryons with positive and negative parities
from lattice QCD simulation with $N_f=2+1$ non-perturbatively
improved Wilson-clover fermions configurations. Their pion mass on
the lattice lies in the range 259-460 MeV with a lattice spacing
$a\sim 0.075$ fm. Many other lattice QCD calculations can be found
in Refs. \cite{Padmanath:2015jea,Brown:2014ena,Basak:2013oya,
Durr:2012dw,Na:2007pv,Can:2015exa}.

As mentioned above, only one doubly charmed baryon state
$\Xi_{cc}^+(3520)$ was reported, whose existence has not been
confirmed yet. For the $\Xi_{cc}^+(3520)$, we collect some model
predictions in Table \ref{mass:doublycharm}. Most of the theoretical
calculations of the mass for this state lie above the SELEX's value.
There also lacks experimental evidence of the triply heavy baryons
although they must exist. However, a large number of $B_c$ mesons
has been observed at Tevatron \cite{Aaltonen:2007gv,Abazov:2008kv}
and LHCb experiments
\cite{Aaij:2012dd,Aaij:2014asa,Aaij:2014jxa,Aaij:2014bva,Aaij:2013cda,Aaij:2013vcx,Aaij:2013gia}.
Hopefully the doubly and triply charmed/bottomed baryons will be
produced at LHC in the near future. The various theoretical
predictions reviewed above will be useful for the future
experimental search of these states.

%% file: section4.1.tex
\section{Candidates for the exotic heavy hadrons}
\label{sec4}

As reviewed in Sec.~\ref{sec3.2}, the theoretical predictions of the
masses of the charmed-strange mesons in the $P$-wave $(0^+, 1^+)$
doublet are around 2.48 GeV and 2.55
GeV~\cite{Godfrey:1985xj,Godfrey:2015dva,Ebert:2009ua,Song:2015nia,Godfrey:1986wj}.
These two values are significantly larger than the experimental
masses of the $D_{s0}^\ast(2317)$ of $J^P=0^+$ and the
$D_{s1}(2460)$ of $J^P=1^+$. This puzzle has stimulated theorists'
extensive interests in exploring their inner structures. Various
exotic assignments were proposed. In this section we review these
efforts. We shall also review the theoretical studies on the
$X(5568)$~\cite{D0:2016mwd}, which consists of four different quarks
$su\bar b\bar d$ (or $sd\bar b\bar u$).

\subsection{The $D_{s0}^*(2317)$ and $D_{s1}(2460)$}
\label{sec4.1}

\subsubsection{Molecular scheme.}

The low mass puzzle of the $D_{s0}^\ast(2317)$ and $D_{s1}(2460)$
inspired various exotic explanations. Among them, the $D^{(*)}K$
molecule interpretation is quite popular. We note that both the
molecular interpretation and the $D^{(*)}K$ couple channel effect
arise from the strong S-wave $D^{(*)}K$ interaction.

The $D^{(*)}K$ molecular interpretation was first proposed in
Ref.~\cite{Barnes:2003dj}, where Barnes, Close and Lipkin found that
a dominantly $I=0$ $DK$ molecular state with some $I=1$ admixture
could explain both the narrow total width of the $D_{s0}^\ast(2317)$
as well as its observed decay to $D^+_s \pi^0$. Later in
Ref.~\cite{Chen:2004dy}, Chen and Li proposed a simple unitive
picture that the $D_{s0}^\ast(2317)$ is a $DK$ molecular state, and
the $D_{s1}(2460)$ is a $D^*K$ molecular state.

Based on the heavy chiral unitary approach, the $S$-wave interaction
between the pseudoscalar heavy meson and the Goldstone boson was
studied in Ref.~\cite{Guo:2006fu} by Guo {\it et al.}. They found a
pole in the charmed sector about $2.312 \pm 0.041$ GeV, which was
interpreted as a $0^+$ $DK$ bound state and regarded as the
$D_{s0}^\ast(2317)$. Besides this, they also predicted a $B \bar K$
bound state $B^*_{s0}$ at about $5.725 \pm 0.039$ GeV. Later in
Ref.~\cite{Guo:2006rp}, Guo {\it et al.} used the same approach to
study the $S$-wave interactions between heavy vector meson and
light pseudoscalar meson. They found a $D^*K$ bound state with a
mass of $2.462 \pm 0.010$ GeV, which was associated with the
$D_{s1}(2460)$. They also predicted a $B^* \bar K$ bound state
($B_{s1}$) with the mass of $5.778 \pm 0.007$ GeV in the bottom
sector.

\begin{table}
\centering
  \caption{Branching fraction of the process $\bar{B}_{s}^{0} \to D_{s0}^{\ast}(2317)^{+} \bar{\nu}_{l} l^{-}$ (to the total decay width of the $\bar{B}_{s}^{0}$) in percentage. Taken from Ref.~\cite{Navarra:2015iea}.}
  \label{sec4:production}
  \begin{tabular*}{8.6cm}{@{\extracolsep{\fill}}lc}
    \hline \hline
     Approach  &   $\mathcal{B}[\bar{B}_{s}^{0} \to D_{s0}^{\ast}(2317)^{+} \bar{\nu}_{l} l^{-}]$ \\
     \hline
     CUM \cite{Navarra:2015iea} &     $0.13$ \\
     QCDSR + HQET \cite{Huang:2004et} &  $0.09 - 0.20$ \\
     QCDSR (SVZ)  \cite{Aliev:2006qy} &  $0.10$  \\
     LCSR         \cite{Li:2009wq}    &  $0.23 \pm 0.11$\\
     CQM          \cite{Zhao:2006at}  &  $0.49 - 0.57$ \\
     CQM          \cite{Segovia:2011dg}   &  $0.44$ \\
     CQM          \cite{Albertus:2014bfa}   &  $0.39$ \\
     \hline \hline
  \end{tabular*}
\end{table}

\begin{figure}
\centering
\includegraphics[width=0.6\textwidth]{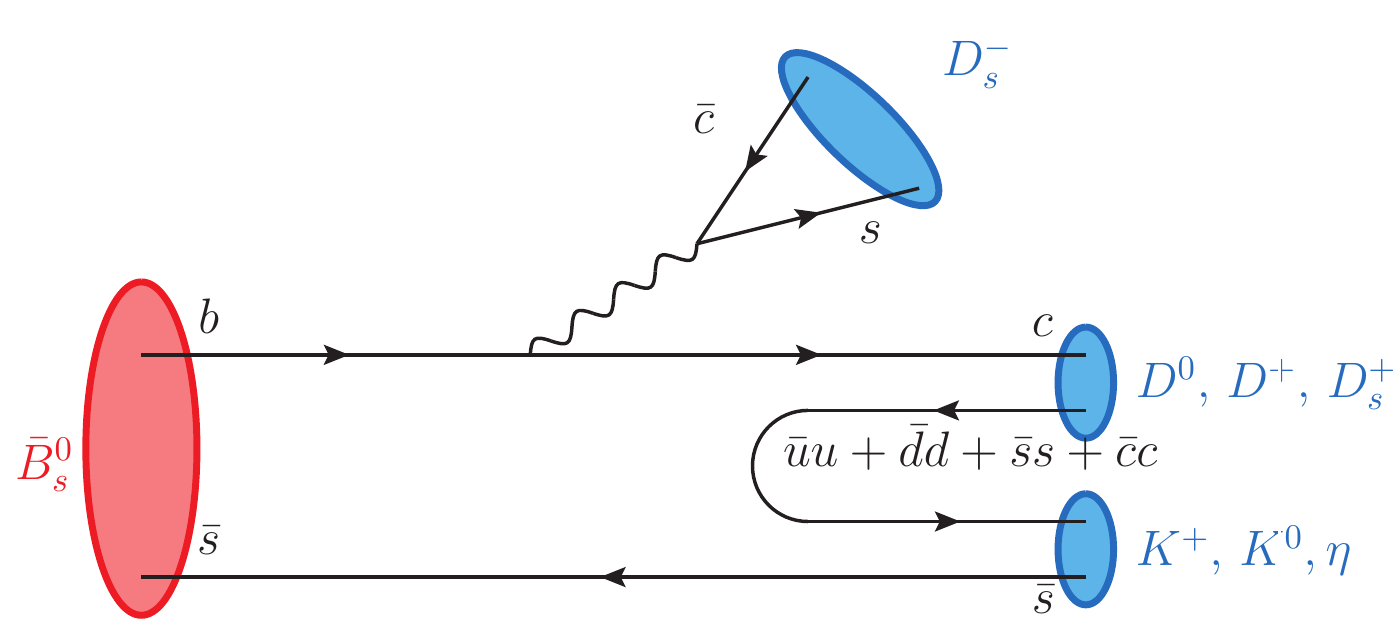}
\caption{(Color online) Mechanism for the $\bar{B}^0_s$ decay into $D_s^-(DK)^+$.
Taken from Ref.~\cite{Albaladejo:2015kea}. \label{sec4:mechanism}}
\end{figure}

\begin{table*}[htb]
\scriptsize
\centering \caption{\label{sec4:radiative} Numerical results for the
radiative decay widths of the $D_{s0}^\ast(2317)$ and $D_{s1}(2460)$
in keV. The first~\cite{Cleven:2014oka} and
fifth~\cite{Faessler:2007gv,Faessler:2007us,Faessler:2008vc} columns
are obtained in the molecule picture. The
second~\cite{Bardeen:2003kt} and third~\cite{Colangelo:2005hv}
columns are obtained when the $D_{s0}^\ast(2317)$ and $D_{s1}(2460)$
are regarded as the conventional charmed-strange mesons. The fourth
column~\cite{Lutz:2007sk} is obtained when the $D_{s0}^\ast(2317)$
and $D_{s1}(2460)$ are generated by coupled-channel dynamics. Taken
from Ref.~\cite{Cleven:2014oka}.
  \medskip}
\renewcommand{\arraystretch}{1.6}
\begin{tabular}{|l|c|c|c|c|c|}
\hline\hline Decay Channel               &
Cleven~\cite{Cleven:2014oka}  & Bardeen~\cite{Bardeen:2003kt} &
Colangelo~\cite{Colangelo:2005hv}    & Lutz~\cite{Lutz:2007sk}   &
Faessler~\cite{Faessler:2007gv,Faessler:2007us,Faessler:2008vc}\\
\hline\hline $\dszero\to D_s^*\gamma$    & $(9.4\pm3.8)$ & $1.74$
& $4-6$             & $1.94(6.47) $     &   0.55-1.41           \\
\hline $\dsone\to D_s\gamma$       & $(24.2\pm10.7)$   & $5.08$
& $19-29 $          & $44.50 (45.14) $  &   2.37-3.73           \\
\hline $\dsone\to D^*_s\gamma$     & $(25.2\pm9.7)$    & $4.66$
& $0.6-1.1 $          & $21.8 (12.47)$    &   --              \\
\hline $\dsone\to \dszero \gamma$  & $(1.3\pm 1.3)$    & $2.74$
& $0.5-0.8$           & $0.13 (0.59)$     &   --              \\
\hline\hline $B_{s0}\to B_s^*\gamma$     & $(32.6\pm20.8)$   &
$58.3$                & --                & --            &
3.07-4.06           \\ \hline $B_{s1}\to B_s\gamma$       &
$(4.1\pm10.9)$    & $39.1$                & --                & -- &
2.01-2.67           \\ \hline $B_{s1}\to B_s^*\gamma$     &
$(46.9\pm33.6)$   & $56.9$                & --
        & --            &   --              \\ \hline
$B_{s1}\to B_{s0}\gamma$    & $(0.02\pm0.02)$   & $0.0061$ & --
        & --            &   --              \\ \hline\hline
\end{tabular}
\end{table*}

Similar results were obtained in Ref.~\cite{Gamermann:2006nm} by
Gamermann {\it et al.}, where they investigated dynamical generation
of the open and hidden charm mesons in a unitarized coupled channel
framework. Their results suggested the $D_{s0}^\ast(2317)$ to be
mainly a $DK$ bound state with no decay modes, except for a tiny one
when allowing isospin violation. More discussions can be found in
Refs.~\cite{Gamermann:2007fi,Gamermann:2007bm,Gamermann:2010ga}.
Later in Ref.~\cite{Navarra:2015iea} the semileptonic $B_s$ and $B$
decays into the $D_{s0}^\ast(2317)$ and $D_{s1}(2460)$ were studied
using a chiral unitarity model in coupled channels, and their
results are shown in Table~\ref{sec4:production}. The $\bar B^0_s
\to D^-_s (DK)^+$ and $B_c \to J/\psi D K$ weak decays, etc. were
similarly investigated in
Refs.~\cite{Albaladejo:2015kea,Sun:2015uva,Albaladejo:2016hae},
where the $D_{s0}^\ast(2317)$ contributes. We show one of the
mechanisms in Fig.~\ref{sec4:mechanism}. In Ref.~\cite{Lutz:2007sk}
Lutz and Soyeur investigated masses and decays of the
$D_{s0}^\ast(2317)$ and $D_{s1}(2460)$. In their studies the
$D_{s0}^\ast(2317)$ and $D_{s1}(2460)$ are also generated by
coupled-channel dynamics, and their radiative decay width are listed
in Table~\ref{sec4:radiative}.

The $S$-wave scattering lengths of the Goldstone boson and heavy
pseudoscalar meson were also systematically studied in
Ref.~\cite{Liu:2009uz}. They found that the $DK$
scattering length is positive, so their interaction is attractive.
However, they also suggested that further exploration of the phase
shifts of the elastic $DK$ scattering was still required in order to
answer whether the $DK$ interaction is strong enough to form a bound
$DK$ molecular state.
The scattering of light-pseudoscalar mesons off charmed and charm-strange mesons
was also studied in Ref.~\cite{Guo:2015dha} by Guo, Meissner and Yao, where
they investigated the $D_{s0}^\ast(2317)$ in a unitarized chiral effective field theory approach.
They analyzed the
light-quark mass and $N_C$ dependence of its pole positions,
and found that the $D_{s0}^\ast(2317)$ pole does not tend to fall down to the real axis for large enough values of $N_C$,
indicating that it does not behave like a standard quark-antiquark meson at large $N_C$.

The opening of a new $S$-wave threshold is frequently accompanied by
an abrupt dip in the magnitude of an amplitude for an already-open
channel. Based on this fact, Rosner sought a unified description of
the underlying dynamics~\cite{Rosner:2006vc}, and suggested that the
$D_{s0}^\ast(2317)$ and $D_{s1}(2460)$ can be viewed as the bound
states of $DK$ and $D^* K$, or as $c \bar s$ states with masses
lowered by the coupling to the $DK$ and $D^* K$ channels,
respectively.

The $D_{s0}^\ast(2317)$ and $D_{s1}(2460)$ have been studied as the
$D^{(*)}K$ molecules in many other models. In
Ref.~\cite{Xie:2010zza} Xie, Feng and Guo used the Bethe-Salpeter
approach and found a bound state of $DK$ which was identified as the
$D_{s0}^\ast(2317)$. A similar approach was used in
Ref.~\cite{Feng:2012zze} where the bound state of $D^*K$ was
identified as the $D_{s1}(2460)$. The scattering amplitude of the
Goldstone bosons off the pseudoscalar $D$-mesons was studied in
Ref.~\cite{Wang:2012bu} in the unitarized heavy meson chiral
approach, where Wang and Wang obtained the $D_{s0}^\ast(2317)$ as a
$DK$ bound state in $(S, I) = (1, 0)$ channel.

Recently in Ref.~\cite{Ortega:2016mms}, Ortega {\it et al.}
performed a coupled-channel computation taking into account the
$D_{s0}^\ast(2317)$, $D_{s1}(2460)$ and $D_{s1}(2536)$ mesons and
the $DK$ and $D^*K$ thresholds within the framework of a constituent
quark model. They obtained a probability of 34\% for the $DK$
component in the $D_{s0}^\ast(2317)$ wave function, and observed
that the meson-meson component is around 50\% for both the
$D_{s1}(2460)$ and $D_{s1}(2536)$ mesons. This method was
recently extended to study the $P$-wave bottom-strange mesons
in Ref.~\cite{Ortega:2016pgg}.

However, the interpretations of the $D_{s0}^\ast(2317)$ and
$D_{s1}(2460)$ as the $DK$ and $D^* K$ molecular states are not
supported in Ref.~\cite{Zhang:2009pn}, where the chiral $SU(3)$
quark model was used. The molecular proposal for the
$D_{s0}^\ast(2317)$ and $D_{s1}(2460)$ was also tested in the
heavy-hadron chiral perturbation theory in Ref. \cite{Mehen:2004uj}.
Their leading order predictions for the electromagnetic branching
ratios are in very poor agreement with the available data, which
disfavored the molecular interpretations for the $D_{s0}^\ast(2317)$
and $D_{s1}(2460)$ mesons.

Assuming the $D_{s0}^\ast(2317)$ and $D_{s1}(2460)$ as the $0^+$
$DK$ and $1^+$ $D^\ast K$ molecular states respectively, their
productions and decay behaviors have been investigated in many
papers~\cite{Cleven:2014oka,Faessler:2007gv,Faessler:2007us,Faessler:2008vc,Faessler:2007cu,Chen:2003jp,Guo:2008gp,Guo:2014ppa,Xiao:2016hoa}.
In Refs.~\cite{Faessler:2007gv}, Faessler {\it et al.} considered
the $D_{s0}^\ast(2317)$ as a $0^+$ $DK$ bound state and calculated
the strong $D^*_{s0} \to D_s \pi^0$ and radiative $D^*_{s0} \to
D^*_s \gamma$ decays using an effective Lagrangian approach. They
evaluated the ratio $R = \Gamma(D^*_{s0} \to D^*_s \gamma) /
\Gamma(D^*_{s0} \to D_s \pi) \sim 10^{-2}$, which is consistent with
the experimental upper limit of $R < 0.059$~\cite{Besson:2003cp}.
Using the same method, they also studied the strong $D_{s1} \to
D_s^* \pi^0$ and radiative $D_{s1} \to D_s \gamma$ decays for the
$D_{s1}(2460)$ meson \cite{Faessler:2007us}. In Ref.
\cite{Faessler:2007cu}, the same authors further analyzed the
branching ratios of $B\to D^{(\ast)}D_{s0}^\ast(D_{s1})$ decays and
calculated the leptonic decay constants $f_{D_{s0}^\ast}$ and
$f_{D_{s1}}$ using the factorization hypothesis.  Their results are
collected in Table~\ref{sec4:radiative} for comparisons. More
discussions for the $D_{s0}^\ast(2317)$ and $D_{s1}(2460)$ decay
properties as the hadron molecules can be found in
Refs.~\cite{Close:2005se}.

\begin{table}[htb]
\centering \caption{\label{sec4:guo} Integrated normalized cross
sections for the inclusive processes $pp \to
D_{s0}^{*}(2317),\,\,D_{s1}(2460),\,\,D_{sJ}(2860)$ and
$D_{s2}(2910)$ at LHC, in units of $\mu$b. The results inside and outside brackets are
obtained using Pythia and Herwig, respectively. Taken from Ref.~\cite{Guo:2014ppa}.}
\begin{tabular}{|c|cccc|} \hline\hline
& $D_{s0}^{*}(2317)$ & $D_{s1}(2460)$ & $D_{sJ}(2860)$ &
$D_{s2}(2910)$ \\\hline
LHC 7 &2.5(0.83) &2.1(0.91) &0.21(-) &0.27(-)\\
LHCb 7 &0.61(0.15) &0.5(0.17) &0.05(-) &0.06(-)\\
LHC 8 &2.9(0.94) &2.4(1.0) &0.24(-) &0.32(-)\\
LHCb 8 &0.74(0.18) &0.61(0.2) &0.06(-) &0.08(-)\\
LHC 14 &5.5(1.6) &4.7(1.7) &0.5(-) &0.65(-)\\
LHCb 14 &1.6(0.35) &1.3(0.38) &0.13(-) &0.17(-)\\
 \hline\hline
\end{tabular}
\end{table}

Considering the $D_{s0}^\ast(2317)$ and $D_{s1}(2460)$ as $DK$ and
$D^\ast K$ hadronic molecules, the partial widths for the radiative
and pionic transitions for the $D_{s1}(2460) \to D_{s0}^\ast(2317)
\pi^0$ and $D_{s1}(2460) \to D_{s0}^\ast(2317) \gamma$ were
evaluated to be about $0.19\mbox{--}0.22$ keV and $3.0\mbox{--}3.1$
keV respectively in Ref. \cite{Xiao:2016hoa}. In addition, they also
estimated the partial width ratio between the $D_{s1}(2460) \to
D_{s0}(2317) \gamma$ and $D_{s1}(2460) \to D_{s}^\ast \pi^0$ decay
modes to be about $(6.6\mbox{--}10.6) \times 10^{-2}$.

In Ref.~\cite{Guo:2008gp}, Guo {\it et al.} constructed the
effective chiral Lagrangian involving the charmed mesons and
Goldstone bosons at the next-to-leading order taking into account
the strong as well as electromagnetic interactions. They evaluated
the decay width $\Gamma(D^*_{s0}(2317) \to D_s \pi^0)$ to be $180
\pm 110$ keV, consistent with the experimental results. A more
systematical study on hadronic and radiative decays of the
$D_{s0}^\ast(2317)$ and $D_{s1}(2460)$ can be found in
Ref.~\cite{Cleven:2014oka}, whose results were in fair agreement with
available data, as shown in Table~\ref{sec4:radiative}. In
Ref.~\cite{Guo:2014ppa} Guo {\it et al.} studied the inclusive
hadroproduction of the $D_{s0}^\ast(2317)$ and $D_{s1}(2460)$ at the
Large Hadron Collider using effective field theory. Their results
are shown in Table~\ref{sec4:guo}.

The $D_{s0}^\ast(2317)$ and $D_{s1}(2460)$ may have many molecular
partner states. For example, its analog for the $bc$ system is a
$BD$ molecule, which was discussed in Ref.~\cite{Lipkin:2003zk}. The
analogous states of the $D_{s0}^\ast(2317)$ and $D_{s1}(2460)$ with
a bottom quark were predicted in Ref.~\cite{Cleven:2010aw} to be
$M_{B^*_{s0}} = 5696 \pm 40$ MeV and $M_{B_{s1}} = 5742 \pm 40$ MeV,
respectively.

\subsubsection{Tetraquark scheme.}

The $D_{s0}^\ast(2317)$ and $D_{s1}(2460)$ were also interpreted as
the $c q \bar s \bar q$ tetraquark states. This scheme was first
investigated in Ref.~\cite{Cheng:2003kg}, where Cheng and Hou
discussed the masses and decay modes of the $c q \bar q \bar q$
($q=u,d,s$) tetraquark states, and found that the isosinglet
$D_{s0}^*(2317)$ is the only narrow one. Its decay was dominated by
the observed isospin violating decay mode and its width was less
than 100 keV. Later in Ref.~\cite{Browder:2003fk} Browder {\it et
al.} suggested that the $D_{s0}^*(2317)$ and $D_{s1}(2460)$ can be
explained by the mixing of the conventional $P$-wave excited $D^+_s$
mesons with the four-quark states.

The influence of the $^\prime$t Hooft interaction on the $cq \bar q
\bar q$ tetraquark mass spectrum was discussed in
Ref.~\cite{Dmitrasinovic:2004cu} by Dmitrasinovic, where the
$D_{s0}^*(2317)$ and $D_{s1}(2460)$ were identified as the
tetraquark candidates. Similar results and conclusions were obtained
in Ref.~\cite{Dmitrasinovic:2012zz} using a relativistic effective
chiral model. In Ref.~\cite{Dmitrasinovic:2005gc} Dmitrasinovic
further argued that the anomalously small mass difference between
the $D_{s0}^\ast(2317)$ and the $D_0^*(2400)$ (with the mass 2318
MeV~\cite{Abe:2003zm}) suggests that they both have a tetraquark
structure.


In Ref.~\cite{Maiani:2004vq} Maiani {\it et al.} proposed their
diquark-antidiquark model and calculated the mass spectrum of
$[cq][\bar s \bar q]$ states. 
They used both ``good'' diquark of spin $S=0$ and ``bad'' diquark of spin $S=1$~\cite{Jaffe:2004ph}
to construct tetraquark states, and found that there are two states with $J^{P} = 0^{+}$,
three states with $J^P = 1^+$ and one state with $J^{P} = 2^{+}$:
\begin{eqnarray}
\nonumber | 0^{+} \rangle &\equiv& | S_{cq}, S_{\bar s \bar q^\prime} ; J_{[cq][\bar s \bar q^\prime]} \rangle = |0_{cq}, 0_{\bar s \bar q^\prime} ; J = 0 \rangle \, ,
\\ \nonumber
| 0^{+\prime} \rangle &=& |1_{cq}, 1_{\bar s \bar q^\prime} ; J = 0 \rangle \, ,
\\
| 1^{+} \rangle &=& {1\over\sqrt2} \left( |0_{cq}, 1_{\bar s \bar q^\prime} ; J = 1 \rangle + |1_{cq}, 0_{\bar s \bar q^\prime} ; J = 1 \rangle \right) \, ,
\\ \nonumber
| 1^{+\prime} \rangle &=& {1\over\sqrt2} \left( |0_{cq}, 1_{\bar s \bar q^\prime} ; J = 1 \rangle - |1_{cq}, 0_{\bar s \bar q^\prime} ; J = 1 \rangle \right) \, ,
\\ \nonumber
| 1^{+\prime\prime} \rangle &=& |1_{cq}, 1_{\bar s \bar q^\prime} ; J = 1 \rangle \, ,
\\ \nonumber
| 2^{+} \rangle &=& |1_{cq}, 1_{\bar s \bar q^\prime} ; J = 2 \rangle \, .
\end{eqnarray}
They further evaluated the masses of the two states with $J^{P} = 0^{+}$ to be 2371 MeV and 2424 MeV,
the masses of the three states with $J^{P} = 1^{+}$ to be 2410 MeV, 2462 MeV and 2571 MeV,
and the masses of the one state with $J^{P} = 2^{+}$ to be 2648 MeV.
Accordingly, they associated the $D_{s0}^\ast(2317)$ and $D_{s1}(2460)$ with the
lowest-lying $0^+$ state and one of the $1^+$ states,
respectively, and at the same time predicted more $[cq][\bar s \bar q]$ states.

The $D_{s0}^\ast(2317)$ was also investigated by Bracco {\it et al.}
using the QCD sum rule approach~\cite{Bracco:2005kt}. They found
that its mass can be reproduced by the four-quark states $(cq)(\bar
q\bar s)$. Similar results were obtained in
Refs.~\cite{Kim:2005gt,Wang:2006uba} using the same approach.

However, the interpretations of the $D_{s0}^\ast(2317)$ and
$D_{s1}(2460)$ as the pure four-quark states are not supported in
Ref.~\cite{Zhang:2006hv} where the chiral $SU(3)$ quark model was
used.

Assuming the $D_{s0}^\ast(2317)$ and $D_{s1}(2460)$ to be the $c q
\bar s \bar q$ tetraquark states of $J^P=0^+$ and $1^+$, their decay
behaviors were studied in various models~\cite{Terasaki:2003qa}. In
Ref.~\cite{Chen:2003jp}, Chen and Li investigated the
$D_{s0}^\ast(2317)$ in $B$ meson decays, and found the ratio
$\mathcal{B}(B \to D_{s0}^\ast(2317) M)/\mathcal{B}(B \to D^{(*)}_s
M)$ ($M = D$, $\pi$ and $K$) to be around either 1 or 0.1,
supporting the $D_{s0}^\ast(2317)$ to be either a $c \bar s$ state
or a tetraquark state, respectively. In
Ref.~\cite{Hayashigaki:2004st} Hayashigaki and Terasaki calculated
the $D^+_s \pi^0$ and $D^{*+}_s \gamma$ decays of the
$D_{s0}^\ast(2317)$ to be around 0.6 keV and 35 keV, respectively.
Their results suggested that its assignment as an iso-triplet
four-quark meson is favored by the severest experimental constraint
on the ratio of the rates for these decays, while assigning it as an
$I = 0$ state (a four-quark or a conventional $c \bar s$) is
inconsistent with this constraint. The partial decay width of
$D_{s0}^\ast(2317) \to D_s^+ \pi^0$ was also calculated in
Ref.~\cite{Nielsen:2005zr} to be in the range of $0.2 \sim 40$ keV.
All these results can be useful for further studies on the
$D_{s0}^\ast(2317)$.

As a $c q \bar s \bar q$ tetraquark state, the production of the
neutral and doubly charged partners of the $D_{s0}^\ast(2317)$ were
studied by Terasaki in
Refs.~\cite{Terasaki:2006qd,Terasaki:2016zbt}, which can also be
useful to verify the tetraquark scheme.

\subsubsection{Conventional charmed-strange mesons with coupled-channel effects.}

Because the $D_{s0}^\ast(2317)$ and $D_{s1}(2460)$ are close to the
thresholds of the $DK$ and $D^\ast K$, respectively, the
coupled-channel effects should be important. In fact, this mechanism
is very probably responsible for their low mass puzzle. These states
can still be categorized into the conventional charmed-strange meson
family, which has been discussed by many theoretical groups.

\begin{figure}
\begin{center}
\includegraphics[width=0.48\textwidth]{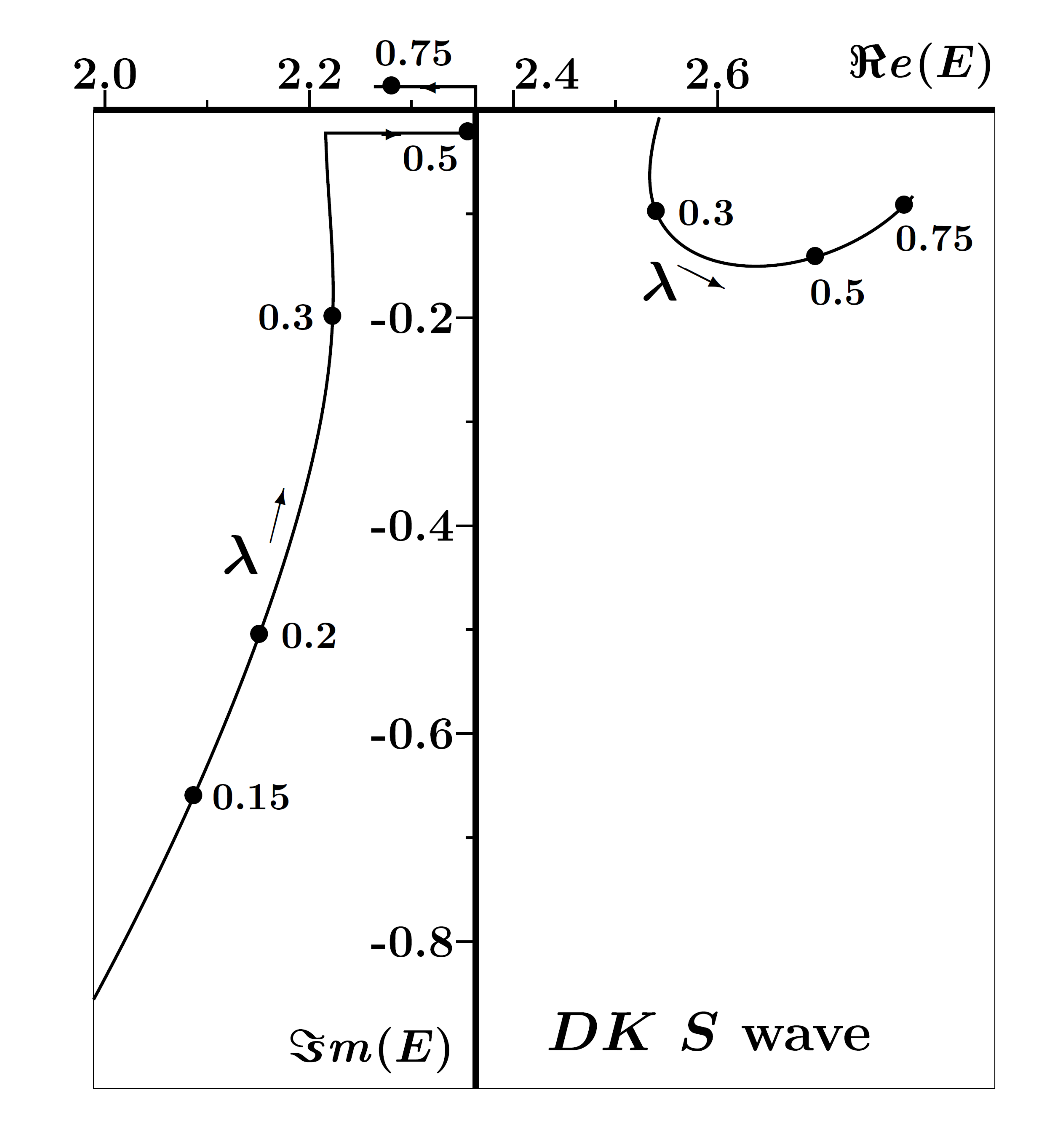}
\end{center}
\caption{The S-matrix poles for the $DK$ $S$-wave scattering as a
function of the coupling constant $\lambda$. The virtual bound
states are displaced slightly downwards, and the real bound states
upwards. Notice that for $\lambda = 0.75$ (physical value) one has a
real bound state. Taken from Ref.~\cite{vanBeveren:2003kd}. }
\label{sec4:couple}
\end{figure}

In Ref.~\cite{vanBeveren:2003kd} Beveren and Rupp described the
$D_{s0}^\ast(2317)$ as a quasi-bound scalar $c \bar s$ state in a
unitarized meson model, and demonstrated how a low-mass scalar
charmed-strange meson can be easily obtained by including its
coupling to the OZI-allowed $DK$ channel. Their obtained S-matrix
poles for the $S$-wave $DK$ scattering are shown in
Fig.~\ref{sec4:couple} as a function of the coupling constant
$\lambda$. Similarly in Ref. \cite{vanBeveren:2003jv}, the same
authors continued to describe the $D_{s1}(2460)$ to be a $J^P=1^+$
$c\bar s$ state, by considering the coupling to the $S$-wave $D^\ast
K$ channel. In Ref. \cite{Coito:2011qn}, the authors further
calculated the mass and width of the $D_{s1}(2460)$ meson in the
Resonance-Spectrum-Expansion model through the coupling of the
$D^\ast K$ channel to the bare $1^+$ $c\bar s$ state.

Later in Ref.~\cite{Hwang:2004cd,Hwang:2005tm}, Hwang and Kim
calculated the mass shift of the $D_s(1^3P_0)$ state by using the
coupled channel effect, and their result suggested that the coupled
channel effect naturally explains the observed mass of the
$D_{s0}^\ast(2317)$. The same result was obtained using the chiral
Lagrangian~\cite{Simonov:2004ar}, by considering the one loop chiral
corrections~\cite{Lee:2004gt}, and by including hadronic loops under
the assumption that these corrections vanish for the ground state
heavy-light mesons~\cite{Guo:2007up}. See also discussions in Ref.~\cite{Zhou:2011sp}.

The mass shifts of the $P$-wave charmed-strange mesons due to their
coupling to the $DK$ and $D^*K$ channels were also studied in
Refs.~\cite{Badalian:2007yr,Badalian:2008zz} by Badalian, Simonov
and Trusov using the chiral quark-pion Lagrangian. They found strong
mass shifts downward about 140 MeV and 100 MeV for the $D^*_s(0^+)$
and $D_s(1^+)$. The two essential factors for these large mass
shifts are: strong coupling of the $0^+$ and $1^+$ states to the
$S$-wave decay channel containing a Nambu-Goldstone meson, and the
chiral flip transitions due to the bi-spinor structure of both
heavy-light mesons. They also predicted masses of the $B^*_s(0^+)$
and $B_s(1^+)$ to be $5695 \pm 10$ MeV and $5730 \pm 15$ MeV,
respectively.

Besides the couple-channel effect, the $D_{s0}^\ast(2317)$ and
$D_{s1}(2460)$ were sometimes investigated together with the
spontaneous breaking of chiral symmetry, which is another possible
reason for its low mass. In Ref.~\cite{Bardeen:2003kt} Bardeen,
Eichten and Hill pointed out that the $(0^+, 1^+)$ spin multiplet was
required in the implementation of $SU(3)_L \times SU(3)_R$ chiral
symmetry in heavy-light meson systems. In Ref.~\cite{Nowak:2003ra}
Nowak, Rho and Zahed pointed out that the $D_{s0}^\ast(2317)$ and
$D_{s1}(2460)$ are consistent with the general pattern of
spontaneous breaking of chiral symmetry in hadrons built of heavy
and light quarks. In Ref.~\cite{Kolomeitsev:2003ac}, Kolomeitsev and
Lutz studied heavy-light meson resonances with quantum numbers $J^P
= 0^+$ and $1^+$ in terms of the non-linear chiral $SU(3)$
Lagrangian where the $D_{s0}^\ast(2317)$ and $D_{s1}(2460)$ can be
reproduced. More discussions can be found in
Refs.~\cite{Nowak:1992um,Bardeen:1993ae,Harada:2003kt,Becirevic:2004uv,Wu:2014era}
especially for the mass splitting of the $(0^+, 1^+)$ doublet.

The $D_{s0}^\ast(2317)$ and $D_{s1}(2460)$ have been studied as
conventional charmed-strange mesons of $J^P=0^+$ and $1^+$ in many
other models. In Ref.~\cite{Dai:2003yg} Dai {\it et al.} calculated
masses of the $(0^+, 1^+)$ and $(1^+, 2^+)$ excited charmed-strange
states using QCD sum rules in the framework of heavy quark effective
theory. Their results suggested that the $D_{s0}^\ast(2317)$ and
$D_{s1}(2460)$ can be interpreted as the $0^+$ and $1^+$ states in
the $(0^+,1^+)$ doublet. This work was developed in
Ref.~\cite{Dai:2006uz}, where Dai {\it et al.} further considered
the contribution of the $DK$ continuum in the formalism of QCD sum
rule, which was found to be significant and largely the reason for
the unexpected low mass of the $D_{s0}^\ast(2317)$. In
Ref.~\cite{Sadzikowski:2003jy}, Sadzikowski calculated the masses of
the $D_{s0}^\ast(2317)$ and $D_{s1}(2460)$ as the charmed-strange
mesons in the MIT bag model, and his results were in a reasonable
agreement with the experimental values. In
Ref.~\cite{Lakhina:2006fy} Lakhina and Swanson studied the quark
mass dependence induced by one loop corrections to the Breit-Fermi
spin-dependent one-gluon-exchange potential and determined the
masses of heavy-light mesons. Their results also suggested that the
$D_{s0}^\ast(2317)$ is a canonical charmed-strange meson. The
$D_{s0}^\ast(2317)$ and $D_{s1}(2460)$ were also studied as
charmed-strange mesons using a potential
model~\cite{Radford:2009bs,Fayyazuddin:2003aa,Liu:2015uya}, within a
covariant light-front approach~\cite{Cheng:2003sm} and with Regge
trajectories~\cite{Zhang:2004cd}, etc.

\renewcommand{\arraystretch}{1.4}
\begin{table*}[htbp]
\footnotesize
\caption{Partial widths of $D_{s0}^*(2317)\to D_s \pi^0$ and
$D_{s1}(2460)\to D_s^*\pi^0$, in unit of keV. Taken from
Ref.~\cite{Song:2015nia}.} \centering
\begin{tabular}{ c c c }\toprule[1pt]
Reference & $\Gamma(D_{s0}^\ast(2317)\rightarrow D_s \pi)$ &$\Gamma(D_{s1}(2460)\rightarrow D^\ast_s \pi)$ \\
\midrule[1pt]
Ref. \cite{Song:2015nia} &11.7& 11.9 \\
Ref. \cite{Lu:2006ry} &32& 35\\
Ref.\cite{Bardeen:2003kt} &21.5&21.5\\
Ref. \cite{Fayyazuddin:2003aa}&16& 32\\
Ref. \cite{Godfrey:2003kg}&$\sim$10&$\sim$10\\
Ref. \cite{Wei:2005ag}&34-44&35-51\\
Ref. \cite{Colangelo:2003vg}&$\simeq6$&$\simeq6$\\
Ref. \cite{Cheng:2003kg}&10-100&--\\
Ref. \cite{Ishida:2003gu} &$155\pm70$&$155\pm70$ \\
Ref. \cite{Matsuki:2011xp}&3.8& 3.9 \\
\bottomrule[1pt]
\end {tabular}
\label{table:singlepion}
\end{table*}

Assuming the $D_{s0}^\ast(2317)$ and $D_{s1}(2460)$ to be the
conventional charmed-strange mesons of $J^P=0^+$ and $1^+$, their
decay behaviors were studied in
Refs.~\cite{Song:2015nia,Bardeen:2003kt,Cheng:2003kg,Fayyazuddin:2003aa,Lu:2006ry,Godfrey:2003kg,Wei:2005ag,Colangelo:2003vg,Ishida:2003gu,Matsuki:2011xp,Wang:2006fg,Wang:2006bs,Wang:2006zw,Wang:2006mf,Fajfer:2006hi,Ke:2013zs,Fajfer:2015zma}
using various methods. The obtained partial widths of
$D_{s0}^*(2317)\to D_s \pi^0$ and $\Gamma(D_{s1}(2460)\rightarrow
D^\ast_s \pi)$ are listed in Table~\ref{table:singlepion}, which are
useful to distinguish various models and understand their underlying
structures.

The radiative decays of the $D_{s0}^\ast(2317)$ and $D_{s1}(2460)$
as the scalar charmed-strange mesons were studied in
Refs.~\cite{Mehen:2004uj,Azimov:2004xk}. Especially in
Ref.~\cite{Colangelo:2005hv}, Colangelo, Fazio and Ozpineci studied
these decays using the light-cone QCD sum rules. They obtained
\begin{eqnarray}
\nonumber \Gamma(D_{s0}^\ast(2317) \to D_s^* \gamma) &=& 4\sim6~{\rm keV} \, ,
\\ \Gamma(D_{s1}(2460) \to D_s \gamma) &=& 19\sim29~{\rm keV} \, ,
\\ \nonumber \Gamma(D_{s1}(2460) \to D_s^* \gamma) &=& 0.6\sim1.1~{\rm keV} \, ,
\\ \nonumber \Gamma(D_{s1}(2460) \to D_{s0}^\ast(2317) \gamma) &=& 0.5\sim0.8~{\rm keV} \, ,
\end{eqnarray}
which are consistent with the experimental values and favor the
interpretation of the $D_{s0}^\ast(2317)$ and $D_{s1}(2460)$ as the
ordinary charmed-strange mesons. These results are shown in
Table~\ref{sec4:radiative} for comparisons. Later in
Ref.~\cite{Wei:2005ag}, Wei, Huang and Zhu calculated their pionic
decay widths in the same framework, and obtained
\begin{eqnarray}
\Gamma(D_{s0}^\ast(2317) \to D_s \pi^0) &=& 34\sim44~{\rm keV} \, ,
\\ \nonumber \Gamma(D_{s1}(2460) \to D_s^* \pi^0) &=& 35\sim51~{\rm keV} \, ,
\end{eqnarray}
which are also consistent with the experimental values and support
their interpretation as the ordinary charmed-strange mesons. More
discussions can be found in Refs.~\cite{Lutz:2007sk,Liu:2006jx}.

More generally, the $D_{s1}(2460)$ can be regarded as a mixture of
the $1^1P_1$ and $1^3P_1$ charmed-strange mesons:
\begin{equation}
 \left(
  \begin{array}{c}
   |D_{s1}(2460)\rangle\\
   |D_{s1}(2536)\rangle\\
  \end{array}
\right )= \left(
  \begin{array}{cc}
    \cos\theta_{1P} & \sin\theta_{1P} \\
   -\sin\theta_{1P} & \cos\theta_{1P}\\
  \end{array}
\right) \left(
  \begin{array}{c}
    |1^1P_1  \rangle \\
   |1^3P_1 \rangle\\
  \end{array}
\right) \, .
\end{equation}
In Ref.~\cite{Song:2015nia}, Song {\it et al.} studied the decay
behaviors of the $D_{s1}(2460)$ systematically using the QPC model.
The single-pion decay $\Gamma(D_{s1}(2460)\rightarrow D^\ast_s \pi)$
depends on the mixing angle $\theta_{1P}$. When the mixing angle
takes the value in the heavy quark limit, i.e.,
$\theta_{1P}=-54.7^\circ$~\cite{Godfrey:1986wj,Matsuki:2010zy,Barnes:2005pb},
the partial width of $D_{s1}(2460)\to D_s^*\pi^0$ was calculated in
Ref.~\cite{Song:2015nia} using the QPC model and is listed in
Table~\ref{table:singlepion}, together with predictions by several
other theoretical
groups~\cite{Bardeen:2003kt,Cheng:2003kg,Fayyazuddin:2003aa,Lu:2006ry,Godfrey:2003kg,Wei:2005ag,Colangelo:2003vg,Ishida:2003gu,Matsuki:2011xp}.

Such a mixing mechanism was also studied via the $S$-wave
intermediate meson loops in Ref. \cite{Wu:2011yb}, where the
propagator matrix was established for this two-state system. The
masses and decay widths of the $D_{s1}(2460)$ and $D_{s1}(2536)$
were pinned down by searching for the pole structures in the
propagator matrix. For the $D_{s1}(2460)$, the pole was identified
at $\sqrt{s}=2454.5$ MeV and the mixing angle was
$\theta_{1P}=-42.5^\circ$.

The productions of the $D_{s0}^\ast(2317)$ and $D_{s1}(2460)$ are
also interesting, which were discussed in the $B$ meson
decays~\cite{Segovia:2011dg,Huang:2004et,Aliev:2006qy,Albertus:2014bfa,Faessler:2007cu,Chen:2003rt,Cheng:2006dm,Segovia:2012yh,Shen:2012mm,Bernlochner:2016bci},
in the $B_s$ decays~\cite{Li:2009wq,Zhao:2006at}, in the $B_c$
decays~\cite{Khosravi:2015tea}, in the $\Lambda_b$
decays~\cite{Datta:2003yk}, in the $\psi(4415)$
decays~\cite{Guo:2005cs}, and in the relativistic heavy ion
collisions (at RHIC)~\cite{Chen:2007zp}, etc. Some of these results
are listed in Table~\ref{sec4:production}.

In Ref. \cite{Segovia:2012yh}, the branching fractions of decays
$B\to D^{(\ast)}D_{sJ}^{(\ast)}$ have been investigated within the
framework of the constituent quark model and in the factorization
approximation. After introducing the finite $c$-quark mass effects,
the authors found that the $D_{s0}^\ast(2317)$ meson could be
described as a pure $c\bar s$ state while the $D_{s1}(2460)$ meson
may have a sizable non-$q\bar q$ component.

Considering the $D_{s1}(2460)$ meson as a $J^P=1^+$ charmed-strange
state, the semileptonic $B_c\to D_{s1}(2460)l^+\l^-$ ($l=\tau, \mu,
e$) and $B_c\to D_{s1}(2460)\nu\bar\nu$ transitions
\cite{Khosravi:2009eb} and the form factors relevant to the
semileptonic $B_{s}\rightarrow D_{sJ}(2460)\ell\nu$
\cite{Aliev:2006gk} were investigated using the three-point QCD sum
rule formalism.

\subsubsection{Lattice QCD simulation.}

Starting from 2003, there are many lattice QCD studies on the
$D_{s0}^\ast(2317)$ and $D_{s1}(2460)$
\cite{Lang:2014yfa,Mohler:2013rwa,Bali:2003jv,Dougall:2003hv,Herdoiza:2006qv,Flynn:2007ki,Liu:2012zya,Mohler:2012na,Moir:2013ub,Torres:2014vna,Kalinowski:2015bwa,Cichy:2016bci}.
Bali presented the lattice results on the scalar $D_s$ meson in the
static limit for the heavy quark in Ref.~\cite{Bali:2003jv}. He
calculated the scalar-pseudoscalar $0^+-0^-$ mass splitting of
$\Delta M=468(43)(24)$ MeV in this limit for $n_f=2$ sea quarks,
which was larger than the value of 338 MeV suggested by a heavy
quark constituent quark model \cite{Bardeen:1993ae}. A finite charm
quark mass correction was also reported, which seems to further
enlarge this discrepancy to support the non-$c\bar s$ interpretation
of the $D_{s1}(2460)$.

In Ref.~\cite{Dougall:2003hv}, the UKQCD Collaboration computed the
spectrum of the orbitally excited $D_s$ mesons in the continuum
limit. Their quenched simulations supported the interpretation of
the $D_{s0}^\ast(2317)$ resonance as a $J^P=0^+$ $c\bar s$ meson.
However, they can't exclude the exotic state possibility due to the
large errors in their calculations. Later, the same collaboration
has also computed the decay constants of $0^+$ P-wave heavy-light
mesons from unquenched lattice QCD at a single lattice spacing
\cite{Herdoiza:2006qv}. For the charm-strange meson, they obtained
the decay constant $f_{D_{s0^+}}=340(110)$ MeV and the
static-strange P-wave decay constant $f_{P_s}^{static}=302(39)$ MeV,
respectively.

In Ref.~\cite{Flynn:2007ki}, the authors analyzed the elastic S-wave
$B\pi, D\pi, DK$ and $K\pi$ scattering from lattice calculations of
the scalar form factors in the semileptonic decays. They extracted
the scattering lengths $m_\pi a=0.179(17)(14), 0.26(26)$ and
$0.29(4)$ for the elastic S-wave isospin-1/2 $K\pi, B\pi$ and $D\pi$
channels respectively. For the $DK$ channel, they found some hints
that there is a bound state which could be identified with the
$D_{s0}^\ast(2317)$ meson.

Liu  {\it et al.} studied the low-energy interactions between light
pseudoscalar mesons and charmed pseudoscalar mesons in
Ref.~\cite{Liu:2012zya}. They calculated the S-wave scattering
lengths of $D\bar K (I=0), D\bar K (I=1), D_sK, D\pi (I=3/2)$ and
$D_s\pi$ using L\"uscher's finite volume technique in full lattice
QCD. Among these channels, the interaction of the iso-scalar $D\bar
K$ is attractive while those of the others are repulsive. This
result supported the interpretation of the $D_{s0}^\ast(2317)$ as a
$DK$ molecule. They also updated a prediction for the isospin
breaking hadronic decay width $\Gamma(D_{s0}^\ast(2317)\to
D_s\pi)=133\pm22$ keV.

In Refs.~\cite{Lang:2014yfa,Mohler:2013rwa}, the authors considered
the $DK$, $D^\ast K$ and $\bar sc$ interpolating operators in the
lattice QCD simulations. They used two different ensembles of gauge
configurations with $N_f=2$ or $2+1$ dynamical fermions and
$m_\pi=266$ or $156$ MeV. A $J^P=0^+$ below threshold state was
established with a binding energy 37 (17) MeV, which was compatible
with the experiment value of 45 MeV for the $D_{s0}^\ast(2317)$
meson. For the $D_{s1}(2460)$, the $N_f=2+1$ simulation obtained a
$J^P=1^+$ strong interaction bound state 44 (10) MeV below the
$D^\ast K$ threshold, which was in agreement with the experiment and
thus identified with the $D_{s1}(2460)$. These obtained energy
levels were later reanalyzed in terms of an auxiliary potential,
employing a single-channel basis $KD^{(\ast)}$ and a two-channel
basis $KD^{(\ast)}, \eta D_s^{(\ast)}$ in
Ref.~\cite{Torres:2014vna}. They obtained similar binding energies
of about 40 MeV with respect to the $KD$ and $KD^\ast$ thresholds,
which were identified with the $D_{s0}^\ast(2317)$ and
$D_{s1}(2460)$ resonances.

\subsubsection{Short summary.}

Various exotic schemes including the molecular states and tetraquark
states have been proposed to explain the low mass puzzle of the
$D_{s0}^\ast(2317)$ and $D_{s1}(2460)$. Up to now, several
experiments scanned the charm-strange meson spectrum below 2.8 GeV
in the $D^{(*)}K^{(*)}$ channels carefully. So far, only four P-wave
states were found. None of the exotic schemes is able to answer
where the traditional $(0^+, 1^+)$ charm-strange mesons in the quark
model are if the $D_{s0}^\ast(2317)$ and $D_{s1}(2460)$ are exotic
states.

The confinement force is flavor independent to a large extent. If
the $D_{s0}^\ast(2317)$ and $D_{s1}(2460)$ are tetraquark
candidates, they should be accompanied by many partner states in the
$SU(3)_F$ multiplets $\mathbf{\bar 3} \otimes \mathbf{\bar 3} \otimes \mathbf{3}
= \mathbf{\bar 3} \oplus \mathbf{\bar 3} \oplus \mathbf{6} \oplus \mathbf{\overline{15}}$. In other
words, one would expect 25 additional tetraquark states below 2.8
GeV. Moreover, one would expect 27 bottomed tetraquark states if we
replace the charm by the bottom. But none of these states has been
observed experimentally up to now.

In short summary, the $D_{s0}^\ast(2317)$ and $D_{s1}(2460)$ are the
P-wave charm-strange mesons in the $(0^+, 1^+)$. The S-wave
$D^{(*)}K$ continuum couples strongly to the P-wave bare quark model
states. As a result, the quark model spectrum is strongly distorted.
The couple channel effects play a very important role in lowering
the quark model energy level. Such a feature is quite common when
the resonances lie around the threshold. For example, the S-wave
${\bar D}^0 D^{\ast 0}$ continuum couples strongly with the P-wave
axial vector charmonium $\chi_{c1}^\prime$. The famous narrow
resonance $X(3872)$ is a mixture of $\chi_{c1}^\prime$ and ${\bar D}
D^{\ast }$. The channel coupling between the S-wave $N{\bar K}$
continuum and P-wave $uds$ quark model state may lead to the
low-lying $\Lambda(1405)$.

%% file: section4.2.tex
\subsection{The $X(5568)$}
\label{sec4.2}

The narrow structure $X(5568)$ was reported by the D\O\, Collaboration
in the $B_s^0\pi^\pm$ invariant mass spectrum~\cite{D0:2016mwd}.
This  charged bottom meson, if it really exists, will be the first
candidate for the fully open-flavor tetraquark state consisting of
four different quarks $su\bar b\bar d$ (or $sd\bar b\bar u$). The D0
Collaboration reported that the spin-parity of $X(5568)$ could be
either $J^P=0^+$ or $1^+$. To date, the $X(5568)$ resonance has
trigged lots of theoretical studies, most of which speculated it to
be a compact diquark-antidiquark tetraquark state.

\begin{figure*}[hbt]
\begin{center}
\includegraphics[width=0.48\textwidth]{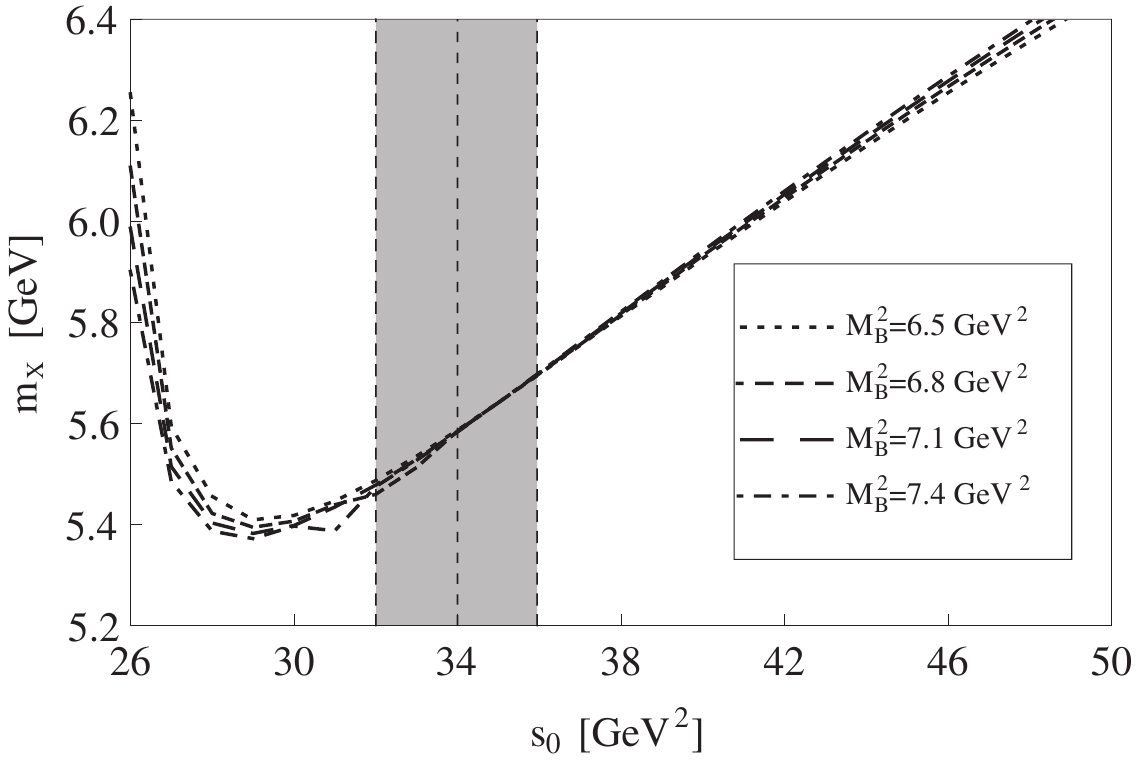}
\includegraphics[width=0.48\textwidth]{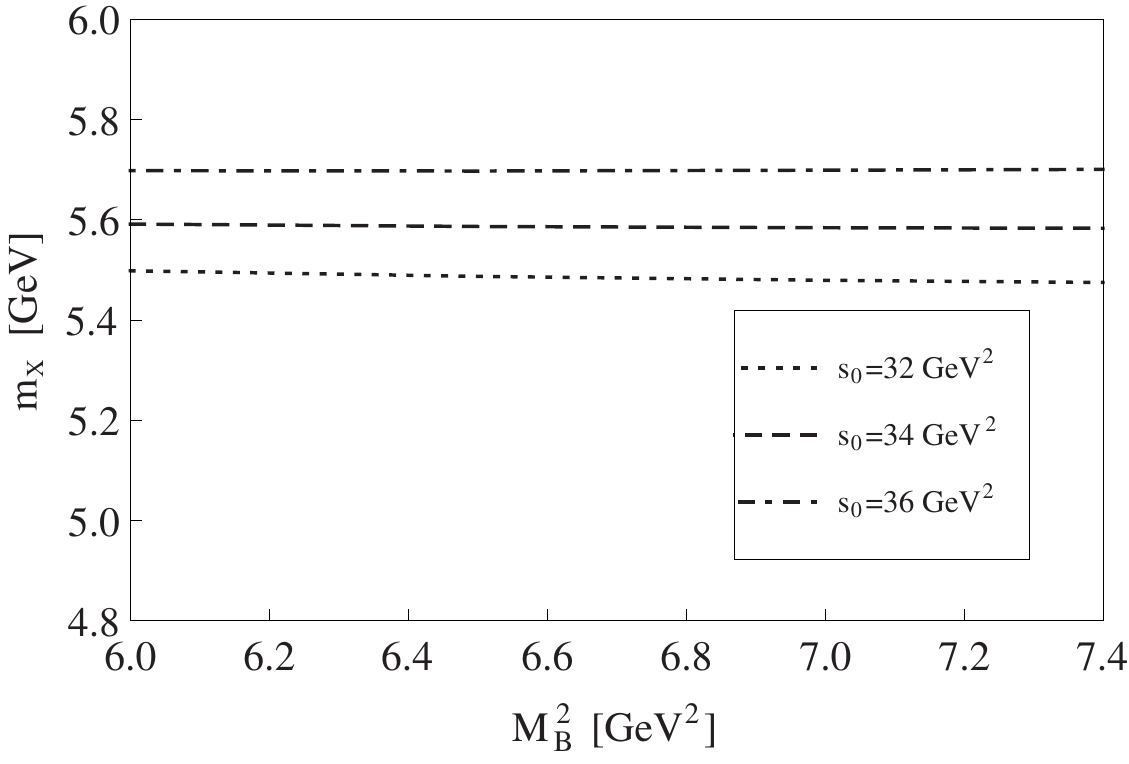}
\\
\includegraphics[width=0.48\textwidth]{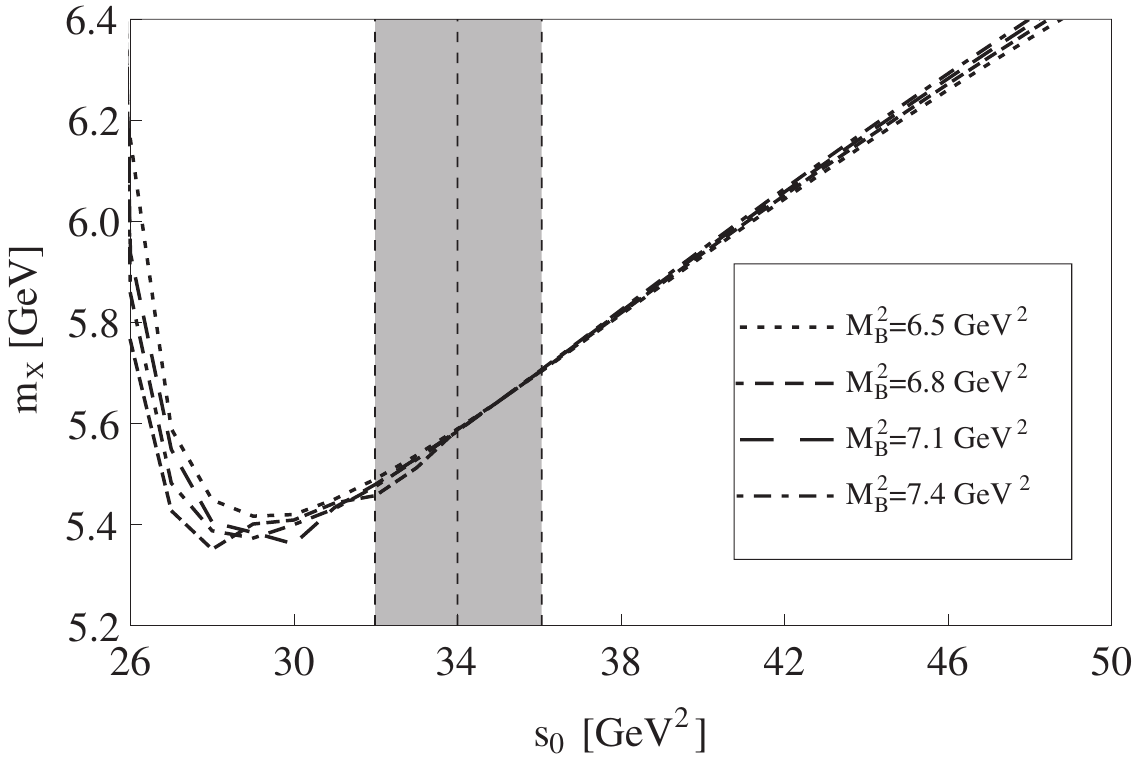}
\includegraphics[width=0.48\textwidth]{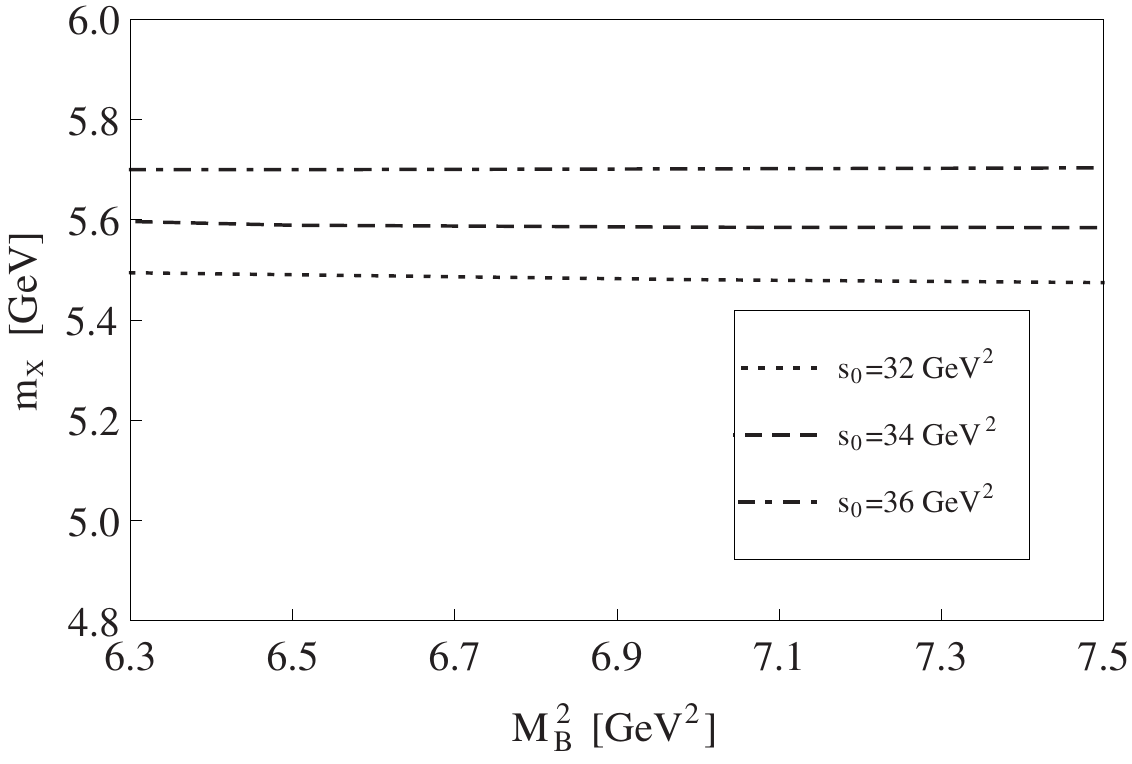}
\end{center}
\caption{Variations of the $X_b$ mass with $s_0$ and $M_B^2$ in the
scalar and axial-vector channels.
Taken from Ref.~\cite{Chen:2016mqt}.
} \label{sec4.1mass}
\end{figure*}

In Ref. \cite{Chen:2016mqt}, Chen {\it et. al.} studied the
$X(5568)$ meson as an exotic open-flavor tetraquark state with
$J^P=0^+/1^+$ in the framework of QCD sum rules. They used the interpolating currents
\begin{equation}
J_{0^+}=s^T_aC\gamma_5u_b(\bar{b}_a\gamma_5C\bar{d}^T_b-\bar{b}_b\gamma_5C\bar{d}^T_a)
\end{equation}
with $J^P=0^+$ and
\begin{equation}
J_{1^+}=s^T_aC\gamma_5u_b(\bar{b}_a\gamma_\mu
C\bar{d}_b^T-\bar{b}_b\gamma_\mu C\bar{d}_a^T)
\end{equation}
with $J^P=1^+$ to calculate the two-point correlation functions.
Both these two currents have antisymmetric color structure
$[\mathbf{\bar 3_c}]_{su} \otimes [\mathbf{3_c}]_{\bar{b}\bar d}$.
After performing numerical analysis, the authors derived stable mass
sum rules in suitable parameter spaces, as shown in Fig.
\ref{sec4.1mass} for the scalar and
axial-vector channels.

They reproduced the mass of the $X(5568)$ meson in both scalar and
axial-vector channels with
\begin{eqnarray}
m_{X_b,\, 0^+}&=&5.58\pm0.14 \mbox{ GeV}\, , \\ \nonumber
\label{sec4.1:X5568mass} m_{X_b,\, 1^+}&=&5.59\pm0.15 \mbox{ GeV}\,
.
\end{eqnarray}
They also discussed the possible decay patterns of the $X(5568)$
state. If the $X(5568)$ meson is interpreted to be a scalar
tetraquark state, its kinematically allowed decay channel would be
the S-wave $B_s^0\pi^+$ and the $B_s^\ast\gamma$ for its neutral
partner. On the other hand, the $X(5568)$ state can decay into
$B_s^\ast\pi^+$. In this case, the quantum number of this resonance
would be $J^P=1^+$. Besides this, its neutral partner may decay into
$B_s^0\gamma$. Thus, the authors of Ref. \cite{Chen:2016mqt}
suggested to search for the neutral partner of the $X(5568)$ in the
radiative decay into $B_s^\ast\gamma$ and $B_s^0\gamma$, which can
be used to determine its spin-parity quantum numbers. Moreover, they
predicted the charmed partner states with quark content $su\bar c\bar
d$ and $J^P=0^+/1^+$:
\begin{eqnarray}
m_{X_c,\, 0^+}&=&2.55\pm0.09 \mbox{ GeV}\, , \\ \nonumber
m_{X_c,\, 1^+}&=&2.55\pm0.10 \mbox{ GeV}\, .
\end{eqnarray}

In the framework of QCD sum rules, the mass of $X(5568)$ was also
studied in Refs.
\cite{Agaev:2016mjb,Zanetti:2016wjn,Wang:2016mee,Tang:2016pcf} by
considering it as a $su\bar b\bar d$ tetraquark state with
$J^P=0^+$. The values of the mass obtained in these works agree with
Eq.~(\ref{sec4.1:X5568mass}) and thus are consistent with the
experimental result~\cite{D0:2016mwd}. However, the mass calculation
disfavors the molecule interpretation of the
resonance~\cite{Agaev:2016urs}. In Refs.
\cite{Agaev:2016ijz,Wang:2016wkj,Dias:2016dme}, the hadronic decay
width of $X(5568)$ was investigated through its strong decay into
$B_s^0\pi^\pm$. They obtained similar results for the decay width of
the $X(5568)$, which are in good agreement with the experimental
value. Moreover, the strong vertices $X_bX_b\rho, X_cX_c\rho$
and the properties for the charmed partner state of $X(5568)$ were
studied in Refs. \cite{Agaev:2016srl} and \cite{Agaev:2016lkl},
respectively.

In Ref.~\cite{Liu:2016ogz} the authors investigated the $X(5568)$ and its partners as
tetraquark states in the framework of the color-magnetic interaction. The tetraquark system was treated as a triquark
plus a heavy antiquark. Adopting a simple chromomagnetic interaction
model, they calculated the color-magnetic interaction matrix
elements for four kinds of tetraquark structures according to the
symmetry of the two light quarks. They found that the $X(5568)$ can
be interpreted as a tetraquark candidate and the other possible
bottom tetraquarks should also exist stably. Very similar
investigations were also performed within a simple quark model with
chromomagnetic interactions in Ref. \cite{Stancu:2016sfd}, in which
a hyperfine interaction parameter $C_{u\bar d}$ was also considered
to improve the calculations. Their result agrees quite well with the
experimental mass of the $X(5568)$.

The spectroscopy of the tetraquarks with one heavy quark
and three light quarks was investigated in a simple quark model by
considering the spin-spin interactions between quarks in Ref.
\cite{Wang:2016tsi}. The orbital angular momenta are vanishing for
the lowest-lying tetraquark states. Using the results for diquark
masses and spin-spin couplings, they obtained the masses of the
tetraquarks with $J^P=0^+, 1^+, 2^+$. They found that lowest-lying
S-wave tetraquark state $su\bar b\bar d$ with $J^P=0^+$ lies about
150 MeV above the $X(5568)$. In Ref. \cite{Ali:2016gdg}, a similar
investigation was performed in the diquark-antidiquark picture to
give predictions about the mass spectrum of the lowest S-wave
bottomed $su\bar b\bar d$ and charmed $su\bar c\bar d$ with
$J^P=0^+, 1^+$. They estimated the lightest scalar $su\bar b\bar d$
tetraquark at a mass of about 5770 MeV, approximately 200 MeV above
the $X(5568)$, and just 7 MeV below the $B\bar K$ threshold. An
S-wave axial-vector $su\bar b\bar d$ tetraquark was predicted around
5820 MeV, which was 250 MeV above the $X(5568)$. The authors
proposed to search for the tetraquark states in the decays of the
$B_c^\pm$ mesons, $B_c^\pm\to X_{b0}^0\pi^\pm$ and $B_c^\pm\to
X_{b0}^\pm\pi^0$~\cite{Ali:2016gdg}.

Some non-resonant schemes have also been proposed to explain the
$X(5568)$ signal. In Ref. \cite{Liu:2016xly}, Liu and Li
investigated the invariant mass distributions of the $B_s\pi$ via
different rescattering processes, such as the triangle rescattering,
the long range interaction process and the weak interaction process.
They found that some bumps arise around the position of the
$X(5568)$ in the triangle rescattering process, which provided a
possibility that the $X(5568)$ signal may be due to some kind of
accumulative effects of the rescattering amplitudes at different
incident energies. If so, the quantum numbers of the $X(5568)$ would
be $J^P=1^-$ because of the P-wave scattering characteristic of the
process $B_s^\ast\pi\to B_s\pi$~\cite{Liu:2016xly}. However, no such
P-wave scattering pole for the $X(5568)$ was dynamically generated
in the unitarized effective field theory \cite{Kang:2016zmv}.

In Ref. \cite{Jin:2016cpv}, the authors studied the production rate
of the $X(5568)$ measured by the D0 Collaboration and found that it
is quite large and difficult to be understood by various general
hadronization mechanisms, such as the string fragmentation model,
cluster model and combination model. They then proposed the
inclusive production formulation for the cross section, and
predicted the distributions and production rates of $X(5568)$ at LHC
energies.

He and Ko classified the symmetry properties of the beauty
tetraquark states $X_b$ ($\bar b q^\prime q^{\prime\prime}\bar q$)
using light quark SU(3) flavor symmetry \cite{He:2016yhd}. These
states containing three light quarks should be in either the {\bf
$\bar 6$} or {\bf $15$} SU(3) flavor representation. They constructed the
leading order chiral Lagrangian to study the possible decays of
$X_b$ into a $B$ meson and a light pseudoscalar octet meson, and
provided search strategies to distinguish whether $X_b$ belongs to
{\bf $\bar 6$} or {\bf $15$}. They predicted a new doubly charged
four quark state if $X_b$ belongs to {\bf $15$}.

However, the existence of the $X(5568)$ was not confirmed by the
preliminary results of the LHCb \cite{Aaij:2016iev} and CMS
\cite{CMS:2016X5568} collaborations. There are also many theoretical
works discussing the difficulties to accommodate such an exotic
structure.

In Ref. \cite{Burns:2016gvy}, Burns and Swanson examined a variety
of explanations for the $X(5568)$ state. They found that the
threshold effect, cusp, molecular and tetraquark are all unable to
give a satisfactory description of the observed state. In their
argument, the threshold model cannot fit the experimental data well.
In the cusp scenario, they were able to fit the data well. However,
this scenario requires the P-wave rescattering with a flavor-blind
interaction, thus predicts the spin-parity of the $X(5568)$ to be
$J^P=1^-$. These unnatural properties are not preferred in
conventional phenomenology. The mass of $X(5568)$ is several
hundreds MeV below the relevant two-body thresholds. However, the
coupling of the $B_s\pi$ or a coupled $B_s\pi-B\bar K$ system are
not strong enough to form the desired states. In the tetraquark
model, they made simple estimates and found that the $X(5568)$ is
too light to be a plausible tetraquark candidate.

In Ref.~\cite{Chen:2016ypj}, Chen and Liu performed a dynamical study of
the interactions of and $B^{(*)} \bar K^{(*)}$ systems
using the one-boson-exchange model. Their study suggested that
the $X(5568)$ can not be assigned to be an isovector $B\bar K$ or $B^* \bar K$
molecular state, but the isoscalar $B \bar K$ and $B^* \bar K$ as well
as the isoscalar $B^* \bar K^*$ molecular states may exist, whose decay behaviors
were also discussed in the present study.

Based on the chiral symmetry and heavy quark symmetry, Guo {\it et.
al.} also proposed several types of models to explain the
structure of the $X(5568)$ \cite{Guo:2016nhb}. Their analyses
supported the conclusions in Ref. \cite{Burns:2016gvy} that none of
the tetraquark model, hadronic molecule, or threshold-effect model
provides a satisfactory description of the signal. They suggested to
search for the $X(5568)$ in the dipion decays of the excited
bottom-strange mesons, e.g., $B^\ast_{s2}(5840)\to B_s\pi\pi$.

In Ref. \cite{Lu:2016zhe}, the authors studied the mass spectra of
open-charm/bottom tetraquark states within the diquark-antidiquark
scenario in the relativized quark model. They calculated the masses
of the scalar and axial-vector diquark and antidiquark by solving
the Hamiltonian with the relativized potential. The masses of the
tetraquark states were then obtained by solving the Schr\"odinger-type
equation. They found the mass of the $sq\bar b\bar q$ state is much
higher than that of the $X(5568)$, which disfavors the tetraquark
configuration of the resonance.

By analyzing a $B_s\pi-B\bar K$ coupled channel system, Albaladejo
{\it et. al.}~\cite{Albaladejo:2016eps} reproduced the spectrum
structure of the $X(5568)$. With the interaction matrix elements
derived from the heavy meson chiral perturbation theory, they found
a pole with the mass and decay width in agreement with the
experimental values. However, if the T-matrix regularization is
employed, a big momentum cutoff $\Lambda\sim 2.8$ GeV will be
required to obtain the same spectrum, which is much larger than a
``natural value" $\Lambda\sim 1$ GeV. The authors thus concluded
that the $X(5568)$ state would not qualify as a resonance
dynamically generated by the unitarity loops.

In the framework of the chiral quark model, Chen and Ping studied
the four-quark system $us\bar b\bar d$ with the quantum numbers
$J^P=0^+$ in both the diquark-antidiquark and meson-meson formalisms
under the SU(3) and SU(4) flavor symmetry~\cite{Chen:2016npt}. To
compute the mass of tetraquark state, they constructed the
tetraquark wave function and hamiltonian in the Gaussian Expansion
Method (GEM) and chiral quark model. The potential energy is
composed of the color confinement, one-gluon-exchange and
one-Goldstone boson exchange. They solved the Schr\"odinger equation
to obtain the masses of the four-quark systems $us\bar b\bar d$,
including the tetraquark system and molecule system. They found that
the masses of the tetraquark states are much higher than that of
$X(5568)$ state while no molecular structure can be formed in their
calculation.

If the $X(5568)$ with $J^P=0^+$ exists, it can decay strongly into
$B_s\pi^+$ only and lies significantly below all the other
thresholds. Such a low mass allows for a more reliable, cleaner and
easier search in lattice QCD. In Ref. \cite{Lang:2016jpk}, the
authors investigated the S-wave $B_s\pi^+$ scattering on the lattice
to search for $X(5568)$ as a scalar exotic resonance. For
completeness, they also considered the $X(5568)$ as a very deeply
bound $B^+\bar K^0$ state, which has a threshold 210 MeV above the
$X(5568)$. However, they didn't find an eigenstate in their lattice
QCD simulation, which does not support the existence of the
$X(5568)$ with $J^P=0^+$.

In the framework of QCD sum rule, Albuquerque {\it et. al.}
investigated the $X(5568)$ state using the molecular interpolating
currents $BK, B_s\pi, B^\ast K, B_s^\ast\pi$ and tetraquark currents
with $J^P=0^+, 1^+$~\cite{Albuquerque:2016nlw}. Their numerical
results did not support the $X(5568)$ as a pure molecule or a
tetraquark state. However, they suggested it to be a mixture of $BK$
molecule and scalar $ds\bar b\bar u$ tetraquark state with a mixing
angle $\sin2\theta\simeq0.15$. There are also some other theoretical
approaches to investigate the $X(5568)$ state
\cite{Esposito:2016itg,Lu:2016kxm}, which supported the negative
results in LHCb \cite{Aaij:2016iev}.

However, the production mechanism of the $X(5568)$ is very different
at the $p\bar p$ and $pp$ colliders. Future experimental efforts are
desirable in the clarification of the situation on the $X(5568)$
state.

%% file: section5.tex
\section{Outlook and summary}
\label{sec5}

In 1976, the first charmed meson was discovered by the Mark I
Collaboration~\cite{Goldhaber:1976xn,Peruzzi:1976sv,Wiss:1976gd}
and the first charmed baryon $\Lambda_c$ was discovered at the
Fermilab~\cite{Knapp:1976qw}. After these observations, many open
charm and open bottom hadrons were observed by the ALEPH, ARGUS,
BNL, CERN R415, CUSB, CUSB-II, DASP, DELPHI, ITEP\&SERP, L3, OPAL,
TPS, TST Collaborations/experiements, etc.

In 2003, two narrow charm-strange states $D_{s0}^*(2317)$ and
$D_{s1}(2460)$ were discovered by the BaBar and CLEO Collaborations,
respectively~\cite{Aubert:2003fg,Besson:2003cp}, which attracted
lots of attentions. After that, many open charm and open bottom
hadrons were observed by the Belle, FOCUS, SELEX, D\O, CDF, LHCb and
CMS collaborations in the past decade. We brief summarize their
statuses here:
\begin{enumerate}
\item All the $1S$ heavy mesons (charmed, charm-strange, bottom and
bottom-strange mesons) are well established.
All the $1P$ charmed and charm-strange mesons were observed experimentally.
There are many other observed excited heavy mesons, which can be accommodated
in the quark model spectrum, although theoretical interpretations are not unique.
More experimental measurements are needed to pin down their classification.

\item All the $1S$ singly heavy baryons (singly charmed and bottom baryons)
are well established, except the $\Omega_b^*$ of $J^P = 3/2^+$. The
singly heavy baryon system is more complicated than the heavy meson
system, and more theoretical and experimental efforts are needed to
understand the excited singly heavy baryons.

\item Only one doubly charmed baryon $\Xi_{cc}^+(3520)$ was reported
by the SELEX Collaboration. However, its existence has not been
confirmed by any other experiments. Most of the theoretical
predictions for the mass of the doubly charmed $ccq$ with
$J^P=1/2^+$ lie above the SELEX's value.

\item There were several candidates for exotic mesons in the
open-charm and open-bottom meson sector. Unfortunately,
$D_{sJ}(2632)$ was not confirmed by subsequent experiments. The
existence of the recently observed $X(5568)$ awaits further
confirmation. It is highly probable that both $D_{s0}^*(2317)$ and
$D_{s1}(2460)$ are the conventional P-wave charm-strange mesons
which are strongly affected by the couple channel effects.

\end{enumerate}

The open-charm and open-bottom mesons provide a wonderful platform
to explore the non-perturbative QCD dynamics in the low-energy
regime and test various theoretical tools and phenomenological
models. In the coming years, more and more excited heavy mesons will
be produced at LHCb, CMS and BelleII. We may expect important
progress in this field in the near future.